\documentclass[preprint]{aastex61}
\usepackage{graphicx}

\def\kms{$\textrm{km~s$^{-1}$}$ \,}

\def\kpc{kpc}

\def\H2{H$_{2}$ \,}
\def\HI{HI}
\def\roH2{$\rho_{\textrm{H}_2}$}
     
\def\MH2{M$_{\textrm{H}_2}$ \,}

\def\Ha{H$\alpha$ \,} 
\def\deg{$^{\circ}$ \,}
 
\def\i{\,{\small I}}

\submitjournal{ApJSupplSer}

\shorttitle{Ionized gas in S0 galaxies} \shortauthors{Sil'chenko et al.}

\begin{document}

\title{The gas kinematics, excitation, and chemistry, in connection with star formation, in
lenticular galaxies.}

\correspondingauthor{Olga Sil'chenko}
\email{olga@sai.msu.su, olgasil.astro@gmail.com}

\author[0000-0003-4946-794X]{Olga K. Sil'chenko} 
\affil{Sternberg Astronomical Institute, M.V. Lomonosov Moscow State University, Universitetsky pr., 13, Moscow, 119234 Russia}
\email{olga@sai.msu.su}

\author[0000-0002-0507-9307]{Alexei V. Moiseev}
\affil{Special Astrophysical Observatory, Russian Academy of Sciences, Nizhnij Arkhyz, 369167 Russia}
\affil{Sternberg Astronomical Institute, M.V. Lomonosov Moscow State University, Universitetsky pr., 13, Moscow, 119234 Russia}
\email{moisav@gmail.com}

\author[0000-0002-4755-118X]{Oleg V. Egorov} 
\affil{Sternberg Astronomical Institute, M.V. Lomonosov Moscow State University, Universitetsky pr., 13, Moscow, 119234 Russia}
\affil{Special Astrophysical Observatory, Russian Academy of Sciences, Nizhnij Arkhyz, 369167 Russia}
\email{egorov@sai.msu.ru}

\begin{abstract}
We present results of long-slit and panoramic spectroscopy of extended gaseous disks in 18 nearby S0 galaxies, mostly in groups. The gas in our S0s
is found to be often accreted from outside that is implied by its decoupled kinematics: at least 5 galaxies demonstrate strongly 
inclined large-scale ionized-gas disks smoothly coupled with their outer \HI\ disks, 7 galaxies reveal circumnuclear polar ionized-gas
disks, and in NGC~2551 the ionized gas though confined to the main galactic plane however counterrotates the stellar component. 
The ionized-gas excitation analysis reveals the gas ionization by young stars in 12 of 18 S0 galaxies studied here; 
the current star formation in these galaxies is confined to the ring-like zones coinciding with the UV-rings. 
The gas oxygen abundance estimates in the rings are closely concentrated around the value of 0.7 $Z_\odot$ %$-0.15$~dex
and do not correlate either with the ring radius nor with the metallicity of the underlying stellar population. By applying
the tilted-ring analysis to the 2D velocity fields of the ionized gas, we have traced the orientation of the gas rotation-plane lines 
of nodes along the radius. We have found that current star formation proceeds usually just where the gas lies strictly in the 
stellar disk planes and rotates there circularly; the sense of the gas rotation does not matter (the counterrotating gas in NGC~2551 forms
stars currently). In the galaxies without signs of current star formation the extended gaseous disks are either in steady-state quasi-polar
orientation (NGC~2655, NGC~2787, NGC~3414, UGC~9519), or are acquired recently through the highly inclined external filaments 
provoking probably shock-like excitation (NGC~4026, NGC~7280). Our data imply crucial difference of the external-gas accretion 
regime in S0s with respect to spiral galaxies: the geometry of the gas accretion in S0s is typically off--plane.
\end{abstract}


\section{Introduction}   

Lenticular galaxies were introduced as a distinct morphological type in the latest version of
the `Hubble fork' -- the galaxy morphology scheme by \citet{hubble}; they were considered by Edwin Hubble
as a transition type between ellipticals and spirals and {\it by definition} had large-scale
stellar disks but lacked spiral arms and intense star formation. Since the global structures
of lenticular and spiral galaxies were very similar \citep[see e.g.,][]{lauri10}, and the only
prominent difference was red color of the disks and absence of noticeable star formation regions within them,
an idea had been formulated rather early that lenticular galaxies were former spirals devoid of gas 
\citep[\it{`they show no trace of gas or anything of the sort'}, ][]{baade}.
However further observations in the 21 cm line had shown that S0 galaxies could be rather rich in
neutral hydrogen \citep[see e.g.,][]{s0_hi}. The origin of this gas and reasons
of the frequent absence of current star formation in the gas-rich S0s remain unclear up to now.

Although lenticular galaxies typically have lower ratio of cold gas mass to their luminosity than
spiral galaxies \citep*{knapp}, it has been found that neutral and even molecular gas is present perhaps in
most of them \citep{welchsage03, welchsage06,welchsage10}, though less than half of gas-rich S0 galaxies 
with extended cold-gas disks experience current star formation of HII-region type \citep{pogge_esk93}. 
The kinematics of ionized gas in S0s is often decoupled from that of their stellar disks that had allowed 
to \citet*{bertola92} to conclude that at least 40\%\ of emission-line S0s acquired their gas from 
some external sources rather recently. Later a factor of environment was found to be important: the study 
by \citet{atlas3d_10} and by \citet{atlas3d_13} showed that among Virgo cluster S0s 
the gaseous and stellar components revealed almost always strict kinematical alignment while the non-Virgo S0s, 
those in groups and in the field, demonstrated decoupled gas kinematics in 50\%\ of all cases. By following this
logic, we have studied strongly isolated S0s which represent the extreme case of sparse environment. We observed them
with the long-slit spectrographs by aligning our slits with the photometric major axes (which are thought to trace the
orientation of the stellar-disk lines of nodes in the case of intrinsically round stellar disks.).
Our results \citep*{isomnras,isosalt} have confirmed that
without external forces to remove the gas, lenticular galaxies demonstrate extended ionized-gas disks in more than 70\%\
of all cases, and about half of them reveal gas counter-rotation with respect to their stellar components.
If the gas has been accreted by S0s from outside, such statistics implies that the directions of gas inflow are 
distributed isotropically. However, in the case of strongly isolated S0s the problem of searching for external gas
source (donors) remains unsolved.

Another interesting question which has intrigued us since then is the incidence of star formation in gas-rich S0s. In our
sample of isolated S0s \citep{isomnras,isosalt} we found that both the gas ionized by young stars and the gas
excited by shocks could be met in the disks, and it was not the amount of the gas that governed this dichotomy.
By comparing the projections of the gas and star rotation velocities onto the line of sight along the 
spectrograph slit (aligned with the stellar-disk major axis), we have suspected that star
formation proceeds when the gas lies strictly in the planes of stellar disks, independently on the gas spin direction --
coincident with the stellar-component spin or antiparallel to it. However, we cannot determine exactly the orientation
of the gas rotation plane by observing it only with the long slit; panoramic spectroscopy is needed. In this
paper we just present the results of our complex study of the ionized gas in 18 S0s for which we have undertaken both long-slit
spectroscopy and velocity mapping with the scanning Fabry-Perot interferometer (FPI) by using the facilities of
the Russian 6-m telescope of the Special Astrophysical Observatory of Russian Academy of Sciences (SAO RAS).

\section{Sample}

\begin{figure*}
\centering
\begin{tabular}{c c}
\includegraphics[width=8cm]{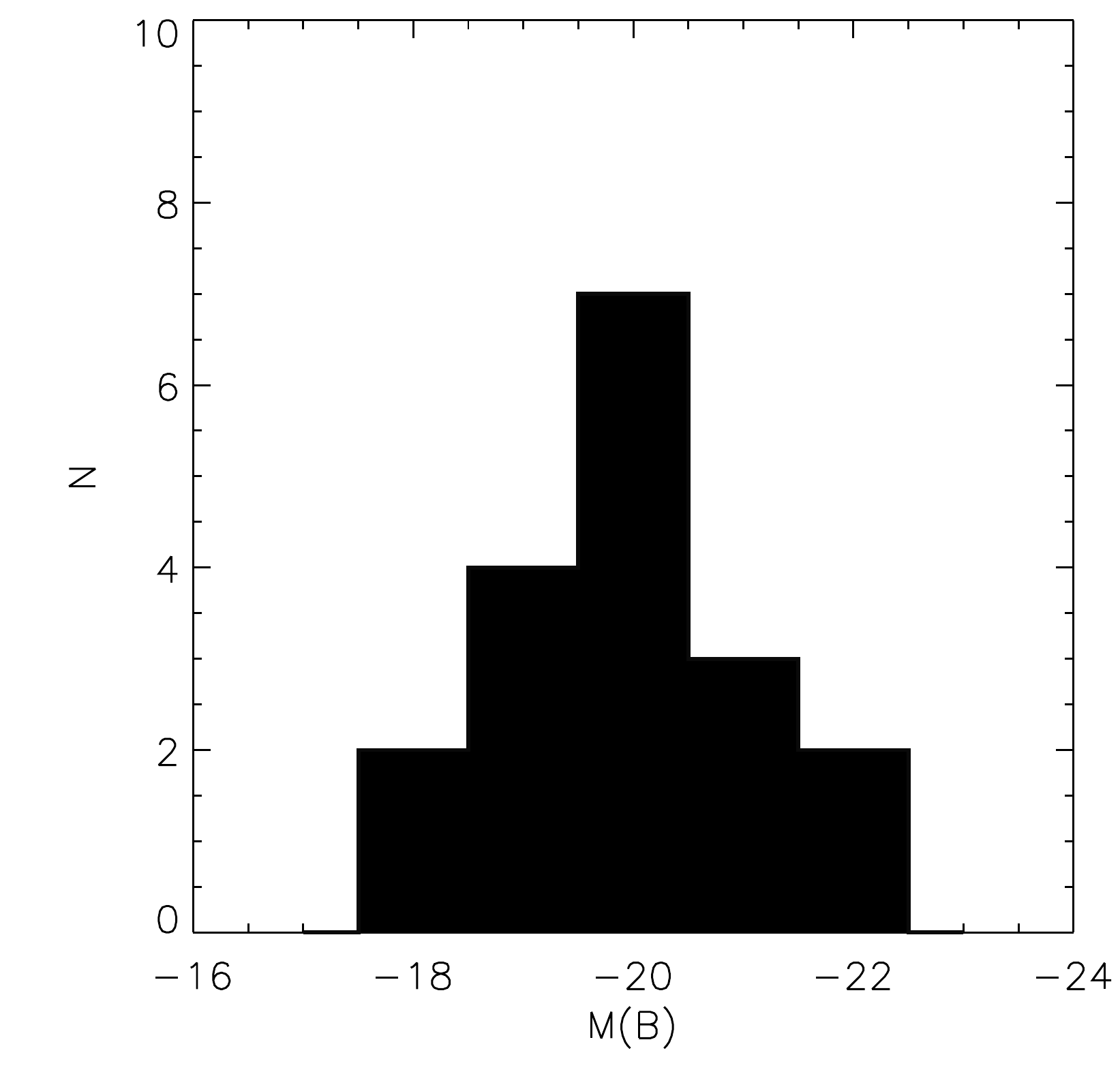} &
\includegraphics[width=8cm]{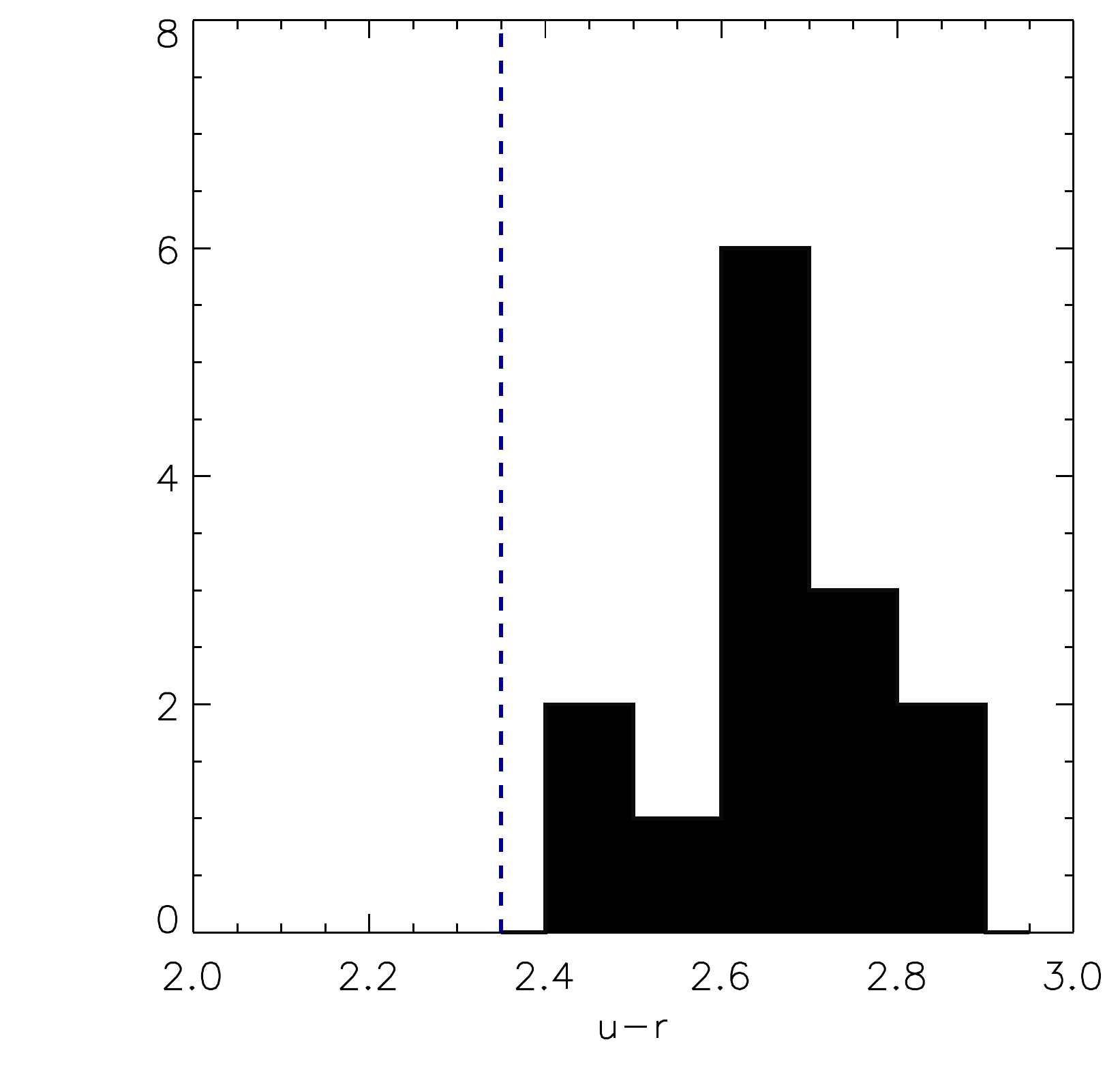} \\
\end{tabular}
\caption{The properties of the sample: the distributions over integrated absolute magnitude assembled from
the HyperLEDA  ({\it left}) and over integrated $(u-r)_0$ color taken from the SDSS/DR9 ({\it right}).
In the right plot we have also shown the boundary separating the red sequence and the blue cloud \citep{sdsscol}
by the vertical dashed line.}
\label{sample}
\end{figure*}

The sample includes 18 S0 and S0/a galaxies with rather extended gaseous disks detected in optical-band emission lines;
the presence of such disks is either known from our previous works, e.g. \citet*{we2551}, \citet*{ringmnras},
\citet*{saltrings}, or is suspected from extended ultraviolet signal reported in \citet{galex}.
To select the targets suitable to be studied with the scanning Fabry-Perot interferometer, we have firstly undertaken
long-slit spectroscopy to trace emission lines toward their full extension; and so on our sample of nearby S0s is
biased toward gas-rich objects. The sample is not {\bf representative over neither galaxy characteristics; it is not volume- or magnitude-limited}.
However, the absolute-magnitude range of the galaxies under consideration covers full range of S0 luminosities, which is
from $M_B=-18$ to $M_B=-22$ according to \citet{vdb_s0} (Fig.~\ref{sample}, left), and the galaxies are homogeneously
red belonging all to the red sequence (Fig.~\ref{sample}, right).
The list of their integrated properties assembled over literature is given in Table~1. The majority of our galaxies, 
13 of 18, demonstrate rings of various size including those seen in the far ultraviolet, according to the GALEX data, so
betraying possible recent star formation. Interestingly, our ringed S0s are mostly unbarred.

\begin{table*}
\scriptsize
\caption{Global parameters of the galaxies}
% %\begin{center}
\begin{flushleft}
\begin{tabular}{lccclrccc}
\hline\noalign{\smallskip}
Name & Type  & 
$M_B$ & $M_H$ & $(g-r)$  & $R_{25}$,$^{\prime \prime}$ &
Environment$^3$ & $M_{\mbox{HI}}$, $10^8\,M_{\odot}$ & $M_{\mbox{H}_2}$, $10^8\,M_{\odot}$ \\
 & (NED$^1$) &  (LEDA$^2$) & (NED) & (SDSS/DR9) & (LEDA) & (NED) & (EDD$^4$) & \citep{atlas3d_4} \\
\hline
IC5285 & S0/a$^2$ & --21.97 & --24.32 & 0.84 & 39.5 & 4 & 37.5 & \\ 
NGC252 & (R)SA(r)0$+$: & --21.42 & --25.13 & 0.78 & 49 & 4 & 28 & 6.29$^{14}$ \\
NGC774 & S0 & --20.17 & --23.69 & 0.86 & 39 & 4 &  &  \\
NGC2551 & SA(s)0/a & --20.04 & --23.02 & 1.04$(=(B-V)_{e,0}^2)$ & 47.5 & 4 & 12 & \\
NGC2655 & SAB(s)0/a & --21.22 & --24.26 & 0.89$(=(B-V)_{e,0}^2)$ & 117  & 4 & 11.4 & 1.3$^5$\\
NGC2697 & SA(s)0$+$: & --18.66 & --22.26 & 0.72$^6$ & 49 & 3 & 5.9 &  1.8 \\ 
NGC2787 & SB(r)0$+$ & --18.45 & --23.34 &  0.95$(=(B-V)_{e,0}^2)$ & 97 &  5 &  9.8$^7$ &  0.18$^8$\\
NGC2962 & (R)SAB(rs)0$+$ & --20.08 & --23.64 & 0.87 & 69 & 4 & 11.0 & $<0.7$ \\  
NGC3106 & S0 & --21.70  & --24.62 & 0.73 & 27 & 4 &  108$^9$ & \\  
NGC3166 & SAB(rs)0/a & --20.30 & --24.34 & 0.81 & 134 & 4 & 4.5$^7$ & 1.7$^{10}$ \\
NGC3182 &  SA(r)0/a$^2$ & --19.81 & --22.84 & 0.77 & 57 & 4 & 0.08$^{11}$ & 2.14\\ 
NGC3414 & S0pec & --20.03 & --23.39 & 0.76 & 81 & 4 & 1.9$^{11}$ & $<0.15$ \\
NGC3619 & (R)SA(s)0$+$: & --19.63 & --23.57  & 0.86 & 117 & 2 & 10.0$^{11}$ & 1.9\\
NGC4026 & S0 & --19.11 & --23.16 & 0.80 & 131 & 2 & 3.2$^{11}$ & 0.88$^7$ \\
NGC4324 & SA(r)0$+$ &  --19.12 & --23.43 & 0.76 & 88.5 & 1 & 16.8 & 0.5 \\
NGC7280 & SAB(r)0$+$ & --18.84 & --22.41 & 0.74  & 60 & 4 & 0.83$^{11}$ & $<0.3$ \\
UGC9519 & S0$^2$ & --18.09 & --21.49 & 0.82 & 24 & 3 & 18.6$^{11}$ & 5.9 \\
UGC12840 & (R)SAB(s)$0^0$ & --21.15 & --24.14 & 0.76 & 36 & 5 & 55$^{12}$ & 6.2$^{13}$\\    
\hline
\multicolumn{9}{l}{$^1$\rule{0pt}{11pt}\footnotesize
NASA/IPAC Extragalactic Database, http://ned.ipac.caltech.edu .}\\
\multicolumn{9}{l}{$^2$\rule{0pt}{11pt}\footnotesize
Lyon-Meudon Extragalactic Database, http://leda.univ-lyon1.fr .}\\
\multicolumn{9}{l}{$^3$\rule{0pt}{11pt}\footnotesize
Environments: 1 -- cluster member, 2  -- rich-group member,
3 -- loose-group member, 4 -- loose-group center,
5 -- field.}\\
\multicolumn{9}{l}{$^4$\rule{0pt}{11pt}\footnotesize
Extragalactic Distance Database, http://edd.ifa.hawaii.edu .}\\
\multicolumn{9}{l}{$^5$\rule{0pt}{11pt}\footnotesize
\citet{ueda}.} \\
\multicolumn{9}{l}{$^6$\rule{0pt}{11pt}\footnotesize
\citet{saltrings}.} \\
\multicolumn{9}{l}{$^7$\rule{0pt}{11pt}\footnotesize
\citet{roberts91}.} \\
\multicolumn{9}{l}{$^8$\rule{0pt}{11pt}\footnotesize
\citet{welchsage03}.} \\
\multicolumn{9}{l}{$^9$\rule{0pt}{11pt}\footnotesize
\citet{eder_hi}.} \\
\multicolumn{9}{l}{$^{10}$\rule{0pt}{11pt}\footnotesize
\citet{wikhenk89}.} \\
\multicolumn{9}{l}{$^{11}$\rule{0pt}{11pt}\footnotesize
\citet{atlas3d_13}.} \\
\multicolumn{9}{l}{$^{12}$\rule{0pt}{11pt}\footnotesize
\citet{haynes11}.} \\
\multicolumn{9}{l}{$^{13}$\rule{0pt}{11pt}\footnotesize
\citet{lisenfeld}.}\\
\multicolumn{9}{l}{$^{14}$\rule{0pt}{11pt}\footnotesize
\citet{coldgas}.}\\
\end{tabular}
\end{flushleft}
\end{table*}

\begin{table*}
\caption{SCORPIO/SCORPIO-2 long--slit spectroscopy of the sample galaxies}
\label{table_obs1}
\scriptsize
 \begin{center}
 \begin{tabular}{lcrcccrl}
\hline\noalign{\smallskip}
Galaxy      & Date             & Grism            & Sp. range, \AA&  Exposure, s & PA(slit), $\degr$  & Seeing, $''$ \\
\hline\noalign{\smallskip}
IC 5285     & 04.10.2015  & VPHG1200R & 5700--7500   & $3 \times 900$ & 93 & 1.3 \\
NGC 252  & 19.11.2011   & VPHG1200@540 & 3700--7200  & $6 \times 900$ & 82 & 1.5 \\
NGC 774  & 08.10.2015  & VPHG2300G & 4800--5500   & $6 \times 1200$ & 170 & 3.6 \\
NGC 2551 & 14.12.2014 & VPHG1200@540&3700-7200&  $8 \times 900$& 13 & 1.5 \\
NGC 2551 & 06.10.2015 & VPHG1200R & 5700--7500   & $3 \times 900$ & 55 & 2.6 \\
NGC 2655 & 15.12.2014 & VPHG1200@540&3700-7200& $2 \times 900 + 6 \times 600$ & 0 & 1.3 \\
NGC 2655 & 15.12.2014 & VPHG1200@540&3700-7200& $3 \times 600 + 8 \times 450$ & 102 & 1.3 \\
NGC 2787 & 30.03.2009 & VPHG2300G & 4800--5500 & $4 \times 1800$ & 109 & 1.5 \\
NGC 2787 & 30.03.2009 & VPHG2300G & 4800--5500 & $3 \times 1200 + 661$ & 72 & 1.5 \\
NGC 2787 & 14.04.2012 & VPHG1200@540&3700-7200& $4 \times 900$ & 108 & 4.0 \\
NGC 2962 & 20.02.2017 & VPHG1200@540&3700-7200& $6 \times 1200$ & 172 & 2.5 \\
NGC 3166 & 28.04.2006 & VPHG2300G & 4800--5500 & $4 \times 1200$ & 87 & 2.8 \\ 
NGC 3166 & 19.02.2017 & VPHG1200@540&3700-7200& $4 \times 1200$ & 87 & 2.5 \\
NGC 3182 & 31.03.2014 & VPHG1200@540&3700-7200& $4 \times 900$ & 153 & 4.0 \\
NGC 3414 & 30.03.2009 & VPHG1800R & 6200--7200 & $4 \times 900$ & 20 & 1.5 \\
NGC 3414 & 31.03.2009 & VPHG1800R & 6200--7200 & $5 \times 1200$ & 150 & 1.3 \\
NGC 3414 & 10.04.2010 & VPHG2300G & 4800--5500 & $7 \times 1200$ & 150 & 3.0 \\
NGC 3619 & 28.02.2014 & VPHG1200@540&3700-7200& $3 \times 600$ & 50 & 3.1 \\
NGC 4026 & 04.04.2009 & VPHG1800R & 6200--7200 & $4 \times 900$ & 178 & 2.8 \\
NGC 4026 & 18.04.2012 & VPHG1200@540&3700-7200& $3 \times 900$ & 176 & 2.5 \\
NGC 7280 & 12.10.2009 & VPHG2300G & 4800--5500 & $9 \times 1200$ & 76 & 1.7 \\
NGC 7280 & 13.10.2009 & VPHG2300G & 4800--5500 & $8 \times 1200$ & 40 & 1.9 \\
NGC 7280 & 05.10.2016 & VPHG1800R & 6200--7200 & $4 \times 900$ & 40 & 1.0 \\
NGC 7280 & 05.10.2016 & VPHG1800R & 6200--7200 & $4 \times 900$ & 76 & 1.4 \\
UGC 9519 & 20.04.2012 & VPHG1200@540&3700-7200& $5 \times 900$ & 75 & 2.0 \\
UGC 12840& 14.11.2014 & VPHG1200@540&3700-7200& $6 \times 900$ & 15 & 1.1 \\
\hline
\end{tabular}
\end{center}
\end{table*}

\section{Observations}

\subsection{Long-slit spectroscopy}

Our long-slit spectral observations were made with the multi--mode focal
reducer SCORPIO \citep{scorpref} and its improved version SCORPIO-2 \citep{scorpio2} installed
at the prime focus of the SAO RAS 6m telescope. Both devices have the same slit size (6\farcm1 in length and 
1\arcsec\ in width) with a scale along the slit of 0\farcs36/px, while for a similar spectral resolution SCOPRIO-2 
provides a twice larger spectral range. With the SCOPRIO-2 we exposed the full optical spectral range, namely 
3700--7200~\AA, by using the volume-phase holographic grating VPHG1200@540 provided spectral resolution  
of about 5~\AA, whereas with the SCORPIO we probed separately the narrower spectral ranges, 4800--5500~\AA,  
6200--7200~\AA, or 5700--7500~\AA, with the  grisms VPHG2300G, VPHG1800R, and VPHG1200R providing the resolution of 
about 2, 2.5, and 5~\AA, correspondingly. The log of the observations is presented in Table~\ref{table_obs1},
the positions of the spectrograph slit against the galaxies images are shown in the Fig.~\ref{fig_fpi}.

\begin{figure}
\centering
	\includegraphics[width=9cm]{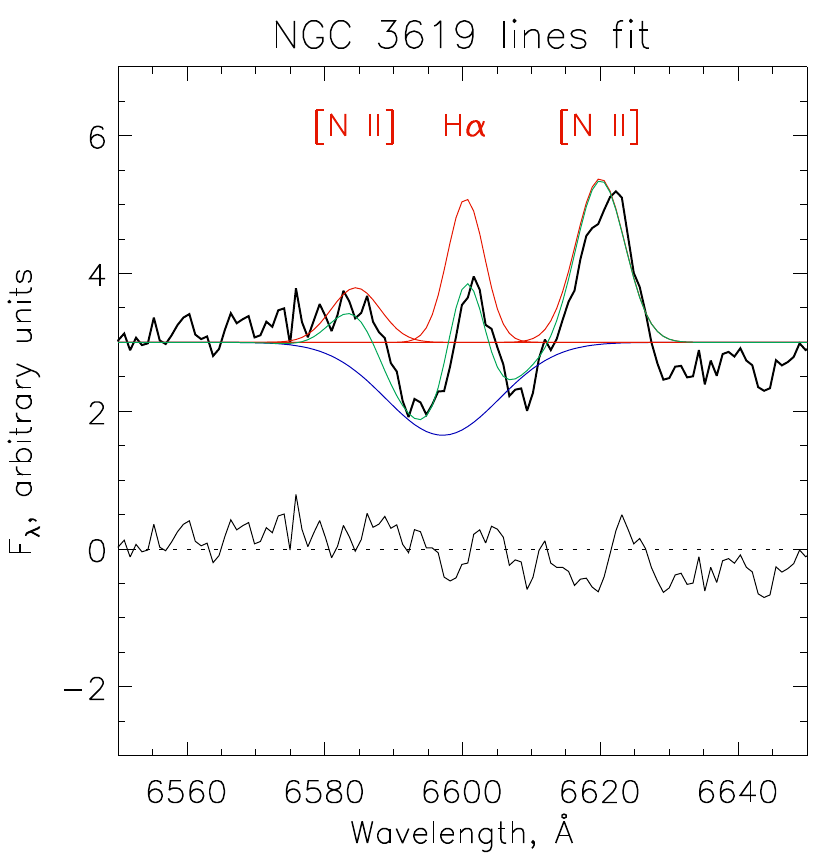}
	\caption{The example of the Gauss decomposition in the red spectral domain of the \Ha+[NII]$\lambda\lambda6548,6583$ emission lines
        of the ionized gas (red colour) and stellar \Ha\ absorption feature (the blue Gaussian) in the observed spectra of NGC~3619 galaxy (the thick black line).
        {\bf The green line shows the sum of all Gaussians.} The thin black line in the bottom shows the residuals after subtraction of the model.
	}
	\label{fig_N3619fit}
\end{figure}

The data were reduced by a standard way using the IDL software package developed in SAO RAS. It is at the edges of the slit
that we derived the sky background to subtract it from the  galaxy spectra, by applying the linear or polynomial (with
the degree of 2--3) fit of the sky background distribution along the slit at every wavelength.
Inhomogeneties of the optics transmittance and variations of the spectral resolution along the
slit were taken into account by using the high signal-to-noise acquisitions of the twilight spectra.
The stellar kinematics was analysed by cross--correlating binned spectra with spectra of K-giant stars
observed the same nights as the galaxies; the description of this technique and references on the original 
papers are presented in our previous papers, e.g. \citet{we2551}. The emission lines, namely, the H$\alpha$,
[NII]$\lambda$6583, and [OIII]$\lambda$5007 first of all, were used to derive ionized-gas kinematics, 
by measuring baricenter positions of the lines; in the central bins where the stellar continuum is strong we
applied gauss-analysis to take into account effects of underlying Balmer absorption lines and also
TiI under [OIII]$\lambda$5007. For the latter purpose, we binned the spectra along the slit to reach signal-to-noise ratio
higher than 50-70, and then made gauss-analysis of the line complexes
[NII]$\lambda$6548,6583+H$\alpha$(emission)+H$\alpha$(absorption), H$\beta$(emission)+H$\beta$(absorption),
[OIII]$\lambda$5007(emission)+TiI$\lambda$5007(absorption)+TiI$\lambda$5015(absorption). With this analysis
we are also able to derive {\bf  the equivalent width of H$\alpha$ line (EW$_{\mathrm{H}\alpha}$) and} the flux ratios of the strong emission lines:
[NII]$\lambda$6583 to H$\alpha$, [OIII]$\lambda$5007 to H$\beta$, [SII]$\lambda$6717 to [SII]$\lambda$6731, and $\Sigma$[SII] to H$\alpha$,
which can be used to diagnose the gas excitation mechanisms with the BPT-diagrams \citep*{bpt} and to determine
the gas oxygen abundances for the emission-line regions where the gas is excited by young stars. 
The example of the emission/absorption lines decomposition near the center of NGC~3619 is shown in Fig.~\ref{fig_N3619fit}.
The sensitivity variations along the wavelength range were corrected  by observing spectrophotometric
standard stars the same nights as the galaxies.

\begin{figure*}
\centering
\begin{tabular}{c c}
\includegraphics[width=8cm]{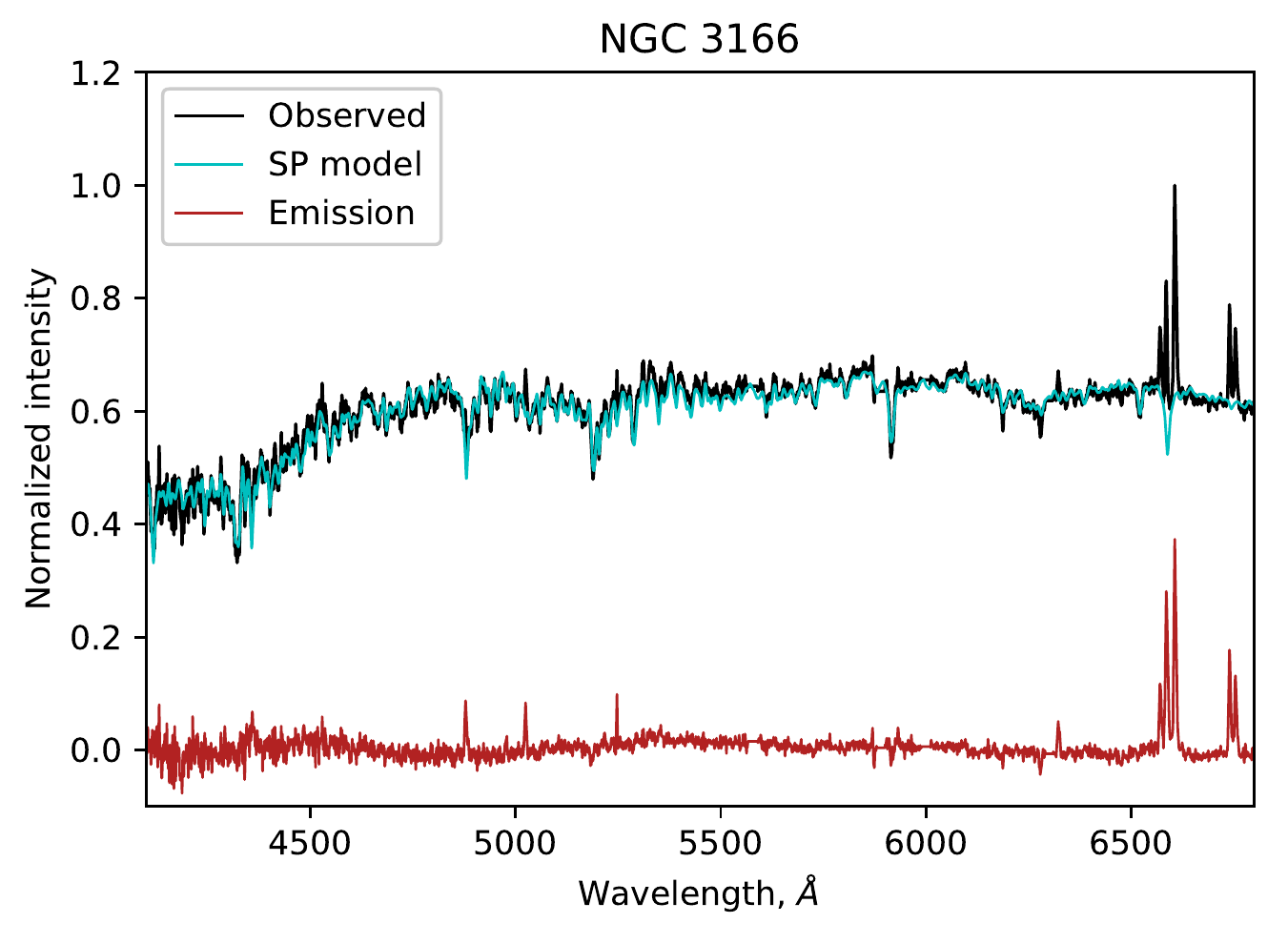} &
\includegraphics[width=8cm]{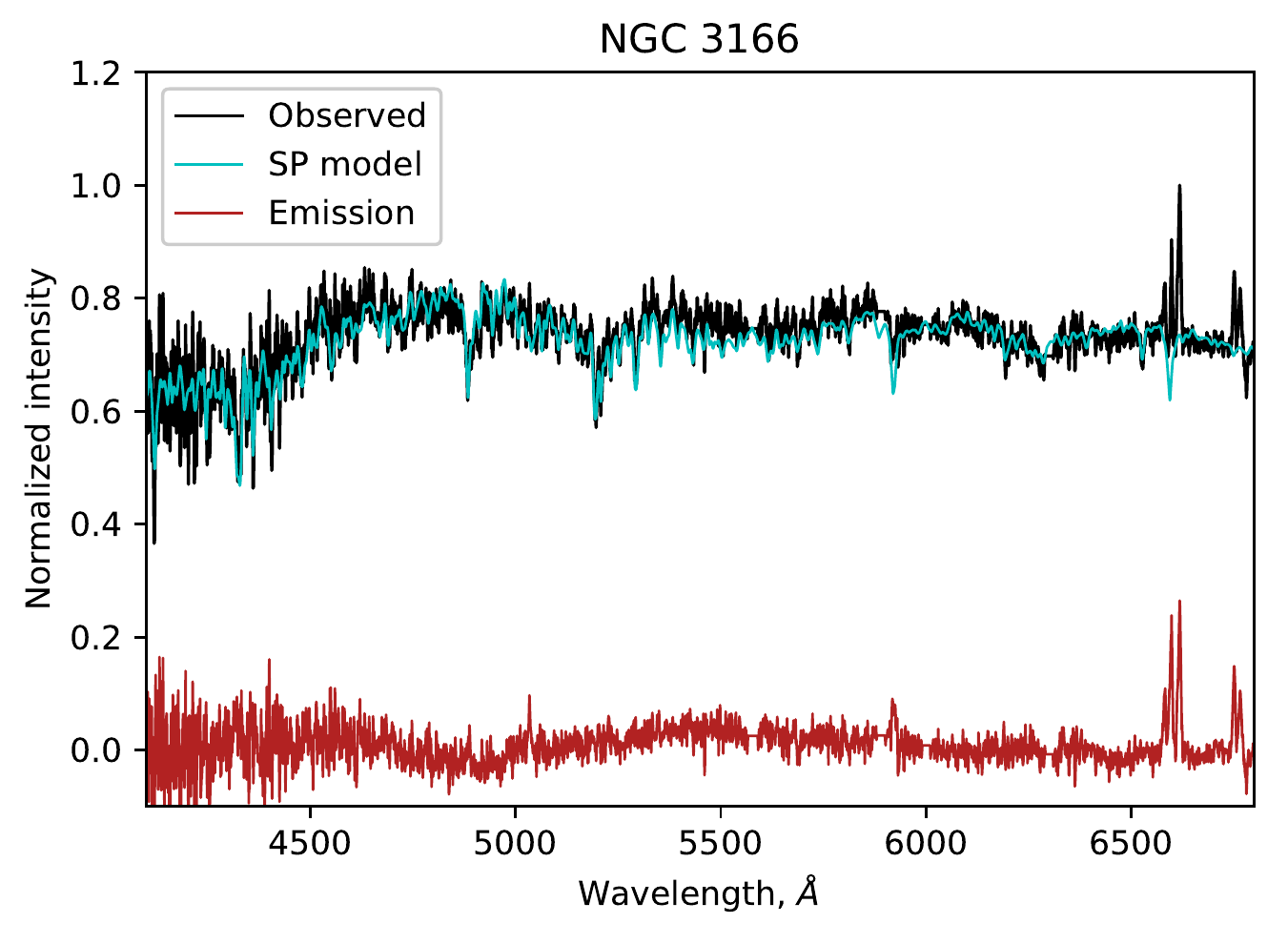} \\
\end{tabular}
\caption{Examples of the pure-emission spectrum extraction in low S/N bins: the emission-line clump
in 14\arcsec\ from the center ({\it left}) and an area of diffuse emission in 23\arcsec\ from the center ({\it right})
in the long-slit spectrum of NGC~3166.}
\label{fig_N3166fit}
\end{figure*}

As it will be noted further, we study the distribution of the above mentioned emission lines
along the slits in more details for two galaxies -- NGC~2551 and NGC~3166. Their spectra demonstrate very extended
emission of diffuse ionized gas together with compact bright clumps. For this particular analysis we consider the spatial
distribution of the line flux ratios measured by fitting single Gaussians to the observed lines at each position along the
slit after subtracting the spectra of stellar population models obtained with the ULySS software \cite{Koleva2009},
as is shown in Fig.~\ref{fig_N3166fit}.

\subsection{Observations with the scanning Fabry-Perot interferometer}

Scanning  Fabry-Perot interferometer (FPI) allows to make 3D--spectroscopic study of the ionized-gas  kinematics over large
field of view, whereas the {\bf narrow interfringe} spectral range contains usually only a single emission line selected by a band--pass
filter with FWHM of about 15--30~\AA. Using the information from the long--slit spectroscopy we have preliminarily chosen
spectral lines in which the most extended ionized gas emission could be expected: H$\alpha$, [NII]$\lambda6583$,
or [OIII]$\lambda5007$. The targets were observed at the SAO RAS 6m telescope with the SCORPIO-2 in the 
FPI mode by using the interferometer providing spectral resolution of FHWM= 1.7~\AA\ ($\sim 78~\kms$) 
and full spectral range (interfringe) of $\Delta\lambda=35$~\AA\ around redshifted [NII] or \Ha line, whereas
in the [OIII] emission line these parameters were FHWM= 2.0~\AA\ ($\sim 120~\kms$) and $\Delta\lambda=20$~\AA.
During the scanning process, we have consecutively obtained few tens of interferograms fixing different gaps 
between the FPI plates covering uniformly the interfringe: 40 frames in the \Ha or 30 frames in the [OIII] spectral ranges.
The field of view was 6\farcm1.  The galaxy NGC~3166 was observed with a previous version of the focal reducer SCORPIO 
in the \Ha emission line using FPI providing  FHWM= 2.5~\AA\ ($\sim 112~\kms$)  and $\Delta\lambda=29$~\AA, 
the number of interferograms was 32. The  log  of the observations including exposures times and mean seeing values
is presented in Table~\ref{table_obs2}. The data were reduced using the software package described in details 
by \citet{Moiseev2002} and by \citet{Moiseev2008}. 

Briefly, the used procedures included standard steps for CCD  data reduction (bias subtraction, flat-field corrections,
sigma clipping of cosmic-ray hits) and some specific steps: removing interference rings relating with airglow lines
using an azimuthal averaging of brightness in each frame with masked galaxies and stars \citep[see][]{Moiseev2002};
correction of atmospheric transparency as well as seeing variations and {\bf astrometric} shifts of individual interferograms basing
on the PSF-photometry of foreground stars;  phase-shift correction (wavelength calibration) basing on the scanning of
some selected lines in the spectra of He-Ne-Ar lamp exposed every evening and morning before and after the observational
night. After the primary reduction the observed data are combined into data cubes, where to each  spaxel of the field of
view a 30, 32, or 40-channel spectrum is attributed.

The data cubes were rotated to the `standard' orientation (north at the top). The astrometry grid was created using the Astrometry.net project
web-interface\footnote{http://nova.astrometry.net/} \citep{Lang2010}. The optical continuum and monochromatic
images in the emission lines, the line-of-sight velocity fields and velocity dispersion maps 
were created from the one-component Voigt profile fitting of the spectra as described in \citet{Moiseev2008}. 
All the velocities  presented in the paper are heliocentric.

\begin{table}
\caption{Log of the scanning FPI observations}
\label{table_obs2}
\centering
\scriptsize
\begin{tabular}{lcccrl}
\hline\noalign{\smallskip}
Galaxy & Date &  Emission line & Exposure, s & Seeing,\arcsec \\
\hline\noalign{\smallskip}
IC 5285     & 08.12.2015 & \Ha & $40 \times 150$   & 2.6 \\
NGC 252   & 15.12.2017 & \Ha & $40 \times 150$   &2.4 \\
NGC 774   & 13.11.2014 &  \Ha & $40 \times 240$  & 1.0 \\
NGC 2551 & 08.12.2015 &  \Ha & $40 \times 180$  & 2.1 \\
NGC 2655 & 07.12.2015 & [OIII]  & $30 \times 220$& 2.0 \\
NGC 2697 & 01.03.2016 & [OIII] & $30 \times 300$ & 4.0 \\
NGC 2697 & 02.03.2016 &  \Ha & $40 \times 180$  &  3.0 \\
NGC 2787 & 09.12.2012 & [NII] & $40 \times 180$  &  1.7 \\
NGC 2962 & 29.02.2016 & \Ha & $40 \times 180$   &  2.2 \\
NGC 3106 & 09.12.2015 & \Ha & $40 \times 120$   & 1.4 \\
NGC 3166 & 14.02.2007 & \Ha & $32 \times 160$   &1.3 \\
NGC 3182 & 28.02.2014 & \Ha & $40 \times 160$   &3.0 \\
NGC 3414 & 15.02.2016 & [OIII] & $30 \times 200$ &2.5 \\
NGC 3619 & 28.02.2014 & [OIII] & $30 \times 280$ &3.1 \\
NGC 4026 & 03.04.2014 & [OIII] & $30 \times 180$ &1.8 \\
NGC 4324 & 29.02.2016 &  \Ha & $40 \times 180$  &1.8 \\
NGC 7280 & 18.11.2014 & [OIII] & $30 \times 180$ &1.1 \\
UGC 9519 & 29.02.2016 & [NII] & $40 \times 150$  & 1.6 \\
UGC 9519 & 03.05.2016 & \Ha & $40 \times 230$   & 1.9 \\
UGC 12840& 11.12.2015 &  \Ha & $34 \times 160 + 6 \times 100$  & 1.2 \\
\hline
\end{tabular}
\end{table}

\subsection{Ancillary data}

By wishing to expand our gas velocity fields and to reach physical interpretation of
what we see, for our analysis we involve some additional data on our galaxies which can be
retrieved in open data archives. In particular, we use SAURON IFU [OIII] emission-line data
collected in the frame of the ATLAS-3D survey \citep{atlas3d_1} to refine gas velocity fields in the
centers of the galaxies, and we expand our gas velocity fields into the outermost parts
of the galaxies by attaching interferometric \HI\ data (GMRT observations from the archive of
\url{http://wow.astron.nl} presented also in \citealt{atlas3d_13} and \citealt{Lee-Waddell}).
The star formation regions distributions over the galactic disks are studied with the GALEX
images \citep{galex} obtained in the far- and near-ultraviolet bands (FUV and NUV).
The general structure of the galactic disks is considered with the photometric data
in the optical bands retrieved from the SDSS (\url{http://www.sdss3.org}) and PanSTARRS
(\url{https://ps1images.stsci.edu/cgi-bin/ps1cutouts}) archives, and also with the
NIR images of the galaxies in 3.6~mkm and 4.5~mkm bands which have been obtained and made
public in the frame of the Spitzer-telescope galaxy survey S4G \citep{s4g}. Finally, four of
our galaxies were observed with the IFU PMAS/PPAK in the frame of the CALIFA survey \citep{califa,califa_s,califa_3},
and for these galaxies we have also used the fully reduced datacubes taken from the CALIFA
archive, \url{http://califa.caha.es}.

\section{Results}

\subsection{Kinematics}

\begin{figure*}[p]
 \includegraphics[width=18cm]{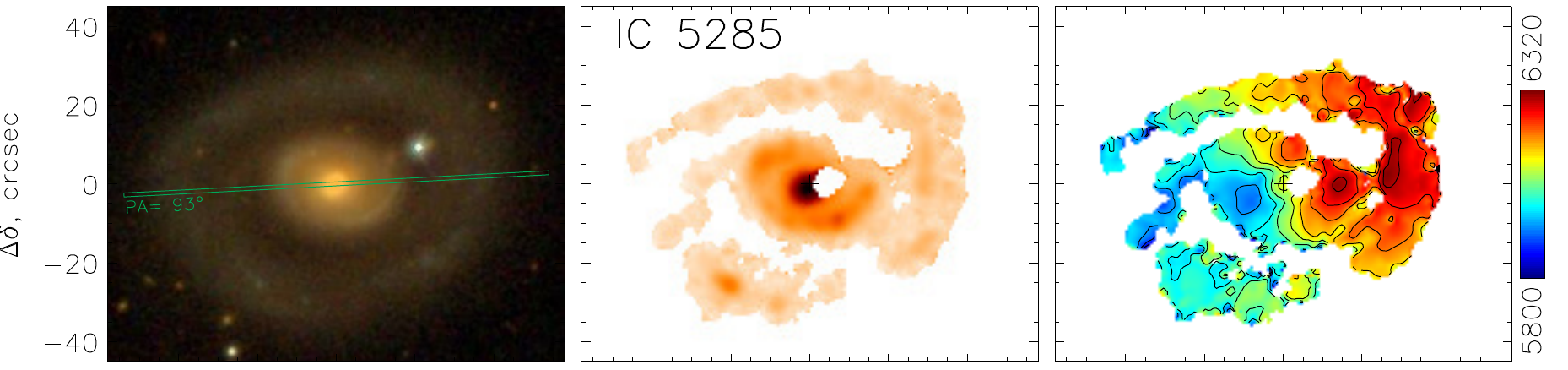}  
 \includegraphics[width=18cm]{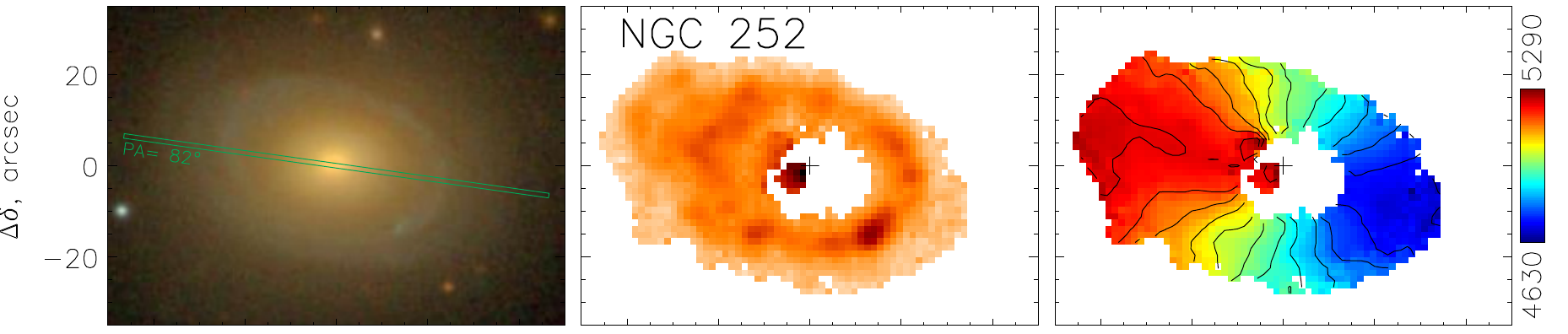}  
 \includegraphics[width=18cm]{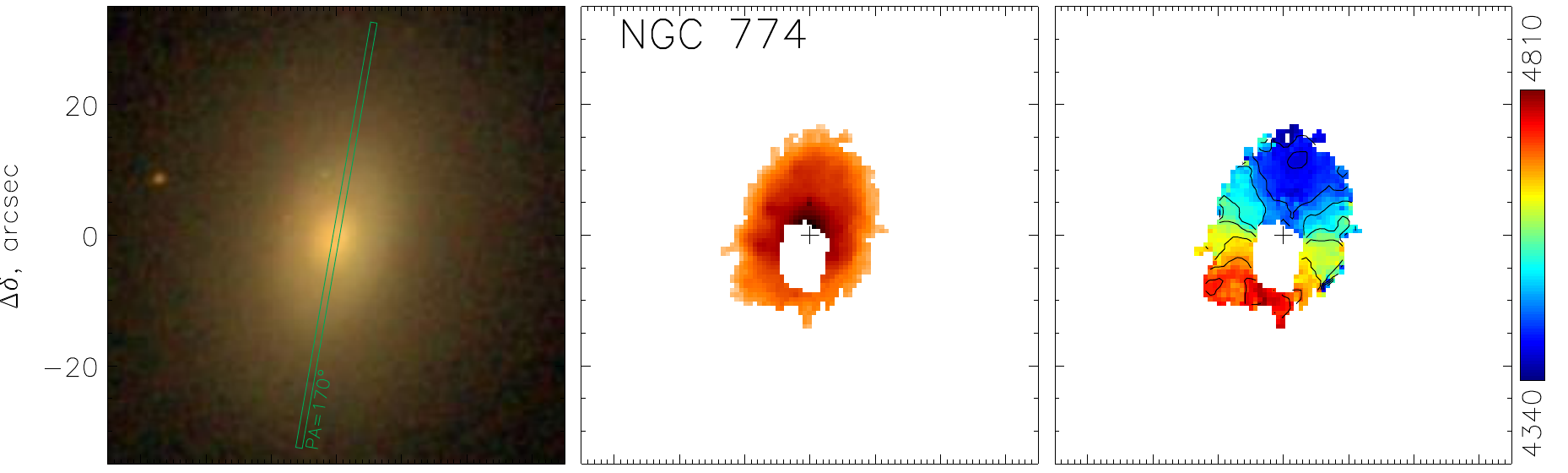}  
 \includegraphics[width=18cm]{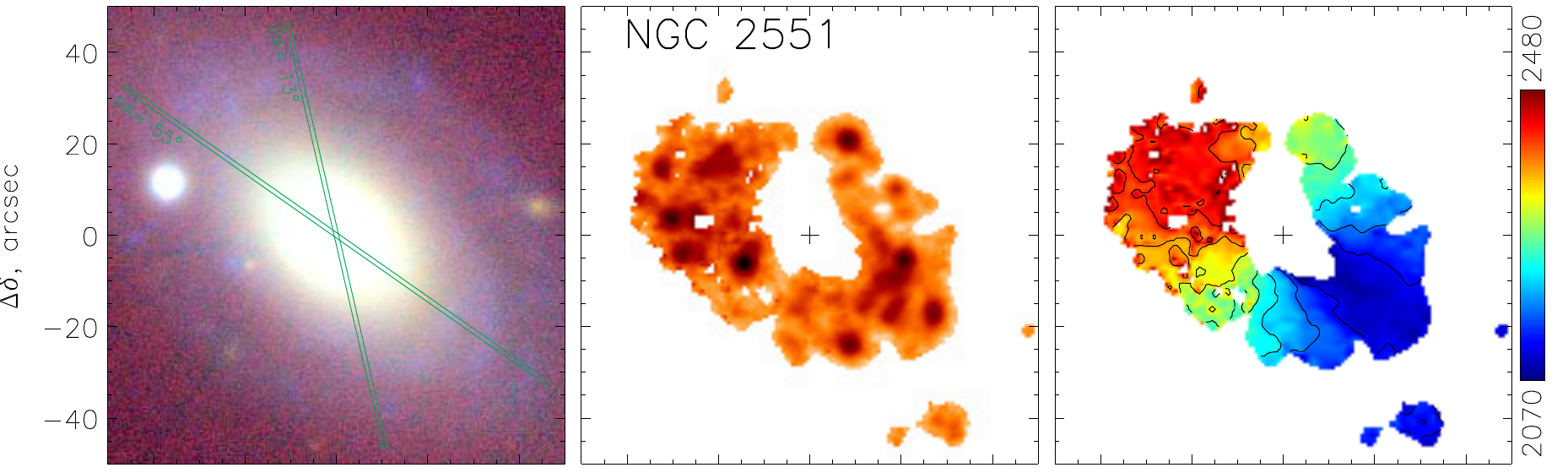}  
\caption{Optical {\bf broad-band} images of the observed galaxies ({\it left}), and the corresponding FPI maps: emission-line images
scaled as square root of the intensity ({\it middle}), ionized-gas velocity fields, with the isovelocity contours  
overplotted through 50~\kms\ step ({\it right}). The color  bar is in \kms. The angular offsets are shown with respect to
photometric centers of the galaxies. The positions of our long slits with marked $PAs$ are overlapped by green. The combined
color images are taken from the SDSS DR14 server; for the targets that were not covered by the SDSS (NGC~2551, NGC~2655, NGC~2697, and NGC~2787)
the PanSTARRS PS1 images are shown, in some cases containing ghosts and color artefacts seen in the original data.}
\label{fig_fpi}
\end{figure*}

\setcounter{figure}{3} 

\begin{figure*}[p]
 \includegraphics[width=18cm]{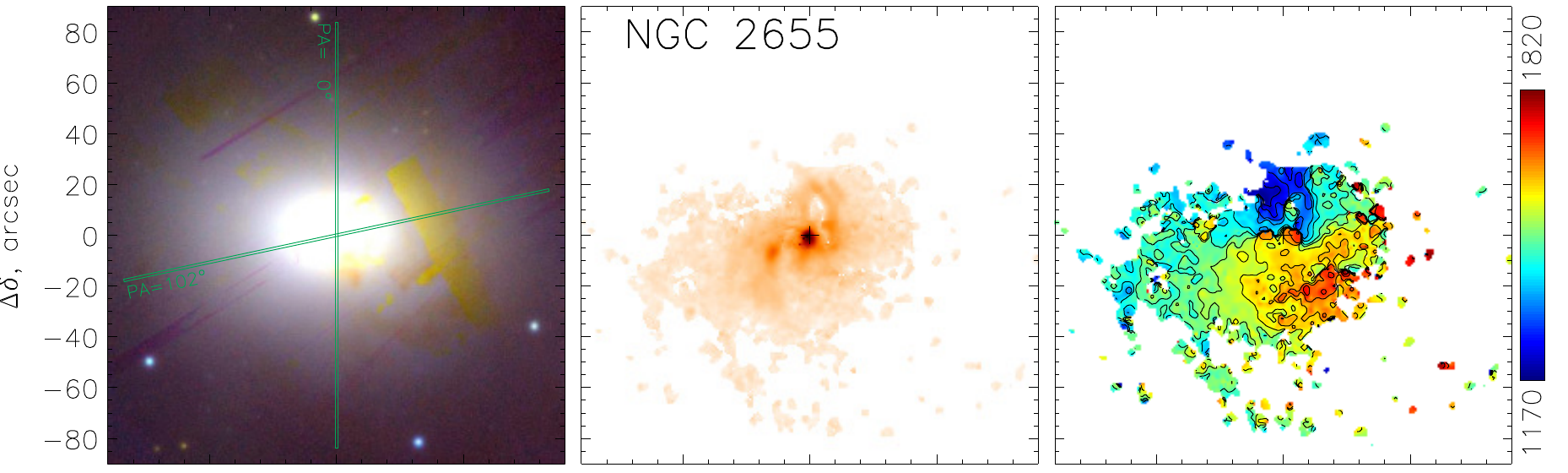}  
 \includegraphics[width=18cm]{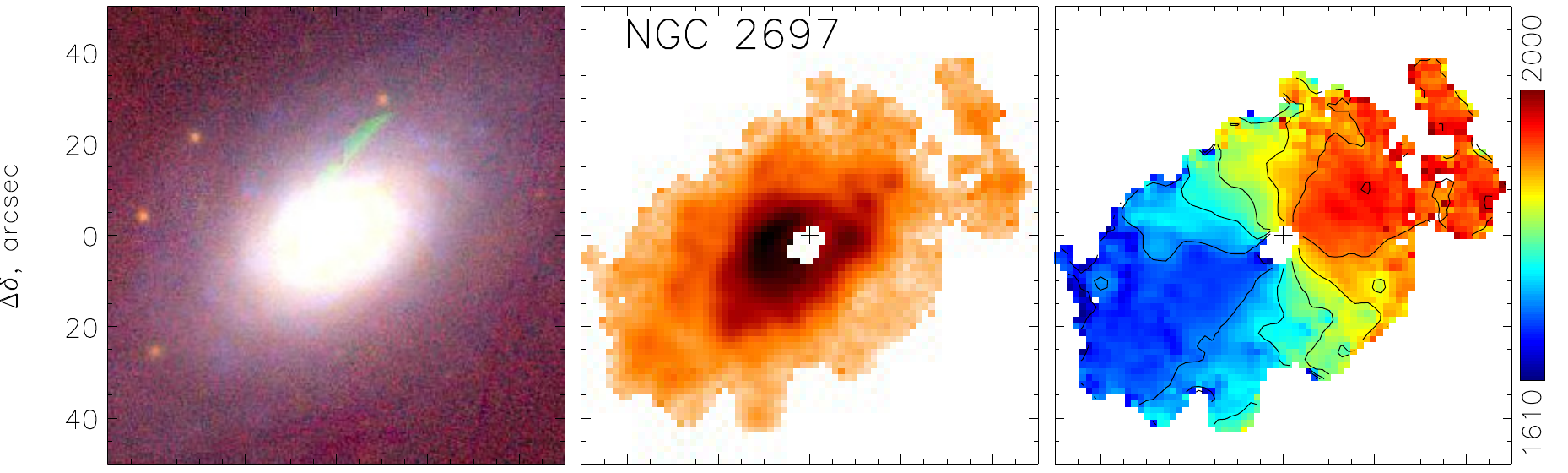}  
 \includegraphics[width=18cm]{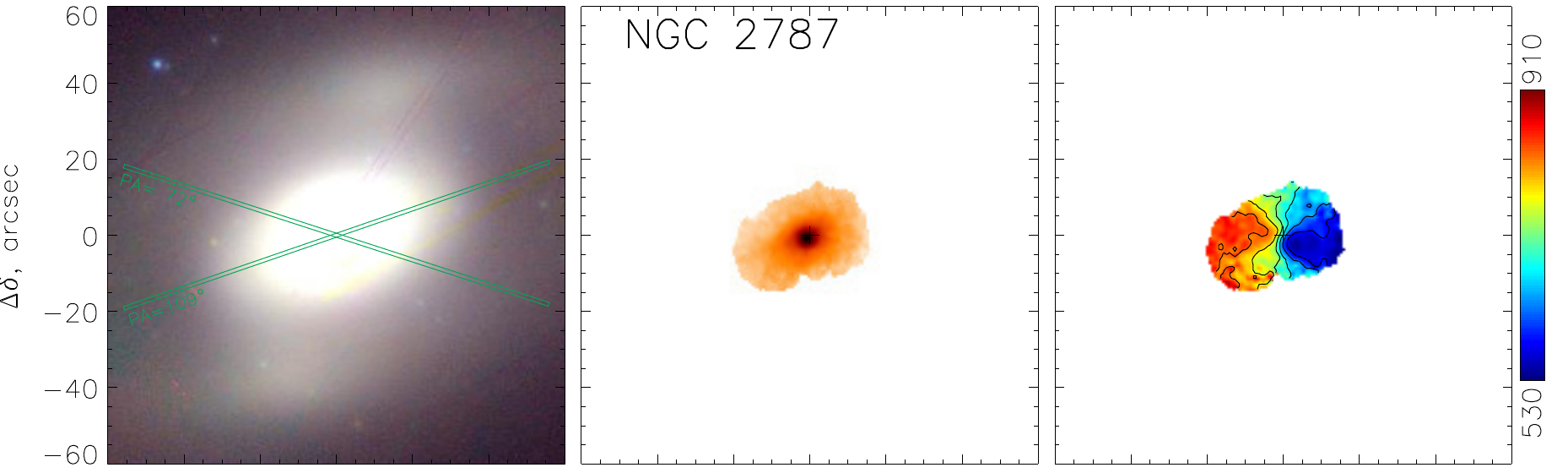}  
 \includegraphics[width=18cm]{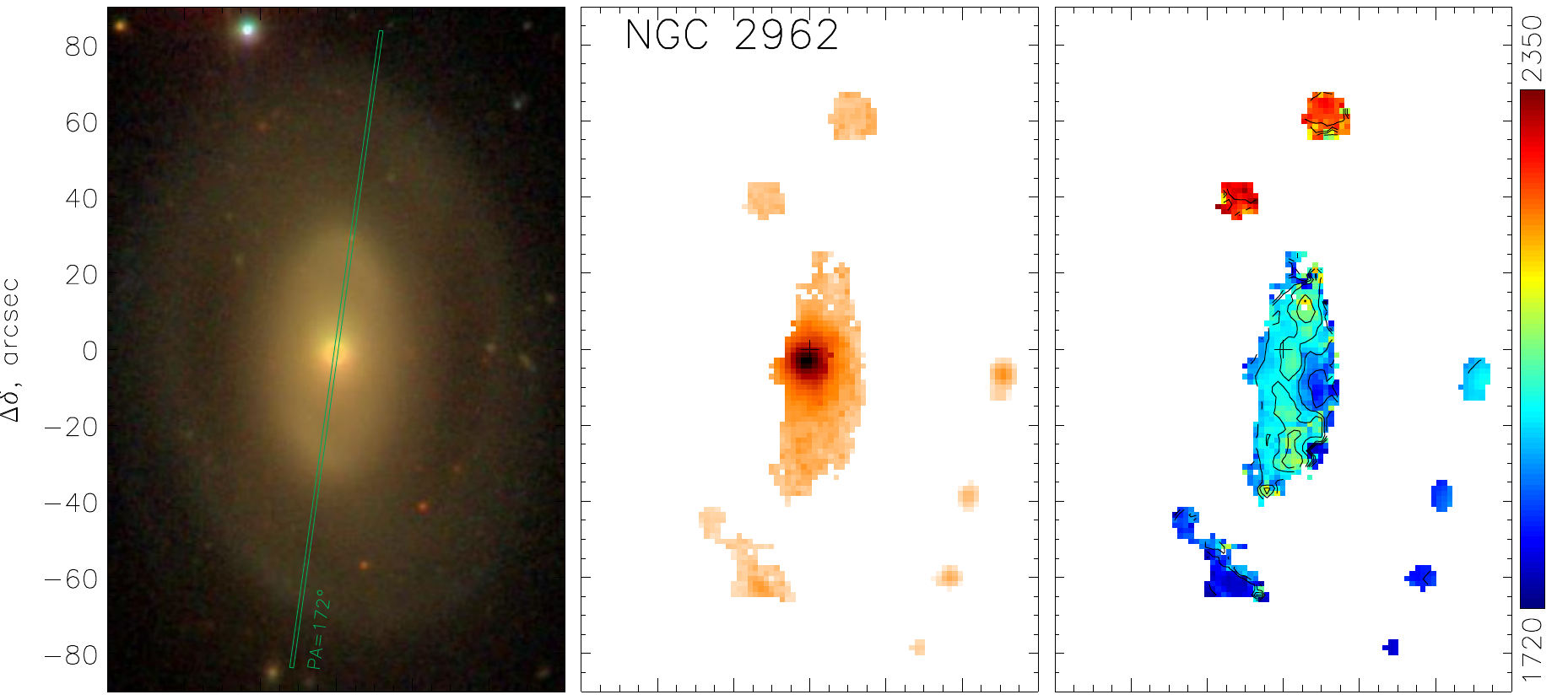}  
\caption{--- continued}
\end{figure*}

\setcounter{figure}{3} 
\begin{figure*}[p]
 \includegraphics[width=18cm]{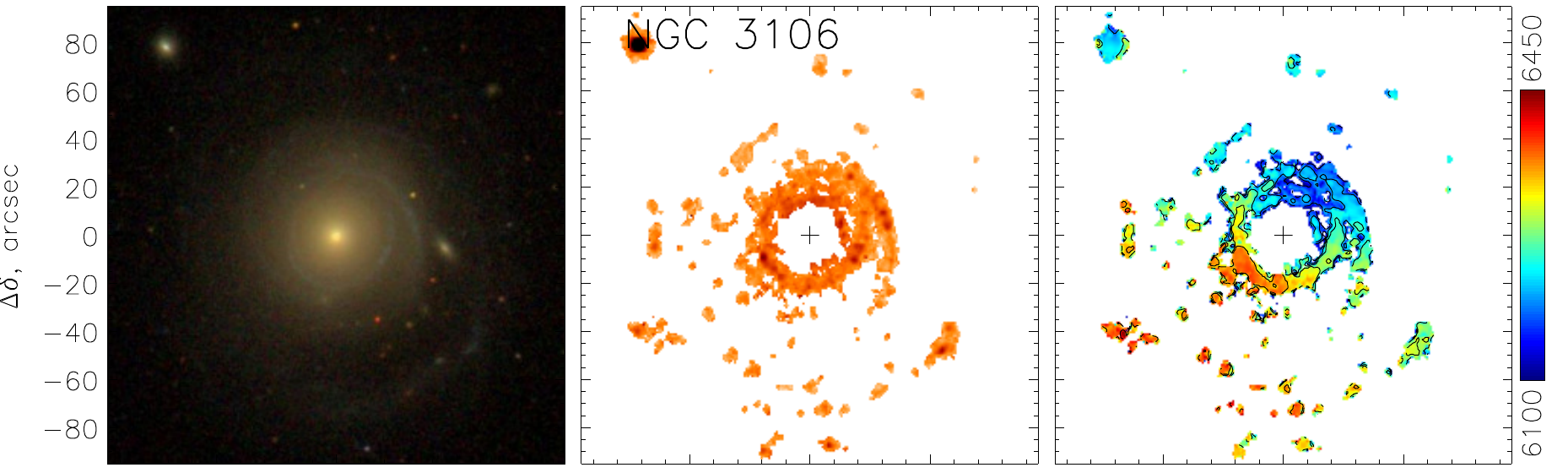}  
 \includegraphics[width=18cm]{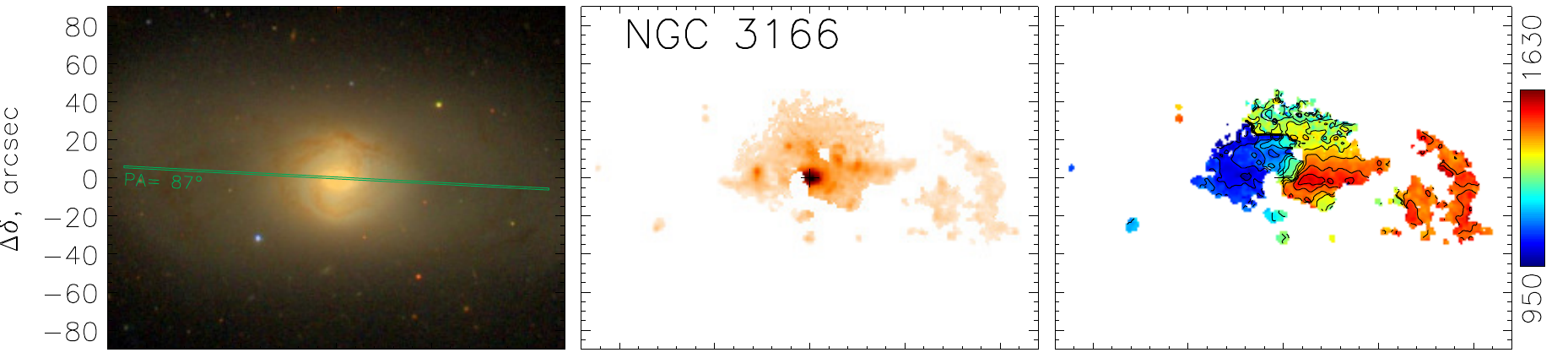}  
 \includegraphics[width=18cm]{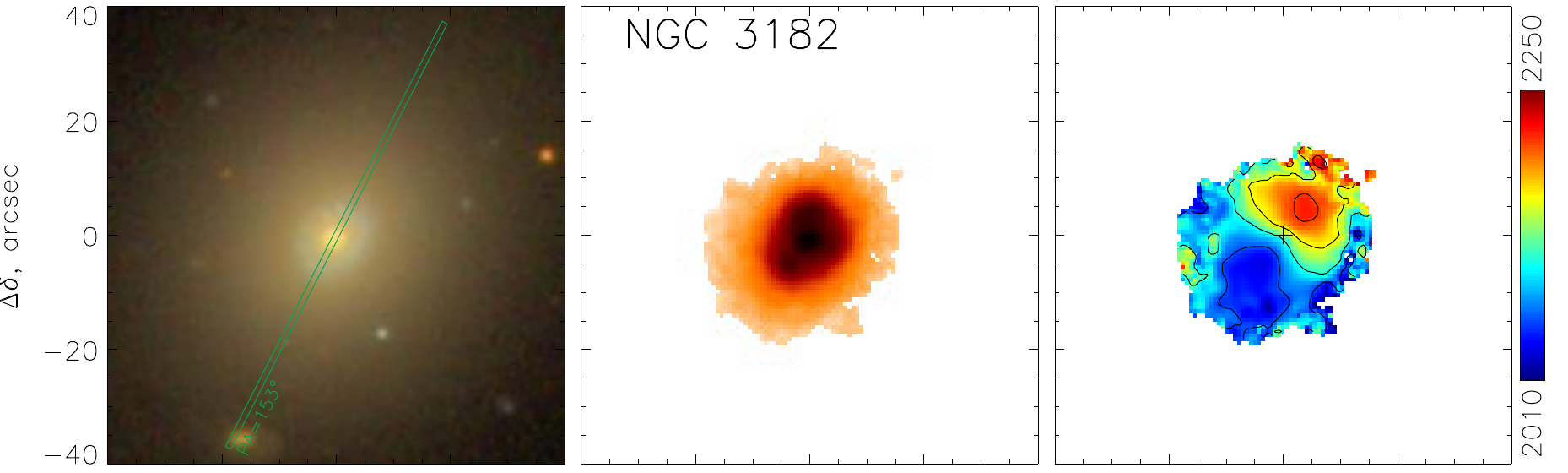}  
 \includegraphics[width=18cm]{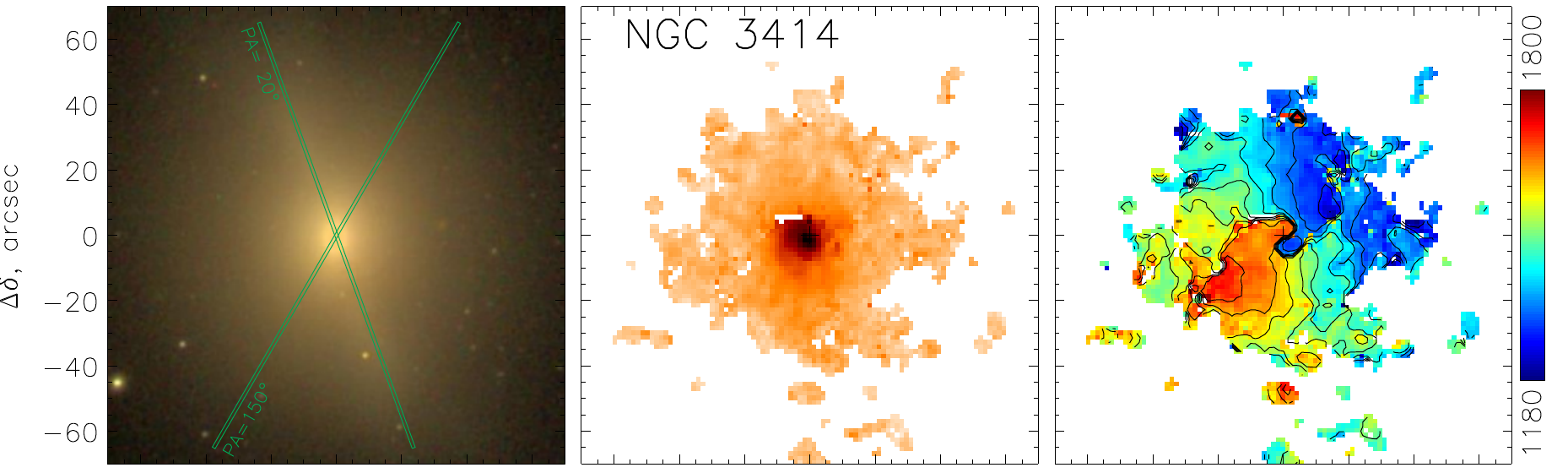}  
\caption{--- continued}
\end{figure*}

\setcounter{figure}{3} 
\begin{figure*}[p]
 \includegraphics[width=18cm]{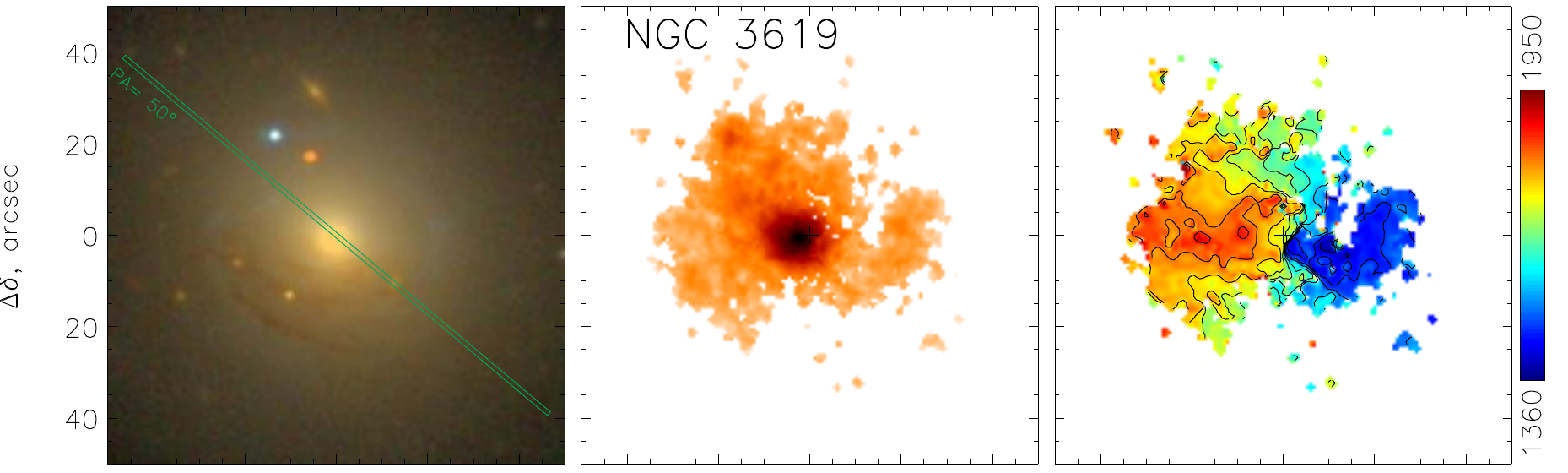}  
 \includegraphics[width=18cm]{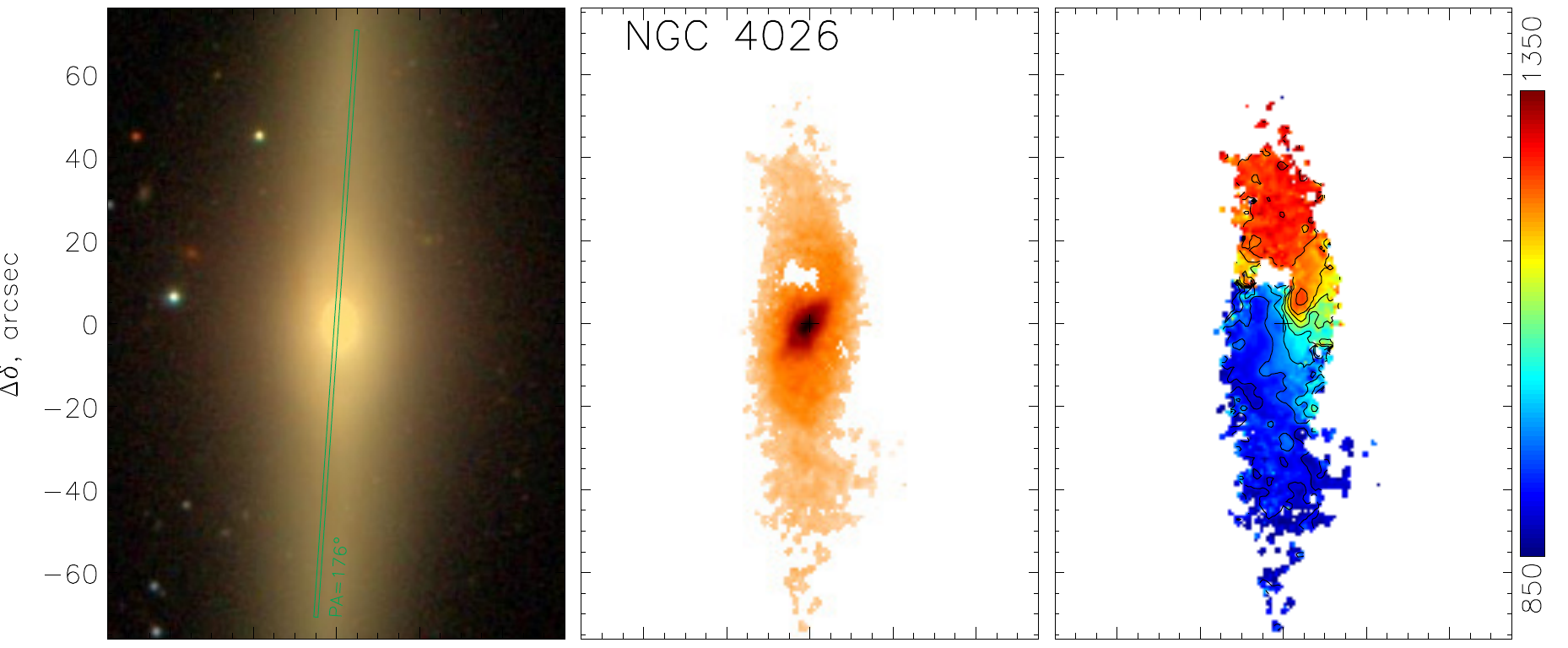}  
 \includegraphics[width=18cm]{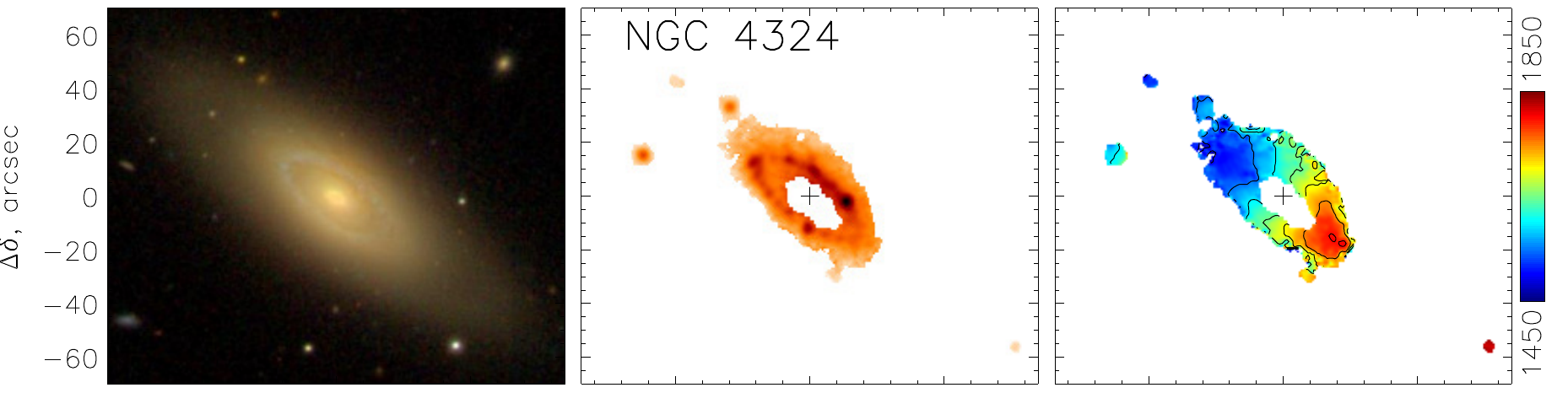}  
 \includegraphics[width=18cm]{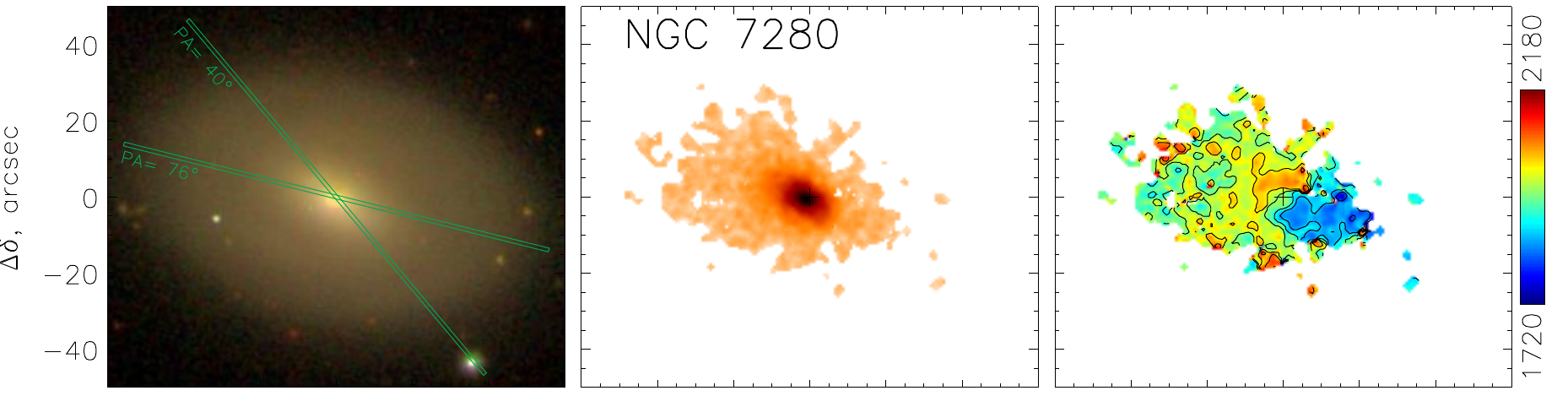}  
\caption{--- continued}
\end{figure*}

\setcounter{figure}{3} 
\begin{figure*}[p]
 \includegraphics[width=18cm]{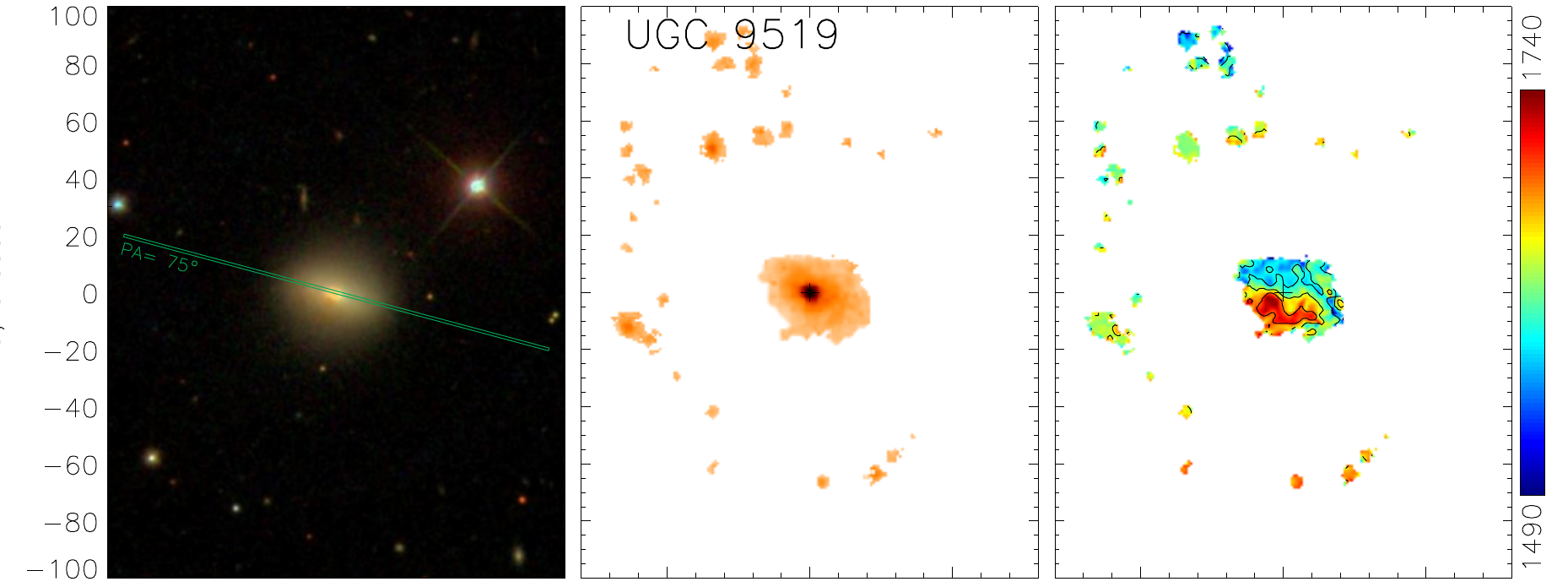}  
 \includegraphics[width=18cm]{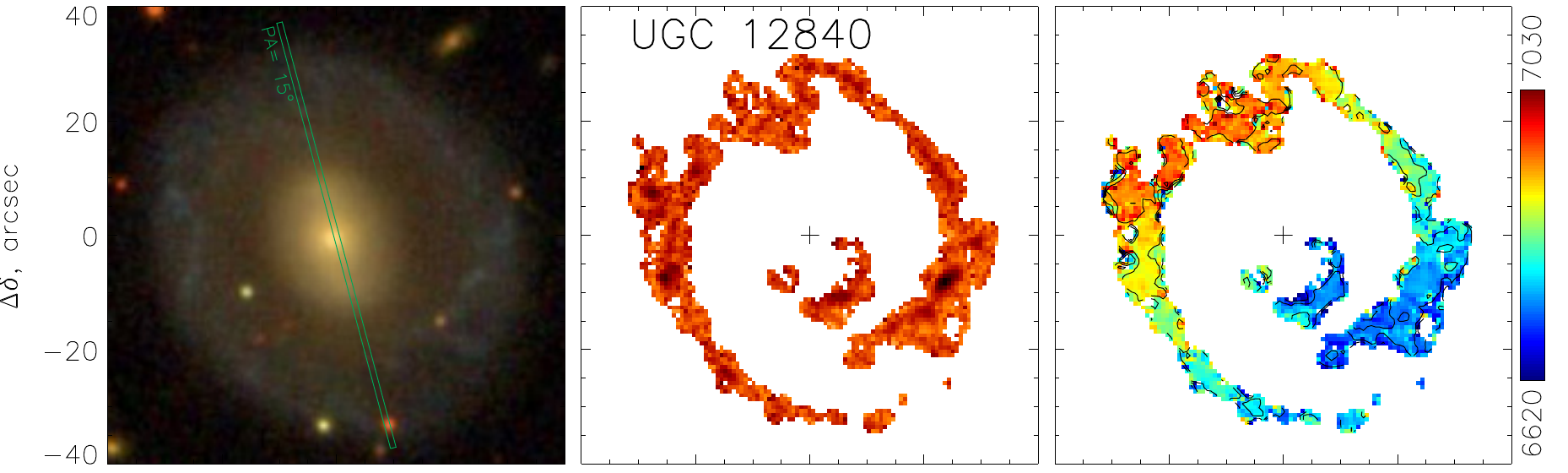}  
\caption{--- continued. For UGC~9519 a combination of the [NII] data for the
radii less than 20\arcsec\ and of the \Ha\ data for the more outer galactic regions
is presented.}
\end{figure*}

\begin{figure*}
\centerline{
 \includegraphics[height=7cm]{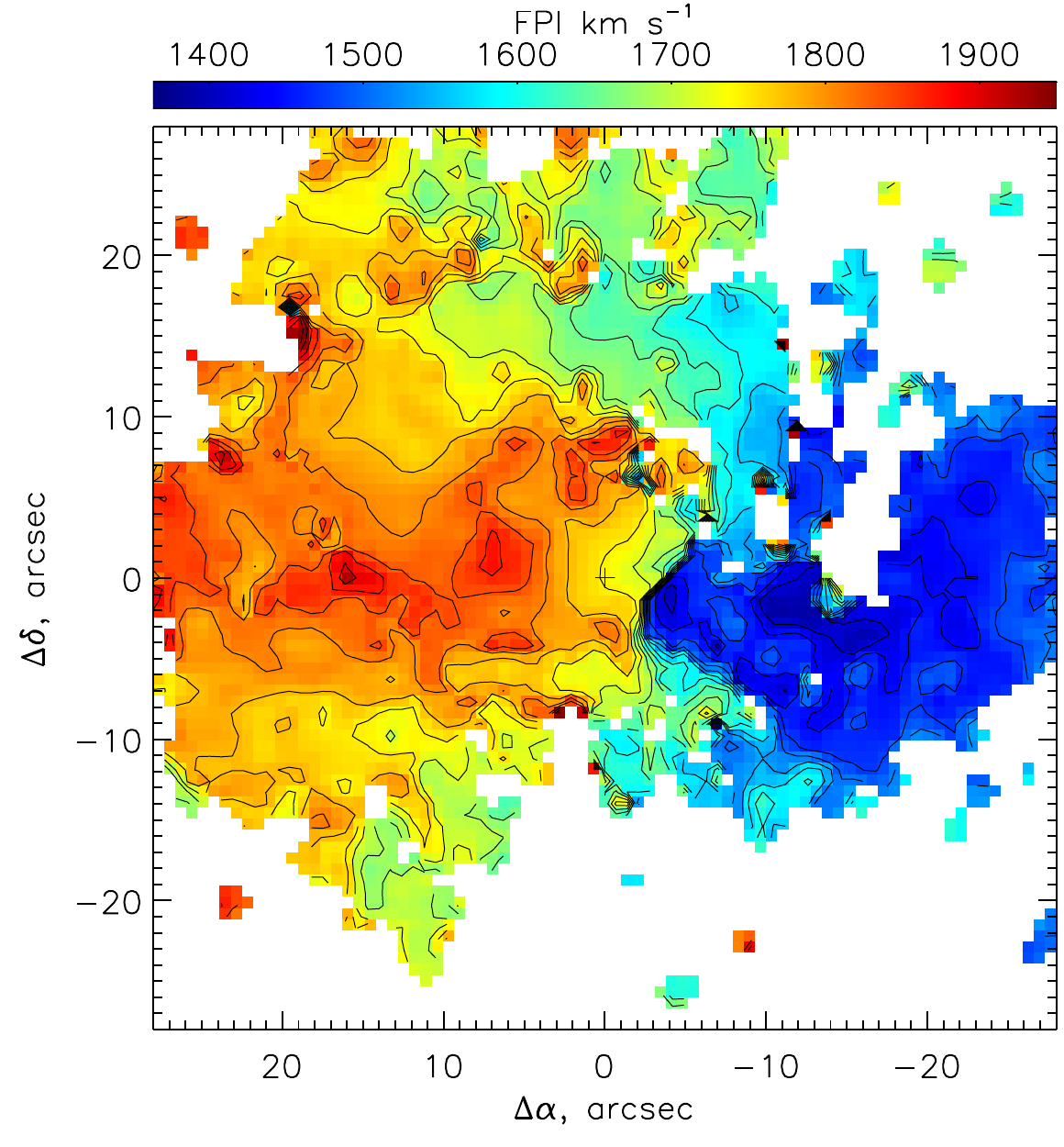}  
 \includegraphics[height=7cm]{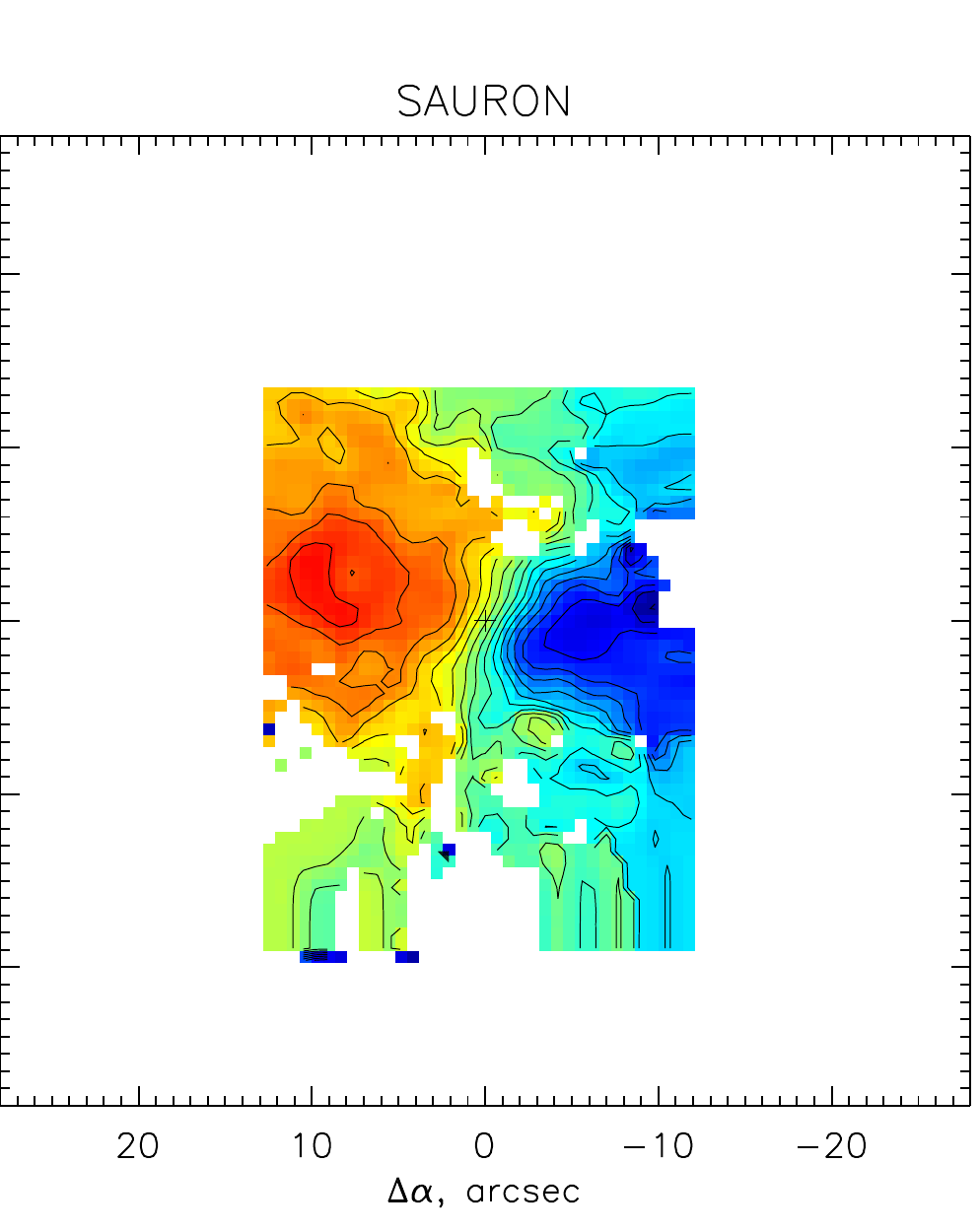}  
 \includegraphics[height=7cm]{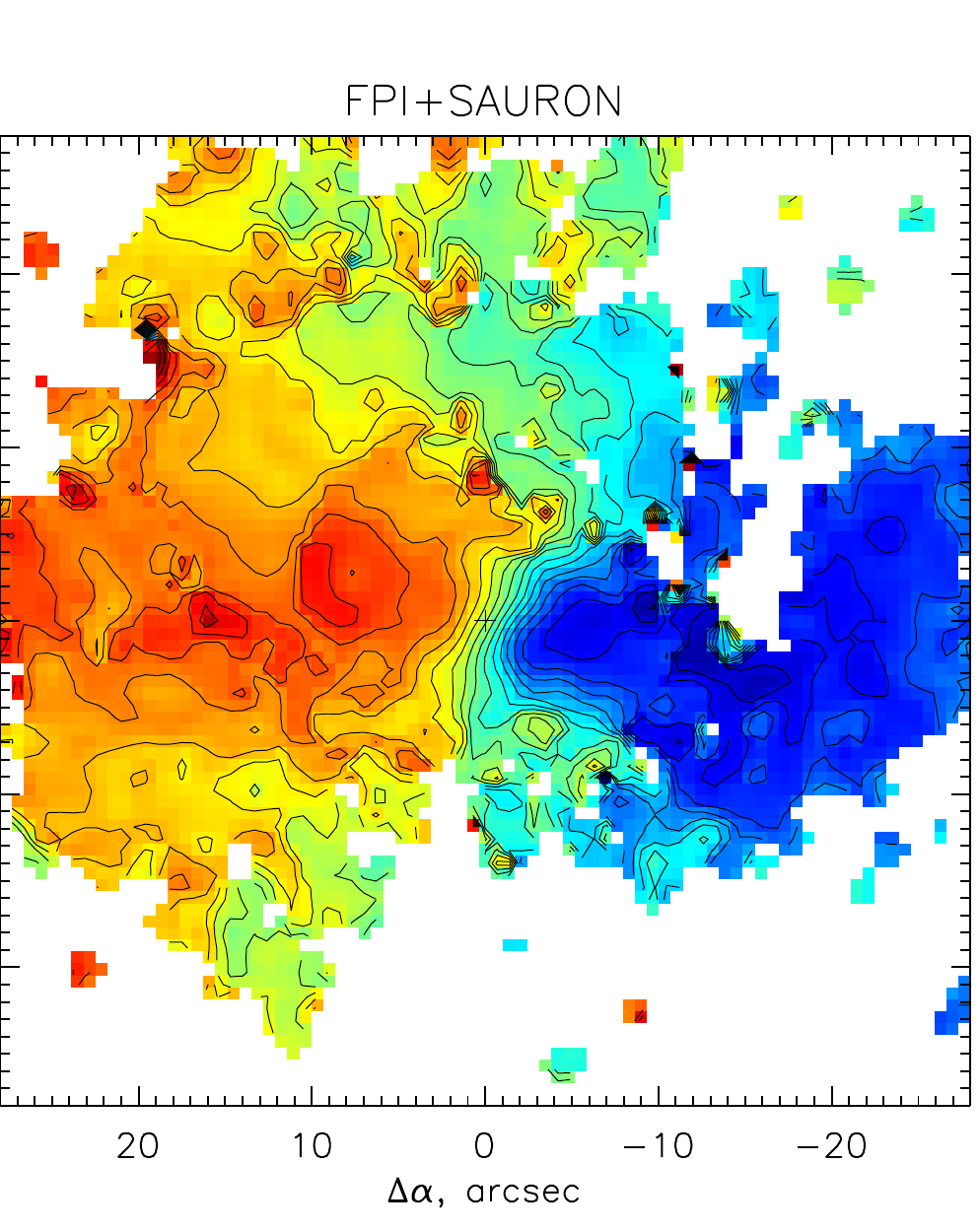}  
 }
\caption{Ionized-gas velocity fields in the [OIII] emission line for the central region of NGC~3619: the original FPI map 
({\it left}), the SAURON data in the same spatial and velocity scales ({\it middle}), and their combination ({\it right}). }
\label{fig_sauron}
\end{figure*}

\begin{figure*}[p]
\centering
\begin{tabular}{c c c}
 \includegraphics[width=6cm]{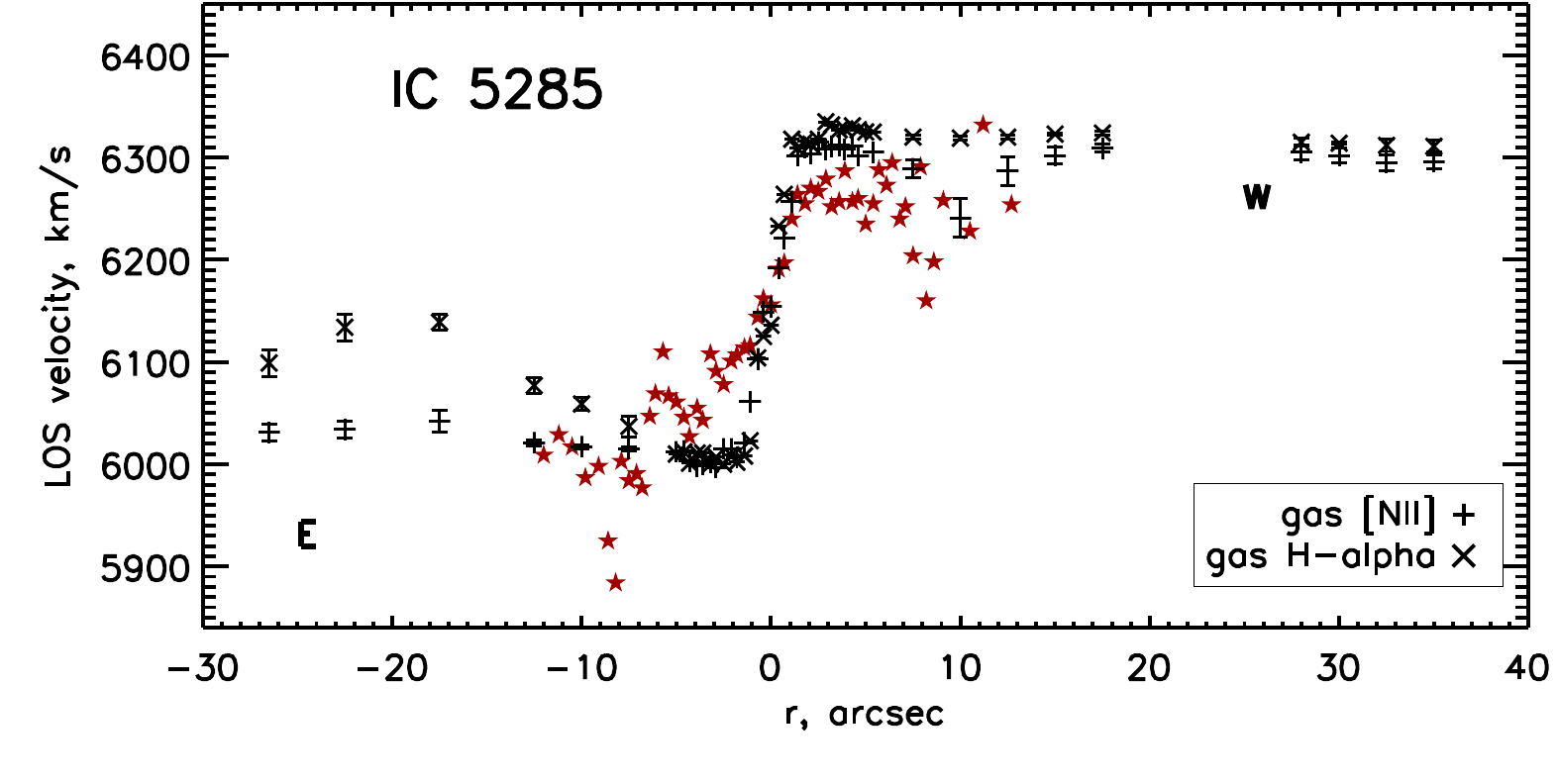} &
 \includegraphics[width=6cm]{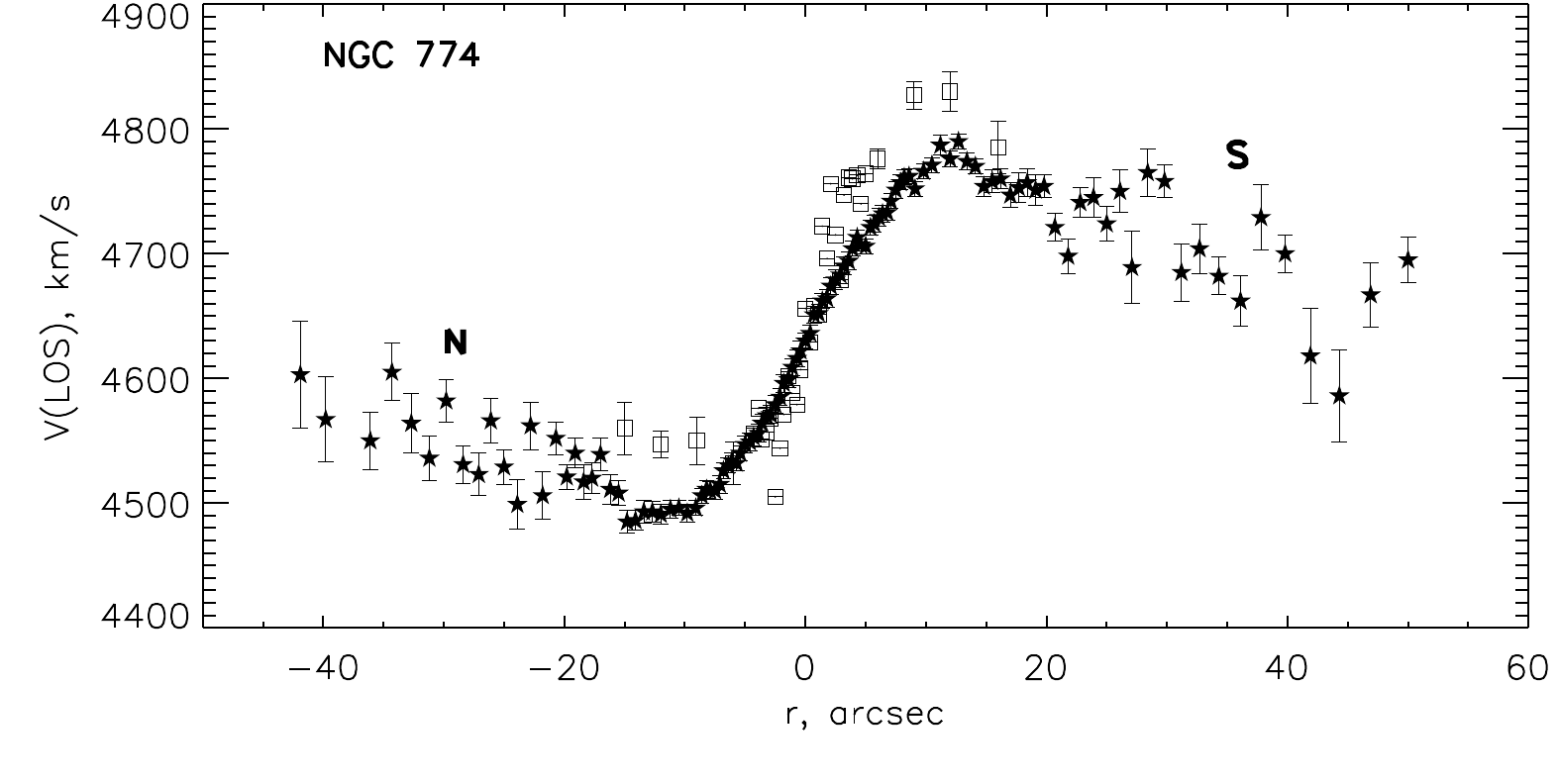} &
 \includegraphics[width=6cm]{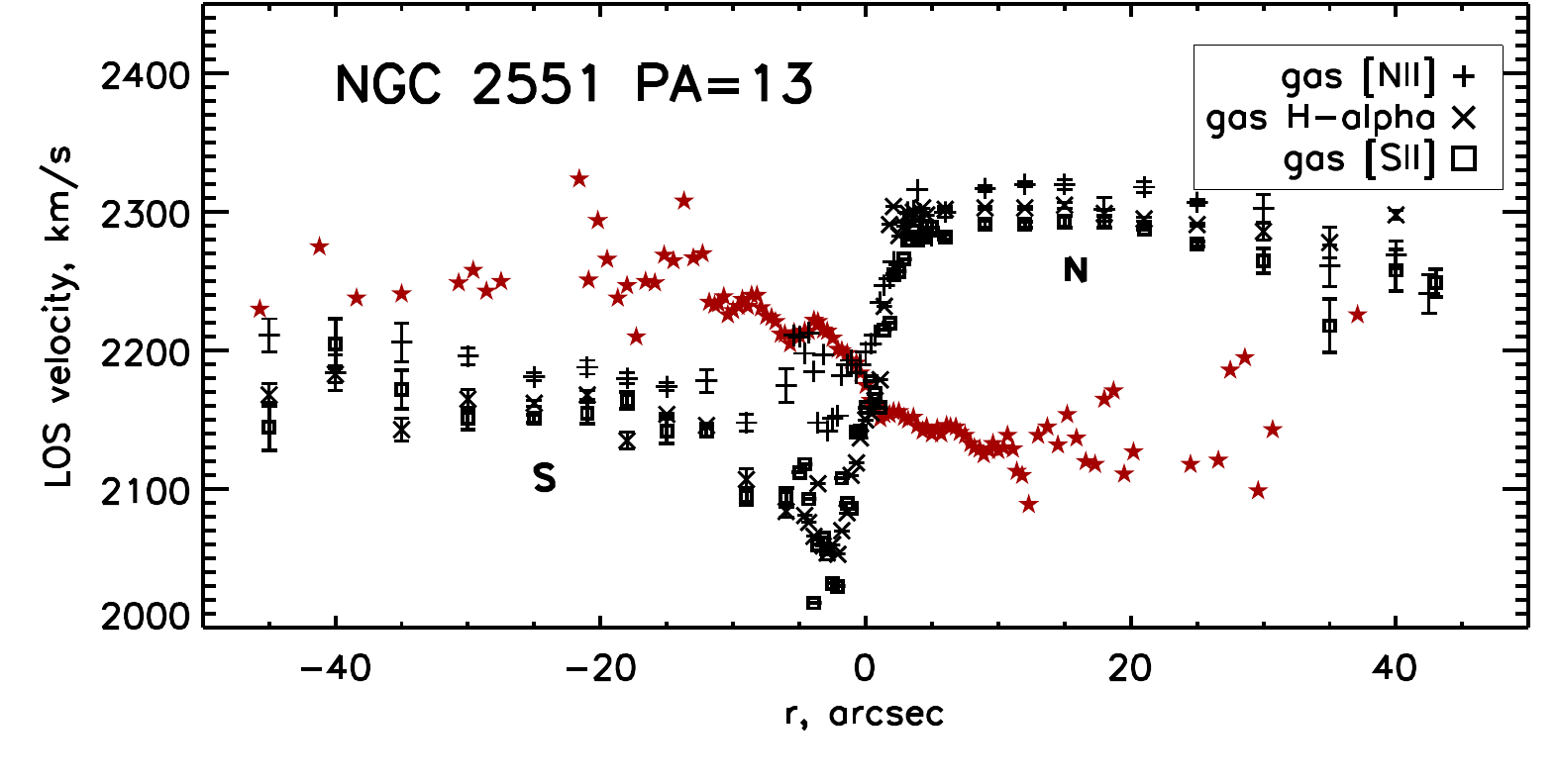} \\
\includegraphics[width=6cm]{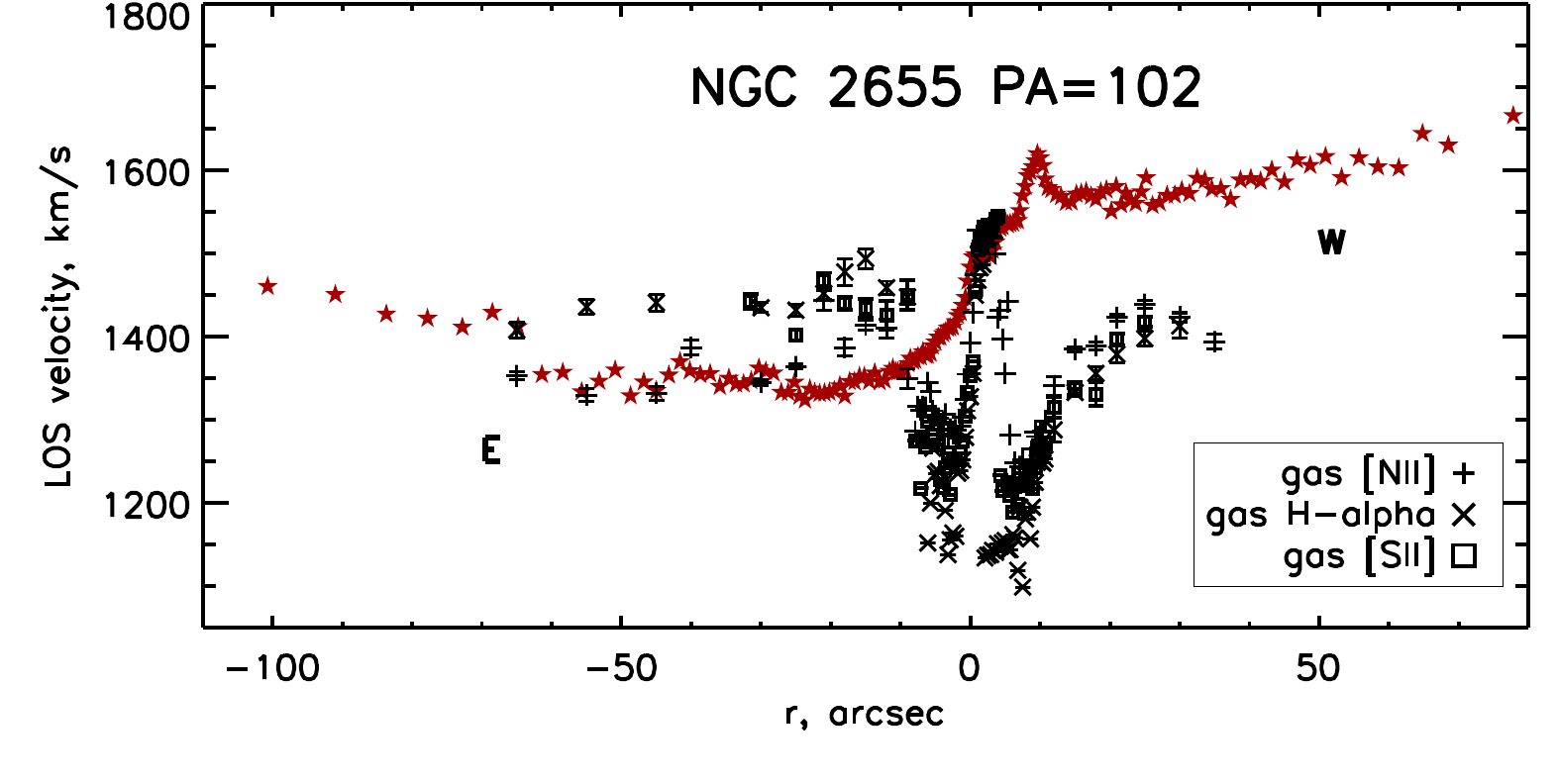} &
 \includegraphics[width=6cm]{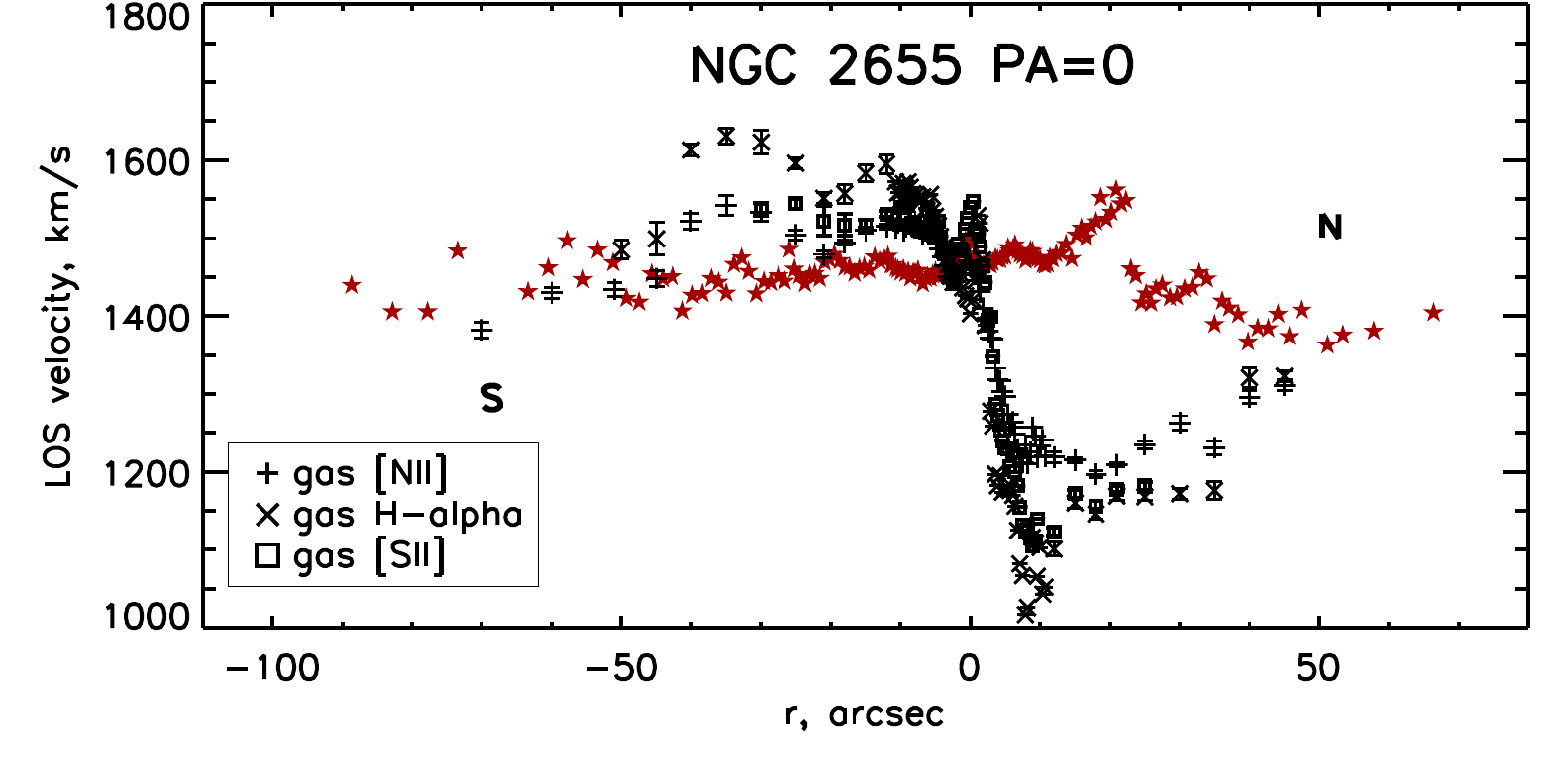} &
 \includegraphics[width=6cm]{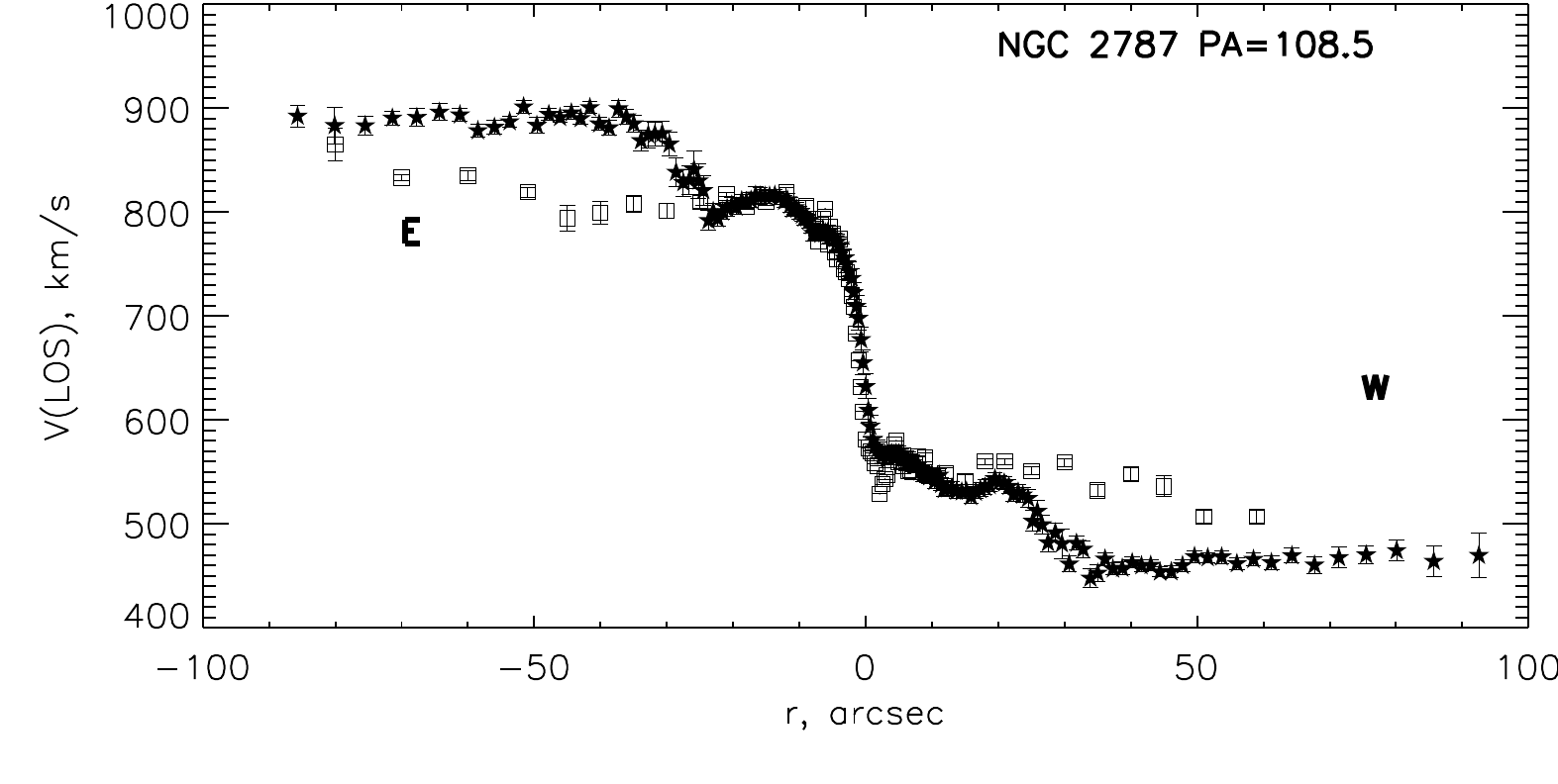} \\
\includegraphics[width=6cm]{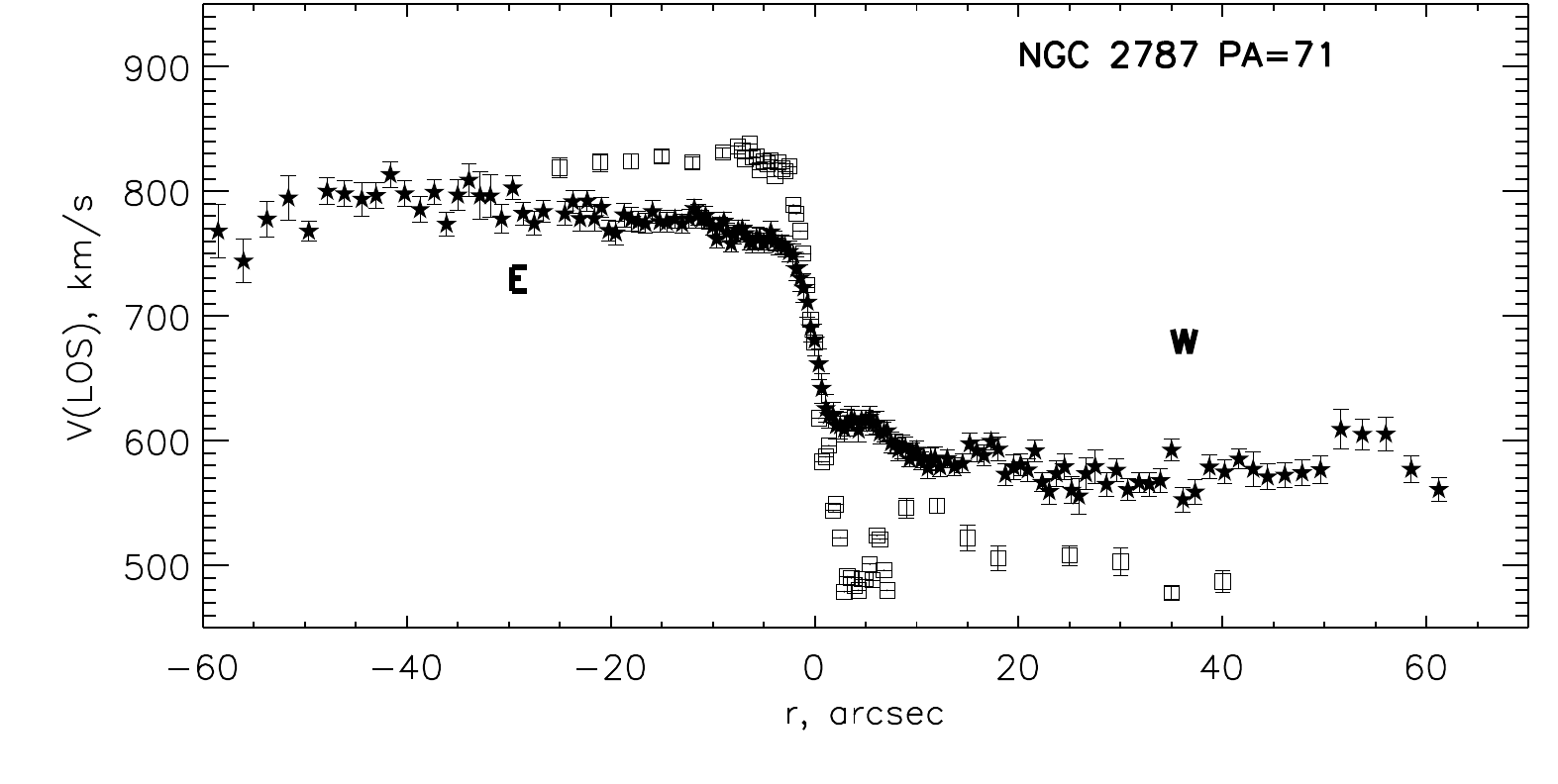} &
 \includegraphics[width=6cm]{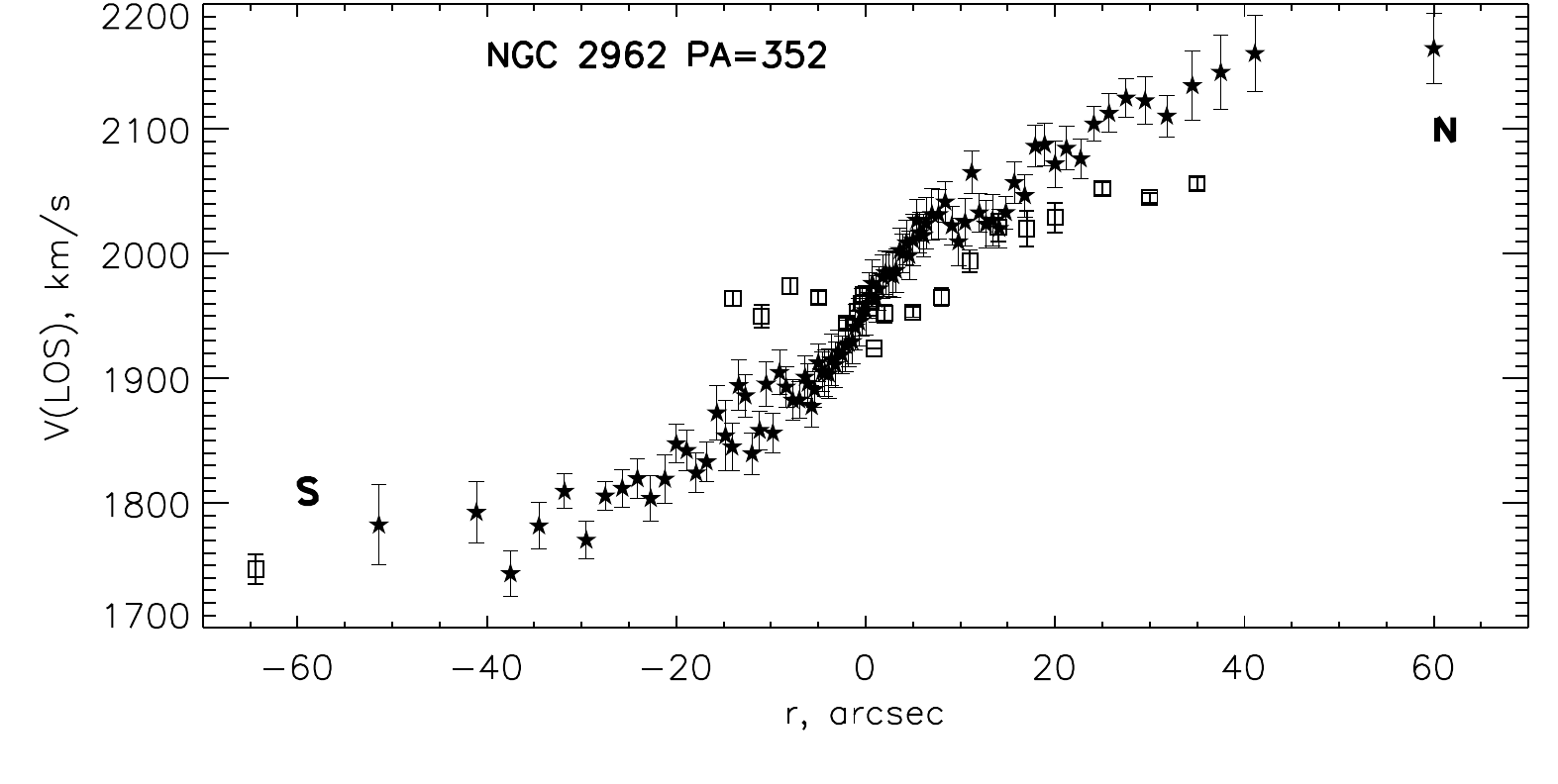} &
 \includegraphics[width=6cm]{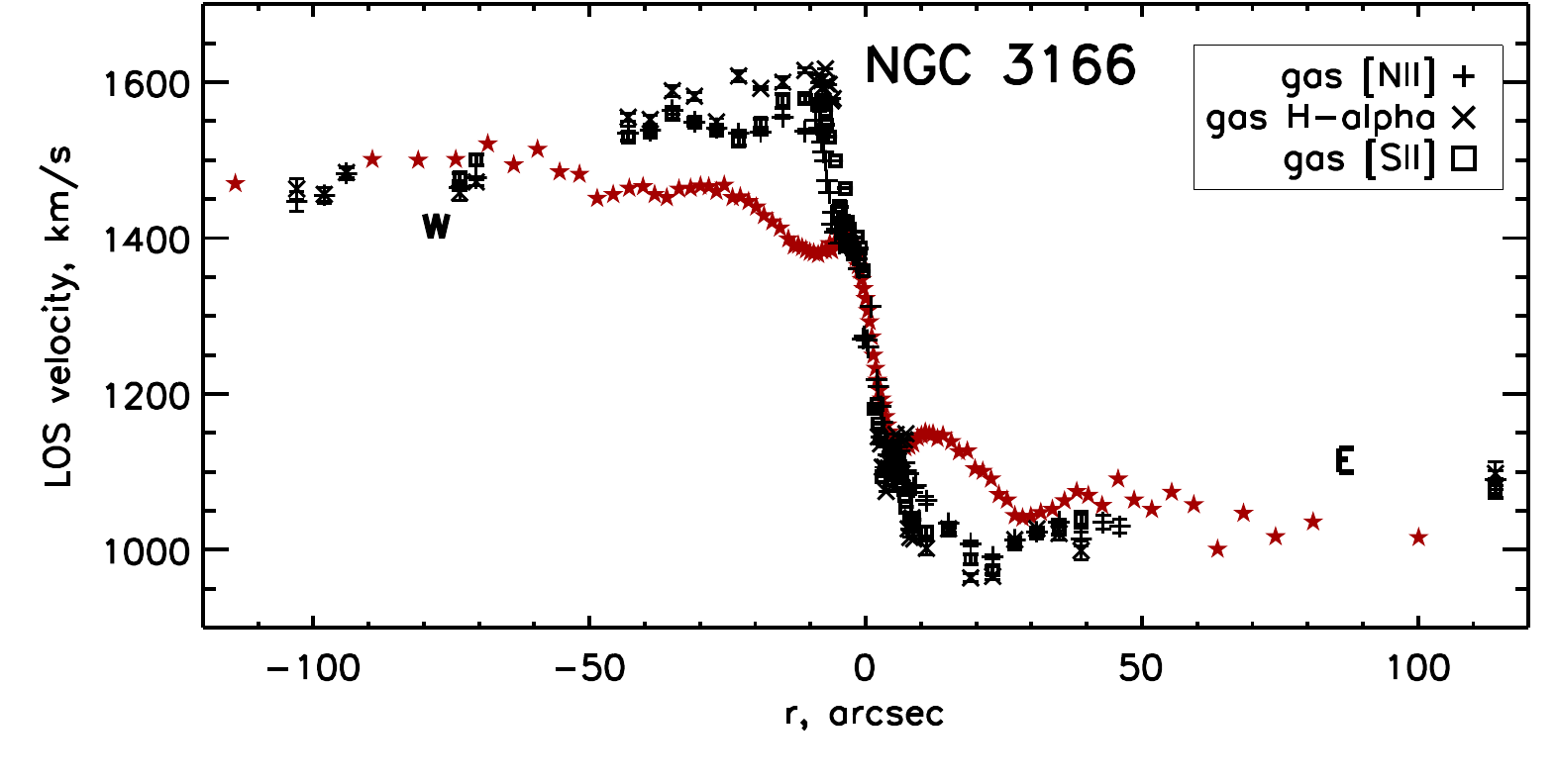} \\
\includegraphics[width=6cm]{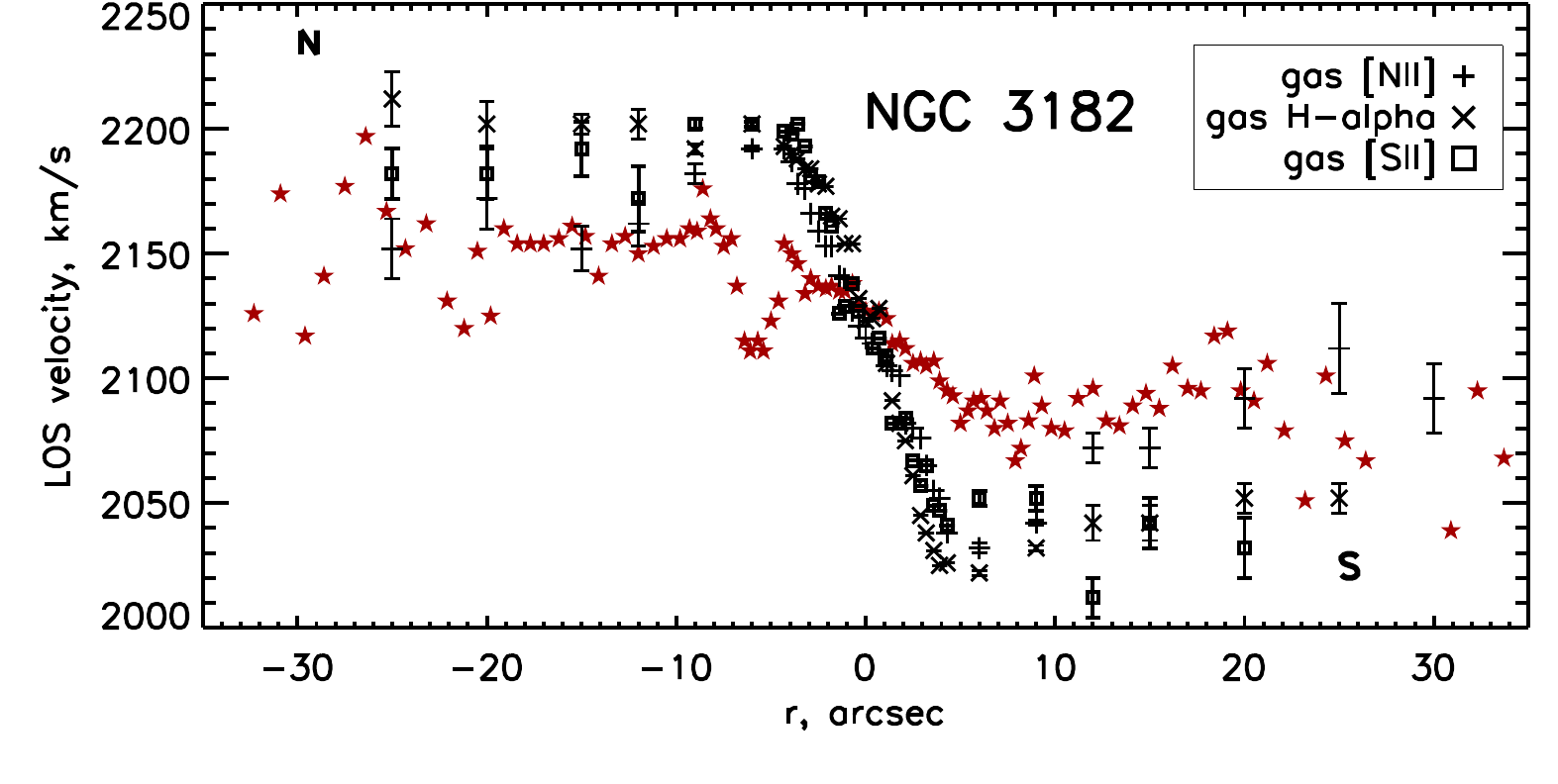} &
 \includegraphics[width=6cm]{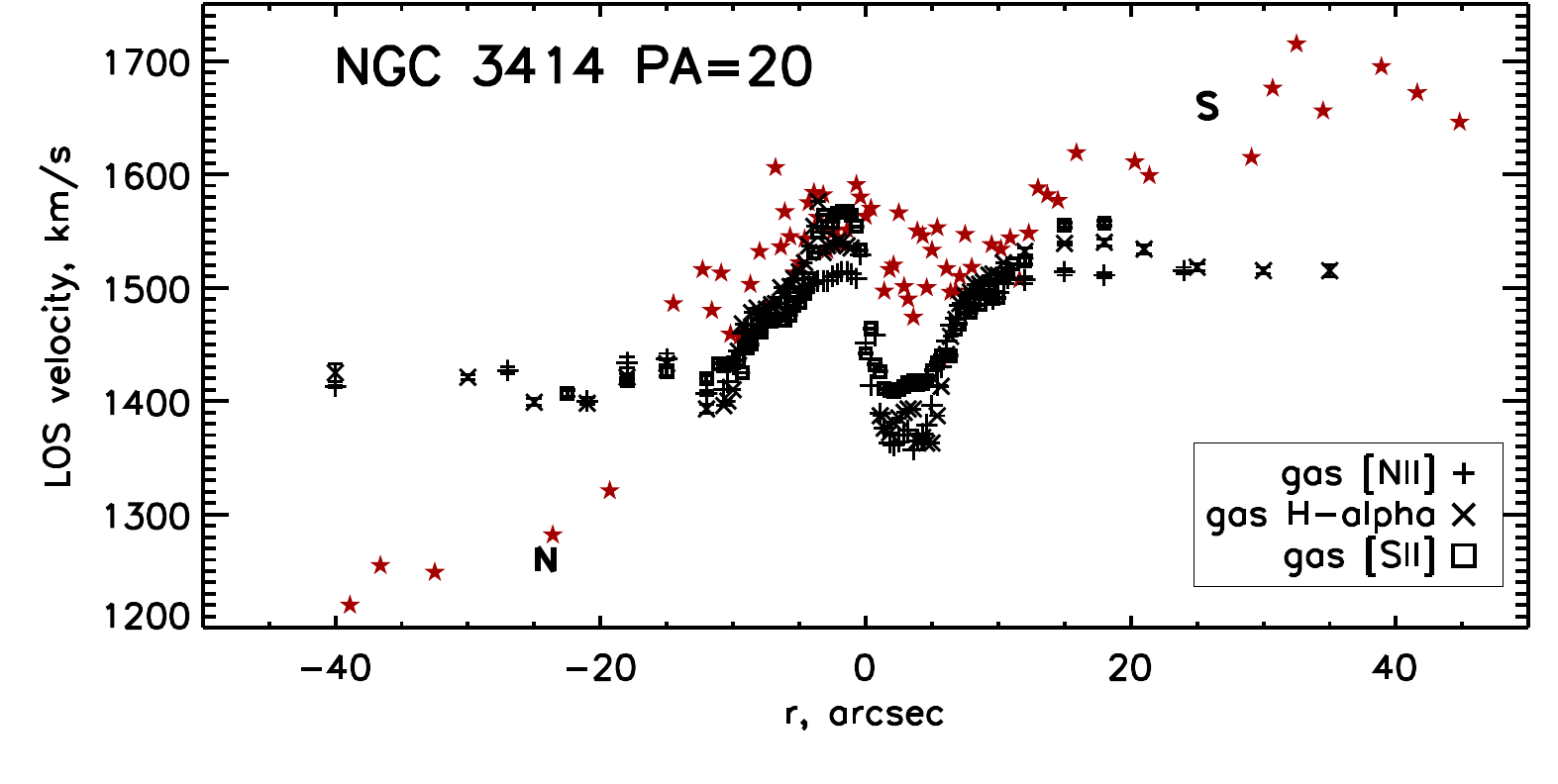} &
 \includegraphics[width=6cm]{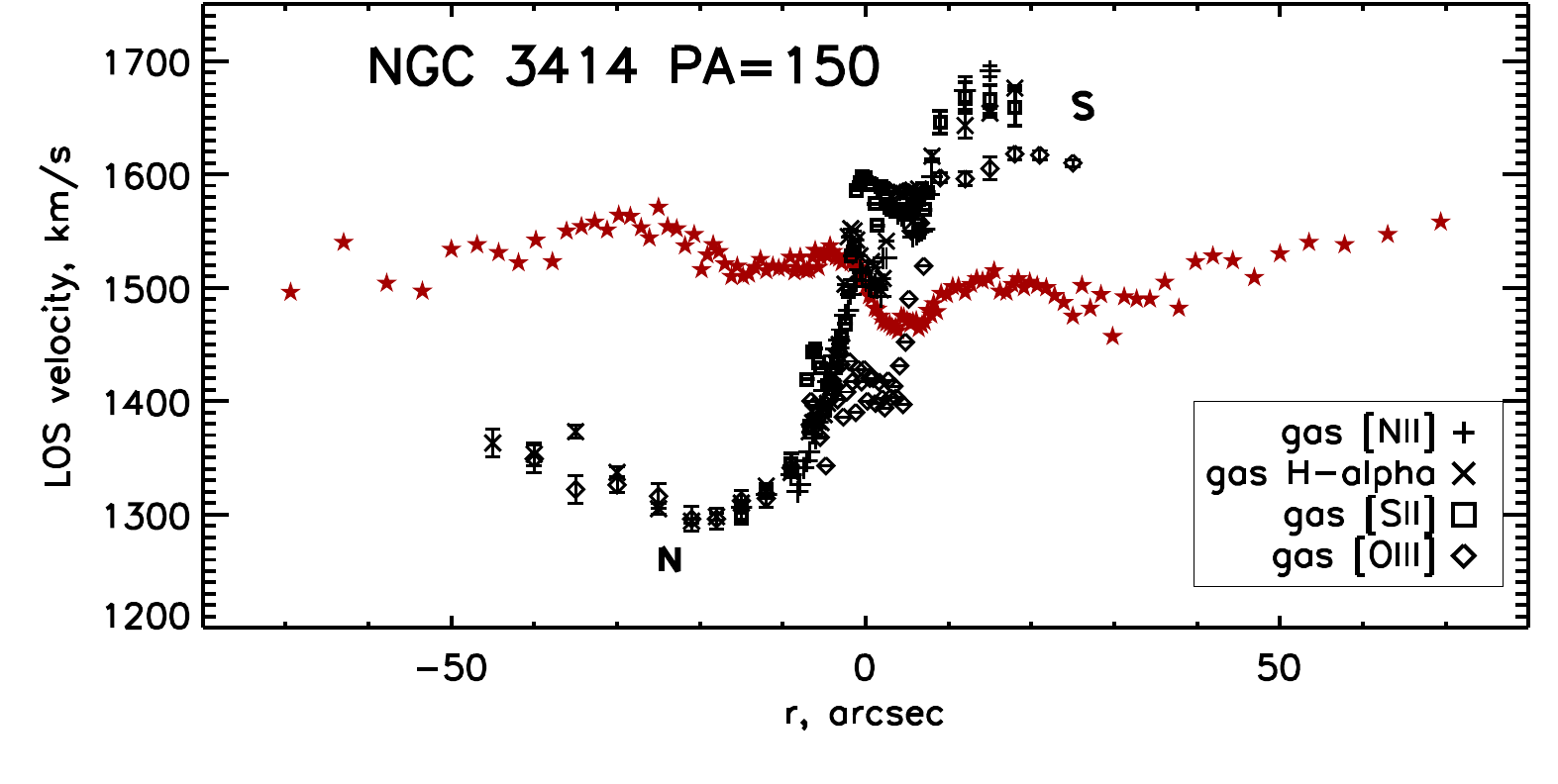} \\
\includegraphics[width=6cm]{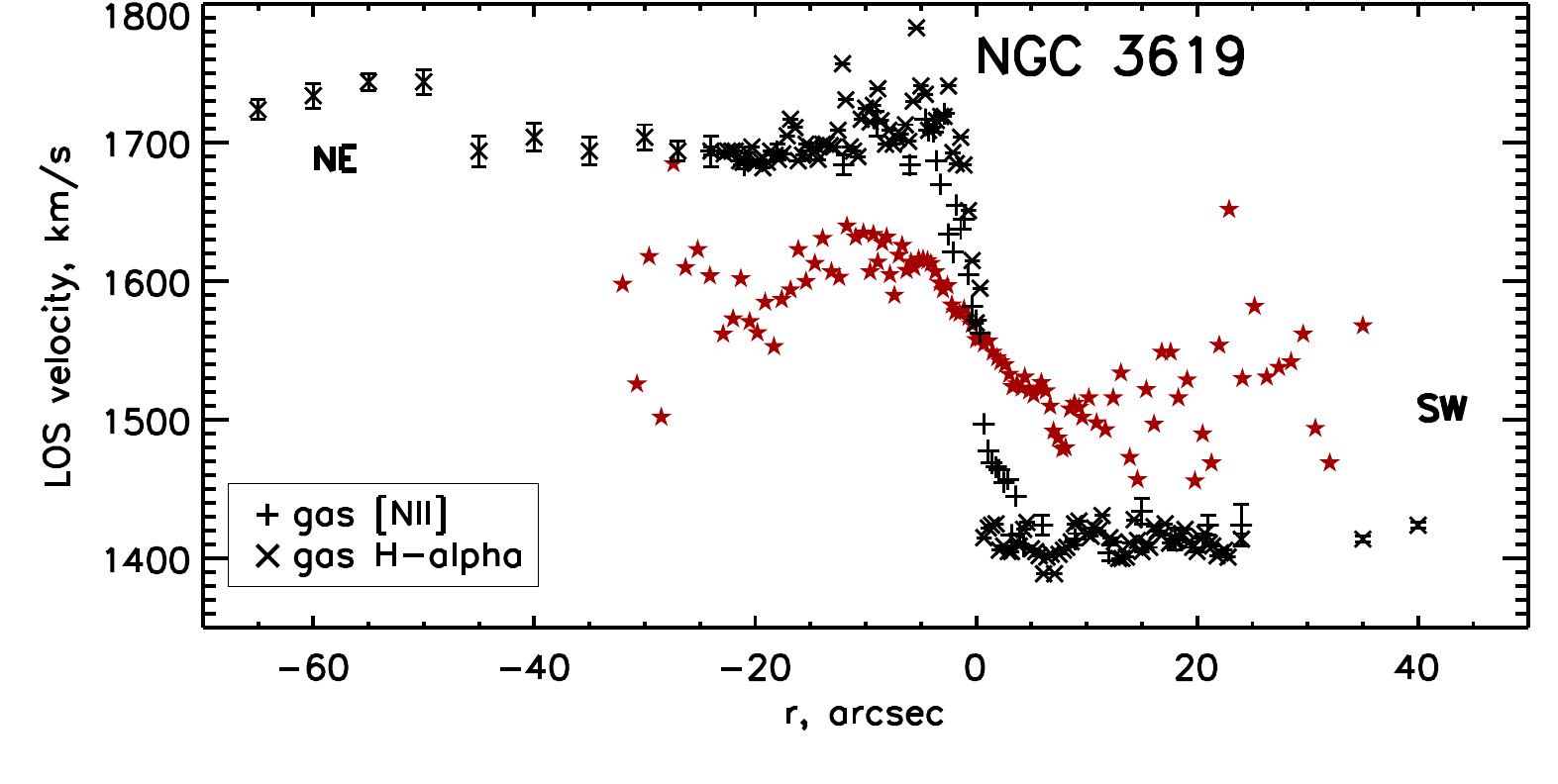} &
 \includegraphics[width=6cm]{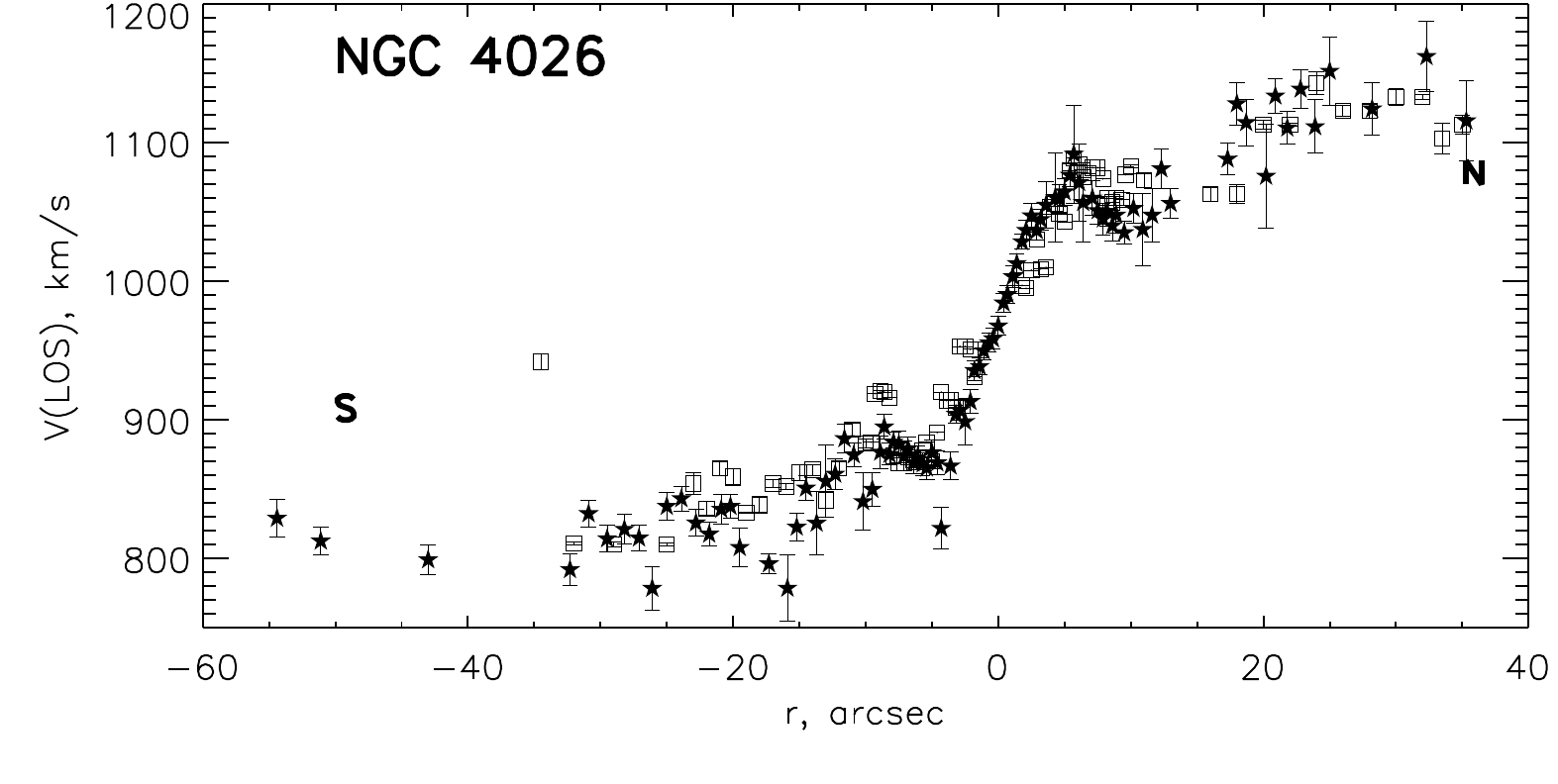} &
 \includegraphics[width=6cm]{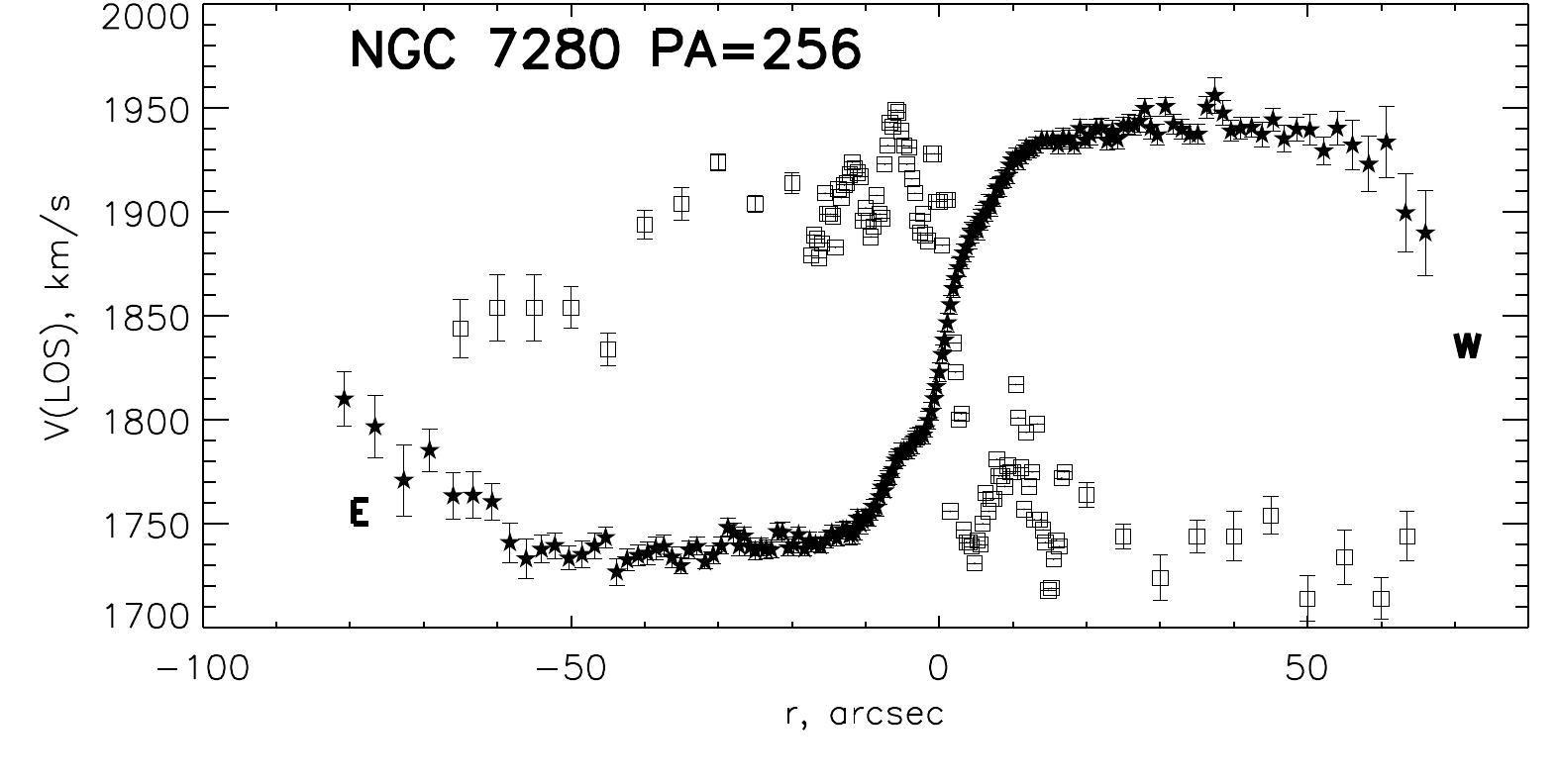} \\
\includegraphics[width=6cm]{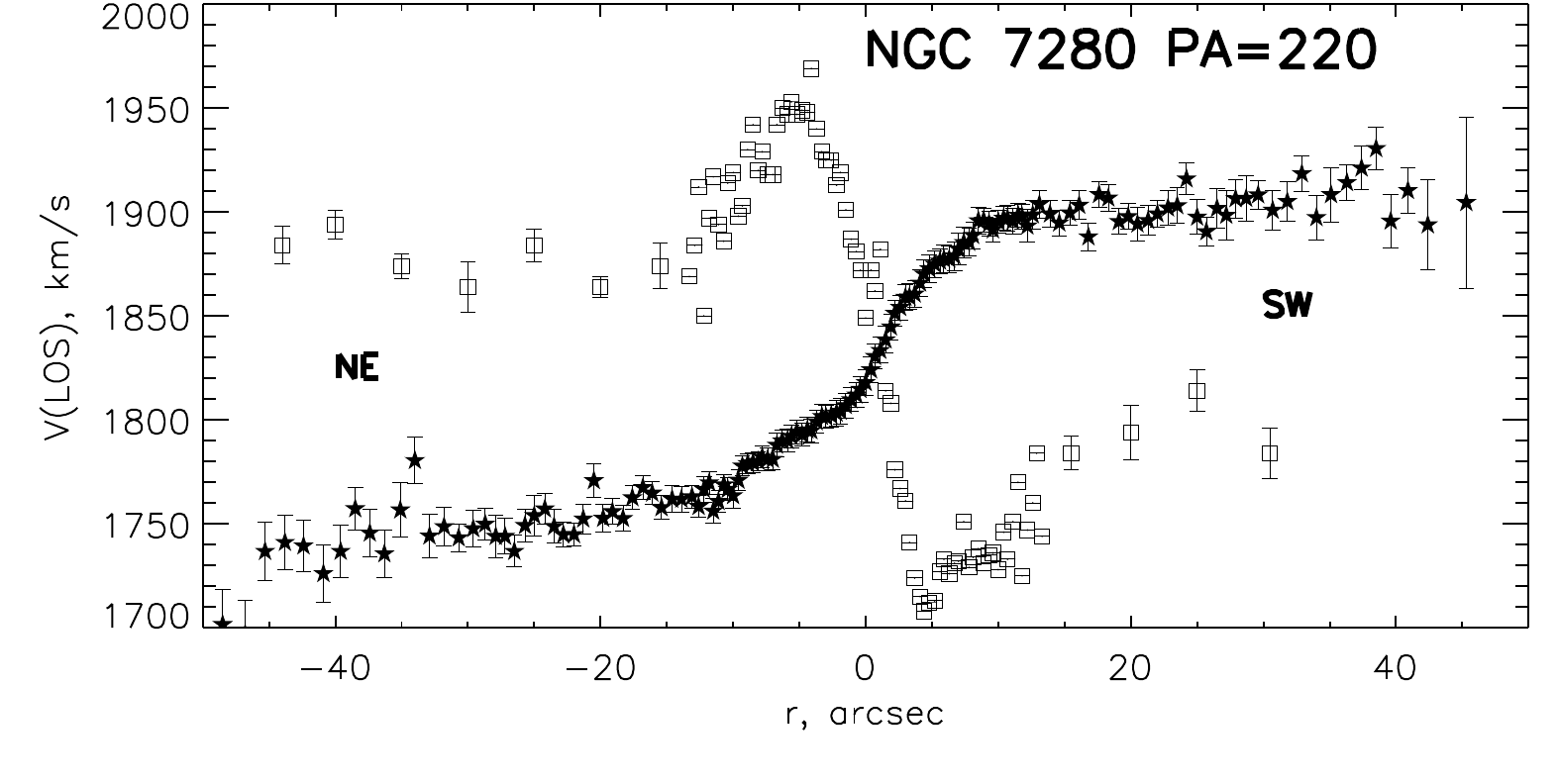} &
 \includegraphics[width=6cm]{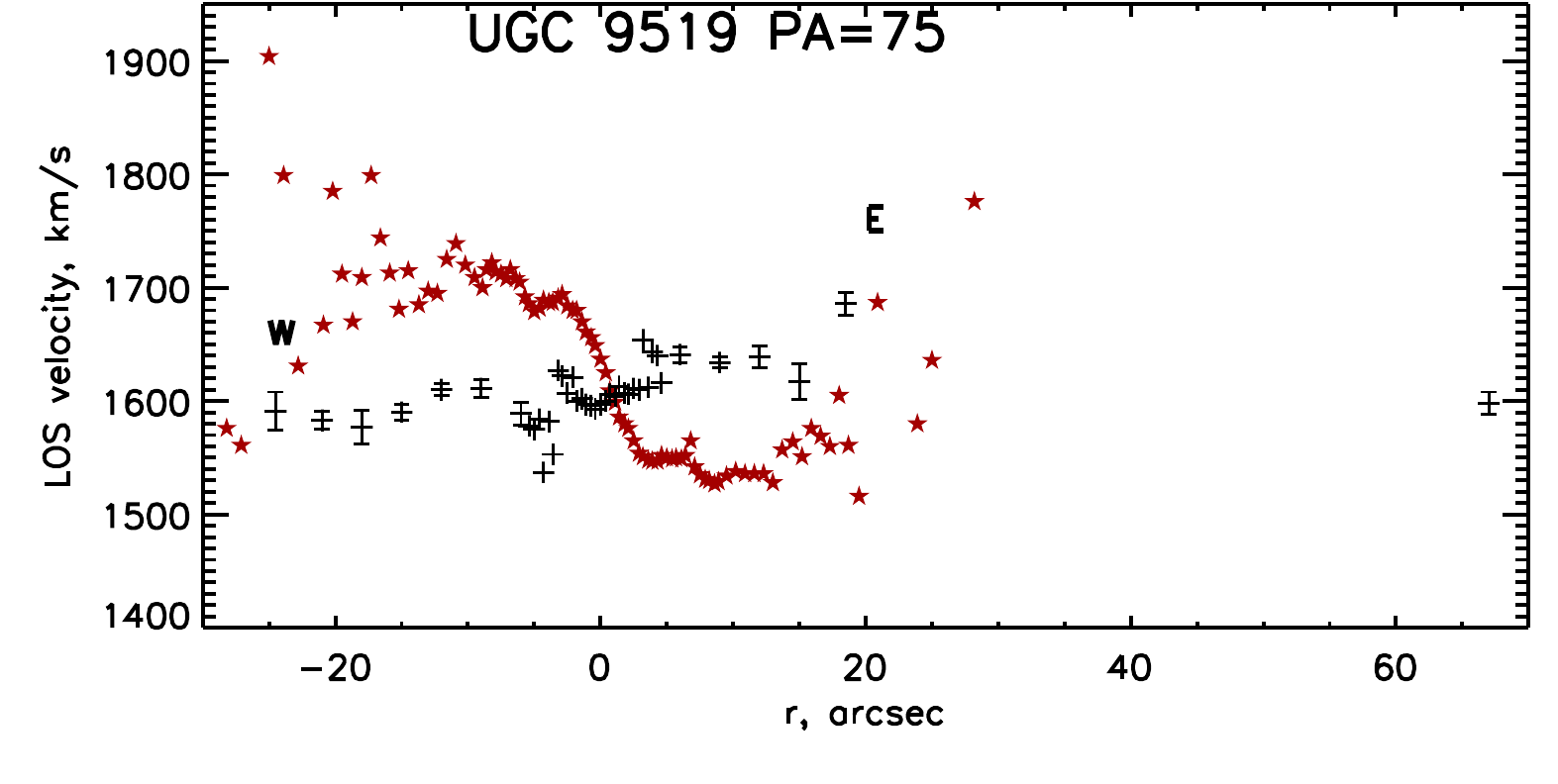} &
 \includegraphics[width=6cm]{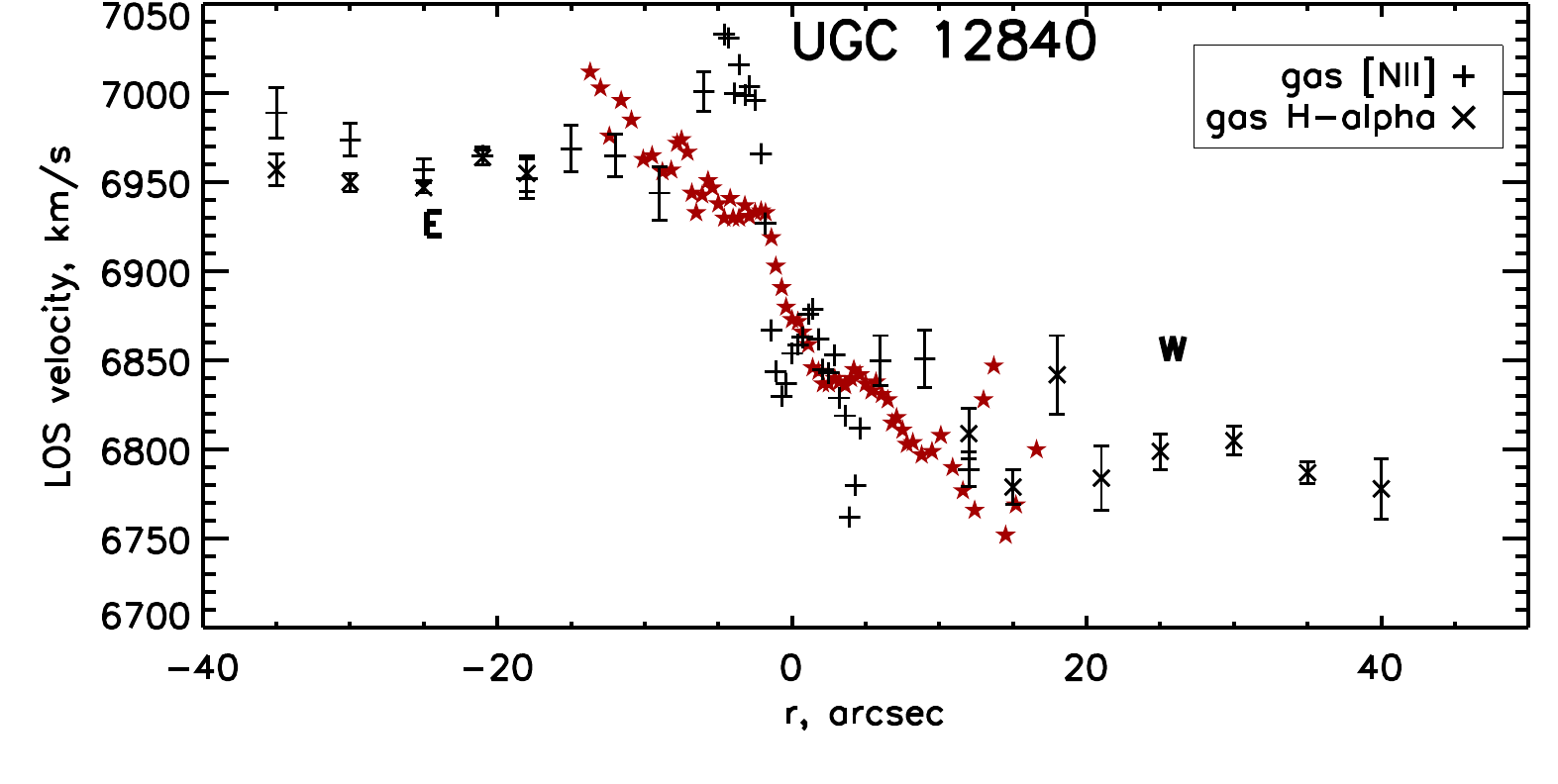} \\
\end{tabular}
\caption{Long-slit line-of-sight velocity profiles obtained with the SCORPIO/BTA;\\
1st row -- IC 5285 (major axis ), NGC 774 (major axis), NGC 2551 ($PA=13$\deg),\\
2nd row --  NGC 2655(major axis), NGC 2655 (minor axis), NGC 2787(major axis), \\
3rd row -- NGC 2787 ($PA=71$\deg ), NGC 2962 (major axis), NGC 3166 (major axis), \\
4th row --  NGC 3182 (major axis), NGC 3414 (major axis), NGC 3414 ($PA=150$\deg),\\
5th row --  NGC 3619 (major axis), NGC 4026 (major axis), NGC 7280 (major axis), \\
6th row -- NGC 7280 ($PA=220$\deg), UGC 9519 (major axis), UGC 12840 (major axis).\\
When only a single emission line is measured (mainly [OIII]$\lambda$5007), it is plotted by
squares, and the stellar component rotation -- by black stars; when several emission lines are independently
measured, they are plotted by various black signs, and the stellar component is indicated by red stars.
The abscissa zero corresponds to the brightest-continuum bin along the slit (to the galactic nucleus).
}
\label{fig_ls}
\end{figure*}

The Figs.~\ref{fig_fpi} and \ref{fig_ls} present the results respectively of our panoramic and long-slit spectroscopy with the 
6m telescope. The optical continuum centers are posed to the coordinate zero points, (0,0) at the both axes. The maps
are smoothed severely to reach the outermost parts of the gaseous disks -- to the typical angular resolution of 2\arcsec--3\arcsec,
so the circumnuclear regions are somewhat poorly resolved. However in some galaxies a certain turn of the kinematical major axis 
(defined as the direction of maximum line-of-sight velocity gradient) in the very center is seen: in NGC~3166,
NGC~3414, and NGC~7280. 

In the central regions of about half of the galaxies their spectral stellar-absorption features are rather deep forcing
the faint ionized-gas emission lines to sink. In those cases the observed gas kinematics can be strongly affected because the small 
spectral range of the FPI observations does not allow us to distinguish the contributions of the nebular and stellar components 
into the integrated spectrum. In  these cases we could not  detect emission lines with the expected velocities  in the
circumnuclear regions which  are blanked in the FPI maps presented in the Fig.~\ref{fig_fpi}. Usually the blanked regions
locate asymmetrically relative the galaxy nucleus (see IC~5285, NGC~774, etc.), because these regions correspond to
the maximal contribution of the bulge luminosity against the distant side of the disk. This situation is invert to the well-known
effect of finding more prominent dust lanes at the nearest side of a galaxy disk.
As concerning the [OIII] data, central parts of some emission-line velocity fields are
available from the surveys undertaken with the SAURON spectrograph \citep{sauron,atlas3d_1}. So for NGC~3619 and NGC~4026
we have succeeded to combine the [OIII] velocity fields derived from the SAURON data cubes and from our FPI data.
Namely, we replaced the central pixels in FPI maps with the SAURON fields if the velocity difference exceeds the limit of about
$50~\kms$. In the resulting fields the nuclear points (where stellar absorption features are significant) possesses to the SAURON data,
while the outer regions retain the FPI measurements with a better spectral resolution.
Figure~\ref{fig_sauron} illustrates this procedure for NGC~3619. The final `combined' velocity fields were 
involved further in our analysis described below. 

The long-slit cross-sections have been mostly obtained with the slit orientation along the photometric
major axis which coincides with the stellar-disk line-of-nodes orientation in the case of an intrinsically round stellar disk.
The round intrinsic shape of stellar galactic disks is statistically confirmed more than once, both from photometric
\citep{rix_zar,andersen_elldisk} and kinematical \citep{franx_dezeeuw,cappellari_rev} arguments. However in some our galaxies
where the kinematical major axes for the gaseous disks derived from the panoramic spectroscopy have been found to
deviate strongly from the stellar-disk lines of nodes (photometric major axes), we have observed two long-slit spectra with
different slit orientations, both being presented in Fig.~\ref{fig_ls}; those galaxies are NGC~2655, NGC~2787, NGC~3414, and
NGC~7280. Below we give some comments on the individual galaxies.

\noindent
{\bf IC~5285.} This luminous S0/a galaxy looks a double-ringed one in continuum, with the outer ring being 
completely detached from the main disk. Interestingly, being mapped in the emission line \Ha,
instead of the outer ring, it reveals a pair of gaseous spirals started from the tips of the inner ring (Fig.~\ref{fig_fpi}). 
Perhaps, the inner ring borders a triaxial bulge because the major axis of the inner ring is turned with respect
to the kinematical major axis; evident non-circular gas motions are seen in the center of the galaxy. In general,
the gaseous disk corotates the stellar component (Fig.~\ref{fig_ls}).

\noindent
{\bf NGC~252.} The galaxy was reported as a ringed one by \citet{iralist}. Later we noted a UV-ring with a radius
about $25^{\prime \prime}$ and reported gas excitation by young stars in the ring \citep{ringmnras}. Now in Fig.~\ref{fig_fpi}
we present an \Ha disk extended up to $\sim 40^{\prime \prime}$ (15 kpc) from the center and rotating with a very
high speed, about 350~\kms. The stars corotate \citep{ringmnras}; the lower visible stellar rotation velocity is perhaps
due to asymmetric drift -- the stellar disk of NGC~252 may be rather dynamically hot. In the very center,
at $R<7^{\prime \prime}$, low-excitation emission line [NII]$\lambda$6583 demonstrates lower rotation velocities than
the \Ha (Fig.~\ref{fig_ls}); it implies the possible presence of a triaxial potential though morphologically the galaxy is
classified as unbarred, SA0$+$.

\noindent
{\bf NGC~774.} It is another S0 galaxy with a UV-ring \citep{galex}; it must possess even current star
formation because recently a core-collapse supernova is detected within the ring (SN2006ee, $13\arcsec$ to SE from the center). 
The ionized gas is asymmetrically distributed over the central part of the galaxy, within $R<20^{\prime \prime}$, or $R<$6~\kpc\ 
(Fig.~\ref{fig_fpi});
beyond the borders of the gaseous disk the stellar rotation velocity curve falls visibly almost to zero (Fig.~\ref{fig_ls}). 

\noindent
{\bf NGC~2551.} In this galaxy the UV-ring is very bright and clumpy both in the FUV and in the NUV, and also is rather wide,
from $R\approx 15^{\prime \prime}$ to $R\approx 30^{\prime \prime}$ (2--5~\kpc) \citep{galex}. The extended 
ionized-gas disk which is coincident radially with the UV-ring (Fig.~\ref{fig_fpi}) counterrotates as a whole with respect to the stellar 
disk \citep{we2551}. Earlier we published the long-slit cross-section along the disk major axis \citep{we2551}; now we present 
a cross-section under $43^{\circ}$ to the line of nodes (Fig.~\ref{fig_ls}), confirming the gas counterrotation, and also the falling
character of the stellar rotation curve.

\begin{figure}
\centering
	\includegraphics[width=9cm]{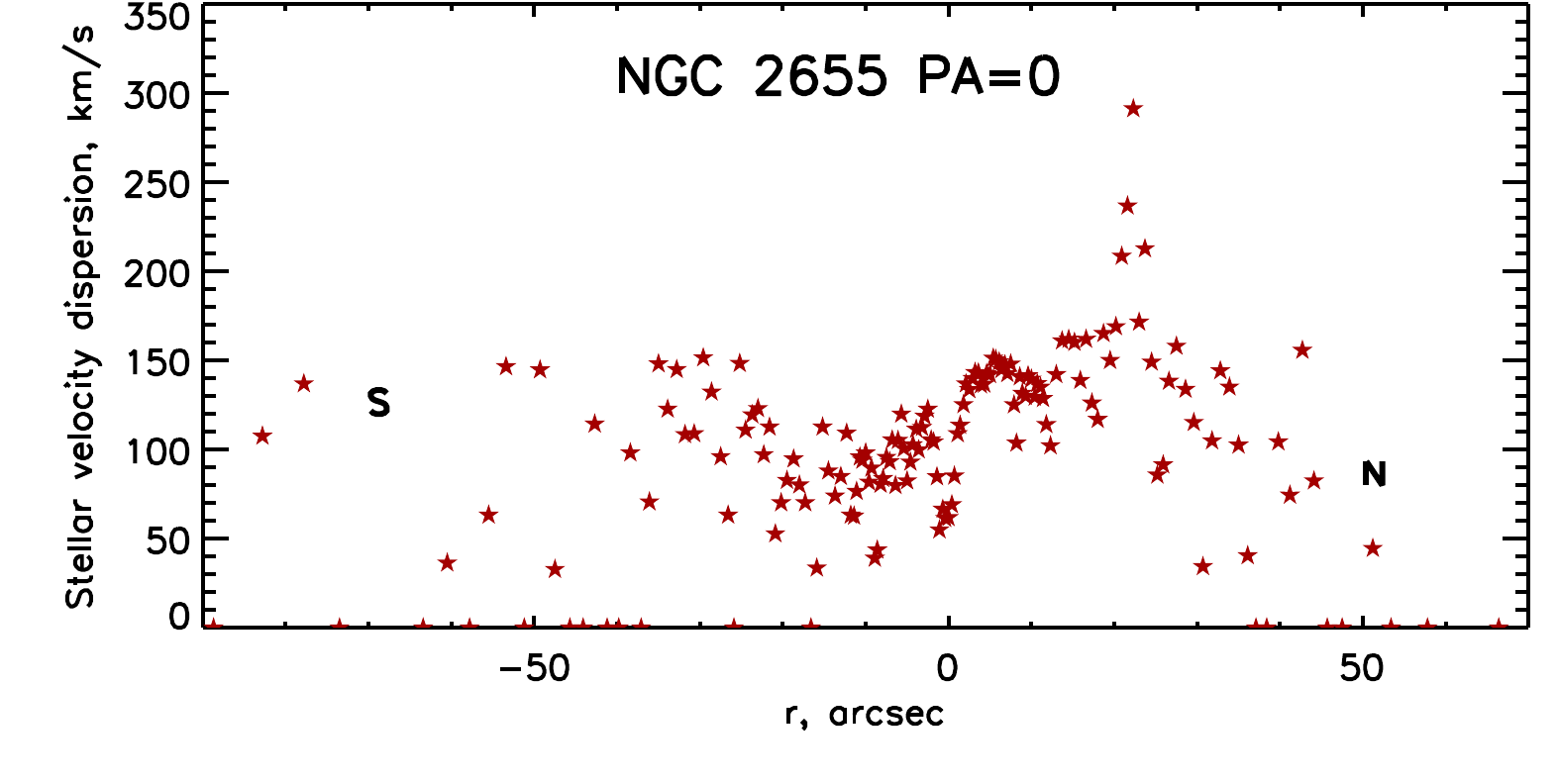}
	\caption{The radial profile of the stellar velocity dispersion along the minor axis of NGC~2655. An increased
        stellar velocity dispersion can be seen to the north of the nucleus.
	}
	\label{fig_N2655sig}
\end{figure}

\noindent
{\bf NGC~2655.} This galaxy is an object of our long-standing interest. Many years ago we started its spectral study
with three long-slit cross-sections \citep{sil2655}; then we noted that the central part of NGC~2655, $R< 5^{\prime \prime}$ (0.8~\kpc),
had quite decoupled gaseous kinematics. Our first interpretation was that it might be an effect of an active nucleus.
Later we had made panoramic spectroscopy with the IFU MPFS of the 6m telescope, and had changed our opinion: in the
center of NGC~2655 we had detected an inner gaseous disk with {\it polar} rotation \citep{polars0} that had then been supported
by a dust lane perpendicular to the isophote major axis \citep{polars0}. This interpretation was later confirmed by discovery of a strongly
warped extended \HI\ disk around NGC~2655 \citep{sparke}. However now, after inspecting the full LOS velocity field of the ionized-gas
component (Fig.~\ref{fig_fpi}) and two orthogonal long-slit cross-sections along the major and the minor axes of the 
isophotes (Fig.~\ref{fig_ls}), we are again forced to change our opinion. The general blue-red asymmetry 
of the gas LOS velocity field is seen in the east-west direction that is fully consistent with the orientation 
of the photometric major axis (stellar-disk line of nodes). The visible effect of the orthogonal turn of the kinematical 
major axis near the center is produced by a compact velocity anomaly just to the north and to the west from the nucleus. 
But when we inspect the long-slit cross-sections made under good seeing conditions (Fig.~\ref{fig_ls}),
we understand that any rotation cannot be the reason of this anomaly: the line-of-sight velocity of the gas drops by some 400~\kms
to the north of the nucleus without any mirrored behavior in the south direction. Interestingly, just in the same place the stellar
velocity dispersion rises enormously (Fig.~\ref{fig_N2655sig}) betraying the presence of a secondary stellar component with strongly
different velocity on our line of sight.  Evidently, at $R=8^{\prime \prime}$ to the north of the nucleus and along the circumnuclear
dust lane to the west of it we see the consequence of minor merging -- the remnant of a gas-rich dwarf satellite destroyed by gravitational
tides and close in projection to the center of NGC~2655.

\noindent
{\bf NGC~2697.} It is a rather low-luminosity S0 galaxy for which no SDSS data were found. We have made our own $gri$ photometry
which results are to be presented elsewhere \citep{saltrings}; in the same paper we analyze also the long-slit
kinematics. Here we note only a well-sampled ionized-gas velocity field (Fig.~\ref{fig_fpi}) demonstrating regular circular gas rotation.

\noindent
{\bf NGC~2787.} It is one more galaxy where we searched for inner polar disk because of the circumnuclear dust lane
perpendicular to the isophote major axis \citep{polars0}. Presently, by exploiting the scanning Fabry-Perot interferometer,
we succeeded to trace the gas rotation in the emission line [NII]$\lambda$6583 up to $R\approx 20^{\prime \prime}$, and we saw 
a turn of the kinematical major axis over this radius range (Fig.~\ref{fig_fpi}). However, in the emission line [OIII]$\lambda$5007, 
by comparing the rotation velocity projections for the stars and for the ionized gas exposed with the long slit in two different 
position angles (Fig.~\ref{fig_ls}), we see also decoupled gaseous kinematics at $R>20^{\prime \prime}$ quite certainly.

\noindent
{\bf NGC~2962.} It is a nearby intermediate-luminosity lenticular galaxy with a detached outer stellar ring at
$R\approx 65^{\prime \prime}-80^{\prime \prime}$ (8--10~\kpc). The galaxy was included into the survey ATLAS-3D \citep{atlas3d_1},
and so panoramic spectroscopy of its central part, $R<15^{\prime \prime}$, with the IFU SAURON had provided us with the
orientations of the stellar and the ionized-gas rotation planes in its inner part. We found that they are
mutually orthogonal \citep{polarsau}, so the galaxy possessed the inner polar disk. However, the major axis of the outer ring
is aligned with the stellar kinematical major axis, and indeed, our scanning Fabry-Perot data (Fig.~\ref{fig_fpi}) confirms that the
outer gas rotates in the main galactic plane. Our long-slit cross-section along the galaxy major axis (Fig.~\ref{fig_ls}) reveals
both effects -- that the rotation axis of the gas within $R=15^{\prime \prime}$ is orthogonal to the rotation axis of the
stellar component and that at $R\approx 50^{\prime \prime}-65^{\prime \prime}$ the gas rotation comes into consistency with
that of the stellar disk. The outer ring of NGC~2962 is well seen in the ultraviolet bands FUV and NUV of the GALEX imaging 
\citep[][ see also our Fig.~\ref{galex}]{marino}.

\noindent
{\bf NGC~3106.} This galaxy, including its gas velocity field, was studied in the frame of the CALIFA survey \citep{califa}.
They have found an `outer' starforming ring in this lenticular galaxy with a radius of some $20^{\prime \prime}$ (8~\kpc) \citep{gomes16}. 
However, our data with a much larger field of view reveals a series of starforming rings in NGC~3106 (Fig.~\ref{fig_fpi}), among which the ring
reported by the CALIFA is the innermost one. The galaxy is close to face-on, nevertheless the rotation is visible in Fig.~\ref{fig_fpi}, and
the orientation of the kinematical major axis is measured quite certainly with our data. 

\noindent
{\bf NGC~3166.} This S0/a galaxy belongs to a tight triplet of disk galaxies embedded into a common \HI\ envelope \citep{haynes81}. 
Many signatures give evidences for the current cold gas accretion onto NGC~3166 from this huge intergalactic gaseous cloud: young age,
$\sim 2$~Gyr, of the nuclear stellar population \citep{sil3166}, large extended molecular-gas reservoir within the galaxy disk 
\citep{nro94}, strongly lopsided structure of the large-scale stellar disk \citep{bournaud05}, multiple twisted dust filaments within 
the central bulge-dominated part of the galaxy \citep{hst3166}. The map in \Ha emission line (Fig.~\ref{fig_fpi}) shows also asymmetric
distribution of the ionized gas, with a prominent \Ha arm at $100^{\prime \prime}$ (11~\kpc) to the west from the nucleus.
The gas isovelocities turn in the very center of the galaxy into a polar orientation. In the outer parts the gas starts to strictly 
corotate the stars only at $R>60^{\prime \prime}$ (Fig.~\ref{fig_ls}), where the large-scale stellar disk begins to dominate 
in the total surface brightness \citep{s0disks}.

\noindent
{\bf NGC~3182.} Though the galaxy surely lacks a bar, it demonstrates a circumnuclear blue (starforming?) ring with a radius of 
$5.5^{\prime \prime}$ (0.8~\kpc). Beyond the ring, different emission lines show different line-of-sight velocity profiles,
the low-ionization [NII]$\lambda$6583 being the less rapid, and as a whole the difference between the projected rotation velocities
of the gas and stars is unexpectedly large for such a low-mass galaxy. We can suspect that this S0 galaxy may possess thick
disk of diffuse ionized gas. Such thick gaseous disks are usually lag the fast equatorial-plane rotation of galactic stellar disks
and demonstrate enhanced low-ionization emission lines [NII] and [SII].

\noindent
{\bf NGC~3414.} A real structure of this galaxy remains a complete mystery up to now. It looks like a round homogeneous body (a spheroid?
a face-on disk?) with an overlaid thin filament-like structure in the orientation of $PA\approx 20$\deg. There were various interpretations
of such a view: it might be a face-on S0 with a very strong bar \citep{bag,chitrejog}, but there were also suggestions that it was a face-on
S0 with a polar ring projected onto the center \citep{whitmore}. In any case, we were not surprised that the gaseous kinematics in NGC~3414 is
completely decoupled from the stellar one (Fig.~\ref{fig_ls}). However, in the very center the gas kinematical major axis turns to stay 
along the thin bar-like structure -- we have obtained the same result from our data with the scanning Fabry-Perot 
in the [OIII]$\lambda$5007 line (Fig.~\ref{fig_fpi}) and from the archive data obtained earlier for the central part of the galaxy 
with the IFU SAURON \citep{sauron_5}.

\noindent
{\bf NGC~3619.} It is another peculiar lenticular galaxy, settling in a rich group and demonstrating a system of outer shells
\citep{ripples} -- possible results of close interaction or even merging. It is rather gas-rich, and the distribution of the neutral
hydrogen is unusually compact for S0s \citep{n3619hi}. In the optical bands, spiral-like dust lanes are projected against the stellar
body to the south from the nucleus implying a circumnuclear gaseous disk settled asymmetrically with respect to the stellar one.
A UV-bright inner ring has been detected in the GALEX survey \citep[][ see also our Fig.~\ref{galex}]{cortese09}. The distribution
of the \Ha emission is also very asymmetric and extended (Fig.~\ref{fig_fpi}); however the ionized gas demonstrates regular rotation with
a possible small kinematical major-axis turn in the very center. The rotation velocities of the gas and star components reveal very different
projections over the full extension of our measurements (Fig.~\ref{fig_ls}), so we may suppose different spatial orientations
of the stellar and gaseous disks in NGC~3619.

\noindent
{\bf NGC~4026.} This edge-on S0 galaxy belongs to a very rich galaxy group, almost cluster, Ursa Major \citep{odenwald}. 
The Ursa Major cluster lacks meantime any hot intergalactic medium so neutral-hydrogen streams survive easily within its volume. 
NGC~4026 has been studied in the frame of the ATLAS-3D survey \citep{atlas3d_1}, and a substantial amount of neutral hydrogen 
has been detected not only within its disk, but also as a narrow external filament connecting just to the galaxy 
center \citep{atlas3d_13}. So we expected decoupled gaseous rotation in this galaxy. Indeed, in the very center 
the [OIII]$\lambda$5007 demonstrates rapidly rotating gaseous disk strongly inclined to the stellar disk --
it is seen in our Fig.~\ref{fig_fpi}, as well as in the IFU SAURON emission-line data \citep{polarsau}. However, 
beyond the central region, the ionized gas is strictly confined to the stellar disk, and its rotation
follows the stellar component one exactly (Fig.~\ref{fig_ls}).

\noindent
{\bf NGC~4324.} This lenticular galaxy inhabits outskirts of the Virgo cluster, and it is very gas rich (see the Table 1). It is also
a part of the ATLAS-3D survey \citep{atlas3d_1} and demonstrates a blue starforming ring with a radius of some 2~\kpc\ which also contains
all the molecular gas detected so far \citep{atlas3d_18}. However, \citet{dup_sch} detected \HI\ up to 2 optical radii in this galaxy that
gave a hope that ionized gas also may be extended beyond the inner starforming ring. Indeed, in Fig.~\ref{fig_fpi} we see clumpy outer starforming
regions up to $R\approx 70^{\prime \prime}-90^{\prime \prime}$ (7.5~\kpc) from the center; most of them take part in the general regular
rotation of the gaseous disk. However, a few -- for example that to the north-east from the galaxy -- do not participate in the rotation
of NGC~4324 and may be gas-rich satellites. 

\noindent
{\bf NGC~7280.} It is one more lenticular galaxy which attracts our attention for many years. Once we had found an inner polar gaseous ring in
the center of NGC~7280 \citep{sil7280}.  Later, in the frame of the ATLAS-3D survey, it has been confirmed that the neutral hydrogen
distribution in this galaxy is indeed aligned with the polar axis \citep{atlas3d_13}. This galaxy was treated as a gas-rich S0 from
the beginning \citep*{chamaraux87}, and star formation was carefully searched for in its disk \citep{pogge_esk93}. But there was no current
star formation found in the disk of NGC~7280. Instead, we have discovered {\it counterrotation} of the ionized gas with respect to the stars
beyond the central region, at $R>10^{\prime \prime}$ \citep{countersau}. Now, with a full map of the ionized-gas velocities (Fig.~\ref{fig_fpi}),
we confirm the central compact polar disk; the outer gas is distributed asymmetrically and counterrotates with respect 
to the stars (Fig.~\ref{fig_ls}).

\noindent
{\bf UGC~9519.} It is a dwarf lenticular galaxy elongated in $PA=75$\deg. However in the frame of the ATLAS-3D survey 
an outer \HI\ ring was found by \citet{atlas3d_13}, well beyond the optical borders of the galaxy, that is elongated 
in the north-south direction. The long-slit cross-section along the isophote major axis (Fig.~\ref{fig_ls}) has confirmed 
the decoupled kinematics of the gas and its probable external origin. To obtain a full picture, we have observed UGC~9519 
with the scanning Fabry-Perot interferometer in two emission lines, \Ha and [NII]$\lambda$6583. Our anticipation has 
justified: the inner part of the galaxy, $R<15^{\prime \prime}$, is well mapped just in the emission line [NII]$\lambda$6583
while in the outer part we resolve the outer starforming ring only in the \Ha emission line; Fig.~\ref{fig_fpi} represents
a combination of the [NII] data for the radii less than 20\arcsec\ and of the \Ha data for the more outer galactic regions. The inner gas
rotates in the nearly polar plane with respect to the stellar disk; the outer ionized gas being related surely to the 
neutral hydrogen detected by \citet{atlas3d_13} is quite decoupled from the main body of the galaxy. 

\noindent
{\bf UGC~12840.} This galaxy is a giant, isolated \citep{kig}, gas-rich lenticular one. Just in {\it this} galaxy \citet{pogge_esk93} 
had found extended star formation in the outer ring. We also detect a giant ring, with the radius of some 12.5~\kpc, emitting 
in the \Ha emission line (Fig.~\ref{fig_fpi}).
Though the galaxy looks almost round, $b/a>0.8$, the projection of the rotation velocity is large, about of $\sim 100$~\kms -- 
and the ionized gas corotates strictly the stellar component. Some non-circular gas motions are possible in the very center of the galaxy.

\begin{figure*}
\centering
\begin{tabular}{c c c}
 \includegraphics[width=5cm]{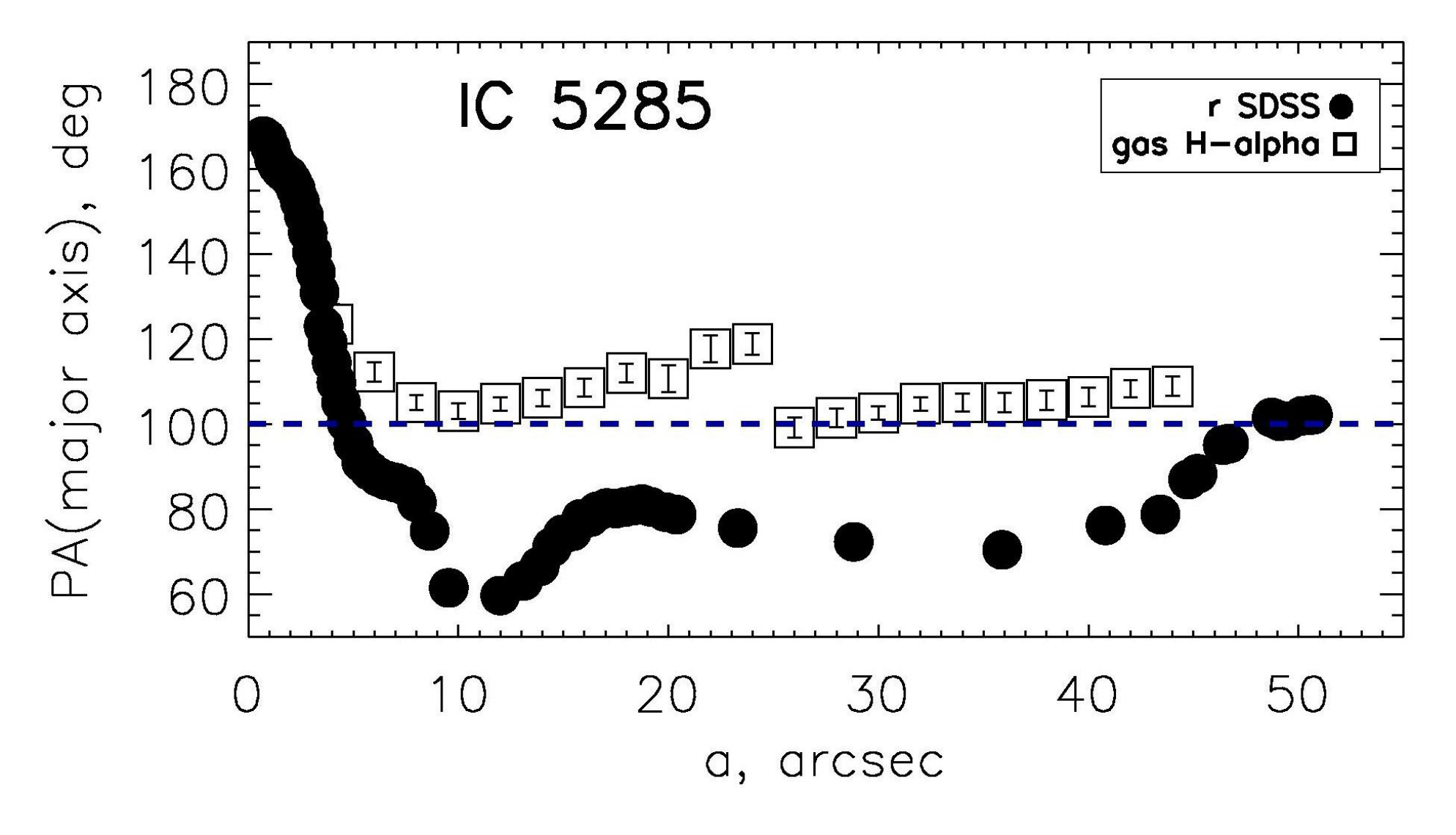} &
 \includegraphics[width=5cm]{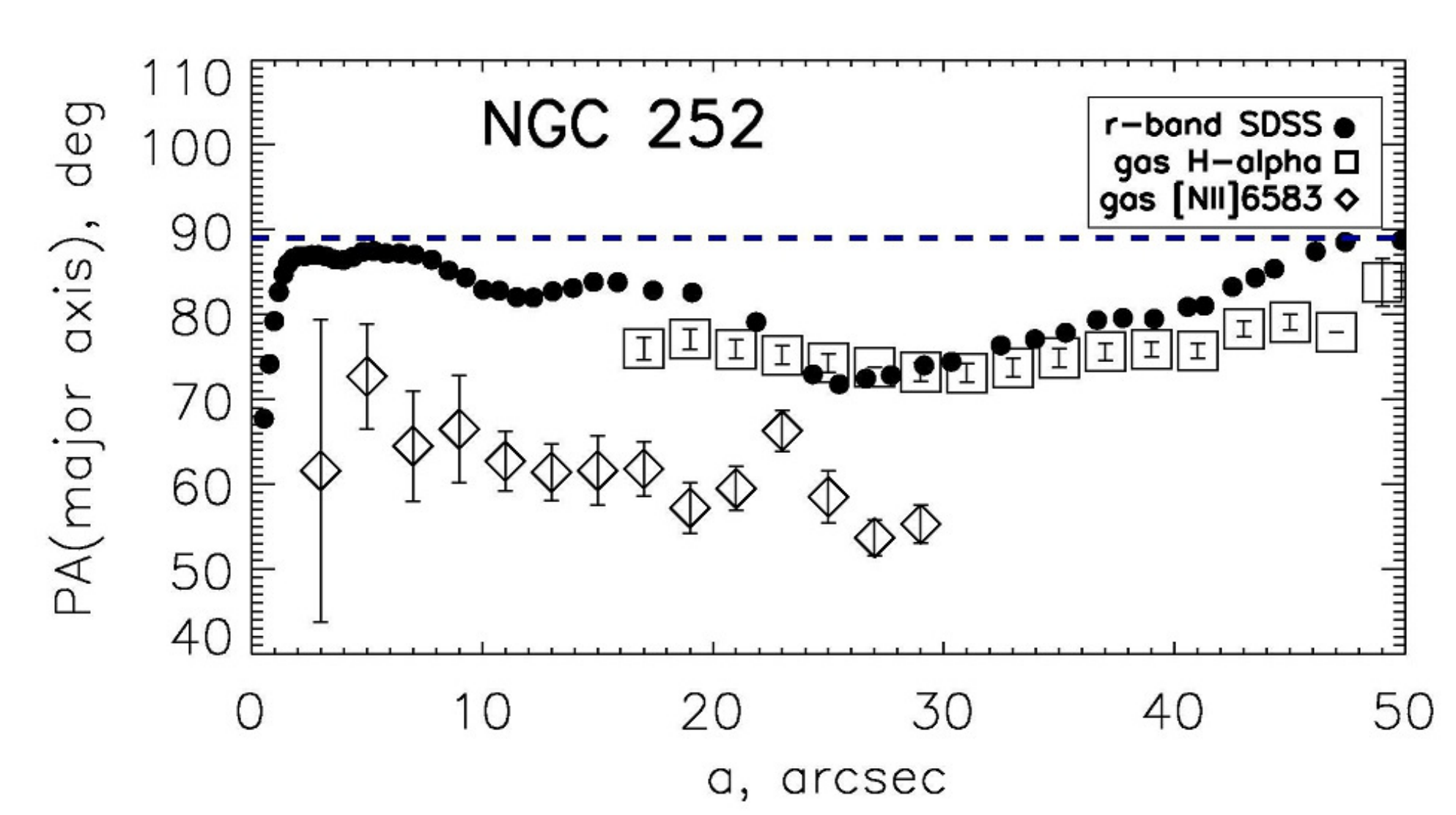} &
 \includegraphics[width=5cm]{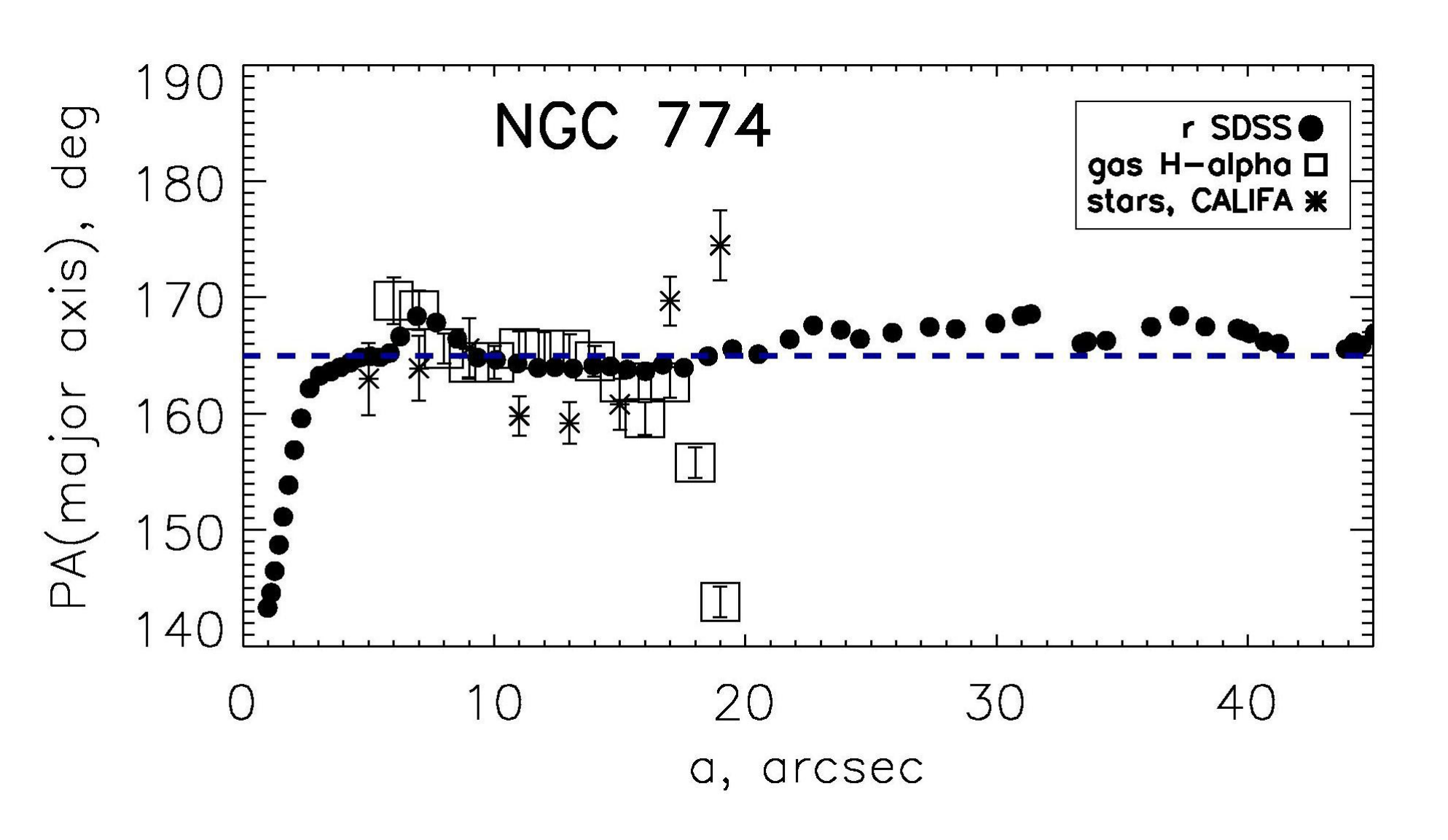} \\
\includegraphics[width=5cm]{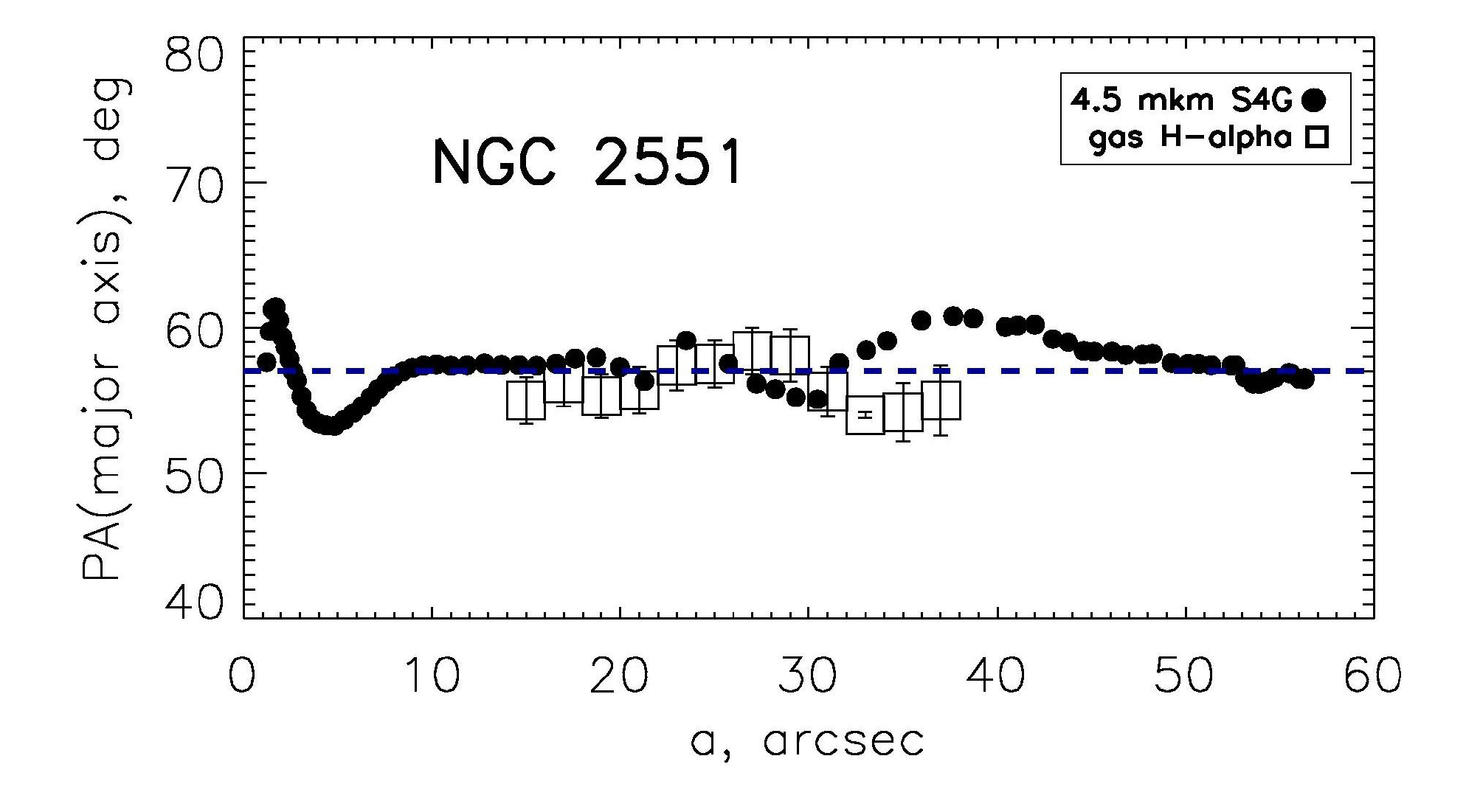} &
 \includegraphics[width=5cm]{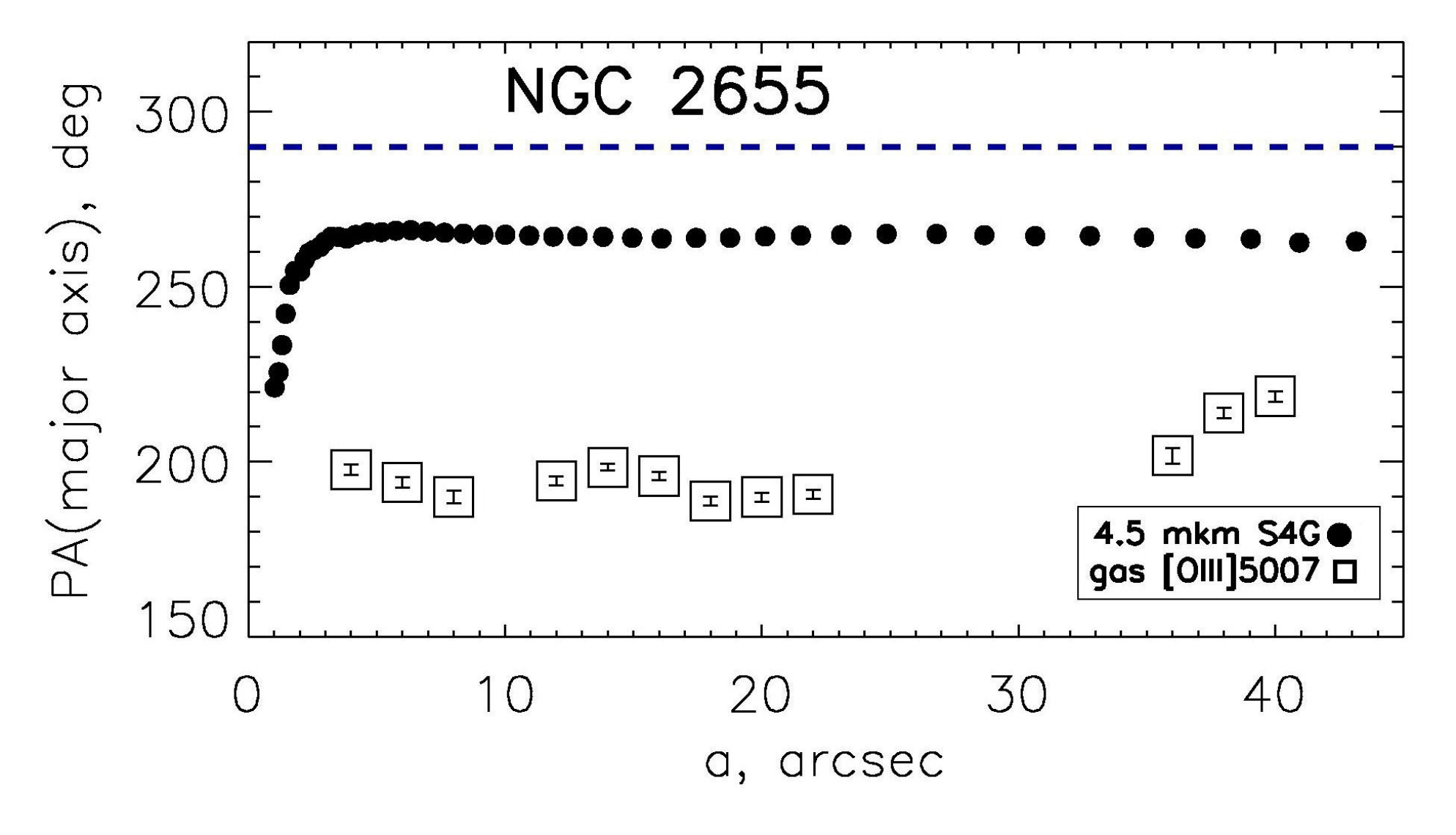} &
 \includegraphics[width=5cm]{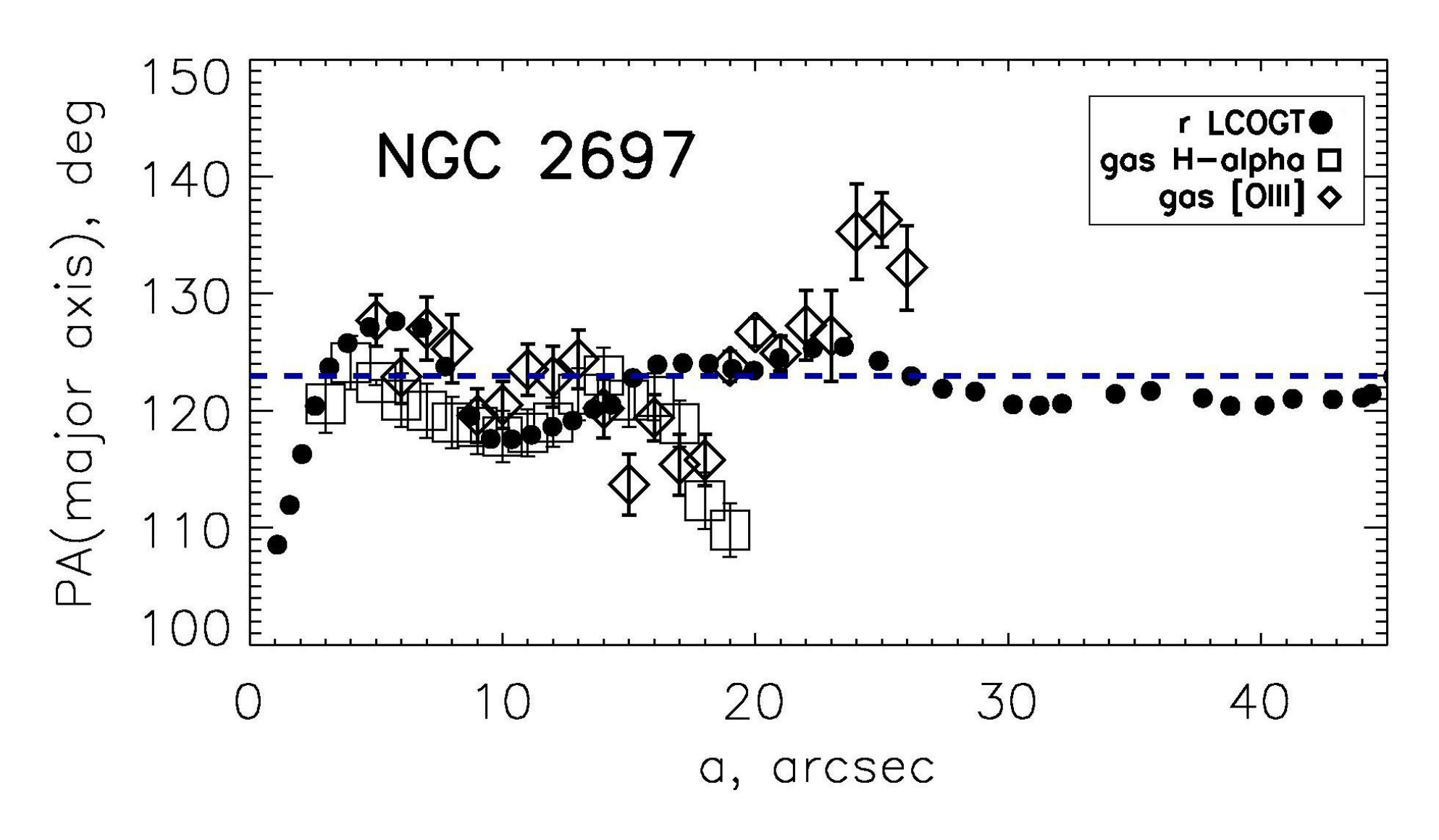} \\
\includegraphics[width=5cm]{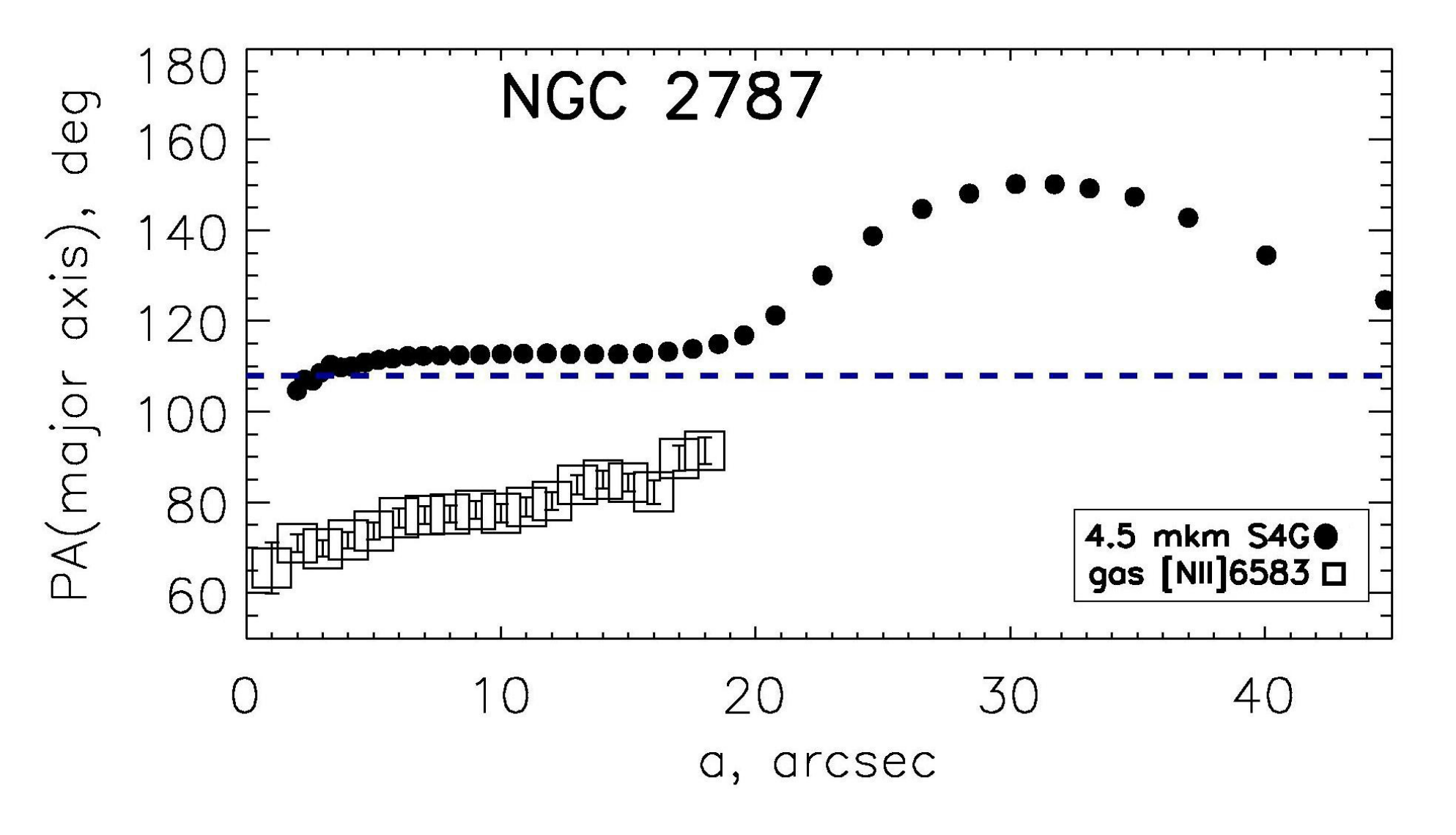} &
 \includegraphics[width=5cm]{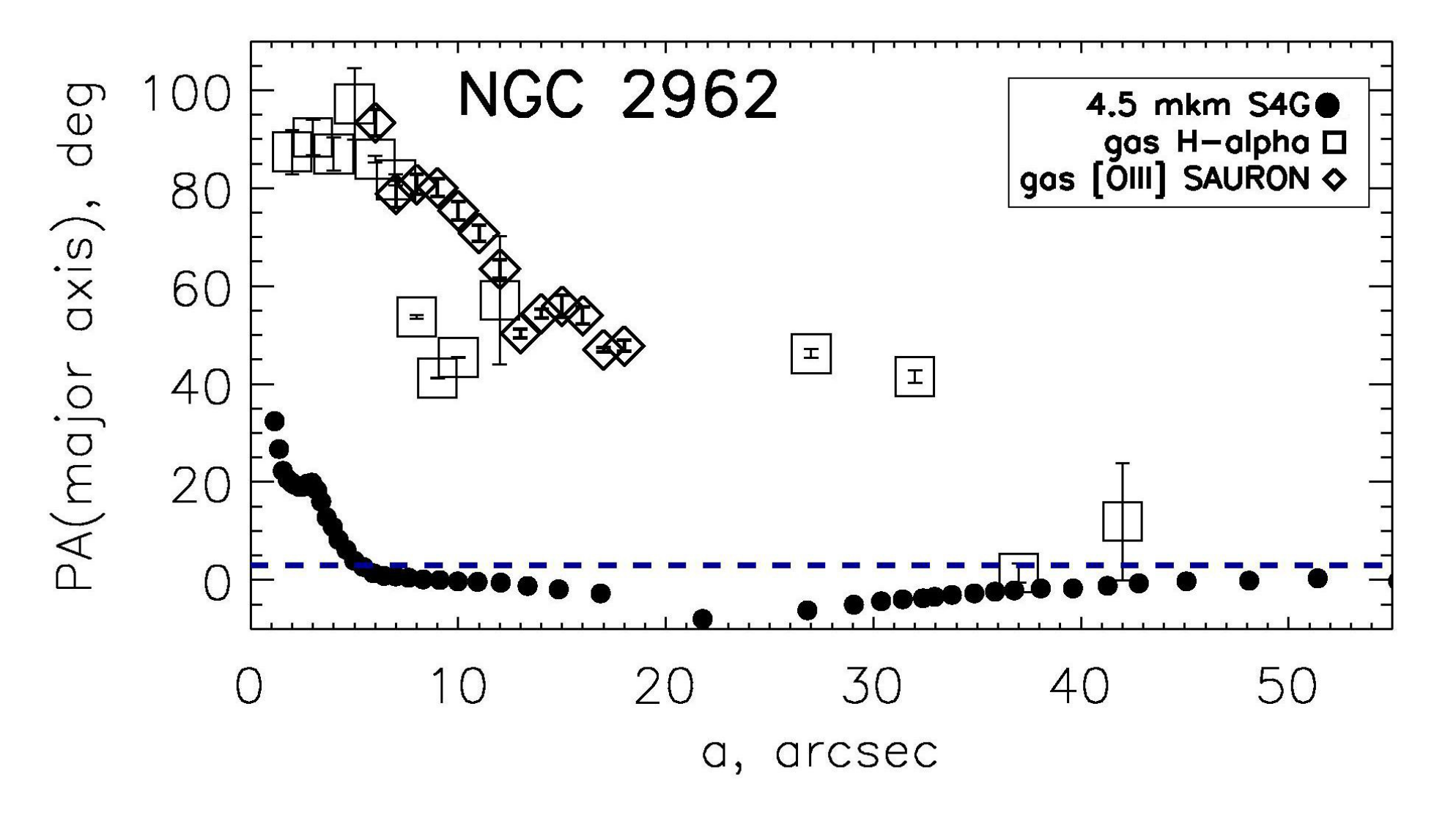} &
 \includegraphics[width=5cm]{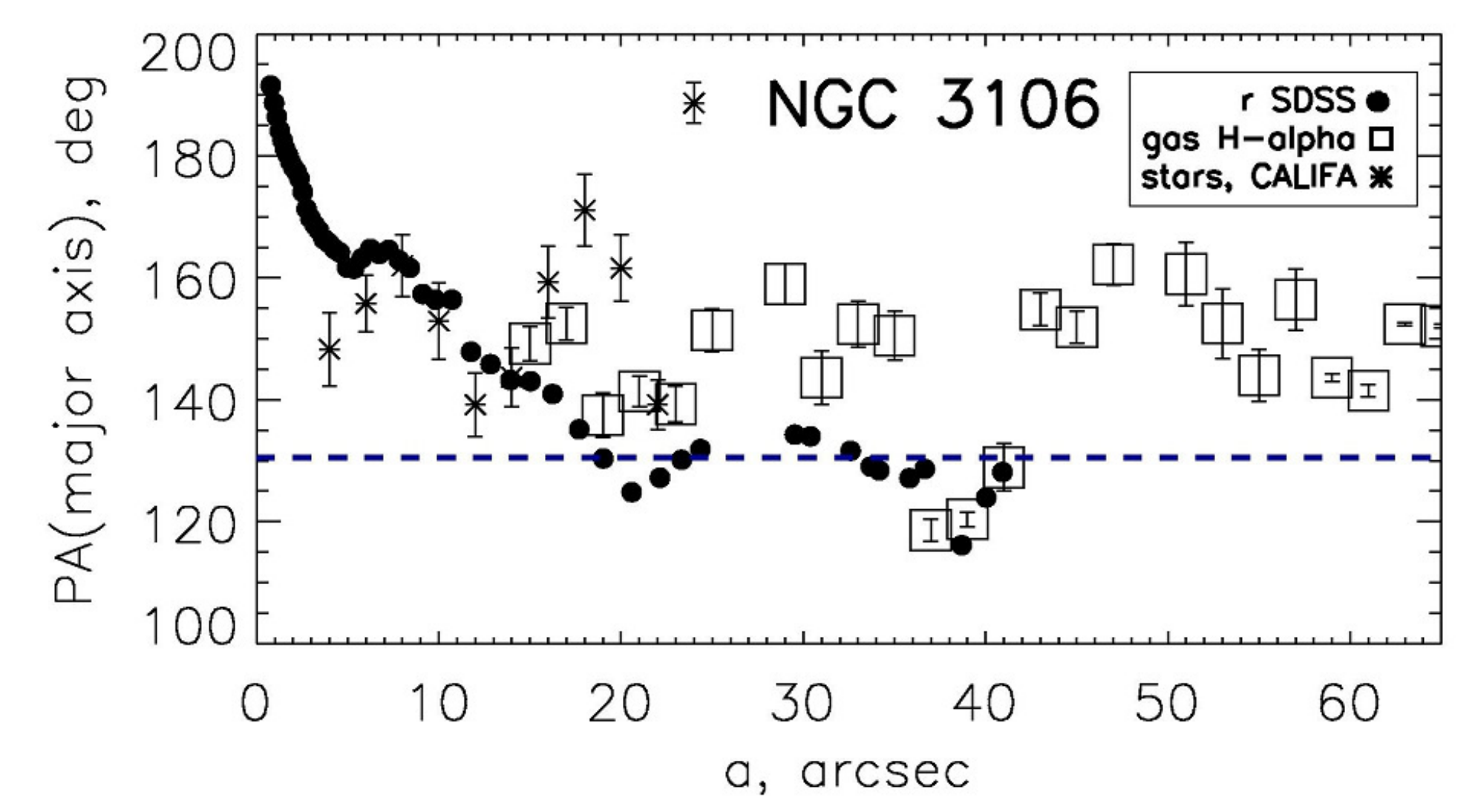} \\
\includegraphics[width=5cm]{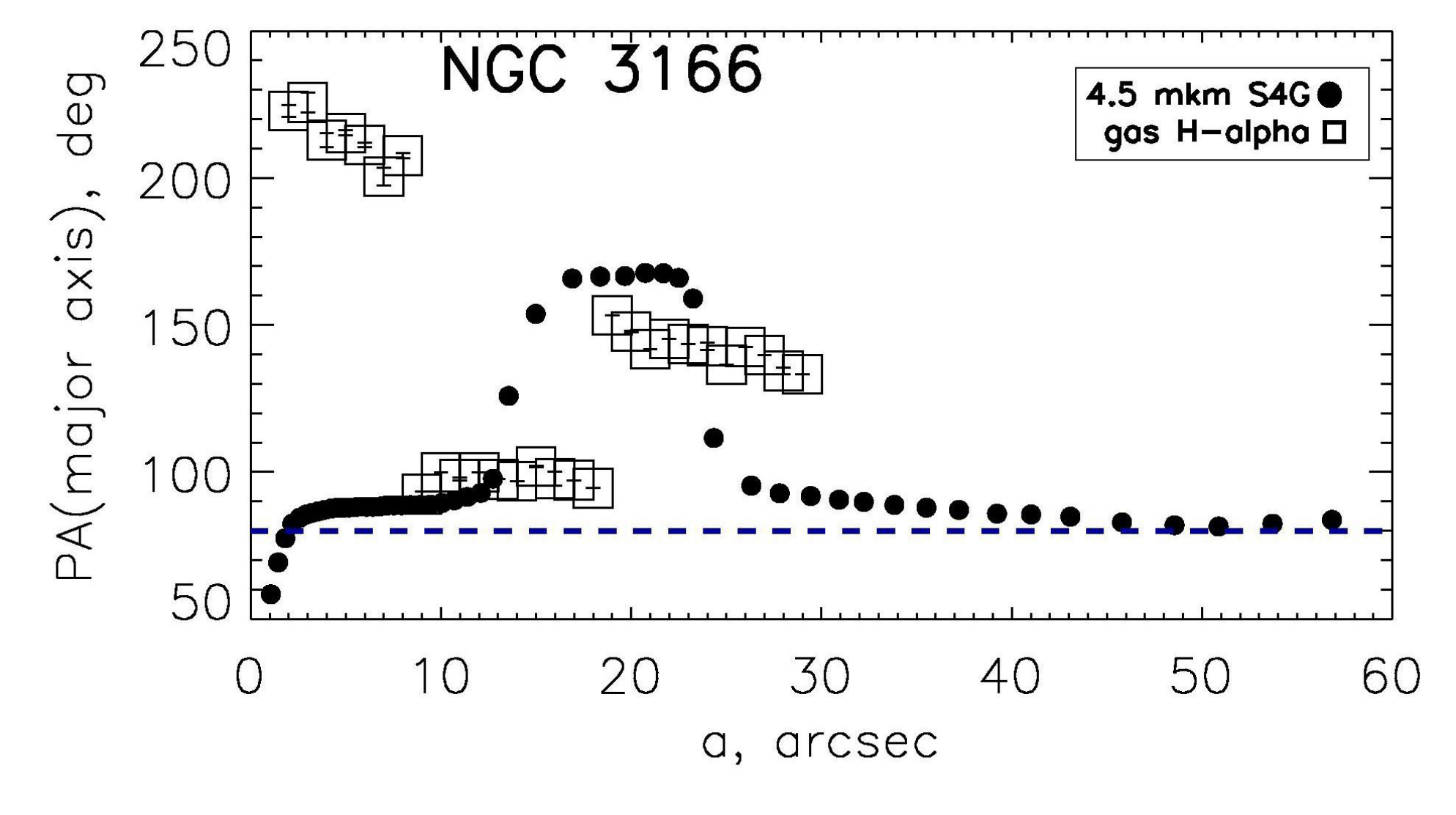} &
 \includegraphics[width=5cm]{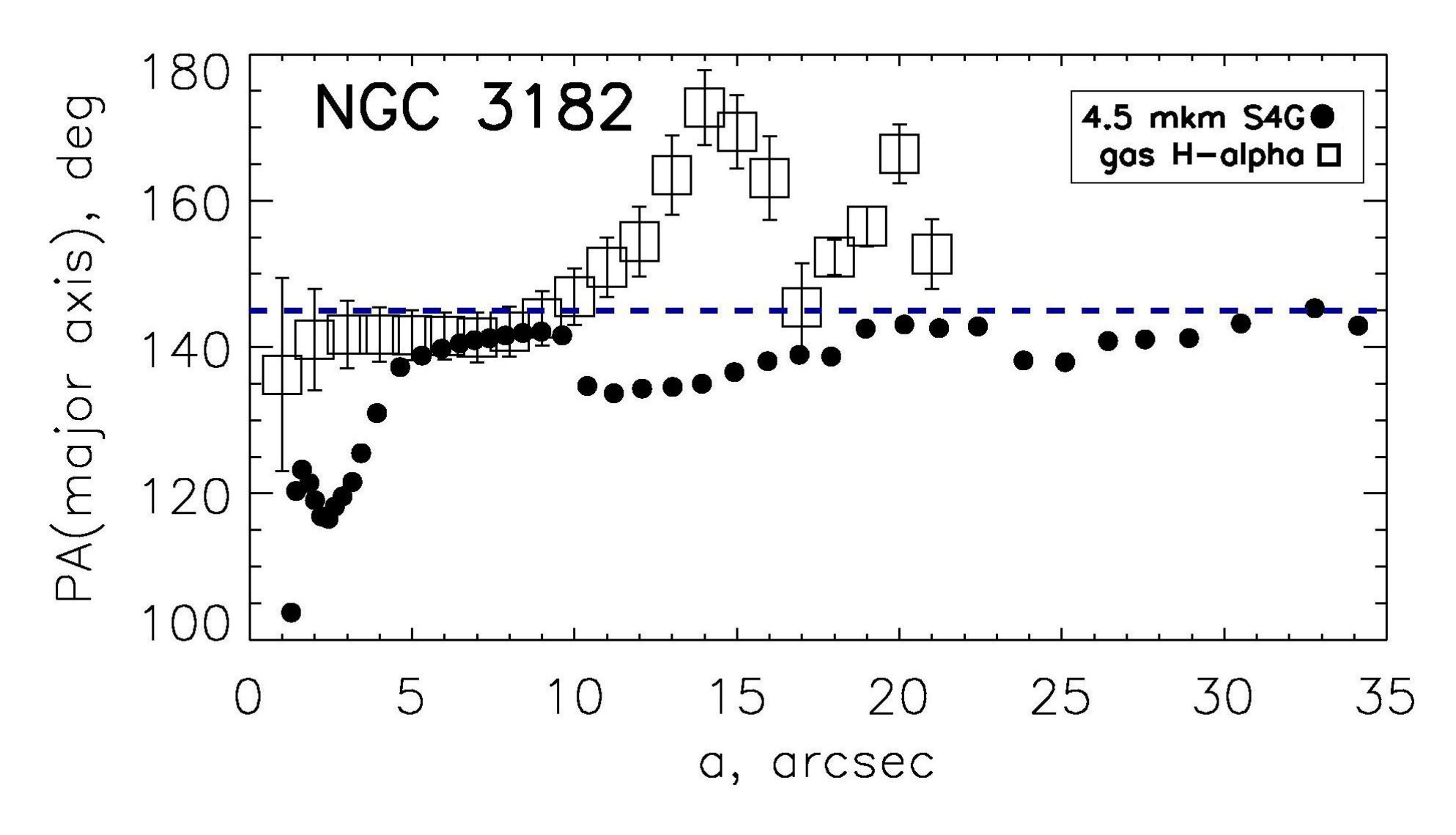} &
 \includegraphics[width=5cm]{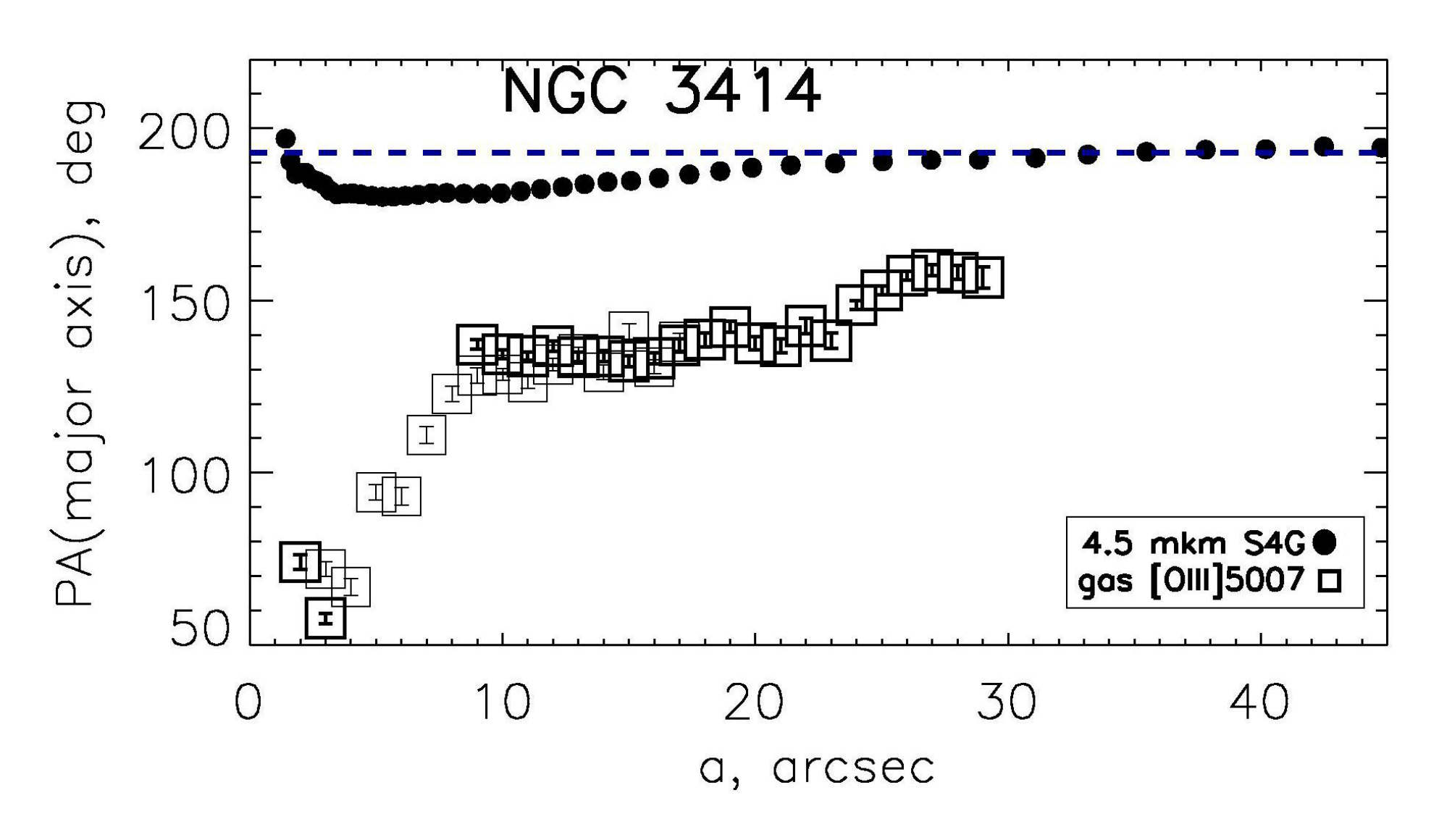} \\
\includegraphics[width=5cm]{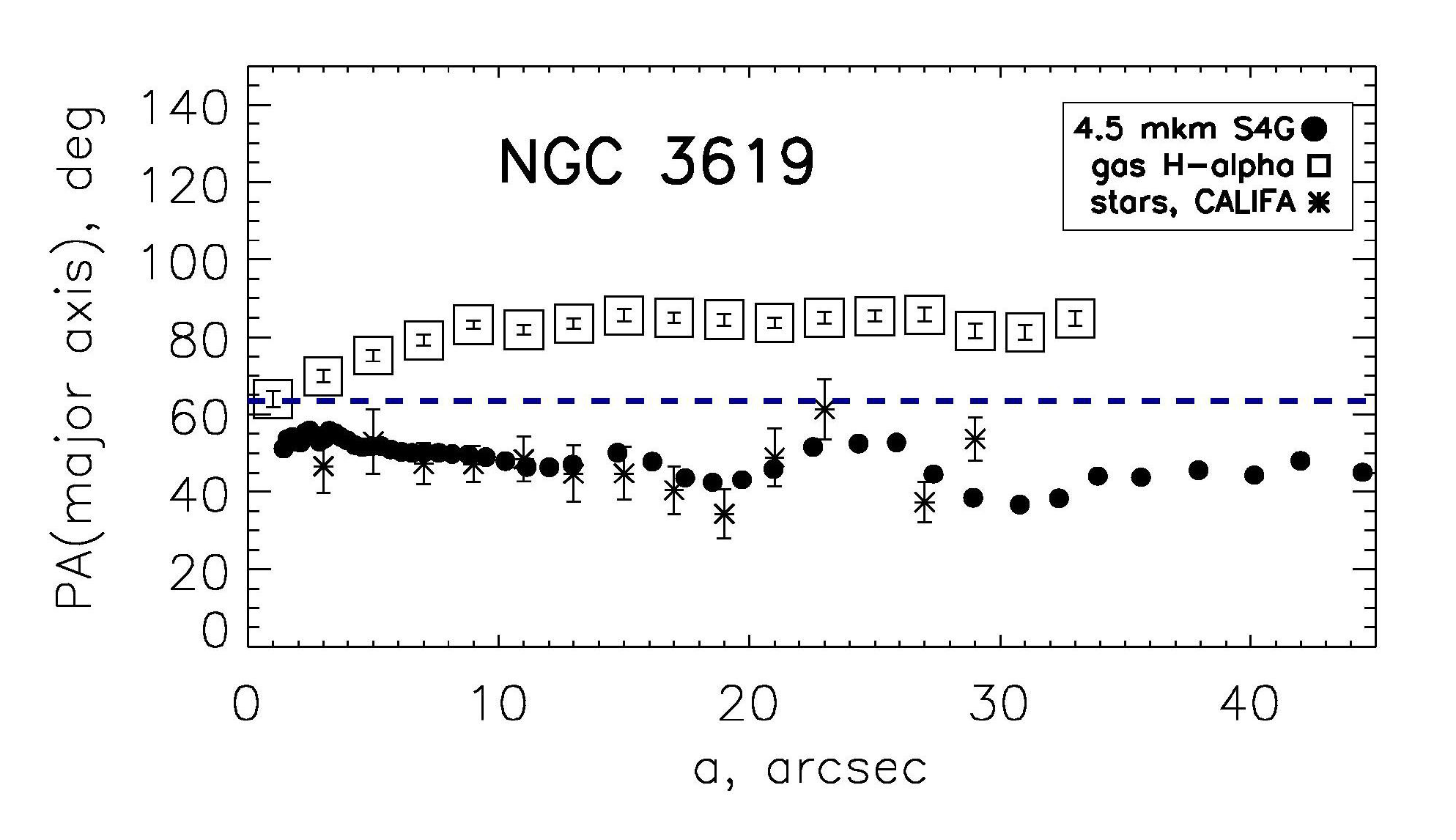} &
 \includegraphics[width=5cm]{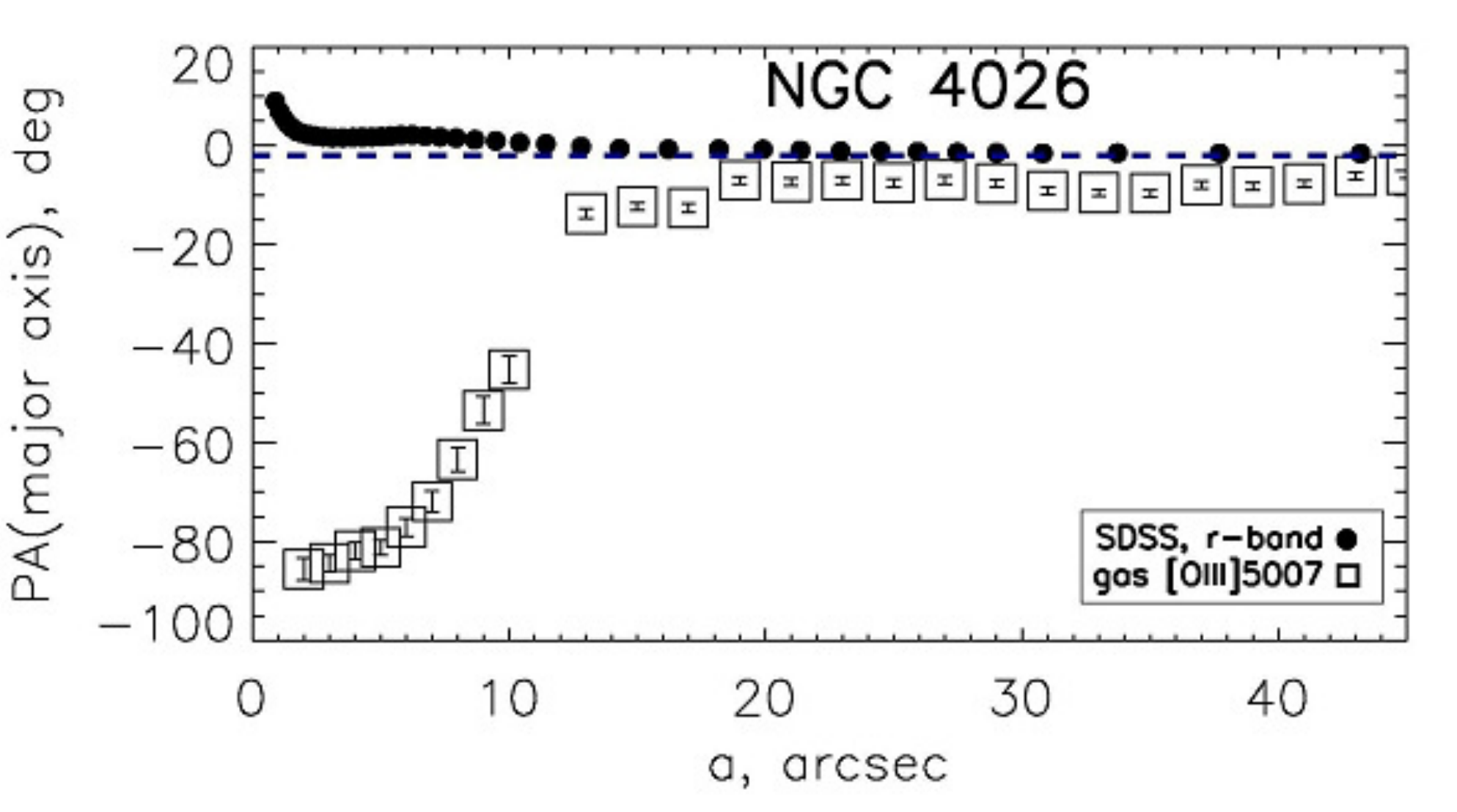} &
 \includegraphics[width=5cm]{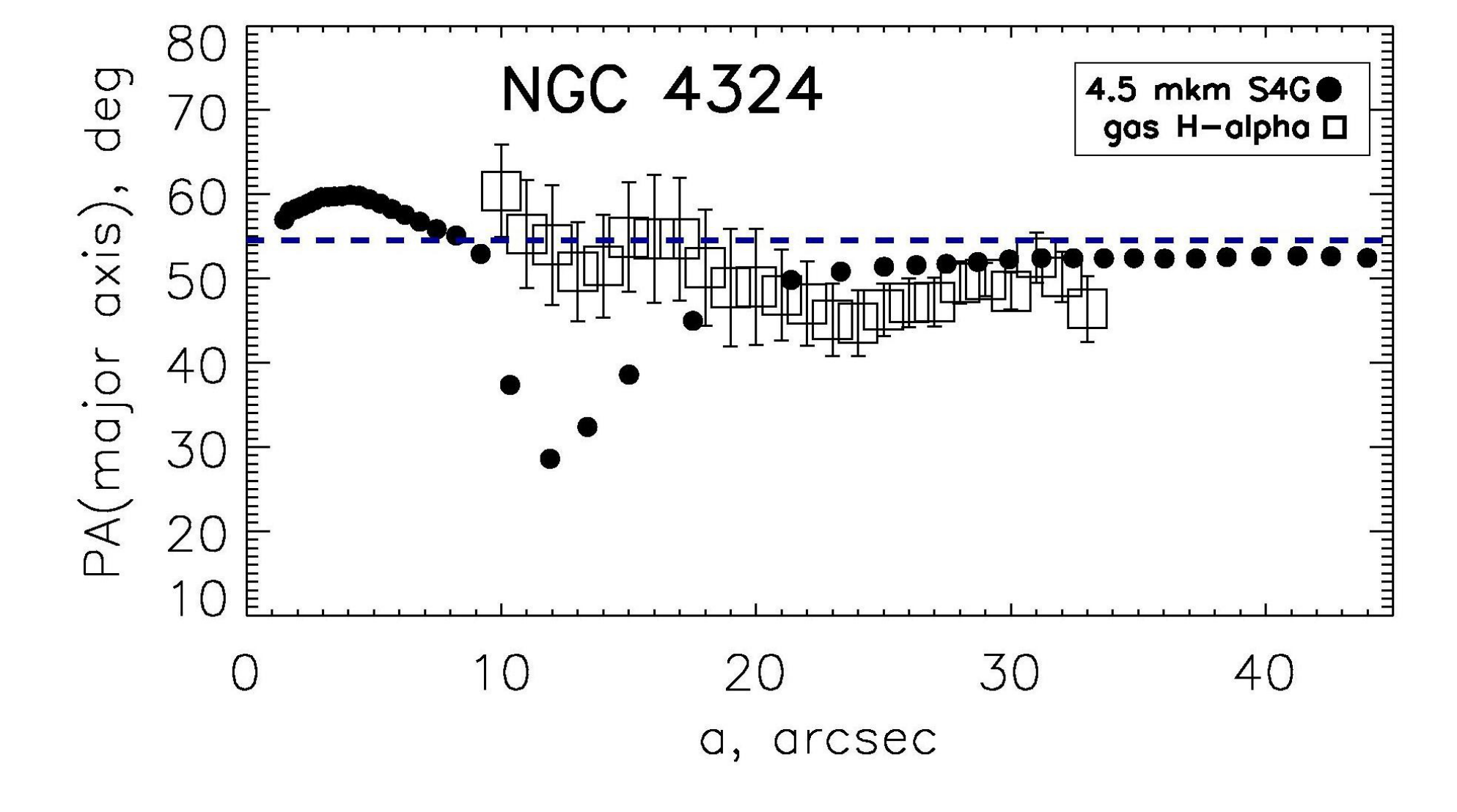} \\
\includegraphics[width=5cm]{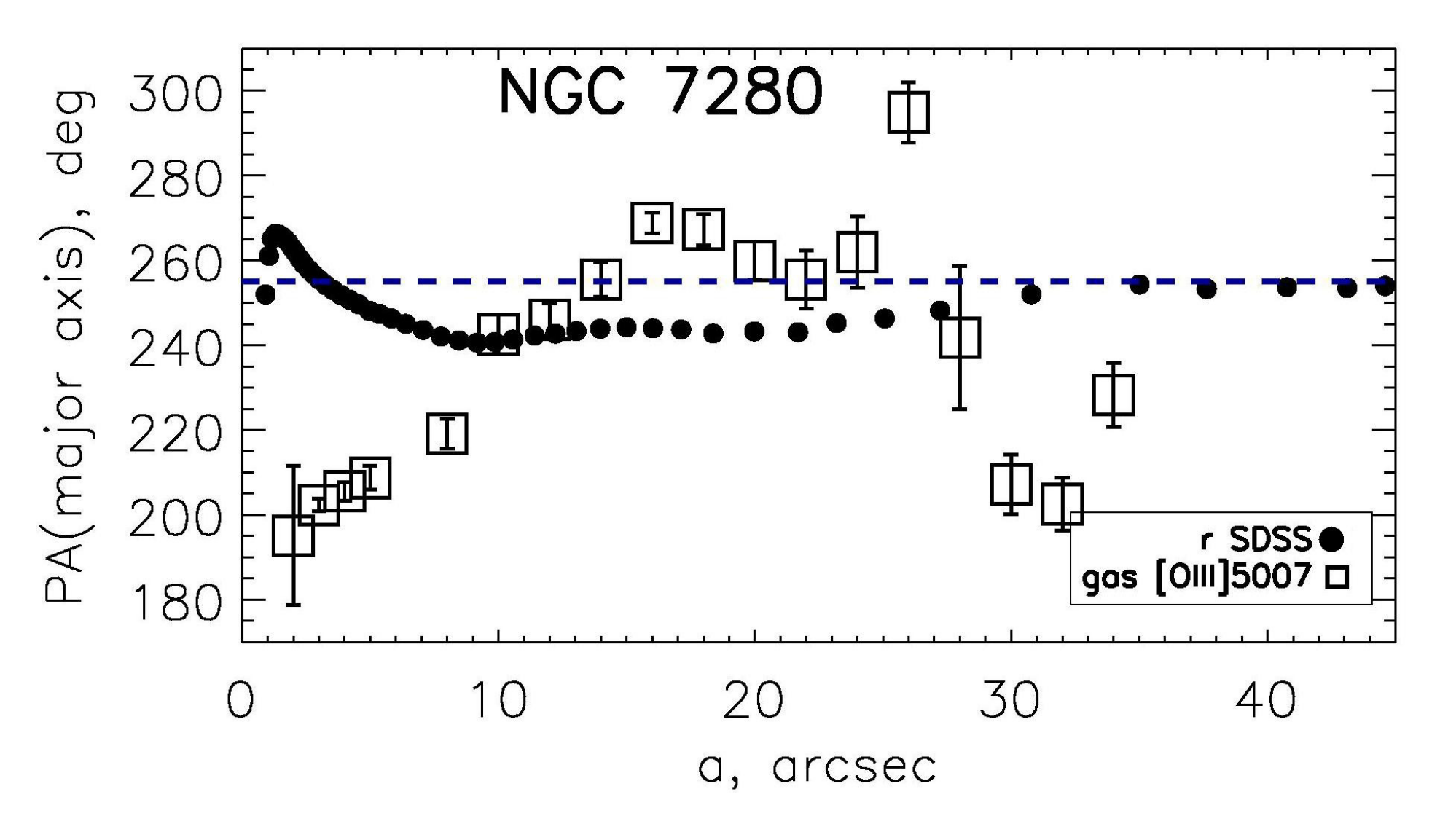} &
 \includegraphics[width=5cm]{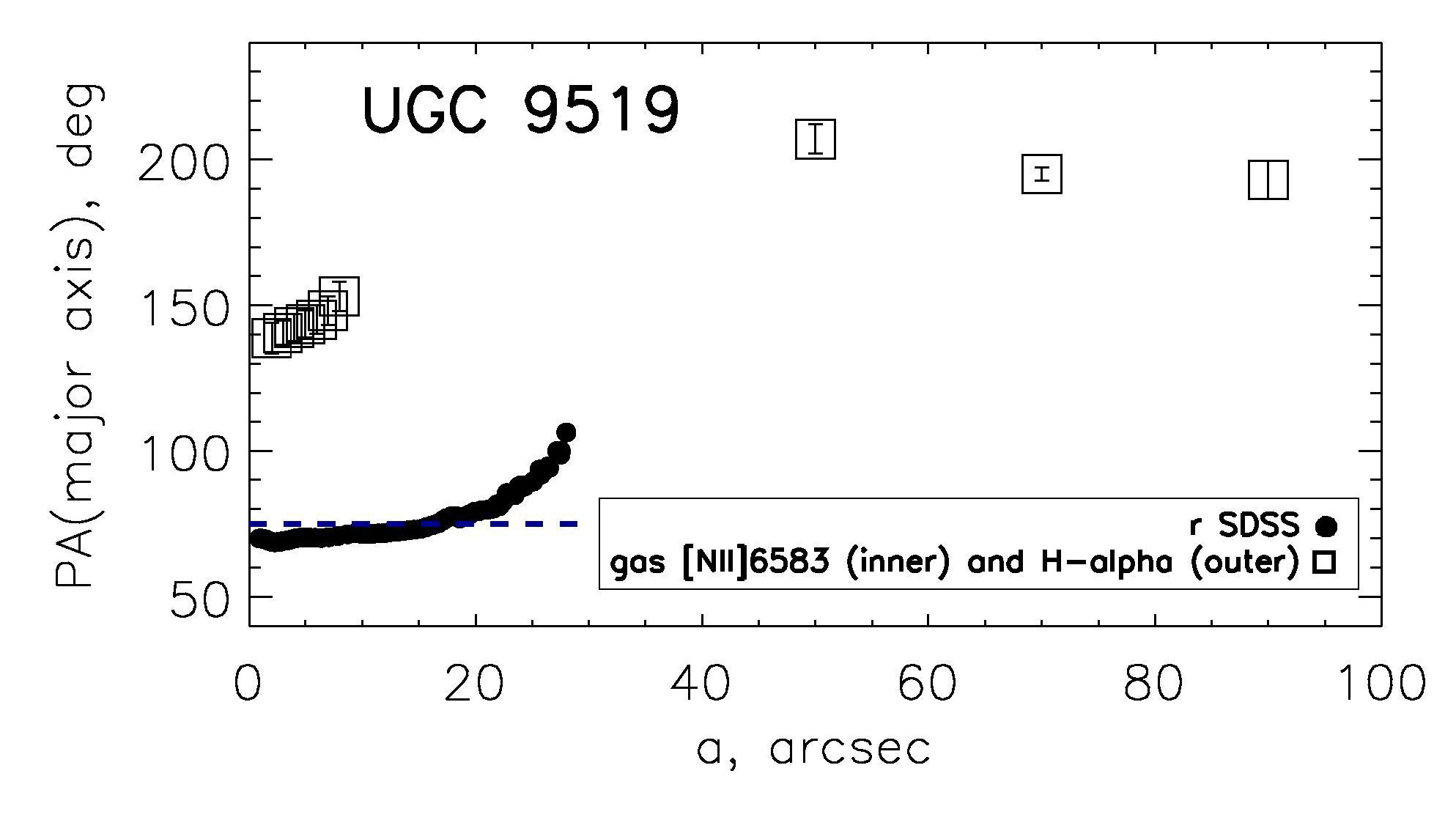} &
 \includegraphics[width=5cm]{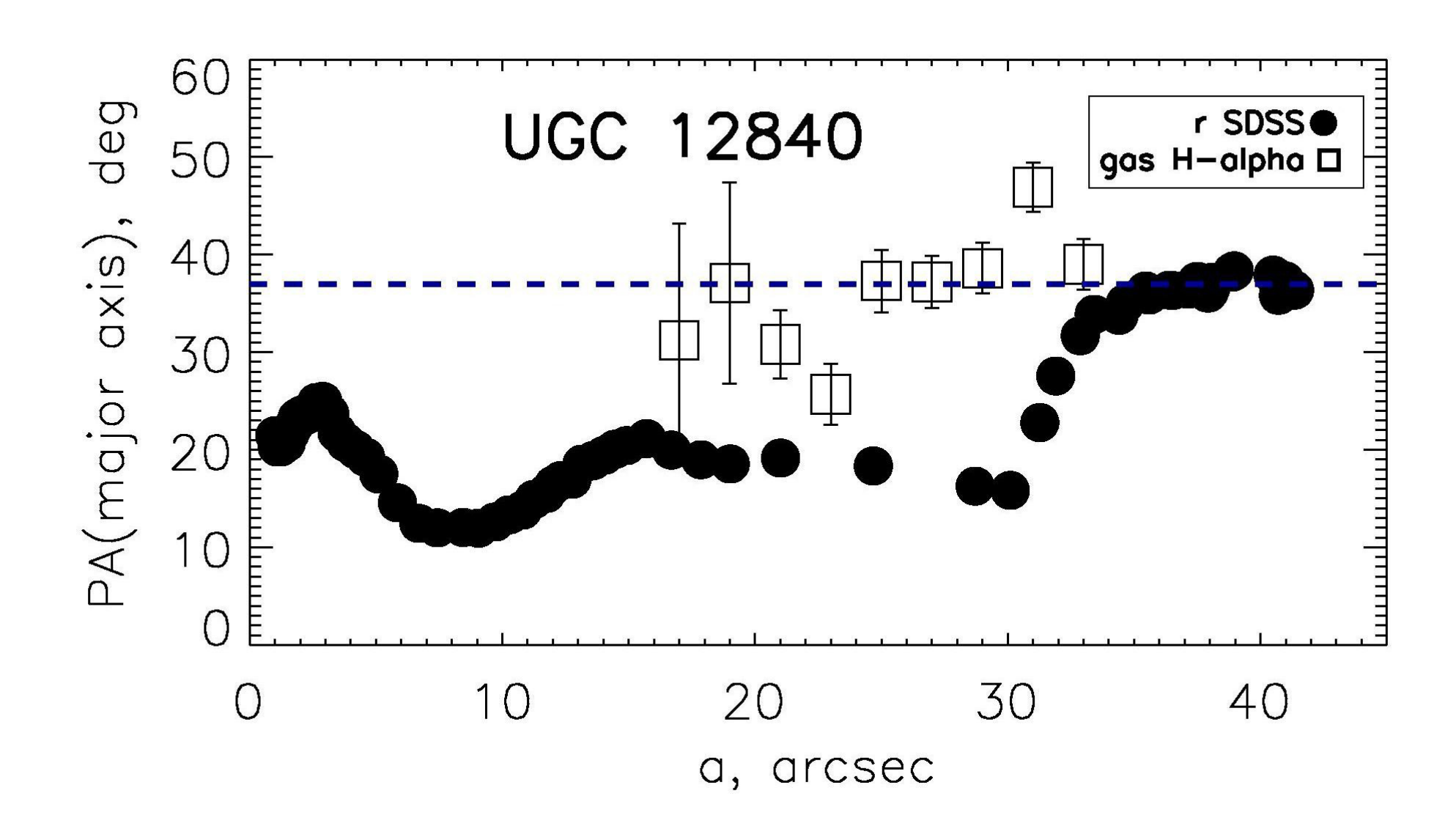} \\
\end{tabular}
\caption{Comparison of the photometric and kinematical major axes orientations;
sometimes the orientation of the kinematical stellar-disk line of nodes is also shown following the analysis
of the SAURON or CALIFA panoramic spectral data. The horizontal dashed lines indicate orientations
of the stellar-disk lines of nodes determined from the position angles of the outermost isophotes.}
\label{fig_pa}
\end{figure*}

\subsection{Tilted-ring analysis to understand the character of the gas motions.}

The profit of two-dimensional velocity fields  representing spatially-resolved projection of rotation velocities at various
distances from the galactic centers onto our line of sight, is that we can check the circular character of the ionized-gas rotation and 
derive spatial orientation of the rotation plane if the rotation is circular {\bf lacking strong radial motions, expansions or inflows}.
Indeed, in the case of circular rotation the maximum
rotation-velocity projection onto the line of sight is observed at the line of nodes; in its turn, the intrinsically round stellar disk looks
like an ellipse with the isophote major axis coincident with the line of nodes. So when the gas is confined to the plane of the galactic disk and
while its rotation is circular, we should see the steepest gradient of the observed line-of-sight velocity (`kinematical major axis') along the
isophote major axis. If the gas rotation is non-circular, that must be the case within a non-axisymmetical potential, the kinematical and
photometric major axes would turn {\it in opposite sense} with respect to the line of nodes \citep{vd97}. Finally, if {\bf we see that}
the gas kinematical major axis does not coincide with the stellar-disk line of nodes, but this deviation is not accompanied by the symmetric
deviation of the photometric major axis {\it within the same radial range}, it means that the rotation plane of the ionized gas stands in the space
differently from the plane of the stellar disk.

By assuming ionized-gas {\it planar  circular} rotation {\bf within some running radial range}, we can derive two angles of its spatial orientation,
{\bf in general varying} along the radius: an inclination of the disk plane to our line of sight and a position angle of its line of nodes tracing {\bf gas plane}
intersection with the sky plane. We have made it by applying to the gas line-of-sight velocity fields tilted-ring analysis \citep{begeman89} adopted for the
ionized-gas FPI data. For this purpose we have used the IDL software DETKA, as described by \citet{moisav04} and \citet{Moiseev2014}. The method is based on
splitting the observed velocity field into elliptical rings (1\arcsec--5\arcsec\ in wide) oriented according to adopted inclination $i_0$ and line-of-sight position
angle of the disk $PA_0$.  In all observed galaxies the center of symmetry of a velocity field lies no further than 1\arcsec--2\arcsec\ from the
photometric nucleus (the center of the inner isophotes), therefore the galaxy nuclei were fixed as the ring  centers.  The $\chi^2$-minimization was used to fit
the observed velocity distribution within every ring by model of quasi-circular rotation with following parameters: kinematical major axis $PA_{kin}$,
inclination $i$, rotation velocity $V_{rot}$, and systemic velocity $V_{sys}$. As a {\bf null} approximation for $PA_0$ and  $i_0$ we
adopted the orientation parameters according to the photometric data (HyperLEDA and isophotal analysis results). During the next steps we fixed often
the $V_{sys}$ derived at the first step. {\bf When the patchy character of a velocity field prevented a stable minimization over three parameters, we
chose to fix firstly the inclination $i$, and then $PA_{kin}$; $V_{rot}$ remained always free. Sometimes we applied an iterative procedure, fixing
$i$ and $PA_{kin}$ by turns.} The $i$ and $PA_{kin}$ were also {\bf always} fixed in the outer regions, where the elliptical rings contain only few points in the
velocity fields. We also masked some local regions notable for their peculiar kinematics, deviating by more than $30-50~\kms$ from the obtained model;
this threshold was individually selected.

Unfortunately, in the most cases the direct tilted-ring fitting described above gives unstable solutions {\bf just}
for $i(r)$ due to noises and {\bf holes} in the ionized-gas velocity fields. The {\bf uncomfortable} problem consists of {\bf relied} unphysical
chaotic changes of the $V_{rot}$ produced by strong {\bf derived} variations of $i$ with radius. \textbf{The well-known  degeneracy between these two quantities was  discussed by numerous authors \citep{begeman89,Kamphuis2015}}.  To avoid this problem we performed the following evaluation
of the $i(r)$ behavior {\bf basing on the first-step tilted-ring analysis results}. Namely, at every $r$ we extracted from the velocity field an elliptical ring
oriented in agreement with $i_0$ and $PA_{kin}(r)$ obtained as described above. Then we fitted the velocities in the ring using only two
free parameters: $i$ and $V_{rot}$, whereas the $PA_{kin}$  and $V_{sys}$ were already obtained {\bf at this radius} and fixed. {\bf The next step was
to repeat the procedure by fixing the obtained $i(r)$ and by leaving $PA_{kin}$ free to refine it}. Finally, we were able to obtain stable fitting results
with a reasonable shape of the rotation curve for the considered galaxies, in their central regions as minimum. The results of this study, including the analysis
of the \HI\ velocity fields available in the references are presented in Sec.~\ref{sec:5.2}--\ref{sec:5.4}.

The comparison of the kinematical major axis orientations running along the radius, with the
stellar-disk lines of nodes is given in Fig.~\ref{fig_pa}. In these plots presenting the tilted-ring analysis results 
we have also overlaid the results of isophote analysis -- the radial variations of the photometric major-axis position angles. 
The lines of nodes are fixed from the outermost isophote major-axis orientation (sometimes taken outside the plotted radial range). In some cases
we have also in hands the orientations of the kinematical major axes for the {\it stellar} components; they are obtained with the same DETKA
software by applying it to the data from the ATLAS-3D survey \citep{atlas3d_1} or from the CALIFA survey \citep{califa_s,califa_3}. In these
cases we can check independently the circular shape of the stellar disk and the axisymmetry of the gravitational potential.

A quick look at the Fig.~\ref{fig_pa} reveals a great variety of behavior of the gas kinematical major axis in the gas-rich lenticular galaxies.
Sometimes the ionized gas lies in the galactic planes and exhibits regular circular rotation 
(NGC~252, NGC~2551, NGC~4324), though the gaseous-disk plane may be warped {\it together} with the stellar plane (NGC~252), 
or the gas may counterrotate the stellar rotation \citep[NGC 2551,][]{we2551}. Sometimes
the gas lies within the decoupled plane homogeneously inclined to the stellar disk (NGC~2655, NGC~3619). Sometimes the central gas is in the polar
plane, but at larger radii it falls into the main galactic plane -- and starts to corotate the stellar disk (NGC~2962, NGC~4026) 
or to counterrotate it (NGC~7280, Fig.~\ref{fig_ls}) or to form a completely detached outer starforming ring (UGC~9519). Sometimes
the central region of a galaxy (its bulge?) looks triaxial, with a twisted isophote major axis, but the ionized gas just within the same radial range
remains to be confined {\bf to the plane of the stellar disk} and demonstrates regular circular rotation (IC~5285, UGC~12840, NGC~3106).
And the last, perhaps the most interesting class of objects: those with the ionized gas confined to the galactic planes and 
demonstrating laminar circular rotation, but {\it only within a limited radial range} --  NGC~774, NGC~2697, NGC~3166, NGC~3182. 
This latter case will be discussed in some more details below, in Section~\ref{sec:5}.

\subsection{Gas excitation and metallicities}

\begin{figure*}
	\centering
	\includegraphics[width=\linewidth]{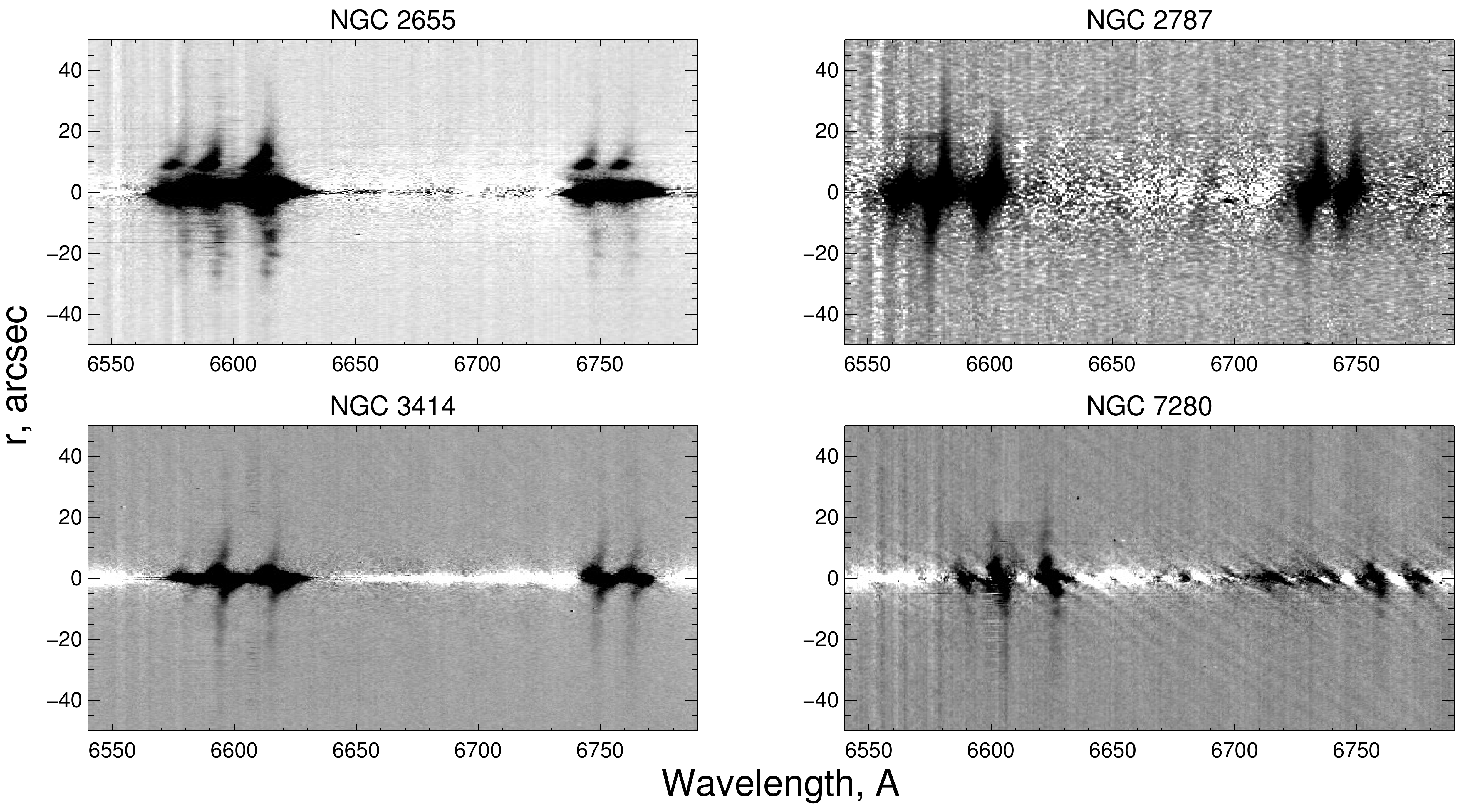}
	\caption{An example of long-slit spectra obtained along the major axes of the four galaxies demonstrating almost homogeneous enhanced emission in [NII] and [SII] lines. Stellar continuum and absorption lines were modelled with ULySS and subtracted.}
	\label{spec2d}
\end{figure*}

\begin{figure*}
\centering
\includegraphics[width=15cm]{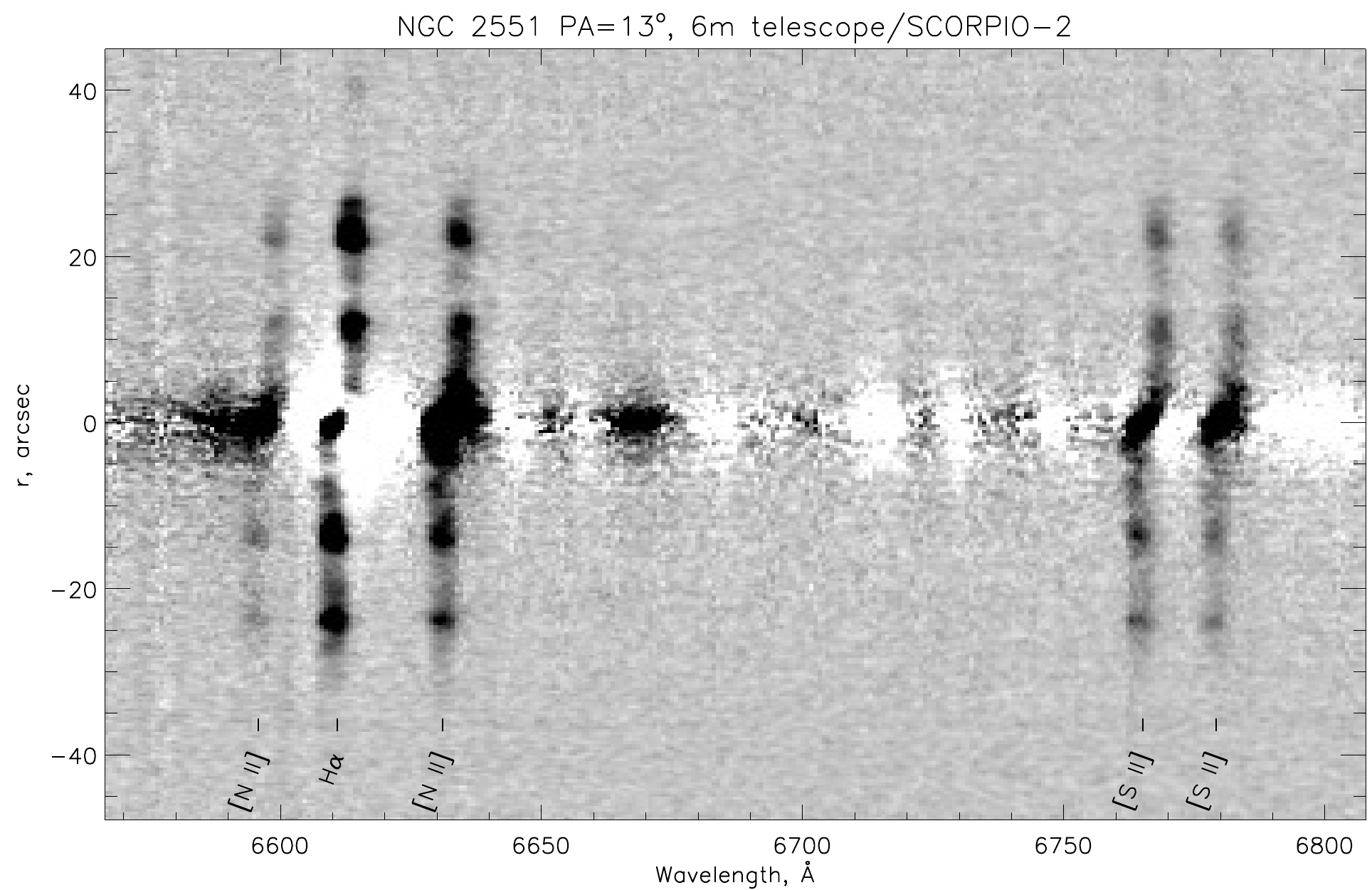}
\caption{An example of the clumpy emission lines in the long-slit spectrum of NGC~2551 -- the S0 galaxy with a broad starforming ring.}
\label{n2551red}
\end{figure*}

\begin{figure*}[p]
\centering
\includegraphics[width=15cm]{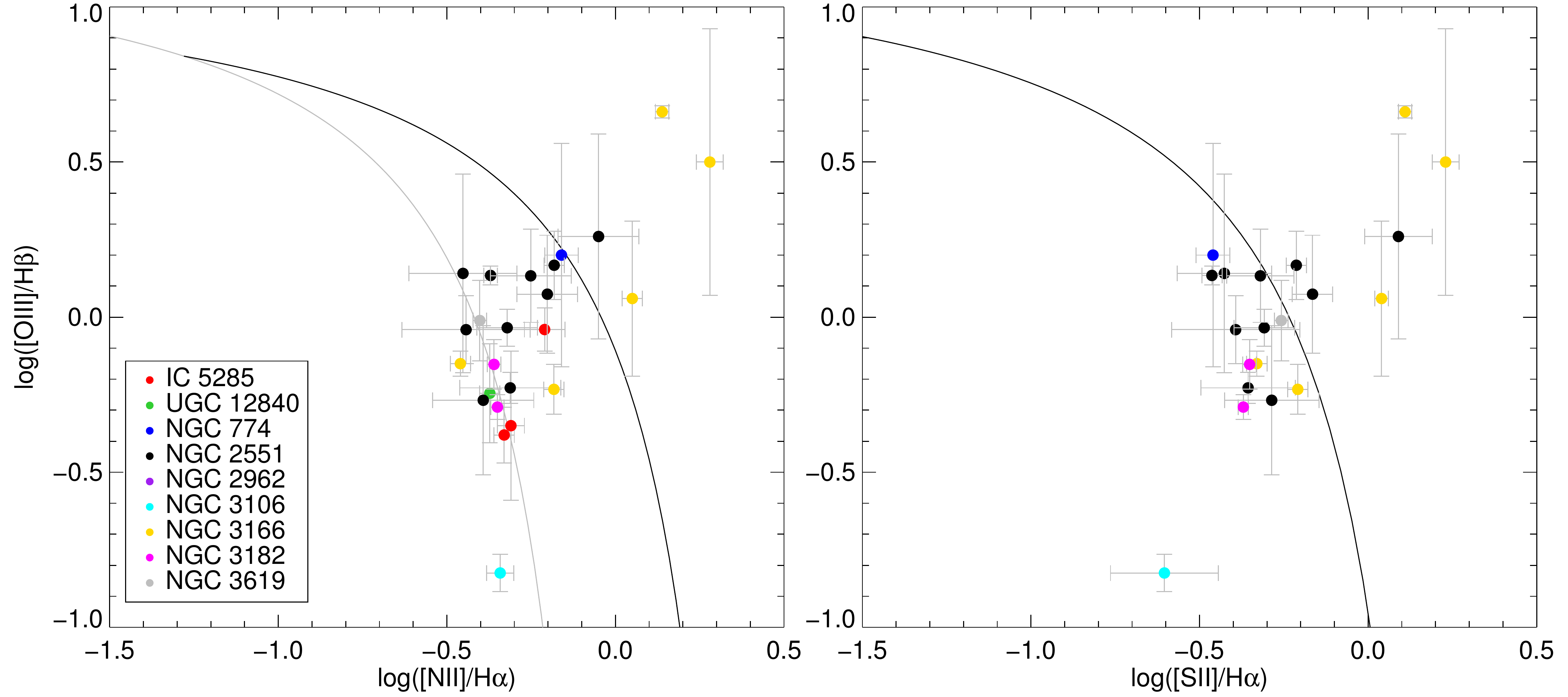}
\caption{BPT-diagrams for some radial zones in 9 S0-galaxies with extended emission. The curves separating 
HII-regions and all other types of gas excitation are from \citet{kewley01} (fat solid one) and from \citet{kauffmann03} (pale one).}
\label{fig_bpt}
\end{figure*}

The source of ionized-gas excitation can be determined by analyzing the flux ratios of low-ionization and high-ionization
emission lines to Balmer emission lines -- at so called BPT (Baldwin-Phillips-Terlevich) diagrams \citep{bpt,vo}.
At least, starforming regions demonstrate quite recognizable emission-line spectra; and the models allow to determine
the chemical composition -- mainly oxygen abundances -- by using the flux ratios of strong emission lines just for the
case of the gas excited by young stars. We have checked the gas excitation in our S0 galaxies by studying the emission lines
in the long-slit spectra obtained with the SCORPIO/BTA; for NGC~774 and NGC~3106 we have used the public spectral datacubes
from the CALIFA survey \citep{califa_3}, for NGC~2697 and NGC~4324 -- additional spectral data obtained earlier
at the 11m SALT of the SAAO with the long-slit spectrograph RSS \citep{saltrings} have been involved into our present
analysis. Weak and {\bf rather} homogeneously distributed along the slit emission lines related to the gaseous disks with strongly
decoupled rotation -- those in NGC~2655, NGC~2787, NGC~3414, and NGC~7280, -- have all demonstrated shock-like excitation: in
their spectra the emission line [NII]$\lambda$6583 is {\bf comparable or even} stronger than the \Ha everywhere along the slit
{\bf (see 2D spectra obtained along major axes of the mentioned galaxies in Fig.~\ref{spec2d})}.  Also similar
pattern of emission-line ratios is {\bf typical for} {\bf diffuse ionized gas (DIG)} {\bf that has allowed to explain diffuse gas excitation
in early-type galaxies by old, strongly evolved stars \citep{pagb94}. }
The other galaxies demonstrate inhomogeneous clumpy emission-line distributions along the slit (see the example of such behavior in Fig.~\ref{n2551red}),
and we have measured the emission-line flux ratios for the individual clumps. We show the BPT diagrams for these measurements in Fig.~\ref{fig_bpt}.
The emission from the regions lying under the theoretical `maximum starburst line' from \citet{kewley01} can be in principle explained through
photoionization by young stars, but composite mechanism of excitation might be responsible for the emission from the clumps lying between 
this line and empirical \cite{kauffmann03} separating sequence.  As it follows from Fig.~\ref{fig_bpt}, half of the clumps in NGC~3166
and some regions in NGC~2551 clearly demonstrate dominating mechanisms of gas excitation different from photoionization by young stars
(e.g., by shocks or by old, post-AGB stars). NGC~774 lies at the border between the HII-region-like excitation and DIG-like or
shock-like excitation. All other resolved clumps lie well below the `maximum starburst line', around the \cite{kauffmann03} HII-fitting
sequence, and their emission can be explained through pure photoionization by young massive stars. So we can determine oxygen
abundances of the gas in these clumps by exploring strong-line methods.

{\bf In general, accurate oxygen abundances in HII regions can be `directly' derived from measurements of electron temperature-sensitive line ratios, such
as [OIII]$\lambda\lambda4959,5007$/[OIII]$\lambda4363$. This is often referred to as
the $T_e$-method. However, in the most cases it is impossible to measure the faint emission lines like [OIII]$\lambda4363$ in distant galaxies.
To avoid this problem many empirical and model-based calibrations relying on strong emission lines only were developed. Despite that, the well-known
problem of discrepancy (up to 0.6 dex) of various metallicity calibrations is still not resolved \cite[see, e.g.,][]{kewley08}}. Here
we have used several empirical methods to derive oxygen abundances $\mathrm{12+\log(O/H)}$, and have compared their results. Because of the limited
spectral range of our data, we selected only those methods that do not demand [OII]$\lambda\lambda3727,3729$ line to be observed:  S-method \citep{pilugin},
O3N2, N2 \citep[both from][]{marino13}, and D16 \citep{dopita16}. Former three of these methods were calibrated using local HII regions
with well-known oxygen abundances determined through the $T_e$-method. The fourth method, D16, is a photoionization model-based one.

Every method used by us has its advantages and disadvantages in comparison with others. Thus, only [NII]/H$\alpha$ ratio is needed 
for the N2 method, but it is applicable only for $\mathrm{8.0<12+\log(O/H)<8.65}$, gives very large scatter (0.16 dex at 1$\sigma$ level) of 
the yielding values, and hence provides only rough estimate of the gas metallicity.
Another popular method -- O3N2 -- uses neighboring emission lines ratios, [OIII]/H$\beta$ and [NII]/H$\alpha$, and hence almost does not suffer
from uncertainties with reddening correction. This method is applicable within $\mathrm{8.17<12+\log(O/H)<8.77}$, and shows large dispersion 
(in average 0.18 dex, but significantly lower for $\mathrm{12+\log(O/H)}>8.4$). {\bf Recent studies have shown that the contamination by a large fraction of
DIG affects the results obtained with O3N2 method in only small extent \citep{Kumari2019, Poetrodjojo2019}.}
The S-method is well-calibrated in the range of $\mathrm{7.1<12+\log(O/H)<8.75}$ and on testing sample yields the results
that agree with estimations made by using the $T_e$-method, with a standard deviation of $\sim 0.05$ dex. In its original form this method suffers
from reddening uncertainties (which are rather high in our data) because all involved emission-line fluxes ([OIII], [SII], [NII]) should 
be normalized to the flux of H$\beta$ line. Finally, the D16 method demands only flux ratios of the lines within the red part
of a spectrum -- [NII]/[SII]  and [NII]/H$\alpha$. This method was developed for application to high-z galaxies, 
and its main advantage is that it doesn't depend on the 
ionization parameter or ISM pressure, and hence it might be used safely also for regions having enhanced [NII]/H$\alpha$ ratios (e.g.,  
lying in the composite excitation zone at the BPT diagrams). However note that it is still calibrated through the photoionization models that
do not include excitation by shocks. This method yields systematically higher oxygen abundances (in comparison with the $T_e$-method) because
of the higher solar abundance used as zero-point in calibration ($\mathrm{12+\log(O/H)}=8.77$ instead of 8.69). Unfortunately, authors of 
the original work do not provide an estimate of typical uncertainty for the D16 calibration.

Taking into account the problems with the metallicity calibrations used in this work, we have applied some corrections to the results 
making them more comparable with each other:
\begin{itemize}
	\item All the values of $\mathrm{12+\log(O/H)}$ obtained with the D16 method were reduced by 0.08 dex to take into
        account the different solar oxygen abundance values assumed;
	\item As an input for the S-method, we have used the [SII]/H$\alpha$ and [NII]/H$\alpha$ line flux ratios and a proposed theoretical Balmer
      decrement H$\alpha$/H$\beta$ = 2.86 \citep{osterbrock}. This correction allowed us to reduce significantly the final uncertainty 
      of the oxygen abundance because of getting rid of uncertainties with the reddening estimates.
	\item All the values beyond the area of applicability of the method used were considered as wrong ones.
\end{itemize} 

\begin{figure*}[p]
	\centering
	\includegraphics[width=0.49\linewidth]{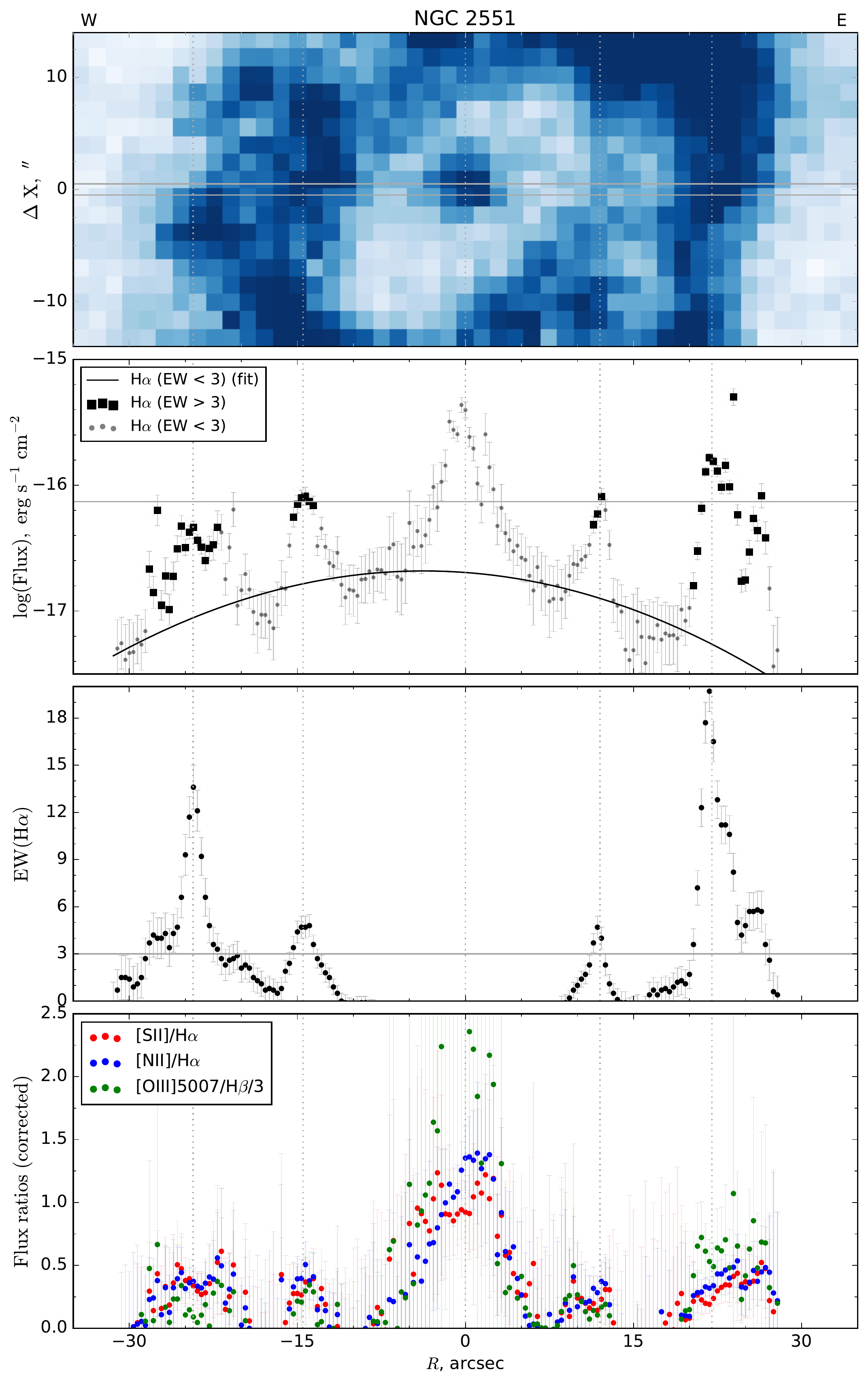}~\includegraphics[width=0.49\linewidth]{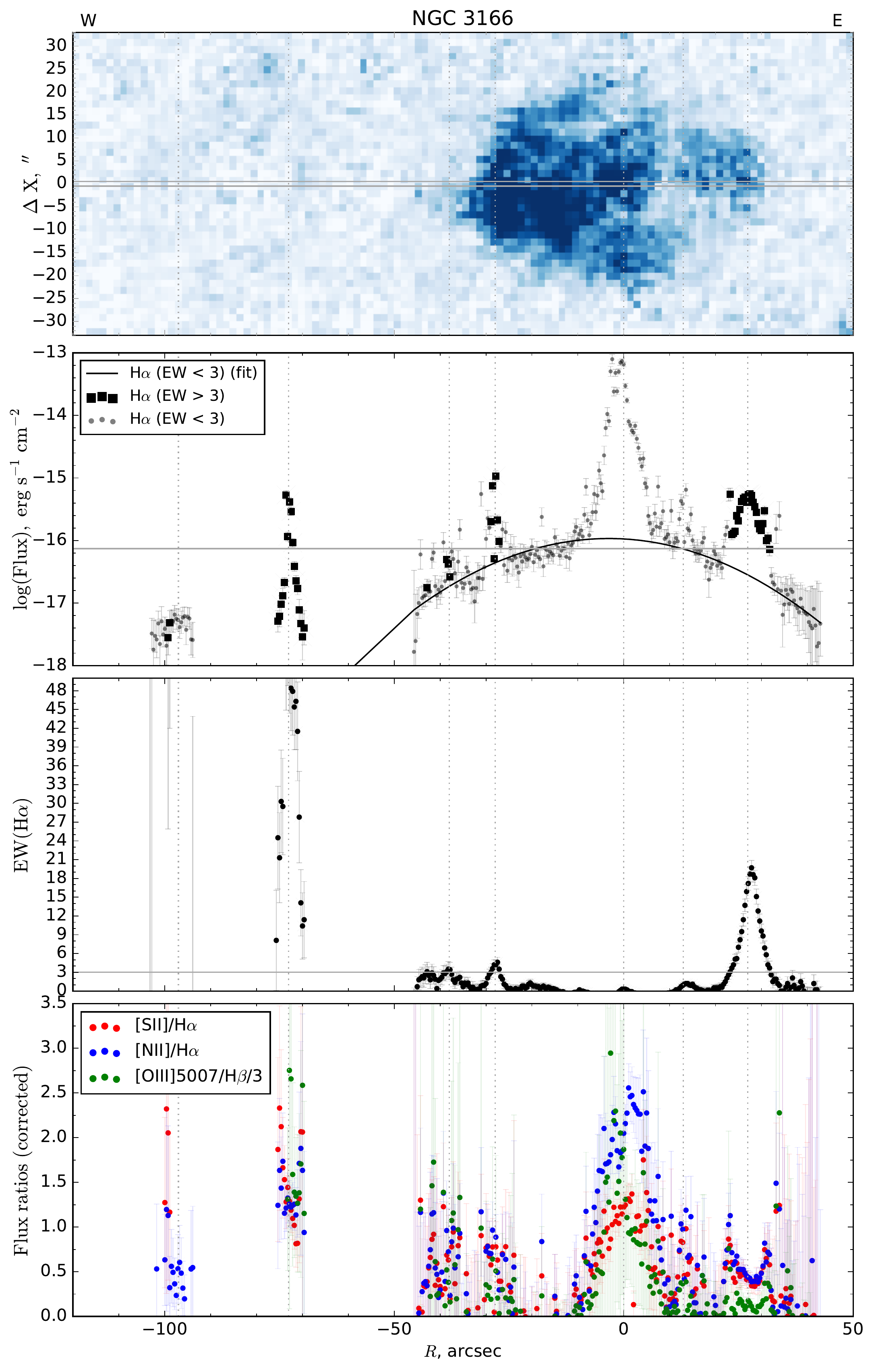}
	\includegraphics[width=0.49\linewidth]{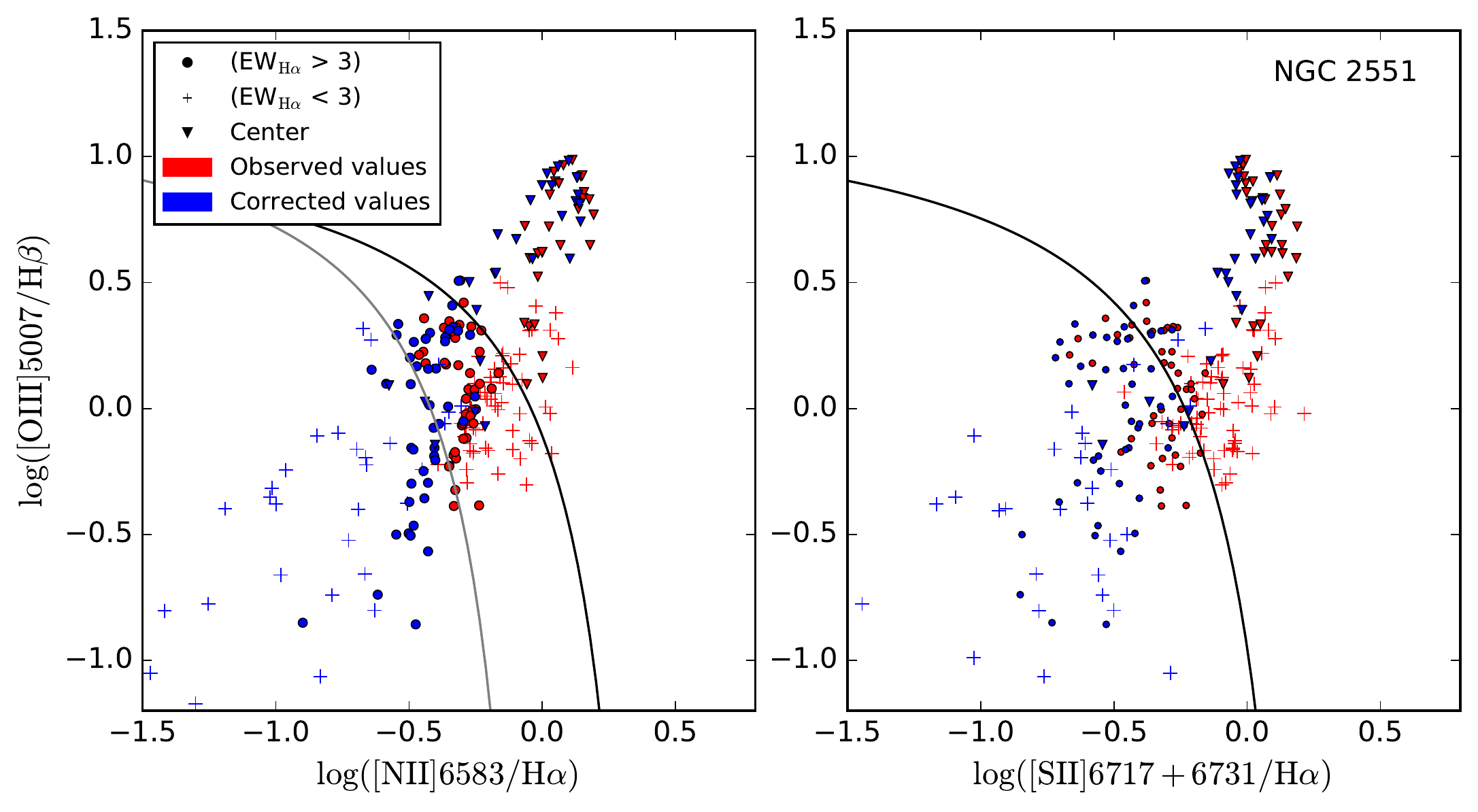}~\includegraphics[width=0.49\linewidth]{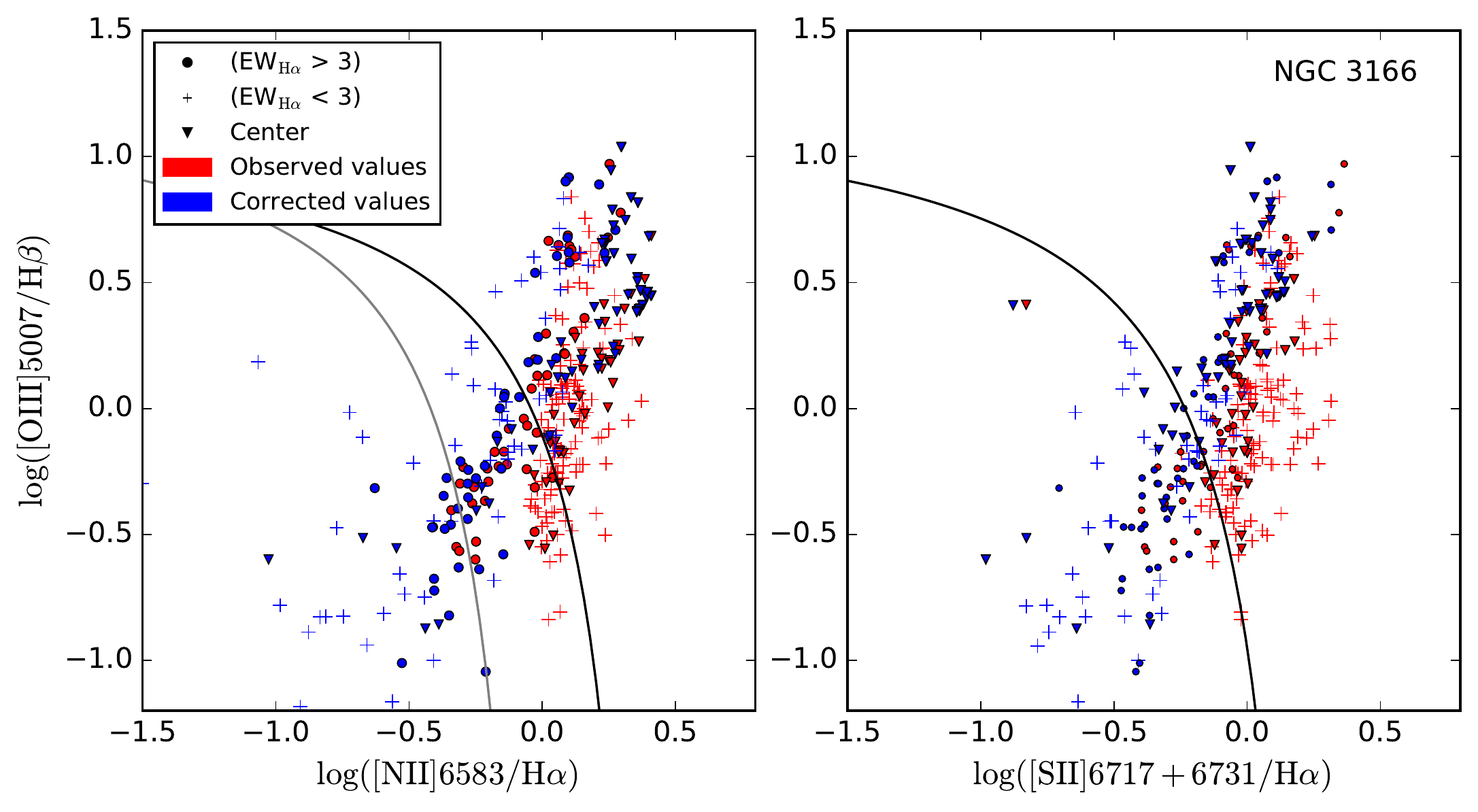}
	\caption{Analysis of the ionization state along the slit for the galaxies NGC~2551 (left) and NGC~3166 (right).
        From top to bottom: position of the slit overlaid on FUV image (GALEX) of the galaxies; distribution of the H$\alpha$
        line flux (horizontal line denotes a level of $F\mathrm{(H\alpha)=7.4\times10^{-17}\ erg/s/cm^{-2}}$ used to identify
        the DIG regions by \citealt{Zhang2017}; {\bf curved line shows the polynomial fit of the DIG flux distribution}); EW(H$\alpha$)
        (horizontal line corresponds to the value of EW(H$\alpha$)=3\AA\
        used to identify the DIG regions by \citealt{Lacerda2018}.); emission lines flux ratios corrected for the DIG contribution;
        BPT-diagrams constructed for individual pixels along the slit. In the BPT-diagram red color shows observed flux ratios
        while blue one -- that corrected for the DIG contribution. Different symbols denote the galactic center and regions with
        $\mathrm{EW(H\alpha) > 3}$ or $\mathrm{EW(H\alpha) < 3}$ \AA. The curves separating
	HII-regions and all other types of gas excitation are from \citet{kewley01} (fat solid one)
        and from \citet{kauffmann03} (pale one)}
	\label{fig_lsfit}
\end{figure*}

\begin{deluxetable}{lccccccc}
	\tablenum{4}
	\tablewidth{19cm}
	\tablewidth{0pt}
	\tabletypesize{\tiny}
	\tablecaption{{  Emission-line ratios in the rings of S0 galaxies}}\label{tab_fluxes}
	\tablehead{
		\colhead{Galaxy name$^a$} &
		\colhead{Radius, $^{\prime \prime}$}  &
		\colhead{EW(H$\alpha$), \AA} &
		\colhead{log([NII]6583/H$\alpha$)} &
		\colhead{log([OIII]5007/H$\beta$)} &
		\colhead{log([SII]6717+6731/H$\alpha$)} &
		\colhead{[SII]6717/6731} &
		\colhead{H$\alpha / \mbox{H}\beta$}
	}
	\startdata
	I5285 & $-12$ & 5.6 & $-0.21\pm 0.06$ & $-0.04\pm 0.07$ & & &    \\
	& $+16$ & 24.3 & $-0.33\pm 0.03$ & $-0.38\pm 0.09$ & & &       \\
	& $+31$ & 31.4 & $-0.31\pm 0.04$ & $-0.35\pm 0.24$ & & &        \\
	U12840 & $-20$ & 20.9 & $-0.37\pm 0.03$ & $-0.21\pm 0.16$ & & & $5.8\pm 0.9$ \\
	N774$^b$  & $9$ & $2.0\pm 0.3$ & $-0.16\pm 0.05$ & $+0.20\pm 0.36$ & $-0.46\pm 0.05$ & $0.76\pm 0.07$ & $>7$ \\
	N2551 & $-29$ & 7.2 & $-0.45\pm 0.16$ & $+0.17\pm 0.32$ & $-0.41\pm 0.14$ & 1.5:: &  $5.1\pm  2.7$ \\
	& $-24$ & 13.8 & $-0.31\pm 0.15$ & $-0.20\pm 0.05$ & $-0.34\pm 0.14$ & $>1.5$ &  $5.0\pm 1.2$ \\
	& $-20$ & 7.0 & $-0.44\pm 0.19$ & $+0.00\pm 0.11$ & $-0.37\pm 0.19$ & $>1.5$ & $6.4\pm 3.2$ \\
	& $-17$ & 4.75 & $-0.39\pm 0.15$ & $-0.24\pm 0.24$ & $-0.27\pm 0.14$ & $1.38\pm 0.16$ & $5.0\pm 3.1$ \\
	& $-14$ & 5.8 & $-0.32\pm 0.09$ & $-0.02\pm 0.06$ & $-0.30\pm 0.09$ & $1.52\pm 0.07$ & $3.8\pm 0.2$ \\
	& $+12$ & 3.9 & $-0.18\pm 0.03$ & $+0.19\pm 0.11$ & $-0.20\pm 0.03$ & $1.29\pm 0.10$ & $4.5\pm 0.7$ \\
	& $+15.5$ & 1.0 & $-0.05\pm 0.12$ & $+0.26\pm 0.33$ & $+0.09\pm 0.10$ & $1.06\pm 0.19$ & $2.15\pm 1.1$ \\
	& $+19$ & 2.8 & $-0.20\pm 0.09$ & $+0.10\pm 0.19$ & $-0.15\pm 0.06$ & $1.22\pm 0.11$ & $4.8\pm 1.9$ \\
	& $+23$ & 14.0 & $-0.37\pm 0.02$ & $+0.14\pm 0.03$ & $-0.46\pm 0.03$ & $1.37\pm 0.08$ & $3.2\pm 0.3$ \\
	& $+27$ & 6.9 & $-0.25\pm 0.12$ & $+0.15\pm 0.15$ & $-0.31\pm 0.10$ & $1.42\pm 0.12$ & $4.0\pm 1.9$ \\
	N3106$^b$ & $21W$ & 22.5 & $-0.34\pm 0.04$ & $-0.80\pm 0.06$ & $-0.59\pm 0.16$ & $1.12\pm 0.15$ & $4.7\pm 0.2$ \\
	N3106$^b$ & $21E$ & 43.5 & $-0.34\pm 0.04$ & $-0.80\pm 0.06$ & $-0.59\pm 0.16$ & $1.12\pm 0.15$ & $4.7\pm 0.2$ \\
	N3166 & $-98$ & $\infty $ & $-0.46\pm 0.03$ & $-0.15\pm 0.04$ & $-0.33\pm 0.03$ & $1.33\pm 0.15$ &    \\
	& $-72$ & 118 & $+0.14\pm 0.02$ & $+0.68\pm 0.02$ & $+0.12\pm 0.02$ & $1.42\pm 0.07$ & $4.1\pm 0.2$ \\
	& $-27$ & 4.2 & $+0.05\pm 0.03$ & $+0.06\pm 0.25$ & $+0.04\pm 0.02$ & $1.14\pm 0.03$ & $>7$   \\
	& $+14$ & 1.7 & $+0.28\pm 0.04$ & $+0.50\pm 0.43$ & $+0.23\pm 0.04$ & $1.14\pm 0.06$ & $>7$  \\
	& $+28$ & 13.3 & $-0.18\pm 0.03$ & $-0.20\pm 0.08$ & $-0.19\pm 0.03$ & $1.36\pm 0.09$ & $5.5\pm 0.4$ \\
	N3182 & $-5.5$ & 13.7 & $-0.35\pm 0.02$ & $-0.29\pm 0.04$ & $-0.37\pm 0.015$ & $1.32\pm 0.05$ & $2.1\pm 0.4$ \\
	& $+5.5$ & 12.2 & $-0.36\pm 0.02$ & $-0.15\pm 0.08$ & $-0.35\pm 0.02$ & $1.31\pm 0.07$ & $3.0\pm 1.0$ \\
	N3619$^c$ & 20 & $8.9\pm 0.5$ & $-0.40\pm 0.02$ & $+0.02\pm 0.13$ & $-0.24\pm 0.02$ & $1.37\pm 0.05$ & $5.3\pm 0.9$  \\
	\enddata
	\tablenotetext{a}{Galaxy ID -- N=NGC, U=UGC, I=IC, P=PGC}
	\tablenotetext{b}{CALIFA data}
	\tablenotetext{c}{CALIFA+SCORPIO data}
\end{deluxetable}

\setcounter{table}{4}
\begin{table*}
	\caption{Excitation and metallicity diagnostic results}
        \label{table_gasoxy}
        \centering
	\scriptsize
	\begin{tabular}{lccccccccc}
	      \hline
	      \hline
		Name & $R, ''$ & $E(B-V)$ & $n_e$, cm$^{-3}$ & $BPT_{N2}$ & $BPT_{S2}$ & \multicolumn{4}{c}{$12+\log(\mathrm{(O/H)})$} \\
		&  &  &  & &  & S  & O3N2 & N2 & D16 \\
		\hline
		IC 5285 & -12.0 &  -- &  -- & -0.17 &  -- &  -- & $8.50 \pm 0.02$ & $8.65 \pm 0.04$ &  -- \\ 
		IC 5285 & 16.0 &  -- &  -- & -0.40 &  -- &  -- & $8.54 \pm 0.02$ & $8.59 \pm 0.03$ &  -- \\ 
		IC 5285 & 31.0 &  -- &  -- & -0.37 &  -- &  -- & $8.54 \pm 0.05$ & $8.60 \pm 0.03$ &  -- \\ 
		\hline
		UGC 12840 & -20.0 & $0.61 \pm 0.13$ &  -- & -0.40 &  -- &  -- & $8.51 \pm 0.04$ & $8.57 \pm 0.03$ &  -- \\ 
		\hline
		NGC 774 & 9.0 &  -- & 1320 & -0.01 & -0.11 & $8.75 \pm 0.04$ & $8.46 \pm 0.08$ &  -- & $8.97 \pm 0.10$ \\ 
		\hline
		NGC 2551 & -29.0 & $0.50 \pm 0.46$ &  -- & -0.28 & -0.11 & $8.52 \pm 0.13$ & $8.41 \pm 0.08$ & $8.53 \pm 0.08$ & $8.55 \pm 0.25$ \\ 
		NGC 2551 & -24.0 & $0.48 \pm 0.21$ &  -- & -0.33 & -0.20 & $8.59 \pm 0.13$ & $8.52 \pm 0.04$ & $8.60 \pm 0.07$ & $8.65 \pm 0.24$ \\ 
		NGC 2551 & -20.0 & $0.70 \pm 0.43$ &  -- & -0.38 & -0.16 & $8.51 \pm 0.16$ & $8.45 \pm 0.05$ & $8.54 \pm 0.09$ & $8.52 \pm 0.31$ \\ 
		NGC 2551 & -17.0 & $0.48 \pm 0.54$ & 50 & -0.42 & -0.14 & $8.50 \pm 0.14$ & $8.51 \pm 0.06$ & $8.56 \pm 0.07$ & $8.48 \pm 0.24$ \\ 
		NGC 2551 & -14.0 & $0.25 \pm 0.05$ &  -- & -0.27 & -0.08 & $8.58 \pm 0.08$ & $8.47 \pm 0.03$ & $8.59 \pm 0.05$ & $8.59 \pm 0.15$ \\ 
		NGC 2551 & 12.0 & $0.39 \pm 0.13$ & 120 & -0.05 & 0.09 & $8.69 \pm 0.03$ & $8.46 \pm 0.03$ &  -- & $8.67 \pm 0.05$ \\ 
		NGC 2551 & 15.5 &  -- & 390 & 0.11 & 0.40 & $8.76 \pm 0.13$ & $8.47 \pm 0.08$ &  -- & $8.54 \pm 0.18$ \\ 
		NGC 2551 & 19.0 & $0.45 \pm 0.34$ & 190 & -0.11 & 0.09 & $8.65 \pm 0.08$ & $8.47 \pm 0.05$ & $8.65 \pm 0.05$ & $8.60 \pm 0.13$ \\ 
		NGC 2551 & 23.0 & $0.10 \pm 0.08$ & 50 & -0.22 & -0.14 & $8.59 \pm 0.02$ & $8.42 \pm 0.02$ & $8.57 \pm 0.03$ & $8.69 \pm 0.04$ \\ 
		NGC 2551 & 27.0 & $0.29 \pm 0.41$ & 20 & -0.12 & -0.02 & $8.65 \pm 0.10$ & $8.45 \pm 0.04$ & $8.63 \pm 0.06$ & $8.69 \pm 0.19$ \\ 
		\hline
		NGC 2962 & -63.0 &  -- & 940 &  -- &  -- &  -- &  -- & $8.47 \pm 0.08$ & $8.21 \pm 0.19$ \\ 
		\hline
		NGC 3106 & 21.0 & $0.43 \pm 0.04$ & 300 & -0.69 & -0.58 & $8.64 \pm 0.06$ & $8.64 \pm 0.02$ & $8.59 \pm 0.03$ & $8.87 \pm 0.19$ \\ 
		\hline
		NGC 3166 & -98.0 &  -- & 80 & -0.44 & -0.14 & $8.47 \pm 0.03$ & $8.47 \pm 0.02$ & $8.53 \pm 0.03$ & $8.44 \pm 0.05$ \\ 
		NGC 3166 & -72.0 & $0.31 \pm 0.04$ & 20 & 0.50 & 0.62 & $-$ & $8.42 \pm 0.01$ &  -- & $8.76 \pm 0.03$ \\ 
		NGC 3166 & -27.0 &  -- & 280 & 0.11 & 0.28 & $-$ & $8.53 \pm 0.06$ &  -- & $8.71 \pm 0.04$ \\ 
		NGC 3166 & 14.0 &  -- & 280 & 0.52 & 0.64 & $-$ & $8.49 \pm 0.09$ &  -- & $8.82 \pm 0.07$ \\ 
		NGC 3166 & 28.0 & $0.56 \pm 0.06$ & 60 & -0.21 & -0.06 & $8.66 \pm 0.03$ & $8.54 \pm 0.02$ &  -- & $8.67 \pm 0.05$ \\ 
		\hline
		NGC 3182 & -5.5 &  -- & 90 & -0.39 & -0.23 & $8.57 \pm 0.02$ & $8.52 \pm 0.02$ & $8.58 \pm 0.03$ & $8.62 \pm 0.03$ \\ 
		NGC 3182 & 5.5 & $0.04 \pm 0.29$ & 100 & -0.35 & -0.17 & $8.56 \pm 0.02$ & $8.49 \pm 0.02$ & $8.58 \pm 0.03$ & $8.59 \pm 0.03$ \\ 
		\hline
		NGC 3619 & 20.0 & $0.53 \pm 0.15$ & 50 & -0.33 & -0.03 & $8.50 \pm 0.02$ & $8.45 \pm 0.03$ & $8.56 \pm 0.03$ & $8.44 \pm 0.03$ \\ 
		\hline
	\end{tabular}
\end{table*}

The results of the excitation analysis and oxygen abundance estimates are {\bf based on the observed flux ratios presented in Table~\ref{tab_fluxes} and}
are summarized in Table~\ref{table_gasoxy}, where different columns correspond
to: name of the galaxy; projected distance of the current clump from the galactic center, in orientation according to Fig.~\ref{fig_ls};
the color excess $E(B-V)$ derived from the observed Balmer decrement; electron density $n_e$ derived from the [SII]6717/6731 flux ratios {\bf \citep{osterbrock}
assuming $T_e=10000K$}; $BPT_{N2}$ and $BPT_{S2}$ parameters -- the shifts from the `maximum starburst line' at the corresponding BPT diagram; the oxygen
abundance $\mathrm{12+\log(O/H)}$ derived using four methods described above. The listed uncertainties of the oxygen abundances derived do not include
systematic uncertainties of the methods, which should be added quadratically. 

As it follows from Table~\ref{table_gasoxy}, for the most of the starforming clumps all the methods yield the values of oxygen abundance
consistent with each other within their uncertainties. The strong discrepancies are observed for the clumps in NGC~2551 
and NGC~3166 demonstrating non-HII-like mechanism of excitation ($BPT_{S2}>0$ and/or $BPT_{N2}>0$) -- the estimated
values of oxygen abundance are not reliable for them. High discrepancy is also observed for the only clump in NGC~774 
locating in the transition zone of the BPT-diagram with the N2 abscissa axis and having the highest electron density
among all studied objects.

\section{Discussion}
\label{sec:5} 

\subsection{Starforming rings in the disk planes of S0 galaxies}

\begin{figure*}
\includegraphics[width=8cm]{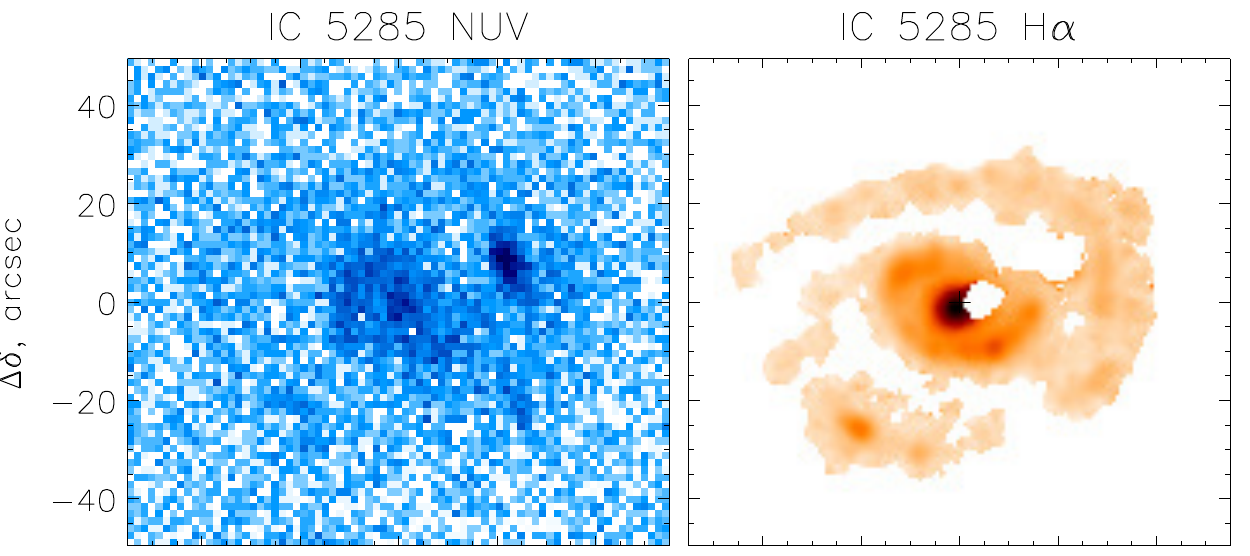} \includegraphics[width=8cm]{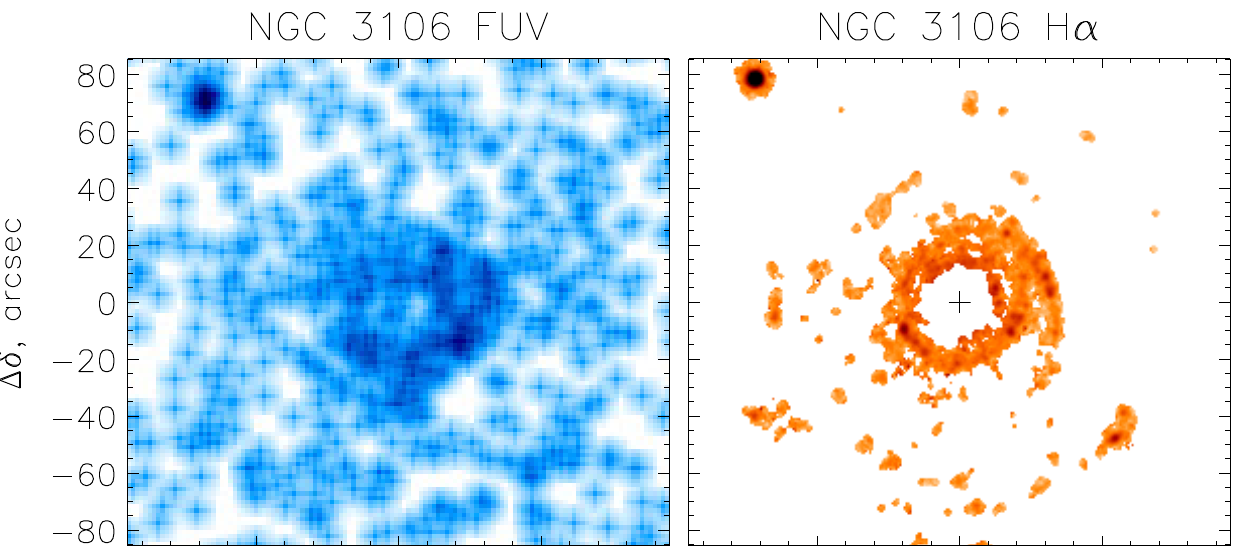}\\
 \includegraphics[width=8cm]{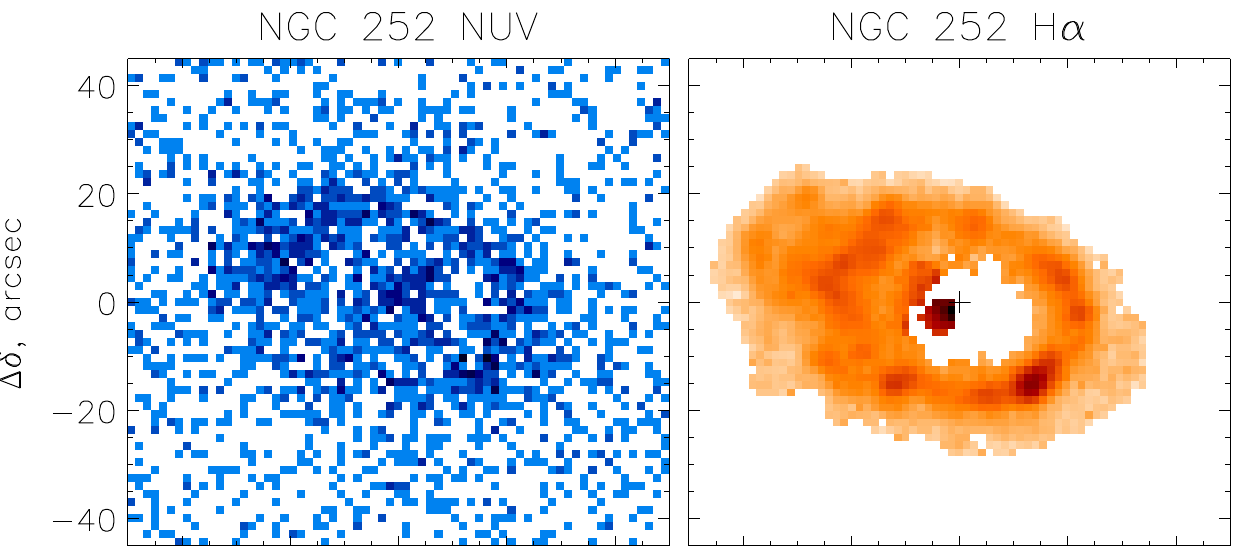} \includegraphics[width=8cm]{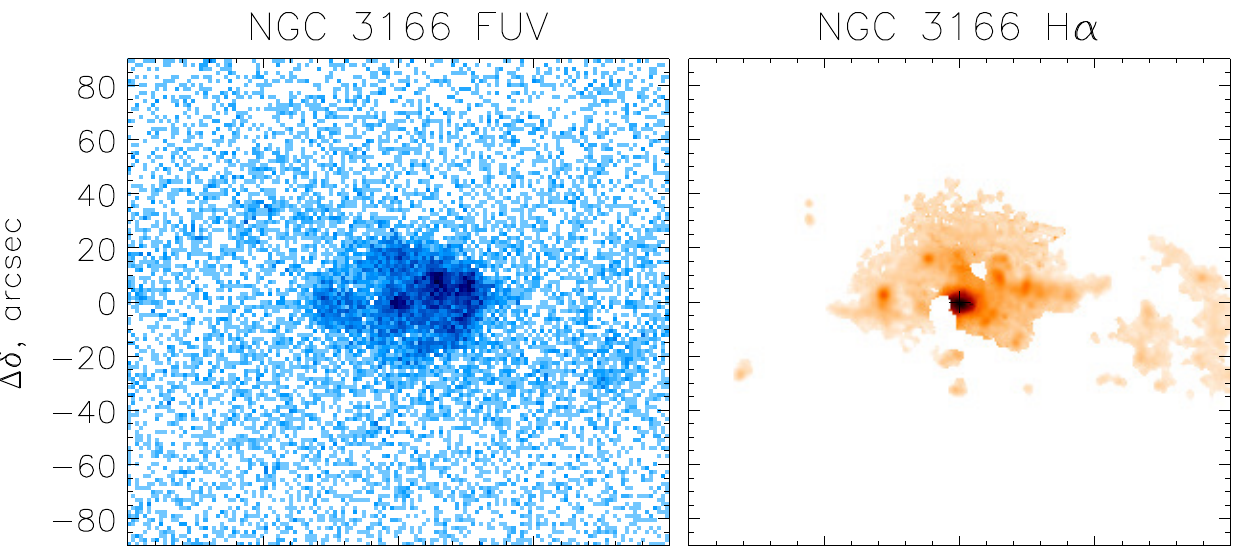}\\
 \includegraphics[width=8cm]{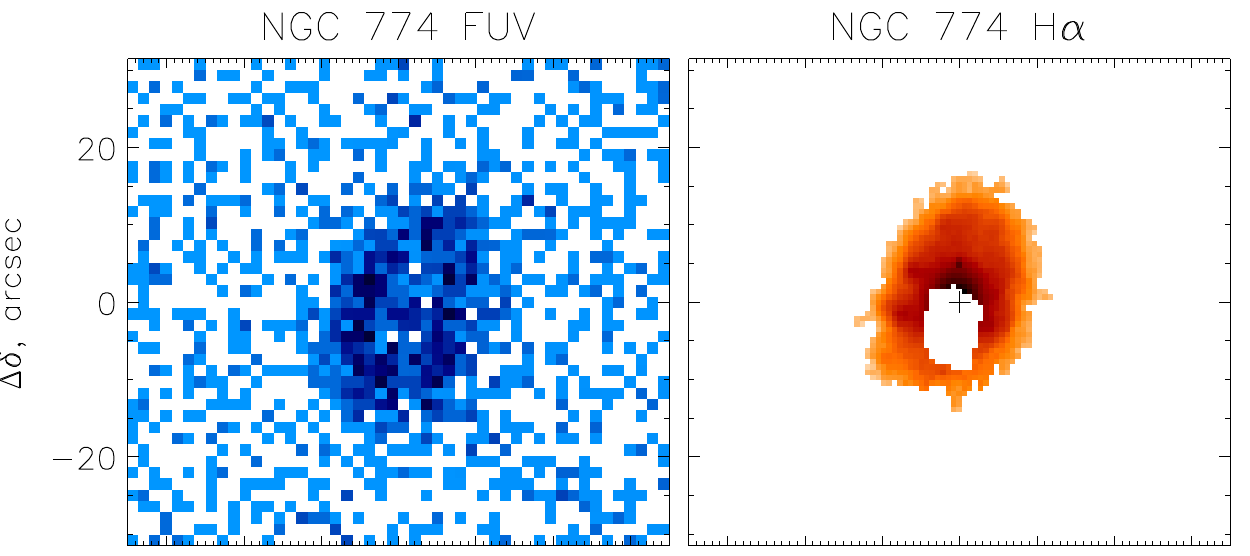} \includegraphics[width=8cm]{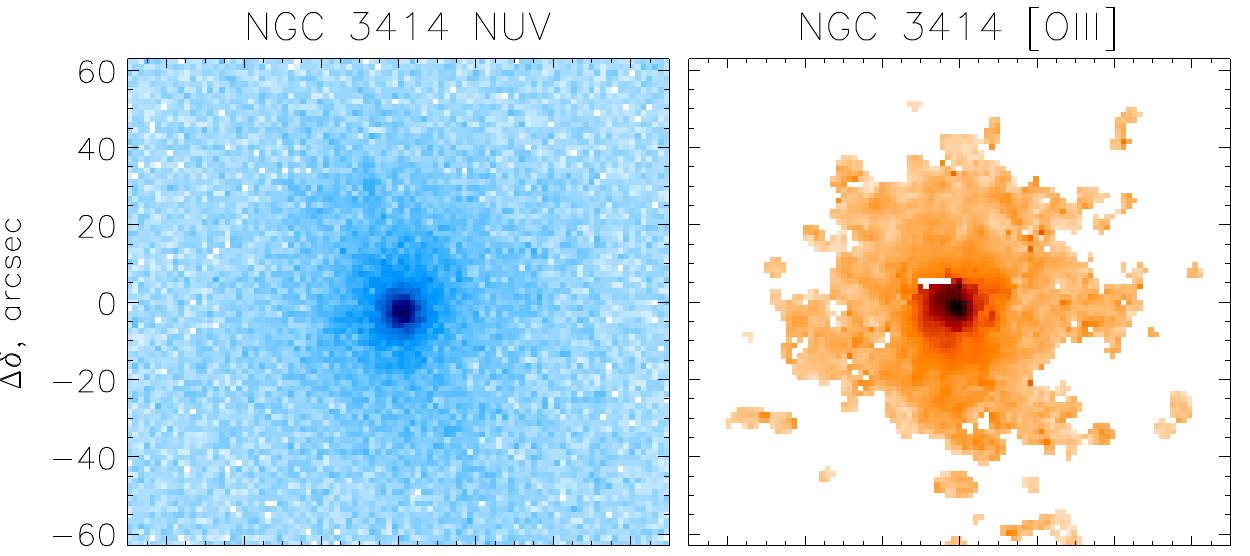}\\
 \includegraphics[width=8cm]{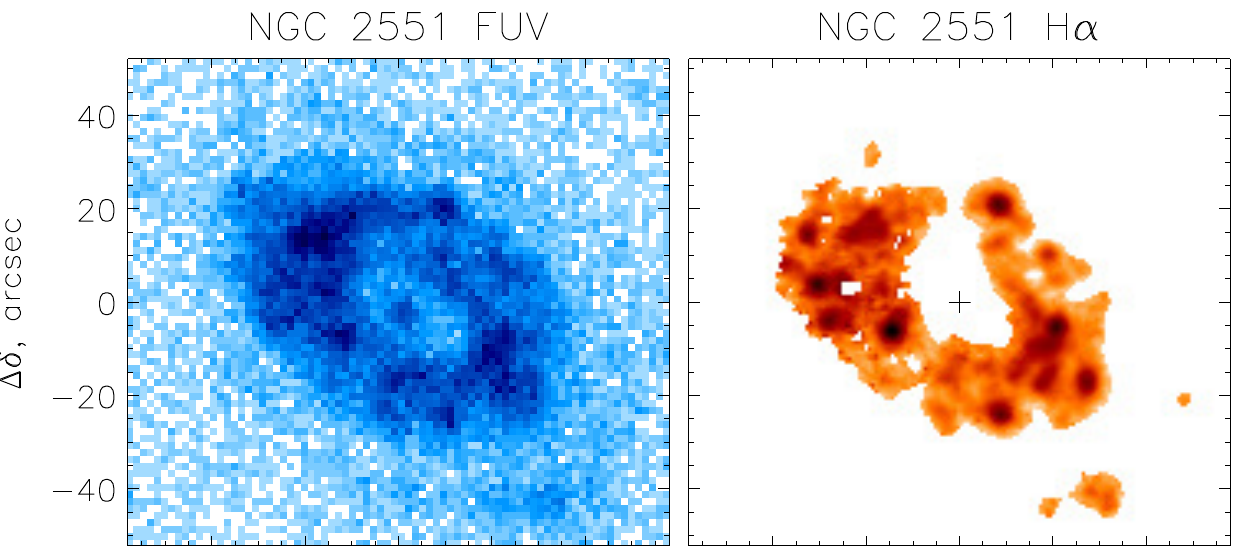} \includegraphics[width=8cm]{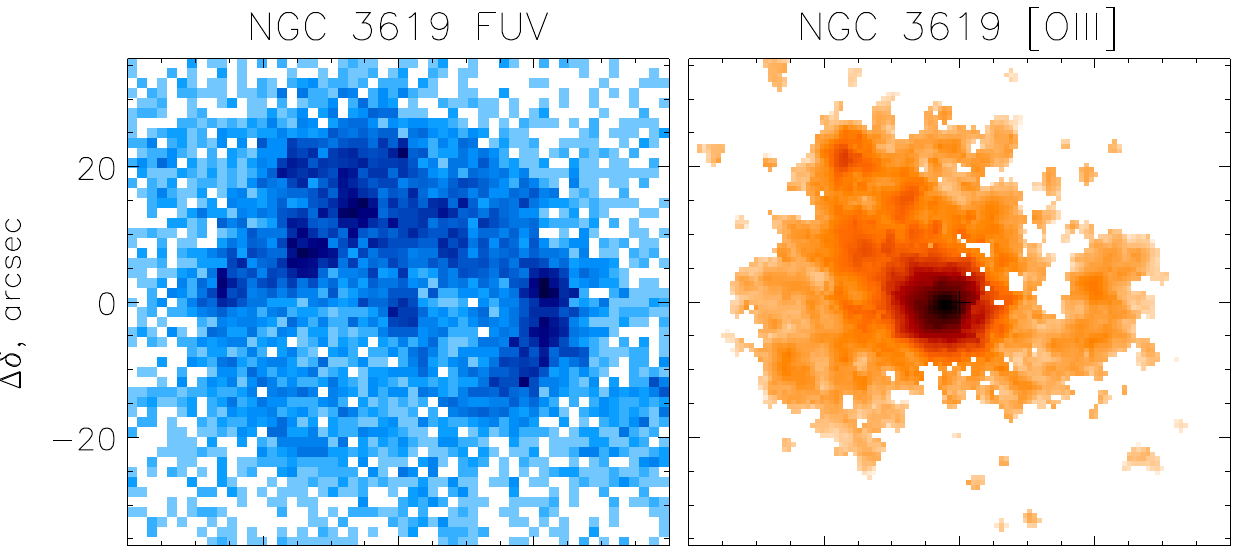}\\
 \includegraphics[width=8cm]{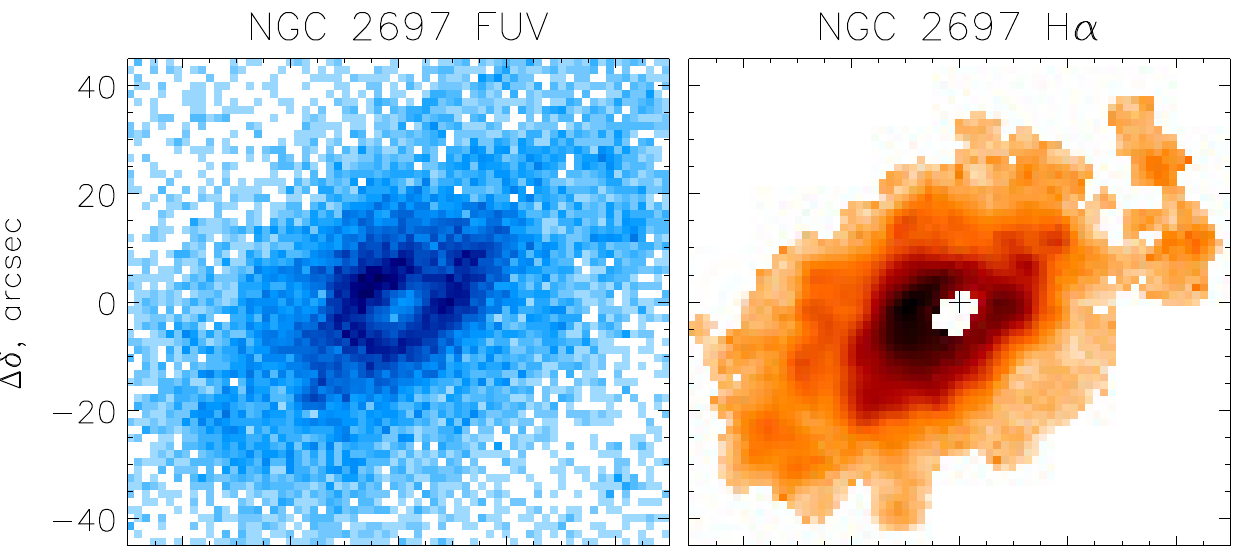} \includegraphics[width=8cm]{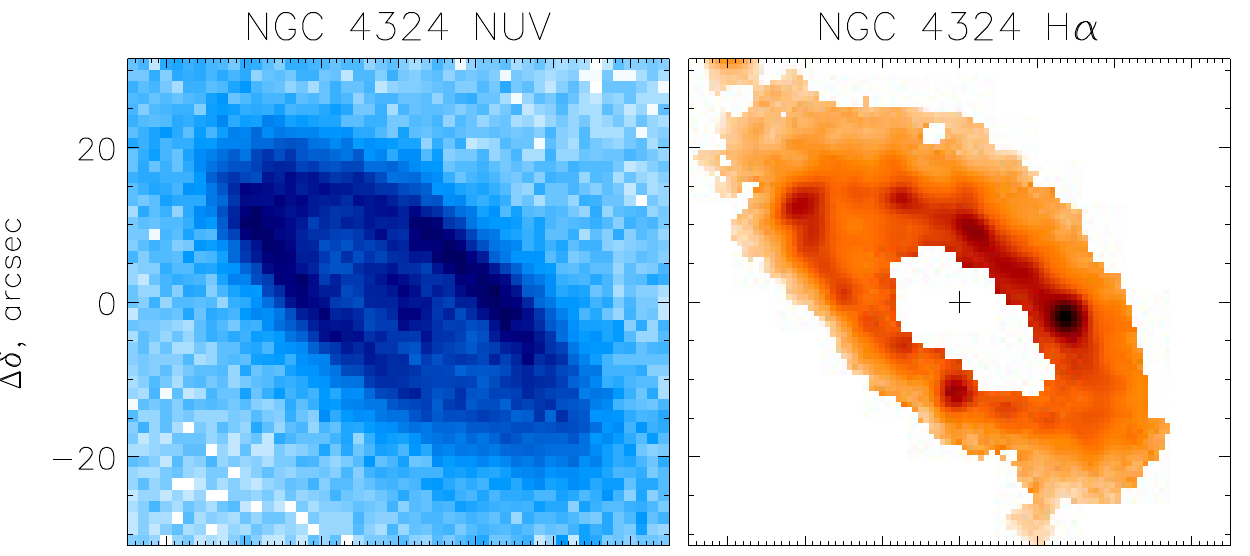}\\
  \includegraphics[width=8cm]{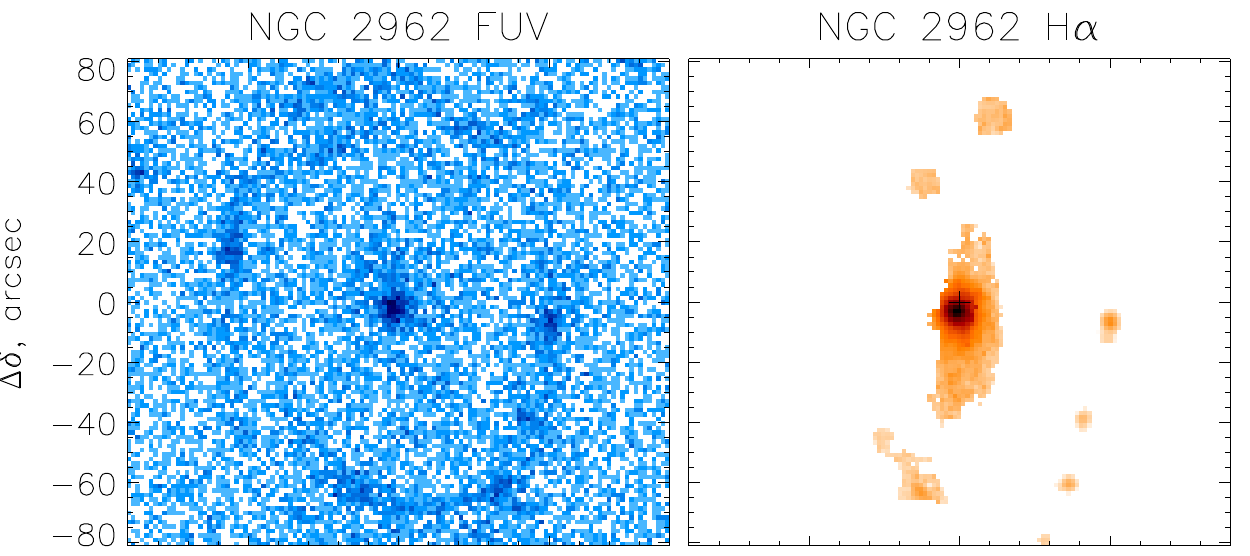} \includegraphics[width=8cm]{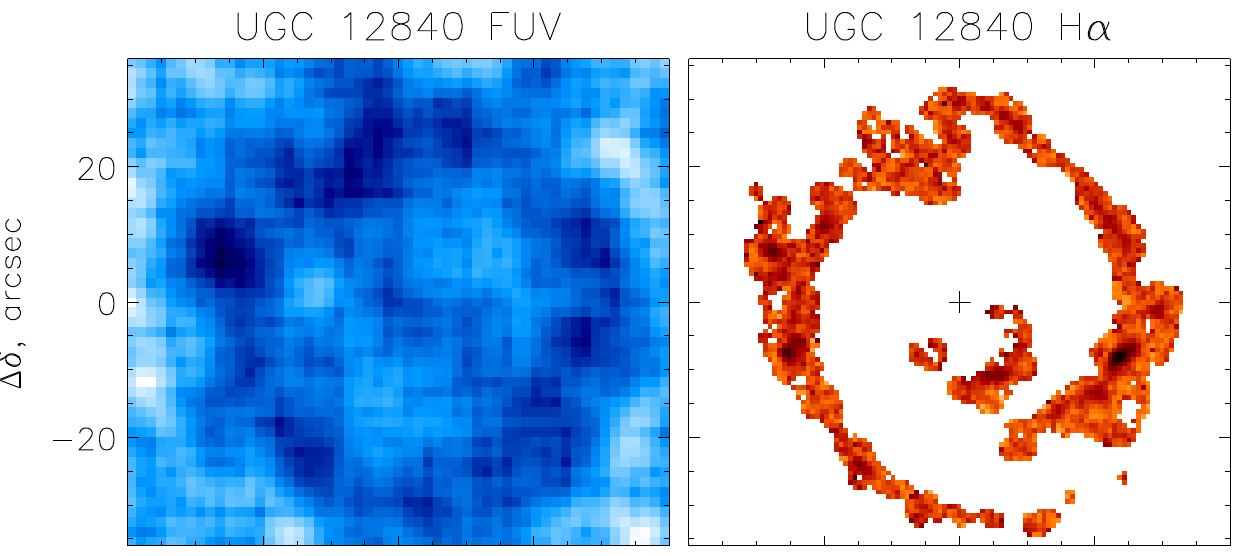}\\
\caption{GALEX images of the sample galaxies observed in the FUV and sometimes in the NUV (if the deep FUV data was not available) bands.
The emission-line images from the SCORPIO/FPI of the 6-m telescope are also shown for comparision. 
The scale is a square root of intensity. The starforming rings are detected in the most cases.}
\label{galex}
\end{figure*}

We have checked the galaxies of our sample in the MAST Archive to see if they are detected in the ultraviolet
with the GALEX. Indeed, the most of them (12 of 18) are detected in the FUV-band of the GALEX survey \citep{galex}, and their UV-morphology 
looks usually ring-like. The UV images of the sample galaxies are shown in Fig.~\ref{galex}. High-contrast ring-like UV structures are clearly
visible even in galaxies without rings of very recent star formation traced by the optical emission lines
(e.g. NGC~774, NGC~3619, etc.).\footnote{The central `holes'  in the FPI maps of some galaxies (IC~5285, NGC~774, etc.) are caused 
by strong \Ha absorption line in the galactic spectra produced by underlying stellar population. We could not take this effect into account, 
because the spectral range of the FPI data is too small.} The UV-rings in NGC~3619 and NGC~4324 
were reported earlier by \citet{cortese09}, this in NGC~2962 -- by \citet{marino}, and the UV-ring in NGC~252 was earlier studied 
by us through long-slit spectroscopy with the SCORPIO of the Russian 6m telescope in the optical spectral range \citep{ringmnras}. 
The results of the gas excitation analysis within the area of UV-bright rings have shown that the UV-bright rings in S0 galaxies
have mostly emission-line spectra typical for the gas excited by young massive stars betraying current star formation. However there 
are two interesting cases where we have resolved a change of the gas excitation mechanism along the radius. In NGC~2551 at
the {\it inner northern} edge of the starforming ring series, $R=12\arcsec -19\arcsec$, the excitation mechanism is composite -- 
the excitation by young stars and by shock waves are both involved, and the electron density is increased just at this radius. 
In NGC~3166 the distribution of the UV surface brightness is not symmetric with respect to the center, it is lopsided with the most 
UV radiation emitted to the west from the nucleus. And the only radial zone in NGC~3166 which demonstrates the HII-region-like emission-line 
spectrum is the western distant (tidal?) arm. All other emission-line rings in NGC~3166 have shock-like or composite gas excitation.
Despite the presence of a large-scale bar in NGC~3166, we would conclude therefore that its ring-like emission-line structures have probably
the tidal  or impact origin being a result of interaction with the neighboring members of the group, NGC~3169 and NGC~3165,
and of cold gas accretion from the common \HI\ group envelope (see below, Subsection 5.2).

We must discuss here that not only in the case of shocks, but also in a case of {\bf DIG domination} the enhanced [NII], [SII], and [OIII] lines
might be observed. Among the sources considered to be responsible for DIG excitation are the radiation of old hot pAGB stars \citep{sokbh1991, pagb94},
the leakage of hard photons from the star-forming regions \citep{hoops03}, and the shock waves (see discussion in \citealt{Zhang2017}). As it was argued by,
e.g., \cite{ff_dig}, low-mass evolved stars might be relevant sources of the gas ionization in S0 galaxies. To consider if the possible contribution
of the DIG affects our conclusion about the presence of shocks in some regions of NGC~2551 and NGC~3166 we made a following
analysis. We identified the regions crossed by our slits that might be related to the DIG. Various criteria of this exists in the literature:
\citet{Zhang2017} proposed to use the H$\alpha$ emission-line surface luminosity as an indicator; after conversion to the observed flux per pixel,
$\mathrm{F(H\alpha) < 7.4\times10^{-17}\ erg/s/cm^{-2}}$ should be considered as DIG excited by old stars. In the same time, according to
\citet{Lacerda2018}, $\mathrm{EW(H\alpha) < 3 }$ \AA\, is expected to be a more precise indicator. In figure~\ref{fig_lsfit} we demonstrate
the obtained radial profiles of the logarithm of the H$\alpha$ line flux, EW(H$\alpha$), and emission lines flux ratios for each galaxy.
In the second panel from the top we marked those regions showing $\mathrm{EW(H\alpha) > 3}$ \AA\, (and hence surely not related to old stars)
by squares, while the regions that might be considered as the DIG are shown by circles. Note that both mentioned criteria can
disagree in particular regions. Since we expect a uniform distribution of the old stars in a galaxy, we may also expect a similar
distribution of their contribution to the ionization. After excluding the areas with $\mathrm{EW(H\alpha) >3}$ and the central part of
the galaxies ($R<5\arcsec$ for NGC~2551, $R<10\arcsec$ for NGC~3166; the ionization conditions in these regions might significantly differ
from the rest of the galaxy, and it is out of scope of our analysis) we have performed a second-order robust polynomial fitting of the
flux logarithm distribution along the slit. The result is shown in the second panel of the Fig.~\ref{fig_lsfit} by solid line. The same fitting
was performed for every emission line analyzed; the obtained smooth models were subtracted from the observed fluxes. After that we have compared
the emission-line flux ratios in the BPT-diagrams (bottom panel of Fig.~\ref{fig_lsfit}) before (red color) and after (blue color) correction for the DIG
contribution. As it follows from these plots, the performed correction changes the position in the BPT-diagrams of the points classified by us as DIG,
but almost does not influence the flux ratios in the regions with $\mathrm{EW(H\alpha) > 3}$ \AA\, that were considered earlier
as shocks or star-forming rings. Hence, we may conclude that the presence of the DIG doesn't change our conclusions
concerning the gas excitation mechanisms in the emission-line clumps. {\bf Note however that we cannot exclude the contribution of the filtered hardened
radiation from HII regions in star-forming rings to the enhancement of [NII], [SII], and [OIII] lines. The DIG component ionized by the filtered radiation
may be more structured than one ionized by low-mass evolved stars considered above. Hence, one may still expect that the DIG contribution is significant
in the studied clumps if there are nearby massive star-forming regions (like it is observed in spiral galaxies, e.g., in M51, \citealt{Greenawalt98} ).
However the models of DIG ionized exclusively by filtered quanta doesn't fully reproduce the enhanced [OIII] emission lines \citep{Weber19}, while at least
for NGC~3166 the ratio of [OIII]5007/H$\beta$ is clearly high in comparison with the clumps related to HII regions (see Figs.~\ref{fig_bpt}, \ref{fig_lsfit}).
So, the ionization by shocks in the clumps above `maximum starburst line' is a preferred mechanism at least for this galaxy.}

\begin{figure*}
\centering
\begin{tabular}{c c}
 \includegraphics[width=6cm]{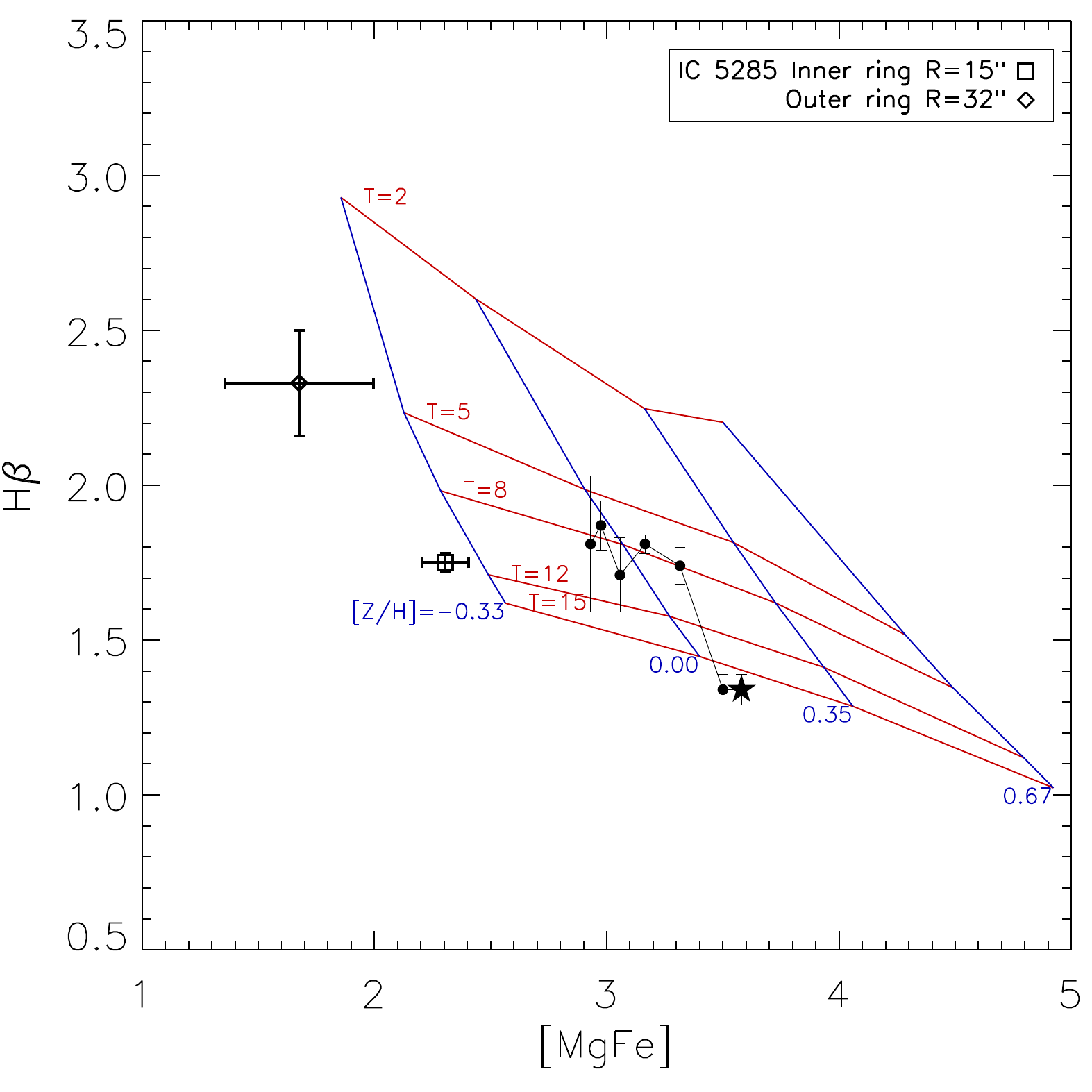} &
 \includegraphics[width=6cm]{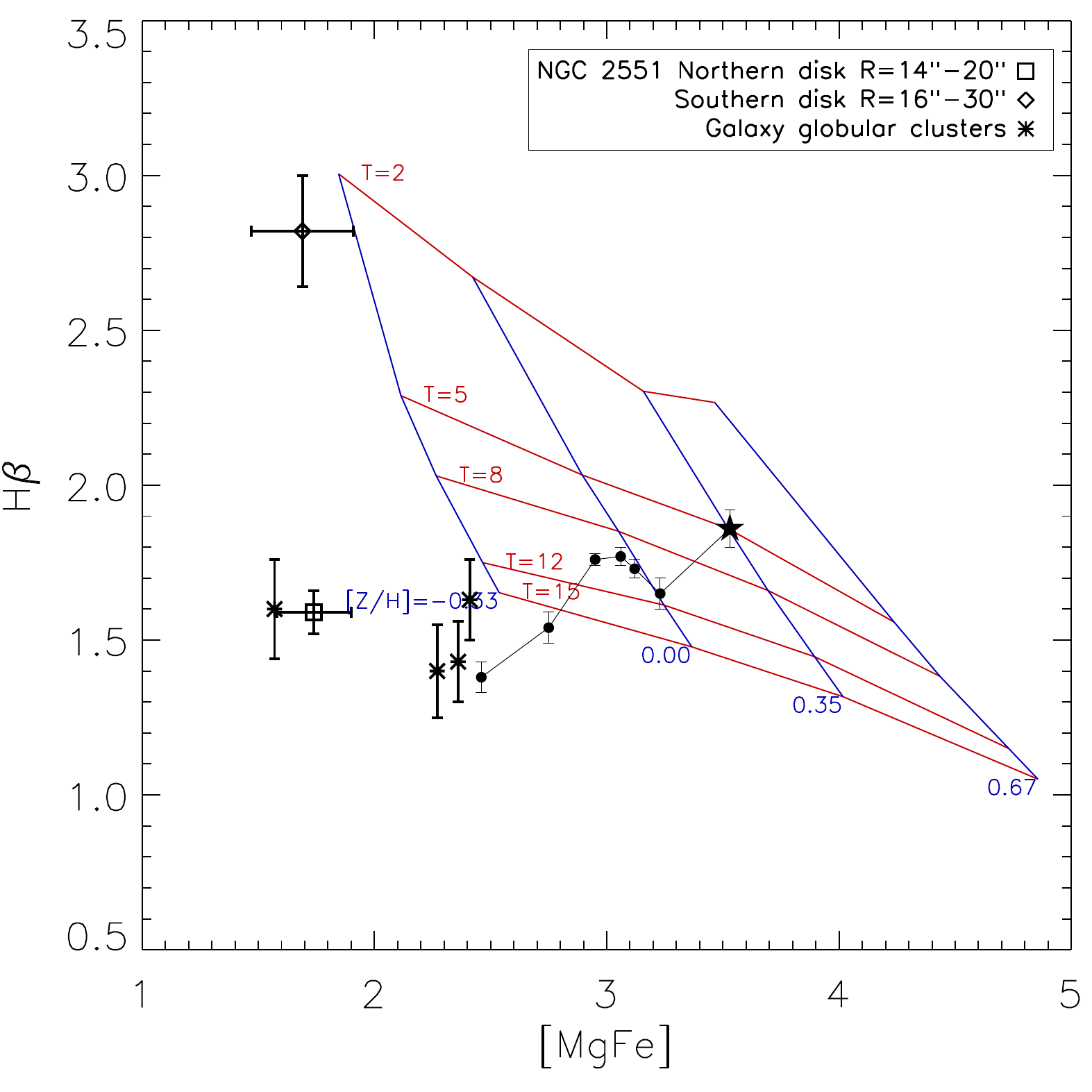} \\
\includegraphics[width=6cm]{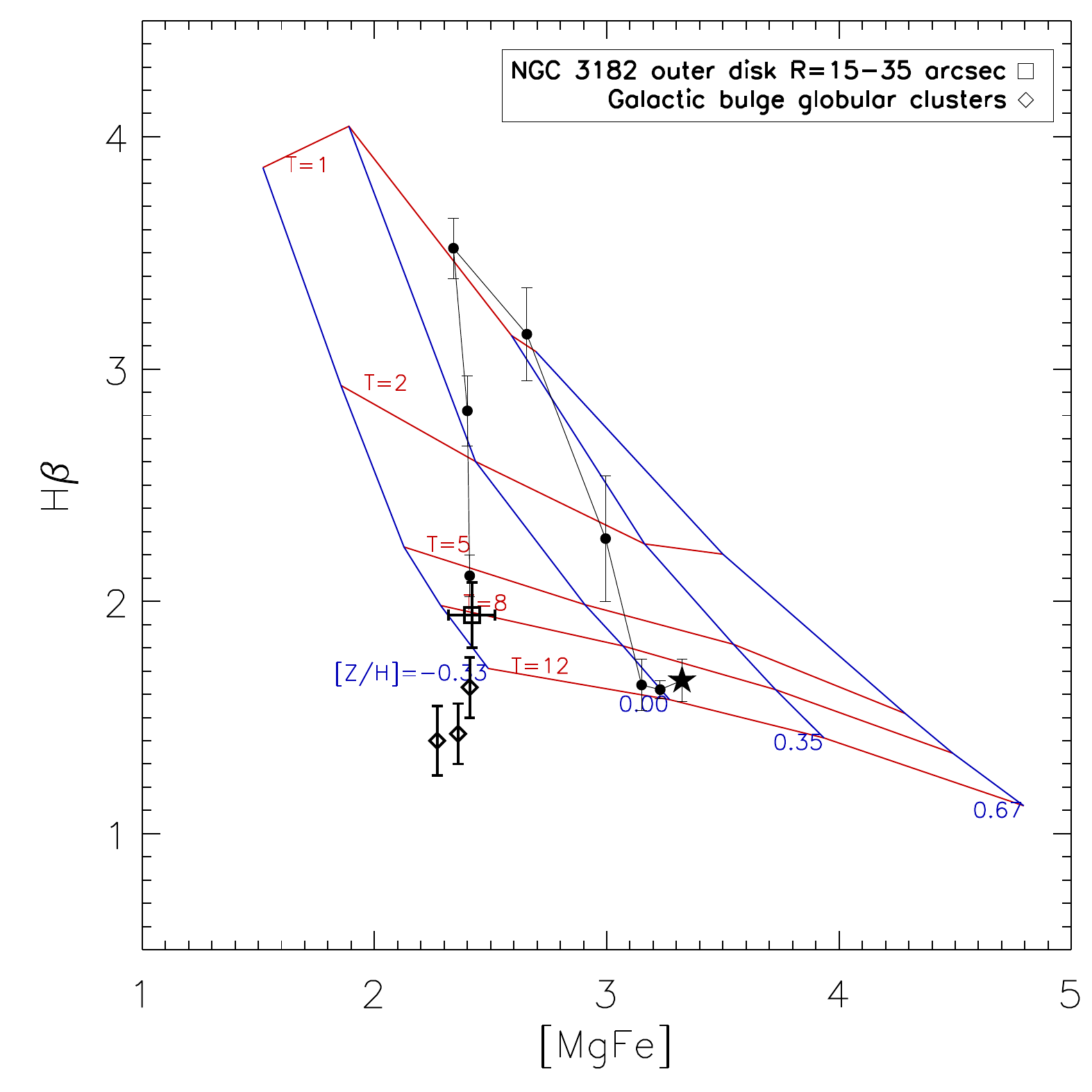} &
\includegraphics[width=6cm]{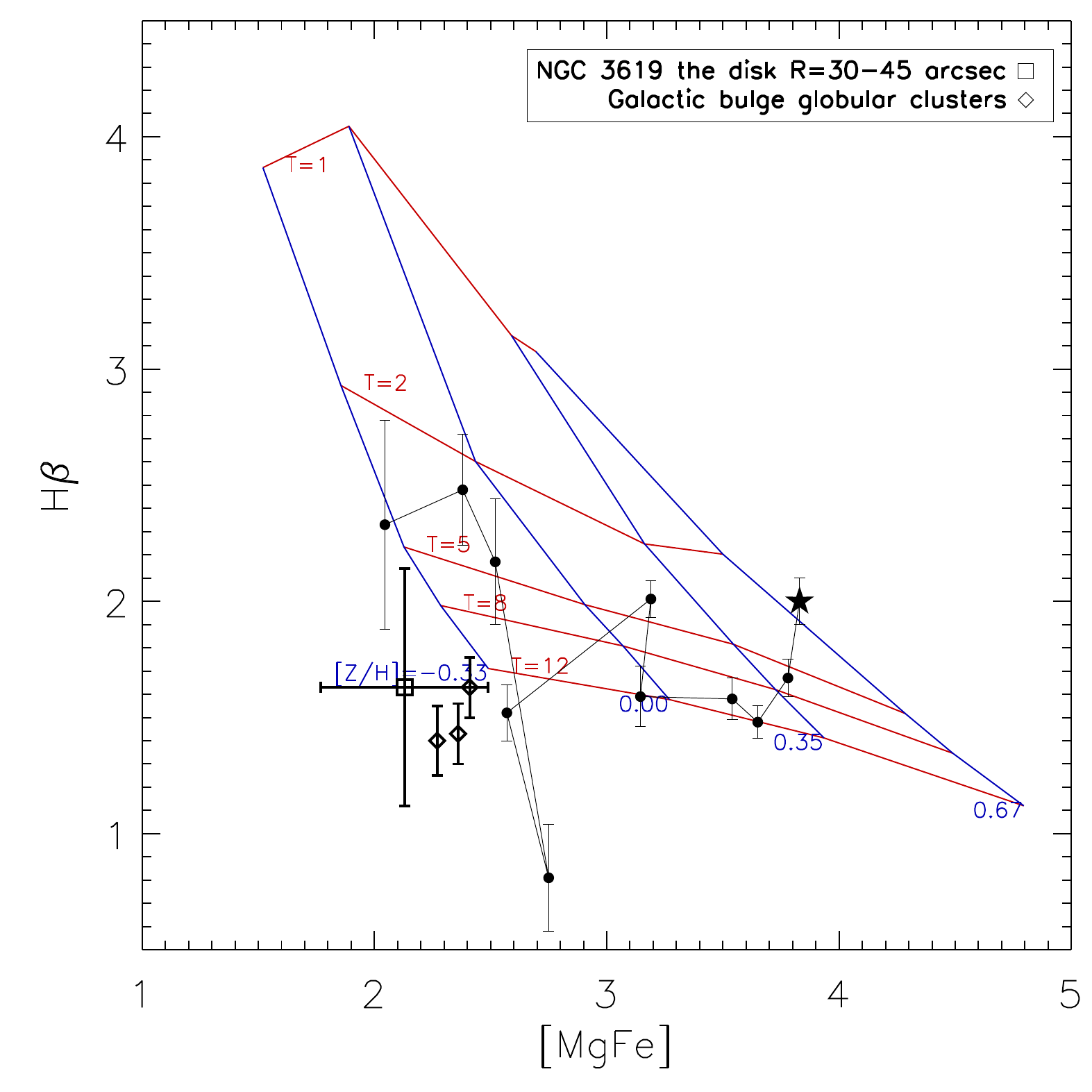} \\
\includegraphics[width=6cm]{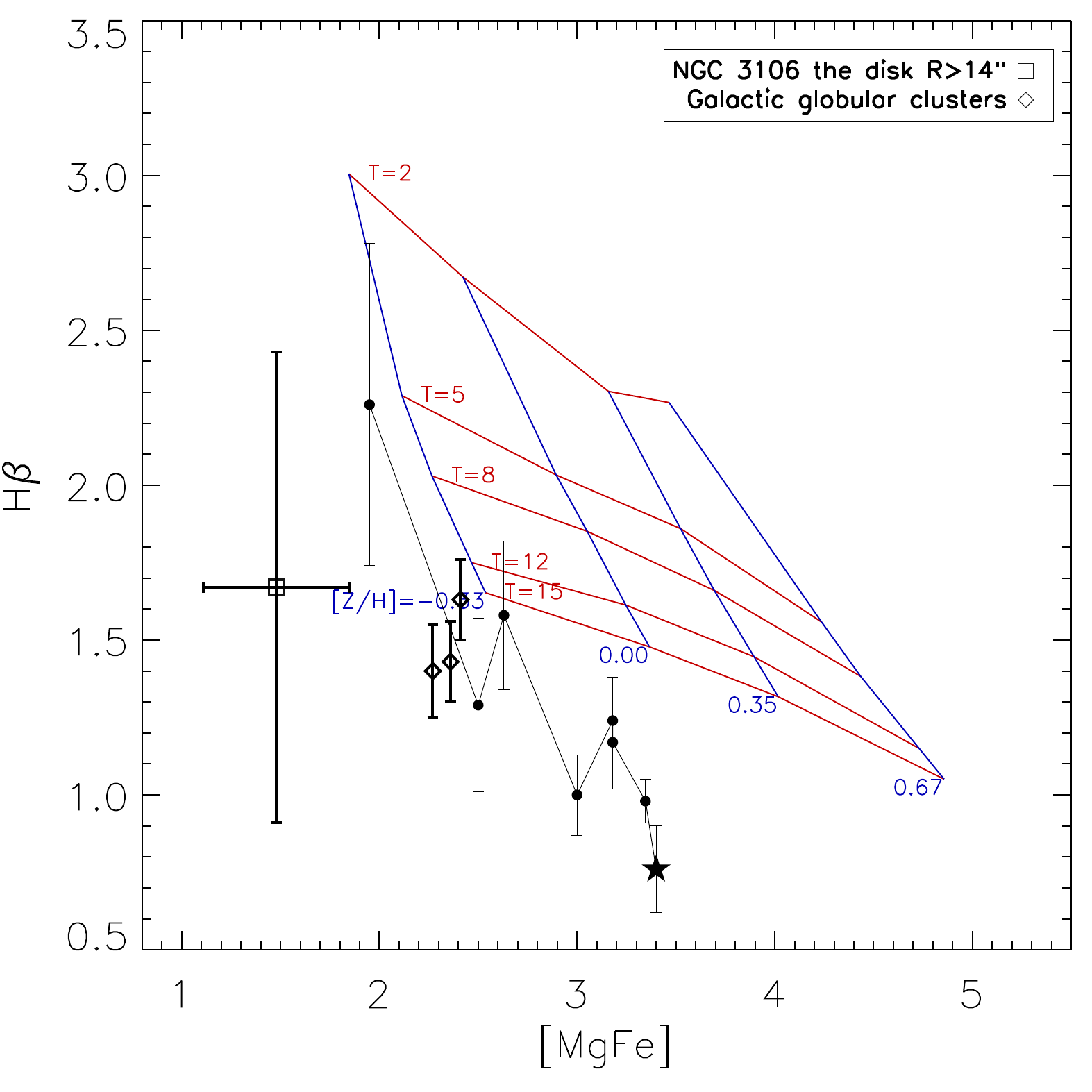} &
\includegraphics[width=6cm]{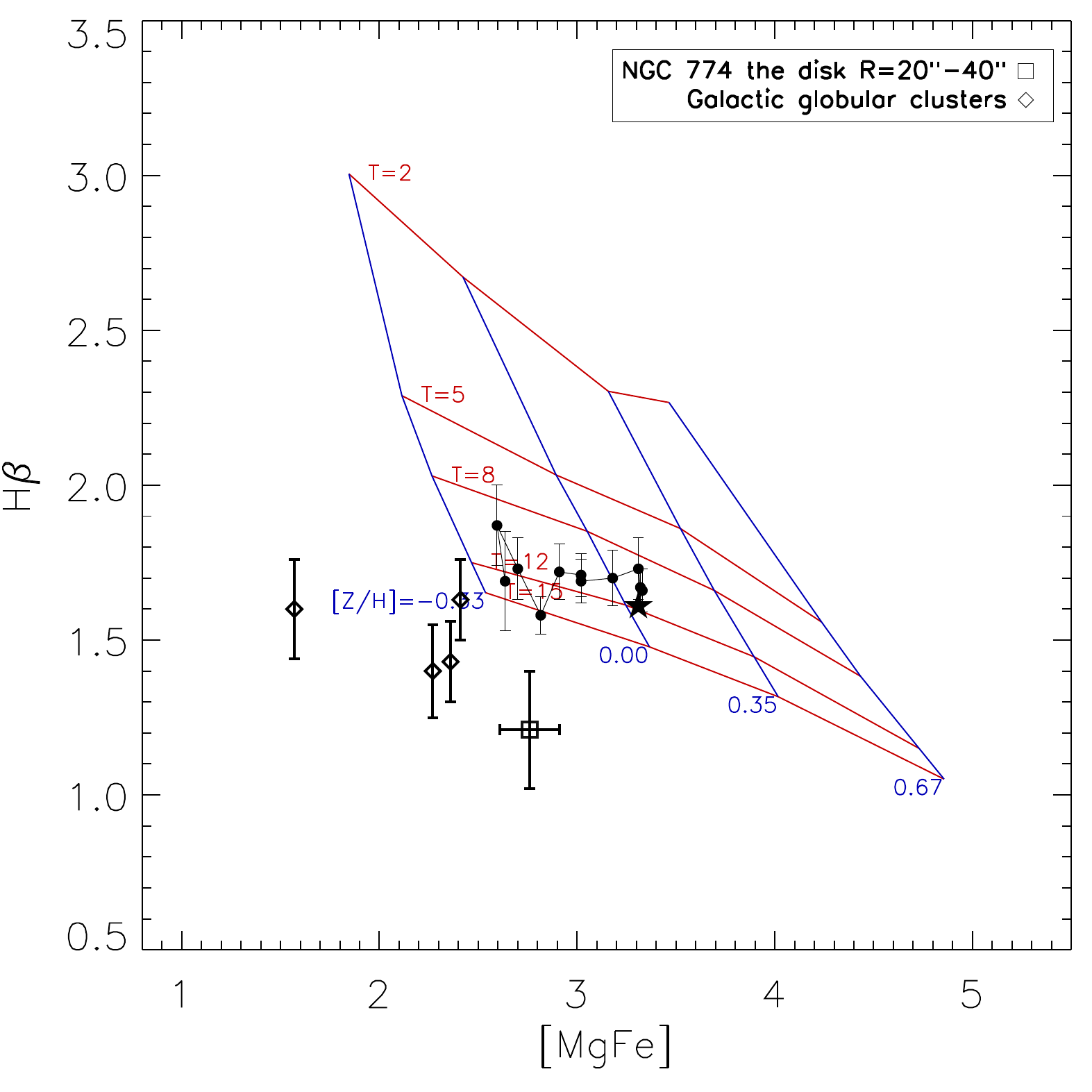} \\
\end{tabular}
\caption{Index--index diagrams for six galaxies. By confronting the H$\beta$ Lick index versus a combined metallicity index involving
magnesium and iron lines, we solve the metallicity-age degeneracy and determine these stellar population parameters with the SSP
evolutionary synthesis models produced by \citet{thomod}. Five different age sequences (red lines)
are plotted as reference frame; the blue lines crossing the model age sequences mark the metallicities of $+0.67$, $+0.35$, 0.00,
--0.33 from right to left. Large black stars correspond to the nucleus for every galaxy, and then we go along the
radius through the points R=1\arcsec, 2\arcsec, 3\arcsec, 4\arcsec, 6\arcsec, 9\arcsec, 12\arcsec $\dots$. A few
globular clusters from \citet{beasley} belonging to the Galactic bulge are also plotted for the reference frame.
}
\label{lickind}
\end{figure*}

\begin{figure*}
\centering
\begin{tabular}{c c}
 \includegraphics[width=8cm]{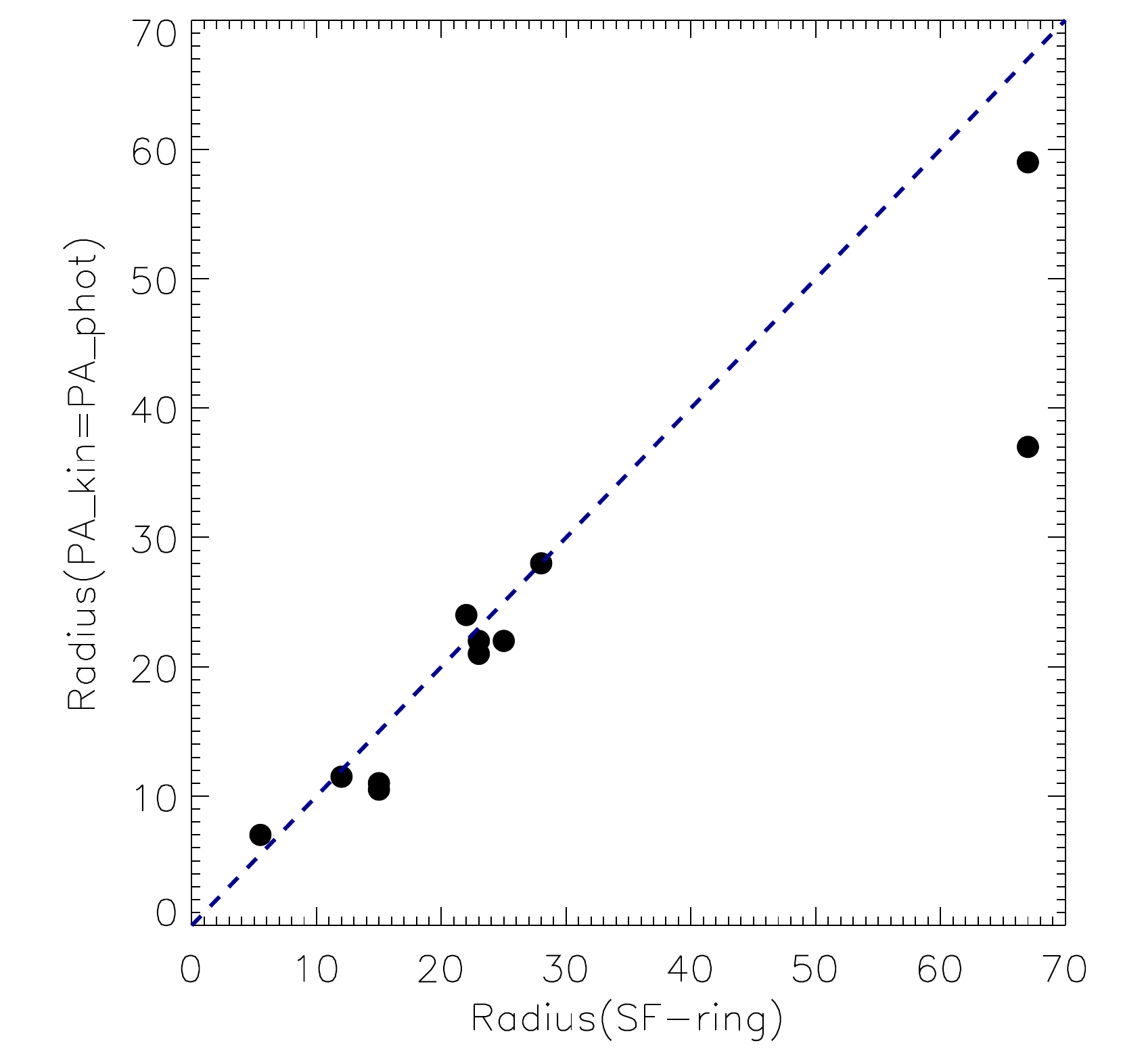} &
 \includegraphics[width=8cm]{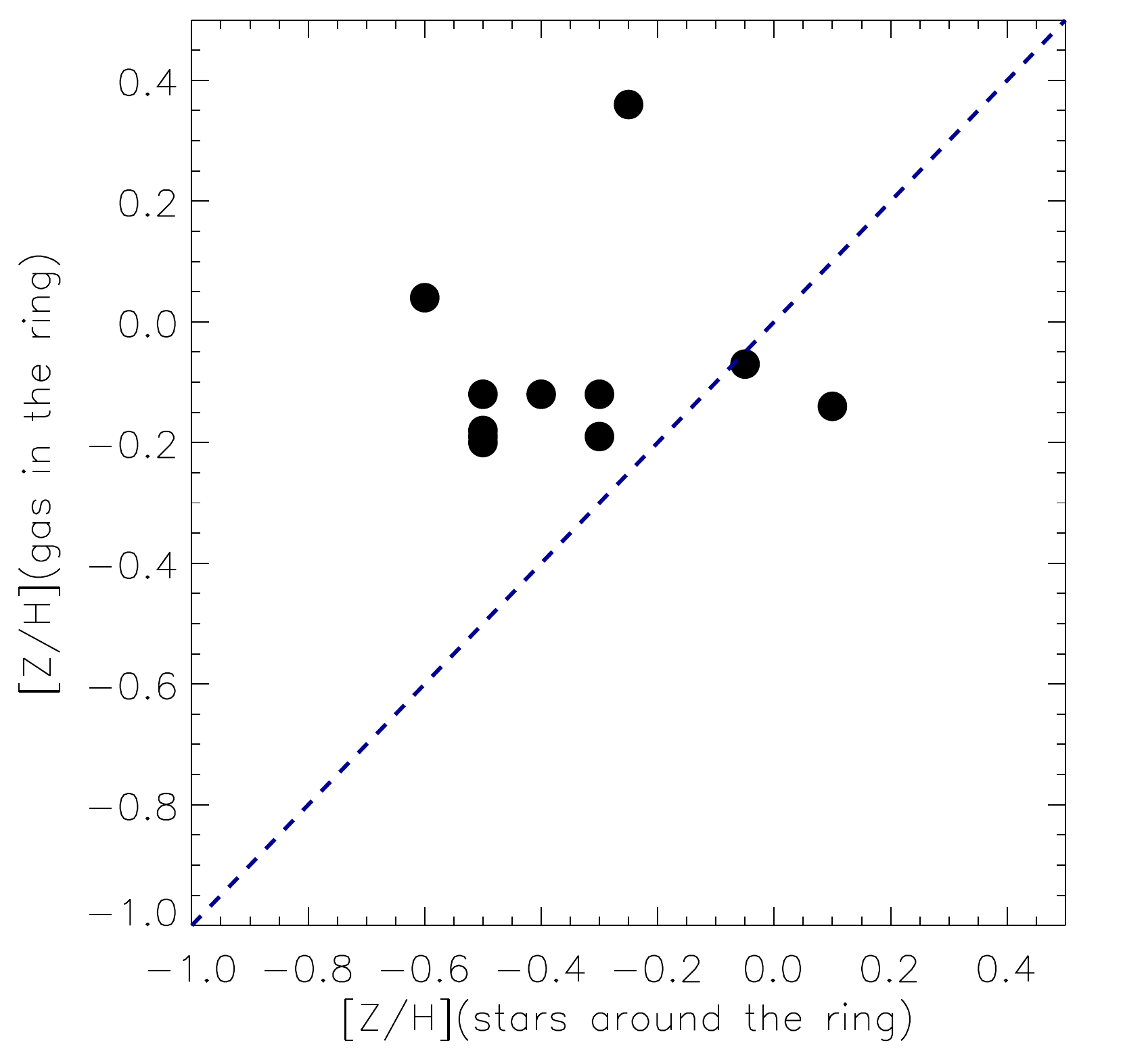} \\
\includegraphics[width=8cm]{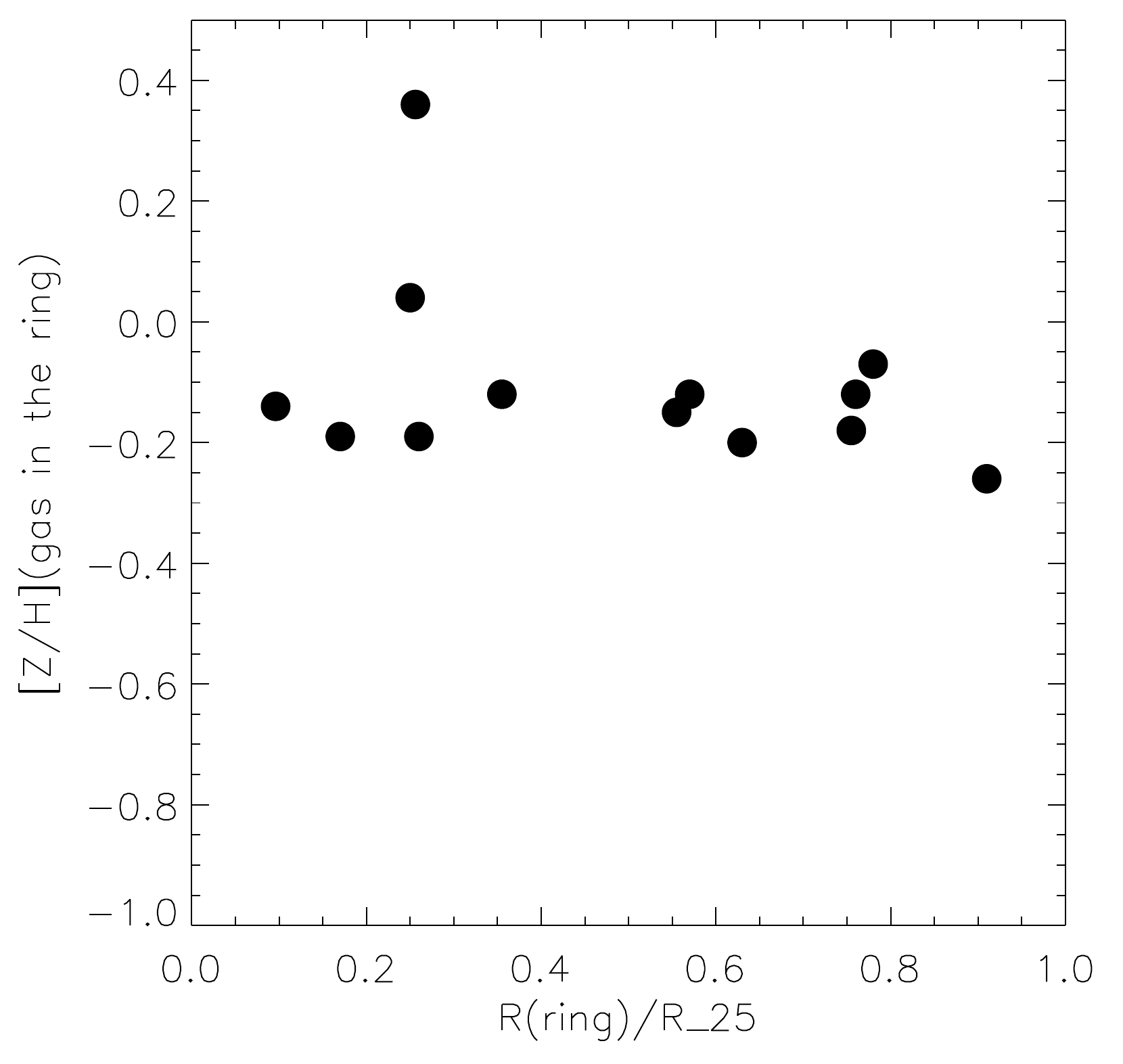} &
\includegraphics[width=8cm]{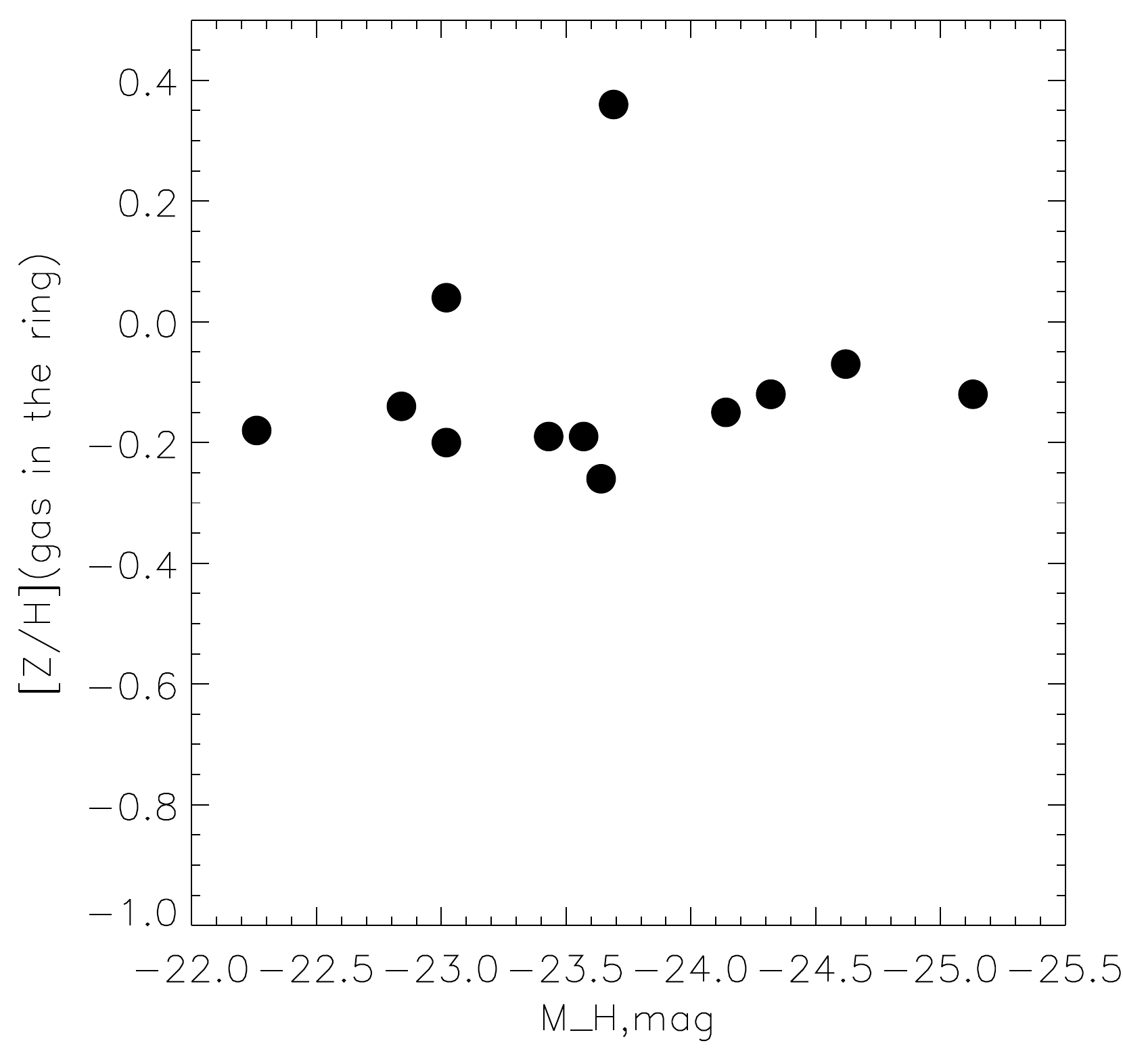} \\
\end{tabular}
\caption{Some correlations for the positions and metallicities of the star-forming rings; we treat here the oxygen abundance
of the gas as a measure of its total metallicity because the oxygen is the most abundant element among those heavier than the helium.}
\label{corring}
\end{figure*}

Figure~\ref{corring} presents some correlations (or the absence of any correlations) related to the UV-rings; for several galaxies
(NGC~2697, NGC~2551, IC~5285) we take into account the multiple UV-rings at various radii. The left upper plot
demonstrates that the radial localization of the UV-bright starforming rings is not casual: the radii of the rings coincide exactly
with the zones where the kinematical major axes are in agreement with the lines of nodes (Fig.~\ref{fig_pa}), or where the gas exhibits laminar
circular rotation in the main planes of the stellar galactic disks. The only point which is not close to the equality line is NGC~2962 where the
outer ring is traced by our scanning Fabry-Perot only partly due to its extreme patchiness, and the outermost radius catched by
the DETKA software is $R\approx 40\arcsec$ -- only two third of the ring radius. From the overall statistics we can conclude that
the necessary condition of the star formation triggering in a ring of S0 galaxy is gas concentration strictly in the main plane
of a galaxy and laminar circular rotation within this plane. The other three plots of Fig.~\ref{corring}
demonstrate a surprising homogeneity of the gas metallicities within the {\it star-forming} rings: they are all concentrated
around the slightly subsolar value, $-0.15$~dex, and do not correlate neither with the radius of the ring nor 
with the metallicity of the underlying stellar population estimated here from the Lick indices measured in our long-slit 
spectra and confronted to the evolutionary synthesis models from \citet*{thomod}.  The metallicities of the stellar
populations are either taken from our earlier works \citep{ringmnras,saltrings}, or calculated particularly for this analysis
from the spectra obtained with the SCORPIO and SCORPIO-2; for the latter galaxies we show the index--index diagrams in Fig.~\ref{lickind}.
The details of the stellar population analysis are analogous to our earlier works and can be inspected in \citet{ringmnras}.
The only point with a supersolar metallicity at the Fig.~\ref{corring} is a starforming ring in NGC~774 which is perhaps dominated by shock-wave
excitation (see the BPT analysis above, Fig.~\ref{fig_bpt}) and which gas oxygen-abundance estimate in the Table~\ref{table_gasoxy} may
hence be unreliable.

\begin{figure*}
\centering
\begin{tabular}{c c}
\includegraphics[width=8cm]{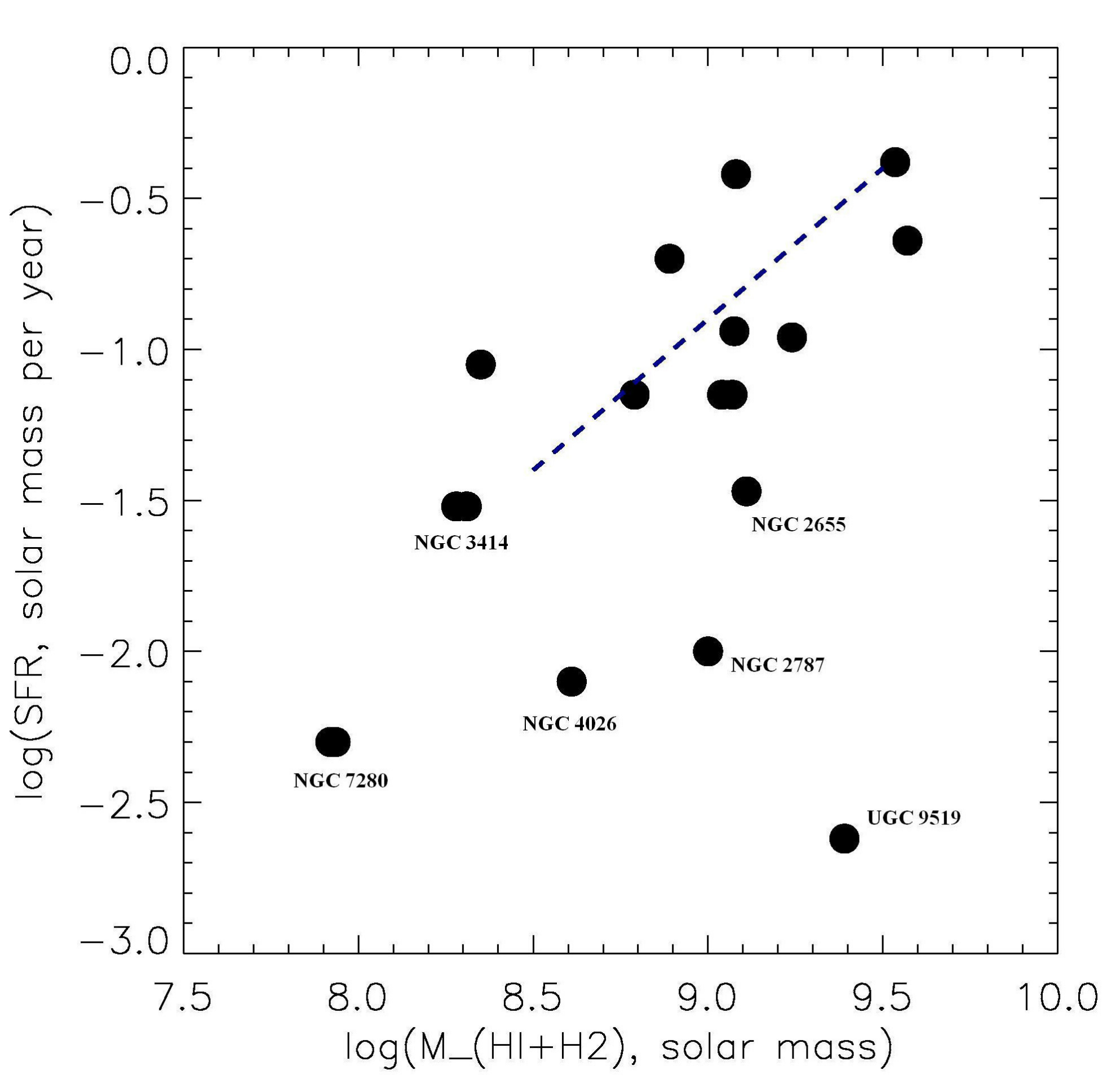} &
\includegraphics[width=8cm]{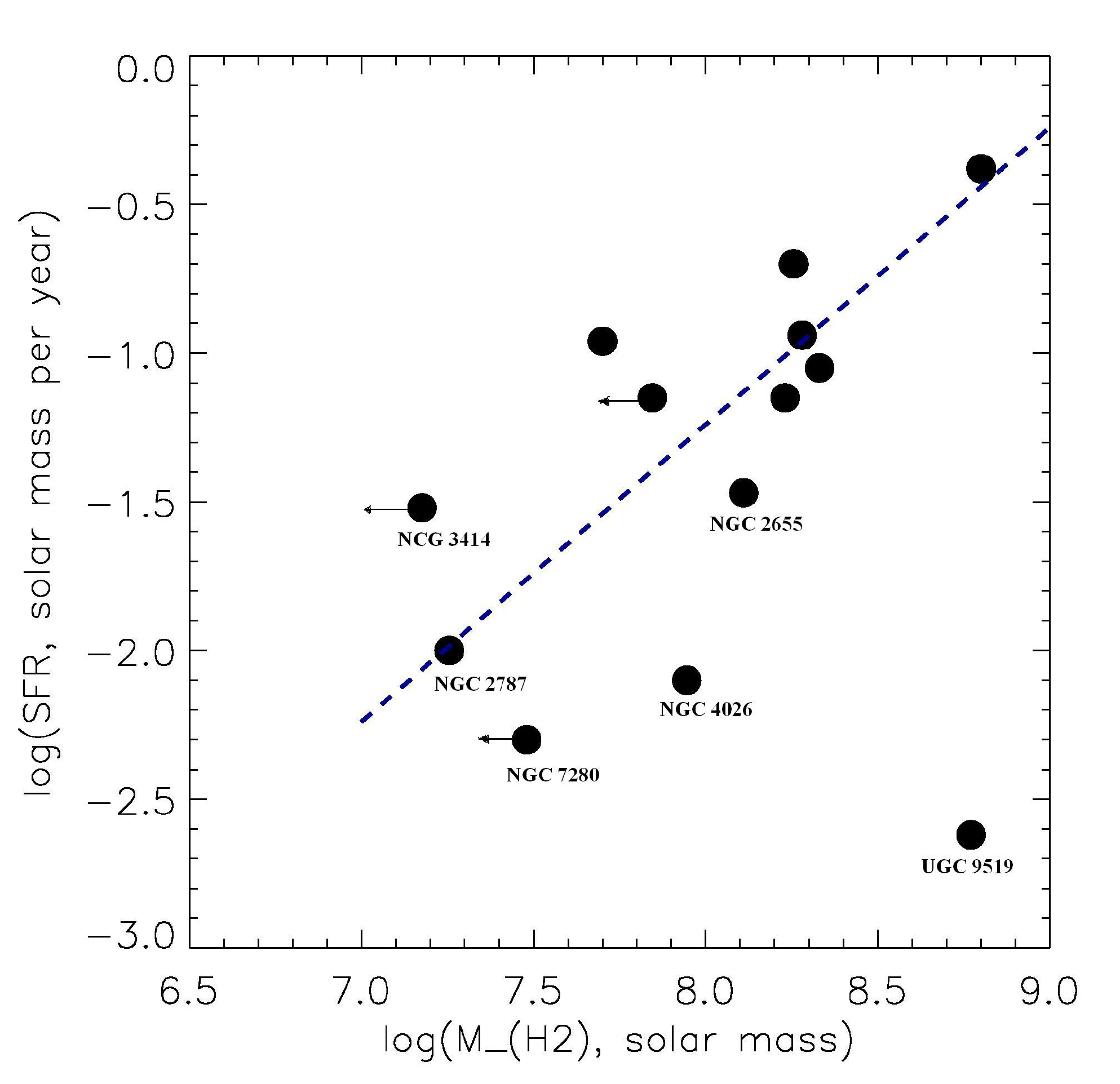} \\
\end{tabular}
\caption{The correlation of the integrated star formation rate with the gas content of the S0-galaxies in our sample.
The slope of the dashed straight line is 1. In the left plot for three galaxies with only upper limits on \H2\ content we give
two signs (circles) for every galaxy -- one for the \H2\ mass equal to the upper limit and the other for the zero \H2\ mass.}
\label{sfr}
\end{figure*}

To illustrate once more our thesis about star formation suppression by gas accretion from highly inclined directions, we have
calculated integrated star formation rates (SFRs) for our galaxies having the integrated FUV-magnitudes in the NED, by using the
calibrations from \citet{lee_sfr}; and then we have compared these SFR estimates with the total gas content in the galaxies,
\HI$+$\H2 , compiled in our Table~1. The comparison -- namely, the Kennicutt-Schmidt relation for our sample -- is shown in Fig.~\ref{sfr}.
One can see that the most galaxies obey the Kennicutt-Schmidt relation with the logarithmic slope of 1, {\bf and this picture is consistent
both for the SFR confronting to the total gas mass, \HI$+$\H2 , or to the \H2\ only}. But several galaxies fall below
this relation, and these are the galaxies with large {\it polar} \HI\ disks (see the subsection~\ref{sec:5.3} and the Table~\ref{table_angles})
or the galaxies where we see outer \HI\ filaments orthogonal to the galactic planes in the \HI\ maps by \citet{atlas3d_13} (see the next subsection).
These galaxies may be very gas-rich, with more than $10^9$ solar masses of neutral hydrogen, as it is observed in NGC~2655 or UGC~9519;
but their current star formation is far less intense than we can expect from the Kennicutt-Schmidt relation. {\bf Interestingly, the star formation
suppression is also detected in UGC~9519 when we consider its \H2\ content only. The molecular gas in UGC~9519 is confined to the central
part of the galaxy; but within this central part it is also distributed in the polar plane \citep{atlas3d_18}!}

\subsection{Accretion seen `by eye': neutral hydrogen around NGC 3166, NGC 4026, and NGC 7280}
 
\label{sec:5.2} 
 
\noindent
{\bf NGC 3166}.  This galaxy is embedded into a large reservoir of \HI\ together with NGC~3169 (to the east from NGC~3166) and NGC~3165
(to the west from NGC~3166), see new Aresibo and GMRT 21-cm observations in  \citet{Lee-Waddell}). Outer streams and HI filaments coinciding
sometimes with the UV-bright structures revealed by the GALEX
are seen to the east-south from the galaxy. The asymmetric {\it inner} structure of NGC~3166 looks even more striking when we inspect
the data on neutral hydrogen and on the far ultraviolet surface brightness distribution within the galaxy (Fig.~\ref{h1} and Fig.~\ref{galex}).
The \HI\ is mostly concentrated {\it to the east} from the nucleus (and in smaller degree in the outer western tidal arm -- the only place where 
the ionized-gas emission-line spectrum reveals excitation by young stars), while the far-ultraviolet radiation -- {\it to the west} 
from the nucleus. Moreover, the $FUV-NUV$ color is extremely red to the east from the nucleus, which may 
be caused by asymmetric dust distribution or by other, than star formation, nature of the UV radiation just over this region.
Here it is the point to mention that already \citet{nro94} noted the stronger horn of the CO-line at receding velocities -- or
to the west from the nucleus where we see now the extended FUV signal. In the situation when the neutral hydrogen is localized in one 
half of the galaxy, and the star formation (UV and H$\alpha$) -- at the opposite side of the galaxy where the molecular gas is also
more prominent, we can only suggest a recent event of the gas acquisition, less than one orbital period ago. The whole
geometry of the accretion implies the cold atomic gas infall under almost right angle into the disk to the east from the galactic center
where shock gas excitation dominates in the emission-line spectrum. Further gas re-distribution over the whole galactic disk
proceeds probably in strongly turbulent regime, because we see the gaseous disk precession near the center (Fig.~\ref{fig_pa}).
The gas compression along the trajectories of orbital rotation provides current formation of molecular clouds and current star formation 
to the west from the galactic center. So the star-forming ring in NGC~3166 is still incomplete.

\noindent
{\bf NGC 4026}. It is another galaxy where a highly elongated \HI\ filament which is almost perpendicular to the galaxy plane
seen edge-on is observed in the wide galaxy outskirts \citep{atlas3d_13}. This time the gaseous stream penetrates NGC~4026 just 
near the center at the right angle. As a result, we see nearly polar rotation of the gas traced by the [OIII]$\lambda$5007 emission 
line up to $R\approx 10\arcsec$ (Fig.~\ref{fig_fpi}). In the outer part of the galaxy, within the area dominated by the large-scale disk,
the ionized gas as well as the neutral hydrogen are confined to the galactic plane (Fig.~\ref{h1}) and rotate
{\it together} with the stellar component of the disk. Interestingly, despite the rather large amount of the gas in the galaxy
and its rather regular distribution and kinematics, no traces of star formation are found all over the galaxy -- no \Ha emission
lines, no far-ultraviolet signal.

\begin{figure*}
\centering
 \includegraphics[width=\textwidth]{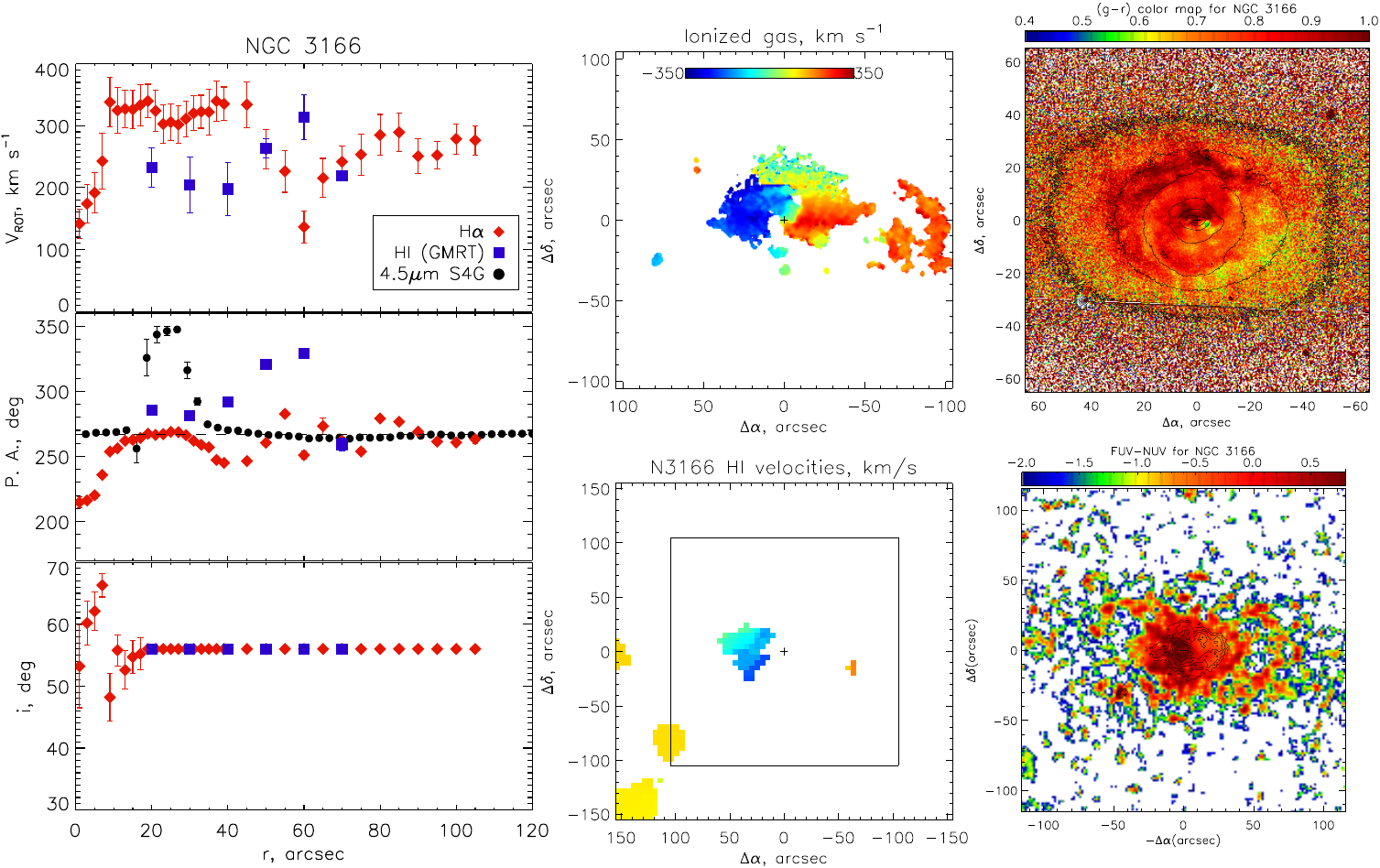} 
\caption{Comparison of the ionized-gas and neutral-hydrogen kinematics for  NGC~3166. The left-hand panels show the radial variations of the 
tilted-ring model parameters for the ionized- and neutral-gas data: the  rotation velocity (top), the $PA_{kin}$ together with
$PA$ of the NIR isophotes from S4G survey (middle), and the inclination (bottom). The ionized gas (FPI)
and \HI\ (GMRT observations from \citet{Lee-Waddell}) velocity fields in the same color scale are shown in the central panels.
The field of view of the FPI map is marked by a square on the \HI\ map. The velocity field are shown after subtraction of the galaxy
systemic velocity.
The {\bf FUV-NUV and} $g-r$ color maps are shown in the right-hand panel. {\bf The isophotes are overlaid: the $r$-band ones onto the
$g-r$ map and NUV ones onto the FUV--NUV map.}
}
\label{h1}
\end{figure*}

\setcounter{figure}{16}
\begin{figure*}
\centering
 \includegraphics[width=\textwidth]{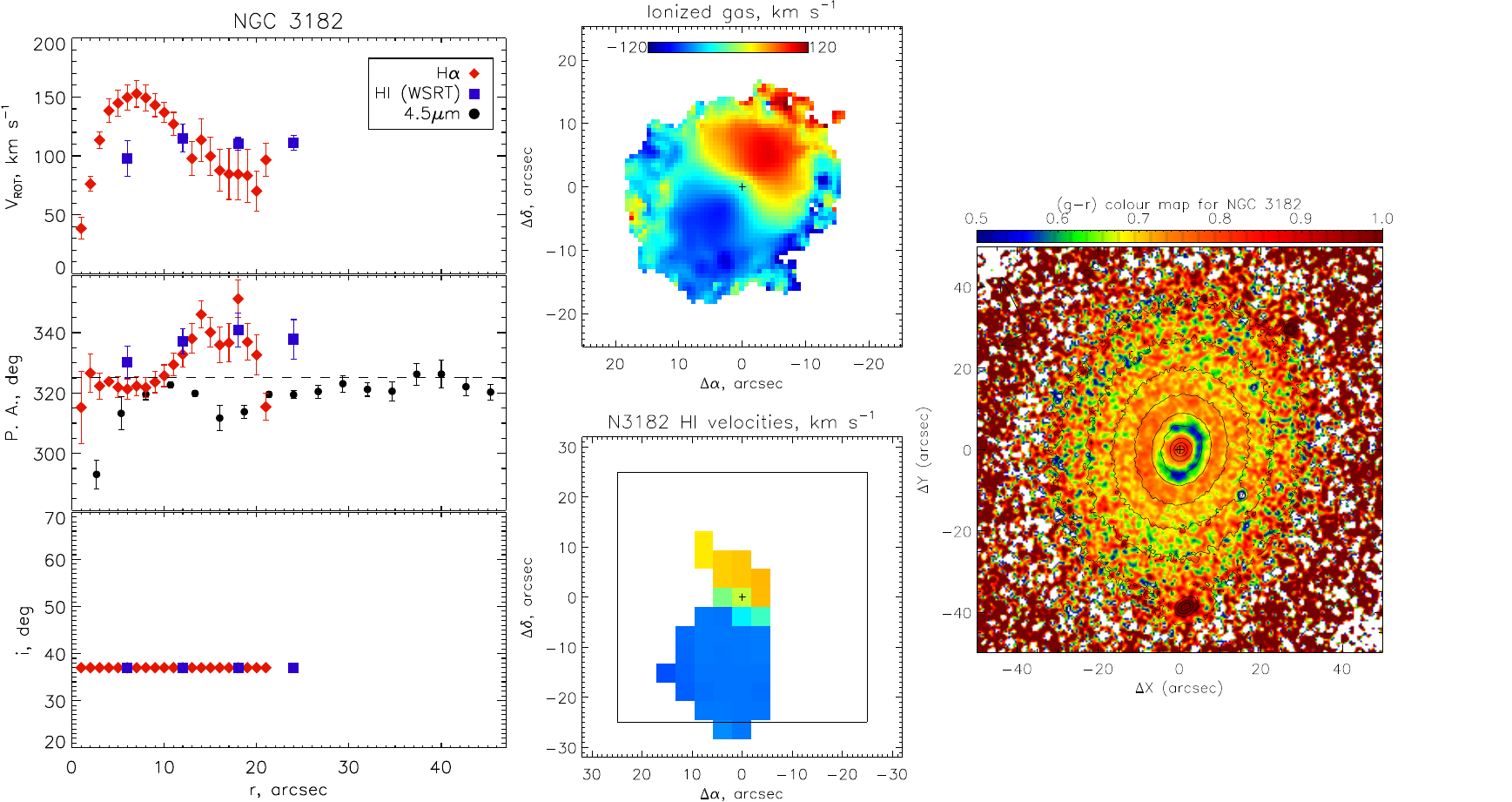} 
  \includegraphics[width=\textwidth]{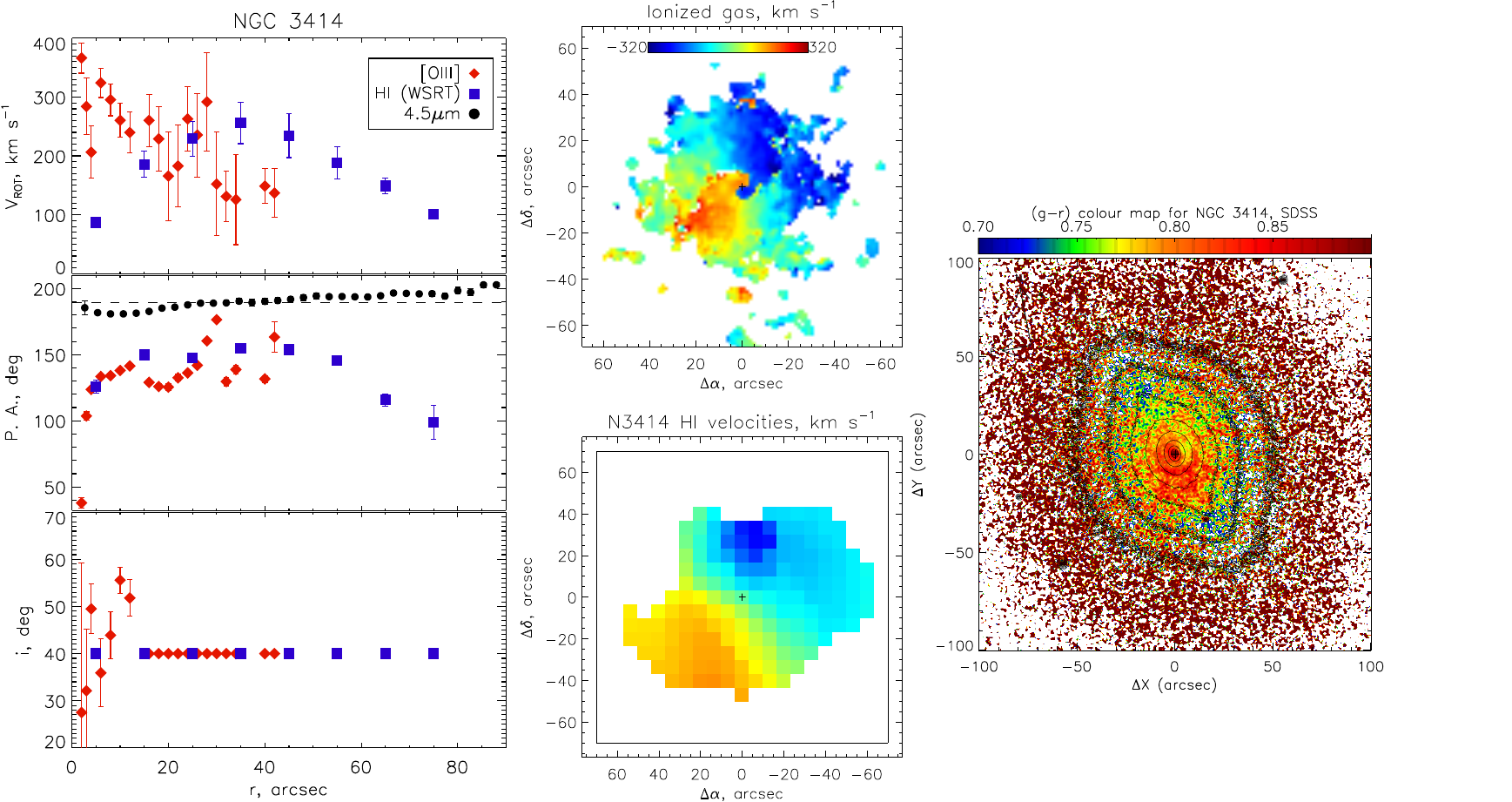} 
\caption{-- continued for NGC 3182 (top panels) and NGC 3414 (the bottom panels).
	The \HI\ maps here and below were obtained at the WSRT and originaly presented in \citet{atlas3d_13}.
         The original FITS were downloaded from WSRT/Atlas3D data archive (\url{http://wow.astron.nl/}).
}
\end{figure*}

\setcounter{figure}{16}
\begin{figure*}
\centering
 \includegraphics[width=\textwidth]{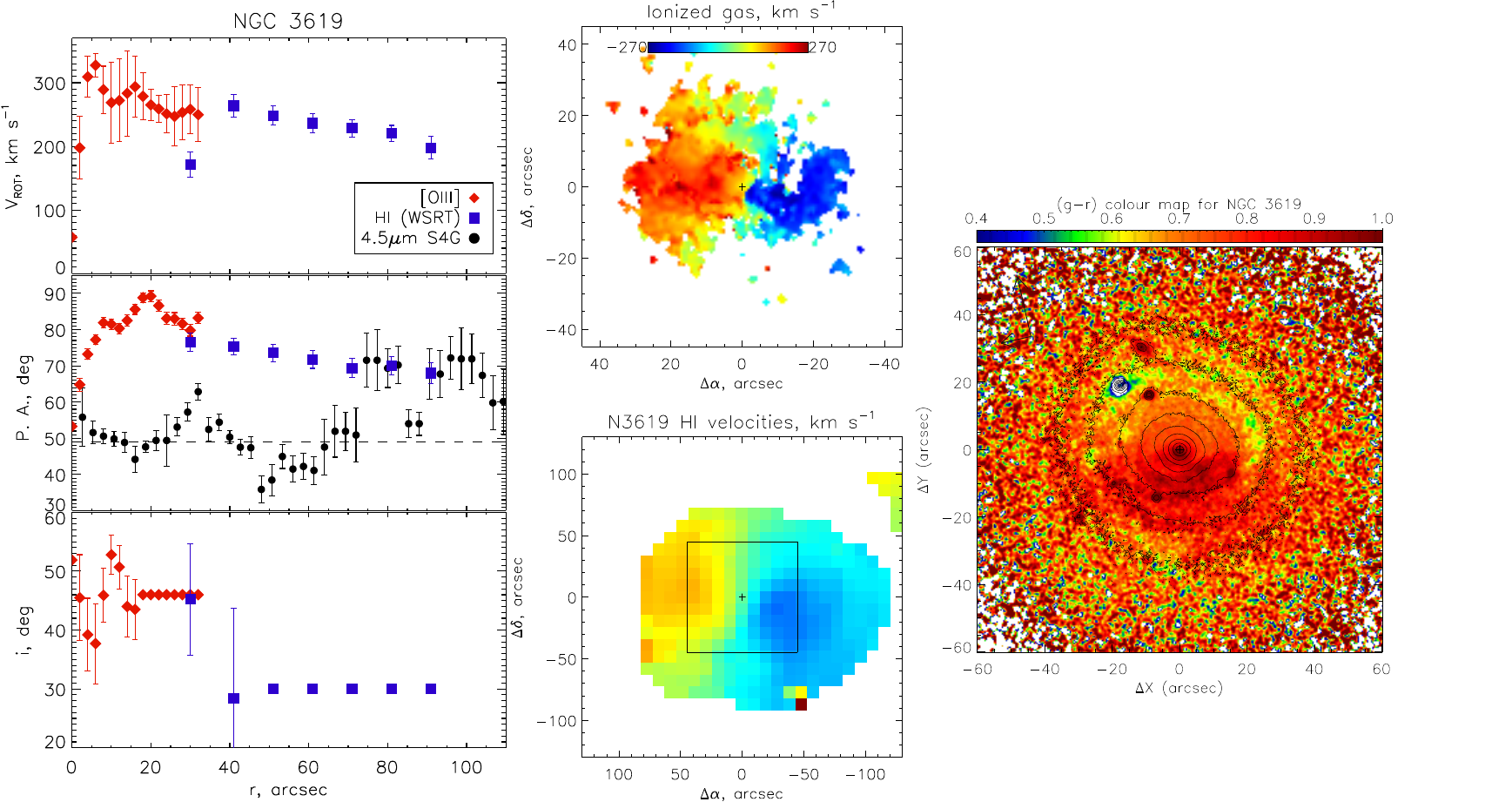} 
  \includegraphics[width=\textwidth]{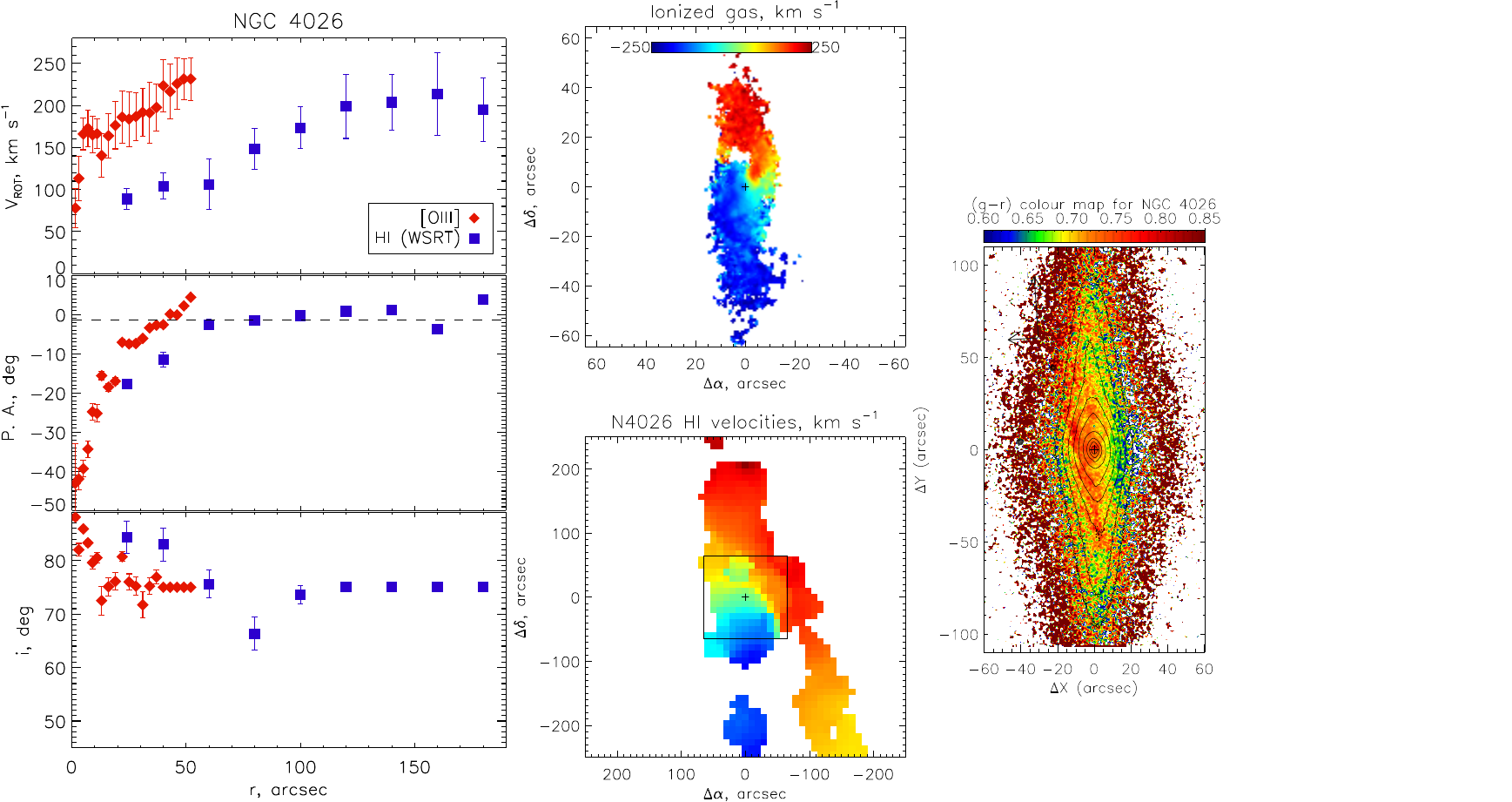} 
\caption{-- continued  for NGC~3619 (top panels)  and  NGC~4026 (bottom panels).
}
\end{figure*}

\setcounter{figure}{16}
\begin{figure*}
\centering
 \includegraphics[width=\textwidth]{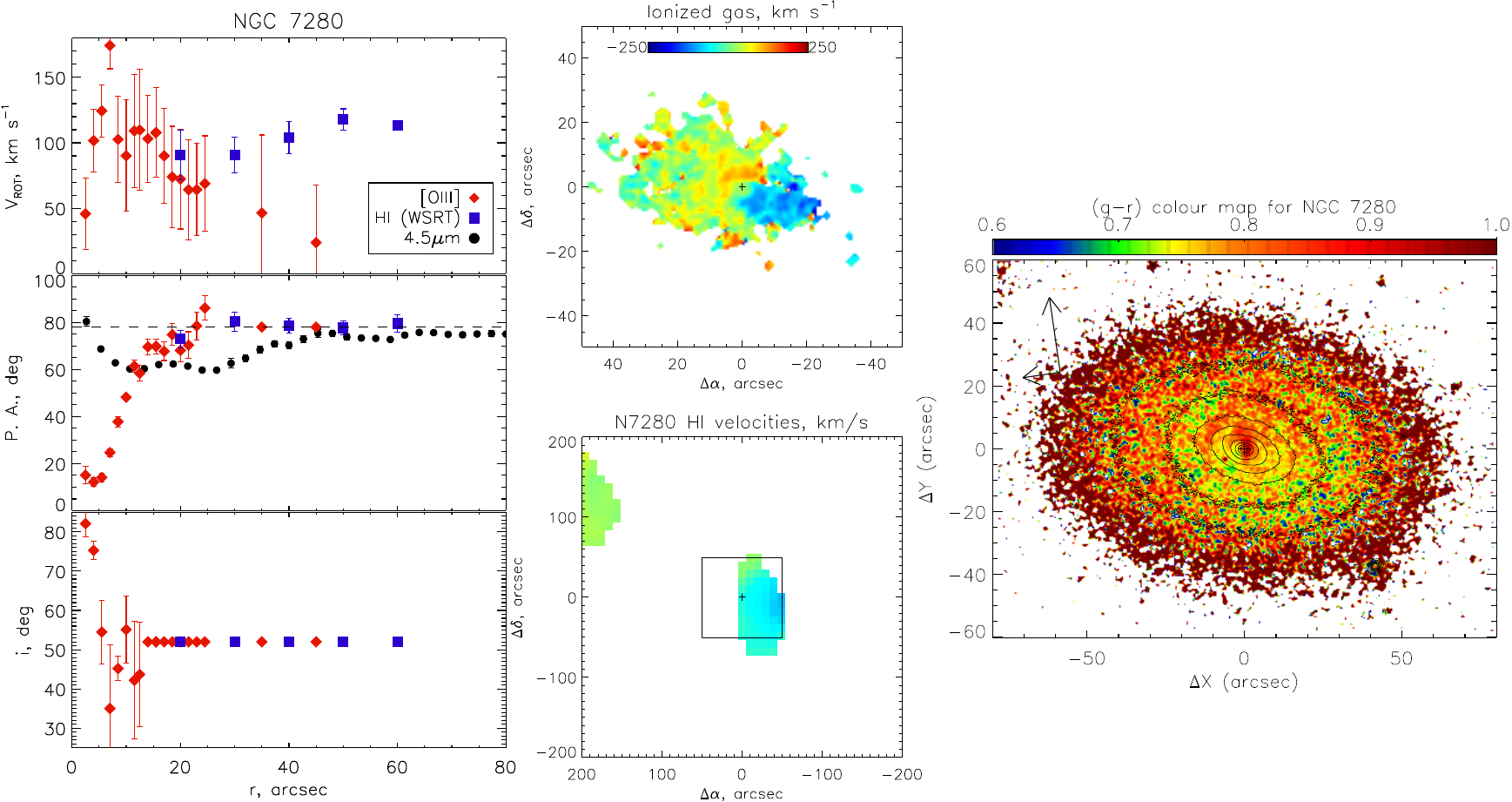} 
  \includegraphics[width=\textwidth]{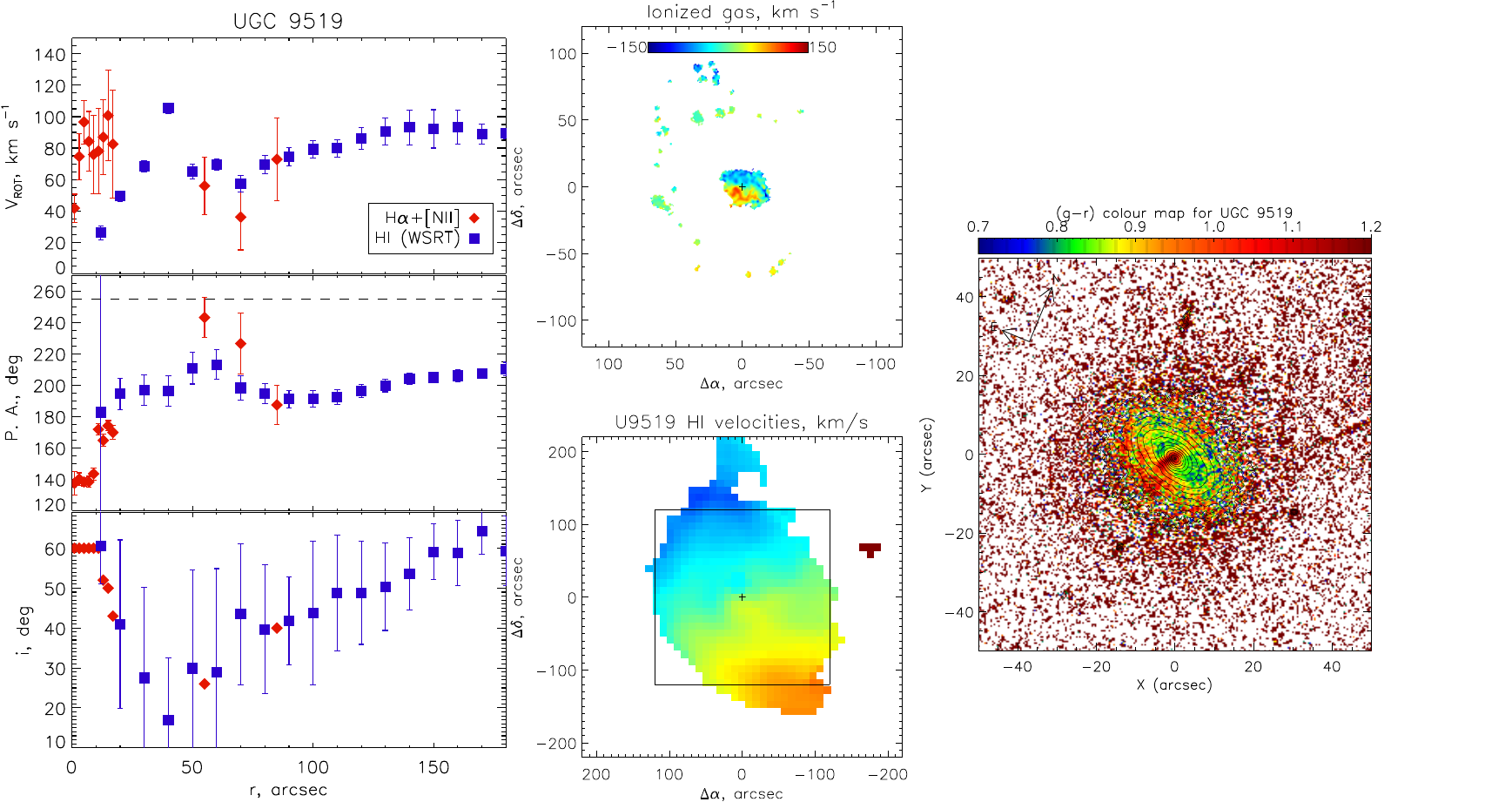} 
\caption{-- continued for NGC~7280 (top panels) and  UGC~9519 (bottom panels).
}
\end{figure*}

\noindent
{\bf NGC 7280}. This galaxy is quite similar to NGC~4026 in all aspects but one. Lacking current star formation and \Ha emission
lines in the disk spectrum, it has a circumnuclear polar disk, seen both in the ionized-gas kinematics (Fig.~\ref{fig_fpi}) and in the
\HI\ distribution \citep{atlas3d_13}; but in the outer-disk area the gas {\it counterrotates} the stars. NGC~7280
is not exactly edge-on, and our surface photometry reveals some signs of triaxiality in its center 
\citep[and see the isophote turn in Figs.~\ref{fig_pa} and \ref{h1}, right]{sil7280};
however the outer gas rotation looks circular though in opposite sense than that of the stars.

\subsection{Extended inclined gaseous disks: neutral hydrogen around NGC 2655, NGC 2787, NGC 3414, NGC 3619, and UGC 9519}

\label{sec:5.3} 

\noindent
{\bf NGC 2655}. The galaxy was mapped in \HI\ and analyzed in detail by \citet{sparke}. The outer \HI\ distribution resembled a
huge disk, with the radius exceeding the optical one by a factor of 2, and the orientation of this disk in its outer parts
was strongly different from the orientation of the stellar disk. However, the \HI\ velocity field obtained with a spatial
resolution about of $40\arcmin$ ($\sim 3.5$~kpc) looked bisymmetrical, and \citet{sparke} had constructed a model of a strongly 
warped, circularly rotating gaseous disk produced probably by multiple minor mergers a few Gyr ago. Indeed, NGC~2655 is a center 
of a group, and several small gas-rich galaxies are still seen not far from it. The {\it multiple} mergers are needed because
the total mass of neutral hydrogen in the galaxy, $2 \cdot 10^9$ solar masses, is too large to be provided by a single satellite
sinking. Our present data revealing strong velocity anomaly close to the center, in some $10\arcsec$ to the north, puts into
doubt a relaxed character of the gas distribution in the galaxy; perhaps, satellite ingestion by NGC~2655 continues just now.
We see no signs of current star formation in and around NGC~2655, and this conclusion coincides with the opinion by \citet{sparke}.

\noindent
{\bf NGC 2787}. The galaxy was mapped in the 21~cm line with the WSRT by \citet{n2787h1}. He reported a \HI\ ring with a radius
of 10~kpc, beyond the optical borders of the galaxy, with regular circular rotation implying the orientation of the line
of nodes of the gaseous disk plane in $PA=140$\deg $\pm 5$\deg. It is almost orthogonal to the inner rotation plane of the
ionized gas (Fig.~\ref{fig_pa}) and also strongly different from the orientation of the stellar disk, $PA_0= 108$\deg. The gaseous disk
isovelocity contour twist in the central part of NGC~2787 may be explained by a strong bar aligned in $PA= 160$\deg; perhaps the
gas when inflowing to the center is settling into a main plane of the triaxial potential orthogonally to the bar major axis. But the outer neutral
hydrogen strongly decoupled from the stellar body is evidently accreted from outside though NGC~2787 is recognized as a completely
isolated S0. Again, any star formation is absent over the whole galaxy extension.

\noindent
{\bf NGC 3414}. An extended, regularly rotating HI disk of this peculiar S0 galaxy has been mapped by \citet{morganti}.
Earlier the galaxy was included into the first survey with the SAURON and later classified as a slow rotator though S0 \citep{sauron_9}.
In fact, the low projected stellar rotation of NGC~3414 is not unexpected because of the disk orientation nearly face-on
\citep{s4gdecomp}. However, the gaseous disk extending up to $R>100\arcsec$ looks much more inclined to the sky plane, and 
so much faster rotating, up to $V_{rot,proj} > 150$~\kms \citep{morganti}. The kinematical major axis and visible elongation
of the large-scale gaseous disk is $PA_{kin,HI}\approx 140$\deg (see Fig.~\ref{h1}) that differs from the stellar isophote major axis,
$PA=12$\deg \citep{s4gdecomp}, and from the thin `bar' elongation in $PA\sim 20$\deg. In the very center, $R<10\arcsec$,
the gas rotation plane starts to turn, and the kinematical major axis traced by the [OIII]$\lambda$5007 emission-line
velocity field aligns with the `bar' (Fig.~\ref{fig_fpi}). In Fig.~\ref{fig_pa} we have plotted the ionized-gas kinematical
major axis orientation by fat black squares (our Fabry-Perot data) and by pale squares (SAURON [OIII]$\lambda$5007 velocity map);
both data sets consistently indicate strong turn of the kinematical major axis by $\sim 90$\deg, toward $PA<50$\deg at $R<3\arcsec$.
This `polar' velocity structure is aligned with the morphological feature in $PA\sim 20$\deg which is sometimes described
as a bar and sometimes -- as a polar ring \citep{whitmore}. Since the stellar photometric and gas kinematical major axes coincide
at $R<3\arcsec$, the data are more consistent with the hypothesis of a polar disk; however this polar disk seen strictly edge-on
lacks any gas at $R>10\arcsec$. Again, over the whole extension of the gas distribution any star formation is absent \citep{sauron_15}.

\noindent
{\bf NGC 3619}. The neutral hydrogen had been mapped in this galaxy as early as in 1989 \citep{n3619hi} with the WSRT.
The whole gas distribution resembled an inner ring embedded into the optical body of the galaxy. The stellar velocity maps
were obtained twice, in the frames of the panoramic spectral surveys ATLAS-3D \citep{atlas3d_2} and CALIFA
\citep{califa,califa_3}, and we have made the tilted-ring analysis of the both velocity fields. The combined analysis
of the stellar kinematics and surface brightness distributions retrieved from SDSS and S4G \citep{s4g} surveys 
implies that the stellar disk of this galaxy is seen nearly face-on: the isophote ellipticity is approximately 0.1
that gives us the inclination of 25\deg under the disk relative thickness of 0.22 \citep{thickness}, and the kinematical
inclination provided by the tilted-ring analysis is also $<30$\deg at $R>10\arcsec$. However the visible cold gas
rotation velocity in the inner region of the galaxy measured through the interferometric mapping of CO \citep{atlas3d_18}, 
\HI\ \citep{n3619hi}, and our \Ha data, is $\sim$200~\kms resulting in kinematical inclination estimate of about 50\deg.
The line-of-nodes position angles are also different: 85\deg -- 90\deg for the gas against $PA_{0,kin}=50$\deg
for the stellar component (Fig.~\ref{fig_pa} and Fig.~\ref{h1}). It means that the gaseous disk in its central part is strongly
inclined with respect to the stellar disk plane; the southern part of the gaseous disk is the nearest to us because we see a red
dust lane projected against the bulge at the color $g-r$ map (Fig.~\ref{h1}). Interestingly, the UV-ring at the
radius of $20\arcsec$ in NGC~3619 is absolutely round that implies the gas compression strictly within the 
plane of the stellar disk. Perhaps, in the outer parts of the disk the gaseous sheet falls completely into the main plane 
of the galaxy -- see the profiles of the left plot of Fig.~\ref{h1} beyond $R>70\arcsec$.

\noindent
{\bf UGC 9519}. Recent deep mapping of the neutral hydrogen in UGC~9519 by \citet{u9519h1} has revealed multi-tiered structure
of cold gas distribution in this galaxy. \citet{u9519h1} identify three main zones of this distribution: the center, $R<25\arcsec$,
the inner ring, $R=40\arcsec - 80\arcsec$, and the outer ring at $R\sim 3\arcmin$, or $>20$~\kpc! Our Fabry-Perot data obtained
in two emission lines, \Ha and [NII]$\lambda$6583, trace the central zone which lacks any star formation and rotates with the
line of nodes in $PA\approx 150$\deg, in polar orientation with respect to the stellar disk, and also the inner ring, up to
$R\approx 80\arcsec$, rotating in a plane with the line of nodes in $PA\approx 20$\deg; so our data are completely consistent with
the \HI\ orientation and rotation (Fig~\ref{h1}). The `inner' ring is detached from the main galactic stellar body; it demonstrates the gas
excitation by young stars (our long-slit data) and high UV surface brightness \citep{u9519h1} so being certainly a starforming ring.
Curiously, the molecular gas is concentrated in the central part of the galaxy \citep{atlas3d_18} where there is no star formation;
it rotates together with the ionized gas, in the polar orientation with respect to the main galactic disk.

\subsection{Long-lived inclined gaseous disks as treated by dynamics}

\label{sec:5.4} 

A problem of possible steady state of inclined gaseous disks was studied more than once as a general problem of galaxy dynamics.
It was provoked partly by impressive phenomenon of large-scale polar rings, partly by understanding inevitable (common) 
gas accretion from an arbitrary direction. So in 1980th-1990th this problem had been solved both analytically, for a case of a spheroidal
potential -- e.g. by \citet{steiman_durisen}, -- and by numerical simulations, for a case of triaxial tumbling potential -- e.g. by 
\citet{colley_sparke}. It was shown that in the case of quasi-axisymmetrical potential the gas acquired under a moderate inclination -- 
say, under 40\deg -- had to settle into a main galactic plane in several precession period, that corresponded for a medium-mass
galaxy to 1~Gyr in the center, $R\sim 1$~kpc, and much longer, $>6$~Gyr, at $R\sim 10$~kpc \citep{steiman_durisen}. \citet{chris_tohline}
considered {\it three} initial directions of gas acquisition, from under 10\deg, 40\deg, and 80\deg\ to the equatorial plane, 
and had found significant differences in the gas behavior: the gas acquired under 10\deg settled into the main plane after 22-25 orbital periods, 
the gas acquired under 40\deg settled after 11-13 orbital period, but after 8 orbital period the disk started to suffer violent gas inflow 
to the center; and the gas acquired at 80\deg conserved
its orientation during the Hubble time but suffered also some radial inflow. Finally, within the triaxial potential the gas could settle
in one of two main planes -- to that orthogonal to the longest axis or to that orthogonal to the shortest axis, and the whole dynamical
evolution allowed to form steady-state polar rings/disks \citep{colley_sparke}.

Recently \citet*{Mapelli2015} have presented the simulation of building large--scale (about 20~kpc in diameter) rings via a minor 
merger between a lenticular galaxy and its gas-rich satellite. The estimated living time of the ring structures has appeared to be less 
than 3~Gyr if the rings form from prograde encounters, whereas it can reach 4--6~Gyr after counter--rotating or non--coplanar 
(inclined by 45\deg or 90\deg) interactions. 

In any case, the dynamical settling of an outer gaseous disk, with a radius of $\sim 10$~kpc, requires a very long time, so for our particular galaxies
described in the previous subsection we can consider them as stationary structures, and so we can try to estimate the accretion direction
by comparing the current spatial orientation of the outer
\HI\ disk with that of the stellar disk. We have made this attempt, and the results are given in the Table~5. The orientations of the
\HI\ disks are taken from the papers mentioned in the previous subsection, and for NGC~3619 and UGC~9519 -- from our own analysis of the
WSRT interferometric velocity fields. The orientations of the stellar disks are taken from the photometric decomposition results of the S4G survey
\citep{s4gdecomp}, and for UGC~9519 we have calculated the orientation parameters of the stellar rotation plane by applying the tilted-ring 
technique to the public SAURON data. The last column gives the estimated angles between the stellar and gaseous planes. For this calculation we
have used the formula which is published by \citet{moisrev}. Two values are given for every galaxy because of the uncertainty with the
nearest side of any disk. Interestingly, most galaxies with the steady-state extended \HI\ disks have a possible variant of their nearly {\it polar}
orientation with respect to the stellar disks. Only NGC~3619 has the neutral-hydrogen disk with the certainly intermediate inclination; it is
the smallest of all, and it contains an inner starforming ring. Perhaps, we could also note some additional consequences of the gas inflow through the
inclined gaseous disks: in the galaxies with the large bars, NGC~2787 and perhaps UGC~9519, the central gas has settled into the planes
orthogonal to the long axes of their {\it bars}.

\begin{table*}
\caption{Mutual inclinations of the stellar and gaseous disks}
\label{table_angles}
% %\begin{center}
\begin{flushleft}
\begin{tabular}{lccccc}
\hline\noalign{\smallskip}
Galaxy & \multicolumn{2}{c}{Stellar disk} & \multicolumn{2}{c}{Outer HI disk} & The angle between the stellar \\
   &  PA(line of nodes) & Inclination &   PA(line of nodes) & Inclination &  and HI disks \\
\hline
NGC 2655 & 110\deg & 20\deg & --60\deg & 60\deg & 80\deg or 40\deg \\
NGC 2787 & 108\deg & 56\deg & 140\deg & 42\deg & 27\deg or 93\deg \\
NGC 3414 & 12.6\deg & 33\deg & 150\deg & 40\deg & 67\deg or 26\deg \\
NGC 3619 & 63.5\deg & 21.5\deg & 73\deg & 30\deg & 10\deg or 51\deg \\
UGC 9519 & 250\deg & 55\deg & 200\deg & 40\deg & 39\deg or 84\deg \\
\hline
\end{tabular}
\end{flushleft}
\end{table*}

\section{Summary}

We have presented the results of our spectral study of gaseous disks in 18 S0 galaxies undertaken with the facilities
of the Russian 6m telescope of the Special Astrophysical Observatory; both panoramic spectroscopy with the scanning
Fabry-Perot interferometer and long-slit spectroscopy over wide spectral range with SCORPIO have been used. The gas in S0s 
is commonly accreted from outside that is implied by its decoupled kinematics: at least 5 galaxies demonstrate extended, strongly 
inclined HI disks smoothly coupled with an inner ionized-gas component, in 3 galaxies we see \HI\ accretion directly, as thin filaments
coming into the galactic disks under right angle; 7 galaxies reveal circumnuclear {\it polar} ionized-gas
rotation, and in NGC~2551 the ionized gas -- and neutral hydrogen too \citep{tang} -- though confined to the main galactic plane 
however counterrotates the stellar component. The ionized-gas excitation analysis at the BPT-diagrams reveals the gas ionization
by young massive stars in 12 of 18 S0 galaxies studied by us; the current star formation in these galaxies is confined to the
ring-like zones coinciding with the UV-rings seen at the galaxy images provided by the GALEX survey. The UV-ring in NGC~774 though
suffering current star formation (a SNII is detected recently in this galaxy) reveals the gas excitation mostly dominated by shock waves. 
We have applied the tilted-ring analysis to the 2D velocity fields of the ionized gas (and also of the neutral hydrogen in some objects).
Tracing the orientation of the gas rotation-plane lines of nodes along the radius, we have found that current star formation proceeds
usually at the radii where the gas lies in the stellar disk planes and rotates circularly; the sense of the gas rotation does not matter
(NGC~2551 forms stars!). In the galaxies without current star formation the extended gaseous disks are either in a steady-state quasi-polar
orientation (NGC~2655, NGC~2787, NGC~3414), or are acquired just now through the infalling highly inclined external filaments provoking
probably shock-wave excitation (NGC~4026, NGC~7280). Our data implies perhaps crucial difference of the external gas accretion regime in S0s
with respect to spiral galaxies: the geometry of the gas accretion in S0s is typically off-plane.

\acknowledgments

This work is based on the spectral data obtained at the Russian 6m telescope of the Special Astrophysical Observatory 
carried out under the financial support of the Ministry of
Science and Higher Education of the Russian Federation. We are grateful to the staff of the Special Astrophysical Observatory 
and especially to Victor Afanasiev for his great contribution into spectroscopy development at the 6m telescope. We thank Michella Mapelli 
for her constructive comments and suggestions, which helped us to improve and clarify our results, 
to Karen Lee-Waddell who has provided us with GMRT maps for NGC~3166, and to Paolo Serra for his help with the WSRT maps.
The observations in 2017 and the
final analysis of the ionized gas properties have been supported by the grant of Russian Science Foundation, project no. 17-12-01335 
`Ionized gas in galaxy disks and beyond the optical radius'. The analysis of star formation in the galactic rings has been supported
by the grant of the Russian Foundation for Basic Researches, no. 18-02-00094a `Outer rings in disk galaxies: star formation, chemical 
composition, origin'. We acknowledge the usage of the HyperLEDA database (http://leda.univ-lyon1.fr). This research
has made use of the NASA/IPAC Extragalactic Database (NED) which is operated by the Jet Propulsion Laboratory, California Institute of 
Technology, under contract with the National Aeronautics and Space Administration. 
To estimate the star formation pattern in the galactic rings, we have used public archive data of the space telescope GALEX.
The NASA GALEX mission data were taken from the Mikulski Archive for Space Telescopes (MAST). STScI is operated by the Association 
of Universities for Research in Astronomy, Inc., under NASA contract NAS5-26555.
This study uses partly data provided by the Calar Alto Legacy Integral Field Area (CALIFA) survey (http://califa.caha.es/).
The CALIFA results are based on observations collected at the Centro Astronomico Hispano Aleman (CAHA) at Calar Alto, operated jointly 
by the Max-Planck-Institut fur Astronomie and the Instituto de Astrofisica de Andalucia (CSIC).
Some figures contain the galaxy images taken from the SDSS and PanSTARRS surveys public databases.
Funding for the SDSS-III has been provided by the Alfred P. Sloan Foundation, the Participating Institutions,
the National Science Foundation, and the U.S. Department of Energy Office of Science. The SDSS-III Web site is http://www.sdss3.org/.
SDSS-III is managed by the Astrophysical Research Consortium for the Participating Institutions of the SDSS-III Collaboration 
including the University of Arizona, the Brazilian Participation Group, Brookhaven National Laboratory, Carnegie Mellon University,
University of Florida, the French Participation Group, the German Participation Group, Harvard University, the Instituto de 
Astrofisica de Canarias, the Michigan State/Notre Dame/JINA Participation Group, Johns Hopkins University, 
Lawrence Berkeley National Laboratory, Max Planck Institute for Astrophysics, Max Planck Institute for Extraterrestrial Physics, 
New Mexico State University, New York University, Ohio State University, Pennsylvania State University, University of Portsmouth, 
Princeton University, the Spanish Participation Group, University of Tokyo, University of Utah, Vanderbilt University, 
University of Virginia, University of Washington, and Yale University. 
The Pan-STARRS1 Surveys (PS1) and the PS1 public science archive have been made possible through contributions 
by the Institute for Astronomy, the University of Hawaii, the Pan-STARRS Project Office, the Max-Planck Society 
and its participating institutes, the Max Planck Institute for Astronomy, Heidelberg and the Max Planck Institute 
for Extraterrestrial Physics, Garching, The Johns Hopkins University, Durham University, the University of Edinburgh, 
the Queen's University Belfast, the Harvard-Smithsonian Center for Astrophysics, the Las Cumbres Observatory Global 
Telescope Network Incorporated, the National Central University of Taiwan, the Space Telescope Science Institute, 
the National Aeronautics and Space Administration under Grant No. NNX08AR22G issued through the Planetary Science 
Division of the NASA Science Mission Directorate, the National Science Foundation Grant No. AST-1238877, the University 
of Maryland, Eotvos Lorand University (ELTE), the Los Alamos National Laboratory, and the Gordon and Betty Moore Foundation.

\end{document}